\documentclass[preprint]{aastex61}

\received{\today}
\submitjournal{AAS}

\shorttitle{H$_{2}$O and dense molecular line emission in ULIRGs}
\shortauthors{Imanishi et al.}

\begin{document}

\title{ALMA Sub-arcsec-resolution 183 GHz H$_{\rm 2}$O and Dense Molecular 
Line Observations of Nearby Ultraluminous Infrared Galaxies}

\correspondingauthor{Masatoshi Imanishi}
\email{masa.imanishi@nao.ac.jp}

\author[0000-0001-6186-8792]{Masatoshi Imanishi}
\affil{National Astronomical Observatory of Japan, National Institutes 
of Natural Sciences (NINS), 2-21-1 Osawa, Mitaka, Tokyo 181-8588, Japan}
\affil{Department of Astronomy, School of Science, Graduate
University for Advanced Studies (SOKENDAI), Mitaka, Tokyo 181-8588, Japan}

\author[0000-0002-6939-0372]{Kouichiro Nakanishi}
\affil{National Astronomical Observatory of Japan, National Institutes 
of Natural Sciences (NINS), 2-21-1 Osawa, Mitaka, Tokyo 181-8588, Japan}
\affil{Department of Astronomy, School of Science, Graduate
University for Advanced Studies (SOKENDAI), Mitaka, Tokyo 181-8588, Japan}

\author[0000-0001-9452-0813]{Takuma Izumi}
\affil{National Astronomical Observatory of Japan, National Institutes 
of Natural Sciences (NINS), 2-21-1 Osawa, Mitaka, Tokyo 181-8588, Japan}
\affil{Department of Astronomy, School of Science, Graduate
University for Advanced Studies (SOKENDAI), Mitaka, Tokyo 181-8588, Japan}

\author[0000-0002-9850-6290]{Shunsuke Baba}
\affil{National Astronomical Observatory of Japan, National Institutes 
of Natural Sciences (NINS), 2-21-1 Osawa, Mitaka, Tokyo 181-8588, Japan}
\affil{Kagoshima University, Graduate School of Science and Engineering, 
Kagoshima 890-0065, Japan}

\begin{abstract}
We present the results of ALMA $\sim$2 mm,  
$\lesssim$1$''$-resolution observations of ten (ultra)luminous infrared 
galaxies ([U]LIRGs; infrared luminosity $\gtrsim$ 10$^{11.7}$L$_{\odot}$) 
at $z <$ 0.15, targeting dense ($>$10$^{4}$ cm$^{-3}$) molecular 
(HCN, HCO$^{+}$, and HNC J=2--1) and 
183 GHz H$_{2}$O 3$_{1,3}$--2$_{2,0}$ emission lines.
Higher HCN to HCO$^{+}$ J=2--1 flux ratios are observed in
some, but not all, AGN-important ULIRGs than in starburst-classified 
sources.
We detect 183 GHz H$_{2}$O emission in almost all AGN-important
ULIRGs, and elevated H$_{2}$O emission is found in two sources
with elevated HCN J=2--1 emission, relative to HCO$^{+}$ J=2--1.  
Except one ULIRG (the Superantennae), the H$_{2}$O emission
largely comes from the entire nuclear regions ($\sim$1 kpc), rather
than AGN-origin megamaser at the very center ($<<$1 kpc).
Nuclear ($\sim$1 kpc) dense molecular gas mass derived from HCO$^{+}$
J=2--1 luminosity is $\gtrsim$a few $\times$ 10$^{8}$M$_{\odot}$, and
its depletion time is estimated to be $\gtrsim$10$^{6}$ yr in all sources.
Vibrationally excited J=2--1 emission lines of HCN and HNC are 
detected in a few (U)LIRGs, but those of HCO$^{+}$ are not.
It is suggested that in mid-infrared-radiation-exposed innermost regions 
around energy sources, HCO$^{+}$ and HNC are substantially less abundant 
than HCN.
In our ALMA $\sim$2 mm data of ten (U)LIRGs, two continuum sources are 
serendipitously detected within $\sim$10$''$, which are likely to be 
an infrared luminous dusty galaxy at $z >$ 1 and a blazar.

\end{abstract}

\section{Introduction} \label{sec:intro}

Ultraluminous infrared galaxies (ULIRGs) with infrared (8--1000 $\mu$m) 
luminosity L$_{\rm IR}$ $\gtrsim$ 10$^{12}$L$_{\odot}$ and luminous infrared 
galaxies (LIRGs) with L$_{\rm IR}$ $\sim$ 10$^{11-12}$L$_{\odot}$ are 
characterized by strong dust thermal emission that is heated by energy
sources \citep{sam96}.
The infrared luminosity is typically much greater than UV-optical 
luminosity, suggesting that the energy sources, either starbursts and/or 
active galactic nuclei (AGNs), are mostly hidden behind dust.
(U)LIRGs are almost exclusively observed as gas-rich galaxy mergers in the 
local universe at $z <$ 0.3 \citep[e.g.,][]{san88,cle96,mur96,duc97}.
Numerical simulations of gas-rich galaxy mergers predict that 
supermassive black holes (SMBHs) grow rapidly in mass through accretion 
and become luminous AGNs at nuclear regions \citep{hop06}.
However, high concentrations of dust and gas in merging (U)LIRGs' nuclei 
\citep[e.g.,][]{dow07,sco15,sak17,ima19} preclude clear distinction of 
hidden energy sources, particularly if compact AGNs are deeply buried 
(=obscured in virtually all directions) because the AGN signatures become 
elusive in extensively used optical spectroscopic classification 
\citep{mai03,ima06}.
Observing at wavelengths of small dust extinction effects is indispensable 
to properly understand the energetic roles of buried AGNs in merging 
(U)LIRGs.

Radiative energy in a starburst originates from nuclear fusion inside stars, 
while in an AGN, a mass-accreting SMBH produces huge radiative energy 
from a compact ($<<$1 pc) accretion disk.
In an AGN, (1) $>$2 keV X-rays and (2) mid-infrared 3--35 $\mu$m emission 
are much stronger than a starburst when normalized to intrinsic UV 
or bolometric luminosity \citep[e.g.,][]{sha11}, 
because in the vicinity of a mass-accreting SMBH, 
(1) UV photons from the accretion disk are upscattered to X-rays by 
inverse Compton process and 
(2) the amount of mid-infrared emitting hot ($>$100 K) dust is much greater 
because of considerably higher UV radiation density than in star-forming 
regions.
Because hard X-rays ($\gtrsim$10 keV) and infrared 3--35 $\mu$m have strong 
penetrating power into dust and gas, spectroscopy in these 
wavelengths can be used as an effective tool to identify luminous buried AGN 
signatures, by distinguishing from starbursts, in gas/dust-rich 
(U)LIRGs' nuclei.
Such spectroscopy of nearby ($z <$ 0.3) (U)LIRGs has been conducted 
systematically and signatures of luminous buried AGNs have been revealed 
in many sources with no obvious optical AGN signatures, 
at $\gtrsim$10 keV hard X-rays \citep[e.g.,][]{ten15,oda17,ric17,yam21} and 
at infrared 3--35 $\mu$m 
\citep[e.g.,][]{gen98,ima06,arm07,ima07a,ima08,nar08,nar09,vei09,nar10,ima10b}. 
The presence of theoretically hypothesized deeply buried AGNs 
(actively mass-accreting SMBHs) in many merging (U)LIRGs has been 
observationally uncovered. 
However, there remain many nearby (U)LIRGs that have 
no significant AGN signatures even in the hard X-rays and infrared.
It is important to distinguish whether (1) they contain no energetically 
important AGNs, or (2) they do, but are elusive even in hard X-rays and 
infrared owing to extremely large extinction 
\citep[e.g.,][]{dow07,mat09,sco17,per21}.
Observations at wavelengths of even lower extinction can provide 
useful information.

(Sub)millimeter at 0.8--3.5 mm is one such wavelengths, with $\lesssim$1/20 
extinction effects than hard X-rays at $\sim$10 keV and 
infrared $\sim$20 $\mu$m \citep{hil83}.
Because energy generation mechanisms are different for a starburst 
(nuclear fusion inside stars) and an AGN (mass-accreting SMBH), 
their effects on the surrounding mass-dominating dense ($>$10$^{4}$ cm$^{-3}$) 
molecular gas at (U)LIRGs' nuclei \citep{gao04} can be different, possibly 
producing different rotational J-transition line flux ratios among 
dense molecular gas tracers at (sub)millimeter. 
(Sub)millimeter dense molecular line flux ratios (J=1--0, J=3--2, and 
J=4--3 of HCN and HCO$^{+}$) characteristic to known optically identified 
luminous AGNs \citep[e.g.,][]{koh05,kri08,izu15,ima16c}, are
also observed in a number of (U)LIRGs that show no discernible AGN
signatures in the optical, infrared, and X-rays 
\citep[e.g.,][]{ima07b,ima09a,pri15,izu16,ima16c,ima18b,ima19}.
Considering that independent (sub)millimeter AGN signatures are observed 
in some of these (U)LIRGs, they can be considered as candidates of 
optically/infrared/X-ray-elusive, but (sub)millimeter-detectable 
extremely deeply buried luminous AGNs \citep{ima18b,ima19}. 

ALMA is an ideal observing facility to apply this (sub)millimeter 
dense molecular line energy diagnostic method to nearby (U)LIRGs, owing to 
very high sensitivity and high spatial resolution which enables to 
(1) pinpoint nuclear ($\lesssim$1 kpc) regions where the putative luminous 
buried AGNs are expected to be present and (2) investigate AGN effects, 
by minimizing the contaminations from surrounding spatially extended 
($\gtrsim$a few kpc) starburst emission.
However, owing to the current ALMA frequency coverage of $>$84 GHz 
(band 3--10), HCN and HCO$^{+}$ J=1--0 lines cannot be observed 
for (U)LIRGs at $z >$ 0.06 where many interesting nearby (U)LIRGs 
are found \citep{kim98}.
Thus, we observed at J=3--2 and J=4--3 lines of HCN, HCO$^{+}$, 
and HNC \citep{ima14,izu16,ima16c,ima18b,ima19}.
A trend of elevated HCN emission, relative to HCO$^{+}$, is observed 
in (U)LIRGs with luminous AGN signatures, compared to those without 
\citep{ima14,izu16,ima16c,ima18b,ima19}.
An enhanced HCN abundance in AGN-affected dense molecular gas 
\citep[e.g.,][]{ala15,sai18,tak19,nak18,kam20,ima20} is one possibility.
However, higher rotational excitation of HCN by warm and dense 
molecular gas in the vicinity of a luminous AGN than in a starburst 
is another scenario \citep{ima18b}, because the critical density of HCN 
is higher by a factor of $\sim$5 than that of HCO$^{+}$ \citep{shi15}.
Addition of different J-transition line data can help distinguish 
the physical origin of the elevated HCN emission in luminous AGNs. 

Since ALMA Cycle 5, band 5 (163--211 GHz or $\sim$1.8 mm) observations
have become available for openuse programs. 
HCN, HCO$^{+}$, and HNC J=2--1 transition lines can be observed in band 5
for nearby (U)LIRGs. 
Because the excitation energy level is lower at J=2 than at J=3 or J=4,
we can (1) discuss dense molecular line flux ratios at J=2--1 
less affected by uncertainty of excitation conditions than at J=3--2 
and J=4--3, and 
(2) J=2--1 lines of HCN, HCO$^{+}$, and HNC may be more reliable 
dense molecular gas mass tracers, by reasonably assuming that J=2--1 
lines are thermalized  and optically thick (i.e., same luminosity 
as J=1--0 in units of [K km s$^{-1}$ pc$^{2}$]) in (U)LIRGs' nuclei, 
where warm and dense molecular gas is highly concentrated 
\citep[e.g.,][]{dow07,sco15,sak17,ima19}.
Further, addition of J=2--1 lines of HCN, HCO$^{+}$, and HNC, to J=3--2 and 
J=4--3 lines, can be used to better constrain physical properties of dense 
molecular gas at (U)LIRGs' nuclei.

In addition, the para-H$_{2}$O 3$_{1,3}$--2$_{2,0}$ line at rest frequency 
$\nu_{\rm rest}$ $\sim$ 183 GHz with an upper energy level of 
$E _{\rm u}$ $\sim$ 205 K. 
(hereafter ``183 GHz H$_{2}$O'') is present in close proximity to the 
HNC J=2--1 line ($\nu_{\rm rest}$ $\sim$ 181 GHz), so they are 
simultaneously observable using ALMA. 
In warm and dense molecular gas illuminated by a luminous AGN, 
it is theoretically predicted that the 183 GHz H$_{2}$O emission line can 
be very bright through maser amplification 
\citep[e.g.,][]{deg77,neu91,yat97,mal02} and/or elevated thermal 
(non-maser) emission owing to enhanced H$_{2}$O abundance caused by 
AGN's X-ray illumination \citep{neu94,meij12}.
Although sensitive observations of the 183 GHz H$_{2}$O lines are difficult 
for very nearby sources at $z \sim$ 0 owing to the poor atmospheric 
transmission of Earth at $\sim$183 GHz, we can observe this line without severe 
Earth's atmospheric H$_{2}$O absorption for (U)LIRGs at $z >$ 0.02, 
rendering this H$_{2}$O line as another potentially good AGN indicator.

In this paper, we present the results of ALMA band 5 (163--211 GHz)
and 4 (125--163 GHz) dense  
molecular (HCN, HCO$^{+}$, and HNC J=2--1) and 183 GHz H$_{2}$O line 
observations of ten (U)LIRGs (Table 1), for which the J=3--2 and 
J=4--3 line data of HCN, HCO$^{+}$, and HNC are available 
\citep{ima13a,ima13b,ima14,ima16b,ima16c,ima17,ima18b}.
Results of a ULIRG, the Superantennae (IRAS 19254$–$7245) at $z =$ 0.0617, 
where strong signatures of AGN-megamaser-origin luminous 183 GHz H$_{2}$O 
emission were found, have previously been published by \citet{ima21}.
In this manuscript, we report our band 4--5 ($\sim$2 mm) 
observational results of ten (U)LIRGs and discuss general properties
of their nuclear dense molecular gas.
Compilation of the J=2--1, J=3--2, and J=4--3 line data of HCN, HCO$^{+}$, 
and HNC, 
and detailed discussion of the physical properties of dense molecular gas 
at (U)LIRGs' nuclei, with the aid of non-LTE modeling \citep{van07}, 
will be presented in a separate paper (M. Imanishi in preparation).
We adopt the cosmological parameters, H$_{0}$ $=$ 71 km s$^{-1}$ Mpc$^{-1}$, 
$\Omega_{\rm M}$ = 0.27, and $\Omega_{\rm \Lambda}$ = 0.73, throughout this paper.
Unless otherwise mentioned, ``H$_{2}$O emission'' indicates ``183 GHz H$_{2}$O 
3$_{1,3}$--2$_{2,0}$ line emission'', while ``molecular line flux ratio'' 
indicates ``rotational J-transition line flux ratio at the vibrational ground 
level (v=0)''.

\section{Targets}

Detailed properties of the observed ten (U)LIRGs in Table 1 are described by 
\citet{ima16b,ima16c,ima18b}.
In short, we selected nearby (U)LIRGs that (1) had different
levels of AGN's energetic contributions to the bolometric luminosity, 
based on optical/infrared/(sub)millimeter spectroscopic energy
diagnostic methods and (2) were expected to show bright molecular
emission lines, to investigate their flux ratios with small statistical
uncertainty. 
We regard NGC 1614 and IRAS 13509$+$0442 as starburst-dominated, 
because there are no luminous buried AGN signatures in the infrared 
as well as (sub)millimeter.
IRAS 06035$-$7102, IRAS 08572$+$3915, IRAS 12127$-$1412, 
IRAS 15250$+$3609, the Superantennae, and IRAS 20551$-$4250 are diagnosed 
to contain luminous obscured AGNs based on 3--35 $\mu$m infrared spectroscopy 
\citep[e.g.,][]{arm07,ima07a,ima08,vei09,nar09,nar10}.
IRAS 12112$+$0305 and IRAS 22491$-$1808 display no obvious luminous 
AGN signatures in the optical and infrared, but possible signatures of 
(sub)millimeter-detected extremely deeply buried luminous AGNs were found 
in IRAS 12112$+$0305 NE (north eastern primary nucleus) and 
IRAS 22491$-$1808 \citep{ima18b}. 
IRAS 12112$+$0305 SW (south western secondary nucleus) is considered 
starburst-dominated \citep{ima18b}. 
Although our sample is not statistically unbiased, nor complete, 
it can provide useful information on the possible variation
of molecular line flux ratios, depending on different AGN's energetic 
contributions.

\begin{deluxetable*}{lcccrrrrccc}[bht!]
\tabletypesize{\scriptsize}
\tablecaption{Basic Properties of Observed (Ultra)luminous Infrared Galaxies 
\label{tbl-1}} 
\tablewidth{0pt}
\tablehead{
\colhead{Object} & \colhead{Redshift} & 
\colhead{d$_{\rm L}$} & \colhead{Scale} & 
\colhead{f$_{\rm 12}$} & 
\colhead{f$_{\rm 25}$} & 
\colhead{f$_{\rm 60}$} & 
\colhead{f$_{\rm 100}$} & 
\colhead{log L$_{\rm IR}$} & 
\colhead{Optical} & 
\colhead{IR/(sub)mm} 
\\
\colhead{} & \colhead{} & \colhead{[Mpc]} & \colhead{[kpc/$''$]}  
& \colhead{[Jy]}
& \colhead{[Jy]} & \colhead{[Jy]} & \colhead{[Jy]}  &
\colhead{[L$_{\odot}$]} & \colhead{Class} & \colhead{Class} \\  
\colhead{(1)} & \colhead{(2)} & \colhead{(3)} & \colhead{(4)} & 
\colhead{(5)} & \colhead{(6)} & \colhead{(7)} & \colhead{(8)} & 
\colhead{(9)} & \colhead{(10)} & \colhead{(11)} 
}
\startdata
NGC 1614 (IRAS 04315$-$0840) & 0.0160 & 68 & 0.32 & 1.38 & 7.50 & 32.12 & 34.32 
& 11.7 & HII$^{a,b}$ (Cp$^{c}$) & SB  \\ 
IRAS 06035$-$7102 & 0.0795 & 356 & 1.5 & 0.12 & 0.57 & 5.13
& 5.65 & 12.2 & LI$^{d}$ & AGN \\
IRAS 08572$+$3915 & 0.0580 & 256 & 1.1 & 0.32 & 1.70 & 7.43  & 4.59  & 12.1 &
LI$^{e}$(Sy2$^{c}$) & AGN \\   
IRAS 12112$+$0305 & 0.0730 & 326 & 1.4 & 0.12 & 0.51 & 8.50 & 9.98 & 12.3 & 
LI$^{e}$ (Sy2$^{c}$) & AGN + SB  \\     
IRAS 12127$-$1412 & 0.1332 & 620 & 2.3 & $<$0.13 & 0.24 &
1.54 & 1.13 & 12.2 & LI$^{e}$ (HII$^{c}$) & AGN \\
IRAS 13509$+$0442 & 0.1364 & 636 & 2.4 & 0.10 & $<$0.23 &
1.56 & 2.53 & 12.3 & HII$^{e}$ (Cp$^{c}$) & SB \\
IRAS 15250$+$3609 & 0.0552 & 243 & 1.1 & 0.16 & 1.31 & 7.10
& 5.93 & 12.0 & LI$^{a}$ (Cp$^{c}$) & AGN \\
Superantennae (IRAS 19254$-$7245) & 0.0617 & 273 & 1.2 & 0.22 & 1.24 & 5.48 &
5.79 & 12.1 & Sy2$^{b,d,f,g}$ & AGN \\    
IRAS 20551$-$4250 & 0.0430 & 188 & 0.84 & 0.28 & 1.91 & 12.78 & 9.95 & 12.0 
& LI or HII$^{d}$ (Cp$^{c}$) & AGN \\
IRAS 22491$-$1808 & 0.0776 & 347 & 1.5 & 0.05 & 0.55 & 5.44
& 4.45 & 12.2 & HII$^{c,e}$  & AGN \\  
\enddata

\tablecomments{
Col.(1): Object name. 
Col.(2): Redshift adopted from ALMA dense molecular line data 
\citep{ima16c}, which are slightly different from the optically derived ones 
\citep{kim98} in some cases.
Col.(3): Luminosity distance (in Mpc). 
Col.(4): Physical scale (in kpc arcsec$^{-1}$). 
Col.(5)--(8): f$_{12}$, f$_{25}$, f$_{60}$, and f$_{100}$ are 
{\it IRAS} fluxes at 12 $\mu$m, 25 $\mu$m, 60 $\mu$m, and 100 $\mu$m,
respectively, taken from \citet{kim98} or \citet{san03} or the IRAS
Faint Source Catalog (FSC).
Col.(9): Decimal logarithm of infrared (8$-$1000 $\mu$m) luminosity
in units of solar luminosity (L$_{\odot}$), calculated with
$L_{\rm IR} = 2.1 \times 10^{39} \times$ D(Mpc)$^{2}$
$\times$ (13.48 $\times$ $f_{12}$ + 5.16 $\times$ $f_{25}$ +
$2.58 \times f_{60} + f_{100}$) (ergs s$^{-1}$) \citep{sam96}.
Col.(10): Optical spectroscopic classification. 
``Sy2'', ``LI'', ``HII'', and ``Cp'' mean Seyfert 2, LINER, HII-region, 
and starburst$+$AGN composite, respectively.
$^{a}$: \citet{vei95}.
$^{b}$: \citet{kew01}.
$^{c}$: \citet{yua10}.
$^{d}$: \citet{duc97}.
$^{e}$: \citet{vei99}.
$^{f}$: \citet{mir91}.
$^{g}$: \citet{col91}.
Col.(10): Infrared and (sub)millimeter energy diagnostic result. 
``AGN'' and ``SB'' mean AGN-important (AGN signatures
significantly detected) and starburst-dominated (no AGN signature), 
respectively.
IRAS 12112$+$0305 consists of two galaxy nuclei, AGN-important primary 
nucleus and starburst-dominated secondary nucleus ($\S$2).
}
\end{deluxetable*}

\section{Observations and Data Analysis} 

We conducted band 5 (163--211 GHz) and 4 (125--163 GHz)
observations of nine ULIRGs 
and one LIRG NGC 1614 (Table 1) in our ALMA Cycle 5 programs 
2017.1.00022.S and 2017.1.00023.S (PI = M. Imanishi).
Data of dense molecular gas tracers, HCN J=2--1 ($\nu_{\rm rest}$ = 177.261 GHz) 
and HCO$^{+}$ J=2--1 ($\nu_{\rm rest}$ = 178.375 GHz) lines, were obtained 
in one observation (called as ``J21a'').
Data of HNC J=2--1 ($\nu_{\rm rest}$ = 181.325 GHz) and 3$_{1,3}$--2$_{2,0}$ line
of para-H$_{2}$O at $\nu_{\rm rest}$ = 183.310 GHz (183 GHz H$_{2}$O) 
were obtained separately (``J21b''). 
Table 2 summarizes our observation details.

\begin{deluxetable*}{lllccc|ccc}[bht!]
\tabletypesize{\scriptsize}
\tablecaption{Log of Our ALMA Observations \label{tbl-2}} 
\tablewidth{0pt}
\tablehead{
\colhead{Object} & \colhead{Line} & \colhead{Date} & \colhead{Antenna} & 
\colhead{Baseline} & \colhead{Integration} & \multicolumn{3}{c}{Calibrator} \\ 
\colhead{} & \colhead{} & \colhead{[UT]} & \colhead{Number} & \colhead{[m]} &
\colhead{[min]} & \colhead{Bandpass} & \colhead{Flux} & \colhead{Phase}  \\
\colhead{(1)} & \colhead{(2)} & \colhead{(3)} & \colhead{(4)} &
\colhead{(5)} & \colhead{(6)} & \colhead{(7)}  & \colhead{(8)} 
& \colhead{(9)}
}
\startdata 
NGC 1614 & HCN/HCO$^{+}$ J=2--1 (J21a) & 2018 September 12 & 41 & 15--1231 & 13 
& J0423$-$0120 & J0423$-$0120 & J0422$-$0643 \\
 & HNC J=2--1/183 GHz H$_{2}$O (J21b) & 2018 September 12 & 41 & 15--1231 & 20
& J0423$-$0120 & J0423$-$0120 & J0422$-$0643 \\
IRAS 06035$-$7102 & HCN/HCO$^{+}$ J=2--1 (J21a) & 2018 August 26 & 42 & 15--782 & 6
& J0519$-$4546 & J0519$-$4546 & J0601$-$7036 \\
 & HNC J=2--1/183 GHz H$_{2}$O (J21b) & 2018 September 21 & 43 & 15--1398 & 15
& J0635$-$7516 & J0635$-$7516 & J0601$-$7036 \\
IRAS 08572$+$3915 & HCN/HCO$^{+}$ J=2--1 (J21a) & 2018 April 1 & 41 & 15--704 
& 19 & J0854$+$2006 & J0854$+$2006 & J0916$+$3854  \\
 & & 2018 September 19 & 41 & 15--1398 & 19 & 
J0854$+$2006 & J0854$+$2006 & J0916$+$3854 \\
 & HNC J=2--1/183 GHz H$_{2}$O (J21b) & 2018 September 19 & 41 & 15--1398 & 27 
& J0854$+$2006 & J0854$+$2006 & J0916$+$3854 \\
IRAS 12112$+$0305 & HCN/HCO$^{+}$ J=2--1 (J21a) & 2018 April 1 & 41 & 15--704 & 11
& J1229$+$0203 & J1229$+$0203 & J1220$+$0203 \\
 & HNC J=2--1/183 GHz H$_{2}$O (J21b) & 2018 September 6 & 43 & 15--784 & 6
& J1229$+$0203 & J1229$+$0203 & J1220$+$0203 \\
IRAS 12127$-$1412 & HCN/HCO$^{+}$ J=2--1 (J21a) & 2017 December 15 & 45 & 15--2517 
& 20 & J1127$-$1857 & J1127$-$1857 & J1215$-$1731 \\
 & HNC J=2--1/183 GHz H$_{2}$O (J21b) & 2017 December 14 & 49 & 15--3321 & 33
& J1127$-$1857 & J1215$-$1731 & J1127$-$1857 \\
IRAS 13509$+$0442 & HCN/HCO$^{+}$ J=2--1 (J21a) & 2017 December 14 & 46 & 15--3083 & 
16 & J1256$-$0547 & J1256$-$0547 & J1359$+$0159 \\
 & HNC J=2--1/183 GHz H$_{2}$O (J21b) & 2017 December 13 & 44 & 15--3083 & 22
& J1256$-$0547 & J1256$-$0547 & J1359$+$0159 \\
  & & 2017 December 24 & 43 & 15--2517 & 22
& J1256$-$0547 & J1256$-$0547 & J1359$+$0159 \\
IRAS 15250$+$3609 & HCN/HCO$^{+}$ J=2--1 (J21a) & 2018 August 31 & 43 & 15--784 & 
14 & J1550$+$0527 & J1550$+$0527 & J1613$+$3412 \\
 & HNC J=2--1/183 GHz H$_{2}$O (J21b) & 2018 August 28 & 44 & 15--782 & 13
& J1550$+$0527 & J1550$+$0527 & J1613$+$3412 \\
 & & 2018 September 20 & 43 & 15--1398 & 13 
& J1550$+$0527 & J1550$+$0527 & J1613$+$3412 \\
Superantennae & HCN/HCO$^{+}$ J=2--1 (J21a) & 2018 September 18 & 43 & 15--1398 & 
6 & J1617$-$5848 & J1617$-$5848 & J1837$-$7108 \\
 & HNC J=2--1/183 GHz H$_{2}$O (J21b) & 2018 September 18 & 43 & 15--1398 & 27
& J1617$-$5848 & J1617$-$5848 & J1837$-$7108 \\
IRAS 20551$-$4250 & HCN/HCO$^{+}$ J=2--1 (J21a) & 2018 August 24 & 45 & 15--500 & 48 
& J2056$-$4714 & J2056$-$4714 & J2109$-$4110 \\
 & & 2018 August 24 & 45 & 15--500 & 48
& J2056$-$4714 & J2056$-$4714 & J2109$-$4110 \\
& HNC J=2--1/183 GHz H$_{2}$O (J21b) & 2018 August 31 & 45 & 15--784 & 37 & 
J2056$-$4714 & J2056$-$4714 & J2109$-$4110 \\ 
& & 2018 August 31 & 45 & 15--784 & 37 
& J2056$-$4714 & J2056$-$4714 & J2109$-$4110 \\
& & 2018 September 20 & 46 & 15--1398 & 37 & 
J2056$-$4714 & J2056$-$4714 & J2109$-$4110 \\
IRAS 22491$-$1808 & HCN/HCO$^{+}$ J=2--1 (J21a) & 2018 September 20 & 46 & 15--1398 &
7 & J2258$-$2758 & J2258$-$2758 & J2236$-$1433 \\
 & HNC J=2--1/183 GHz H$_{2}$O (J21b) & 2018 September 20 & 46 & 15--1398 & 8
& J2258$-$2758 & J2258$-$2758 & J2236$-$1433 \\
\enddata

\tablecomments{ 
Col.(1): Object name. 
Col.(2): Observed molecular line. 
Col.(3): Observation date in UT. 
Col.(4): Number of antennas used for observations. 
Col.(5): Baseline length in meters. Minimum and maximum baseline lengths are 
shown. 
Col.(6): Net on source integration time in minutes.
Cols.(7), (8), and (9): Bandpass, flux, and phase calibrator for the 
target source, respectively.
}
\end{deluxetable*}

We employed the widest 1.875 GHz mode with 1920 channels for each spectral 
window. 
For the J21a observations, we included HCN J=2--1 and HCO$^{+}$ J=2--1 
lines in two spectral windows in one sideband (LSB or USB).
The vibrationally excited v$_{2}$=1, l=1f (v$_{2}$=1f) HCN 
(hereafter HCN-VIB) J=2--1 line at $\nu_{\rm rest}$ = 178.136 GHz and 
vibrationally excited v$_{2}$=1f HCO$^{+}$ (HCO$^{+}$-VIB) J=2--1 line 
at $\nu_{\rm rest}$ = 179.129 GHz were also included in this sideband. 
Further, two spectral windows were set in another sideband to observe 
continuum and serendipitously detected lines.
For the J21b observations, we covered the HNC J=2--1 and 183 GHz H$_{2}$O lines 
and vibrationally excited v$_{2}$=1f HNC (HNC-VIB) J=2--1 line 
at $\nu_{\rm rest}$ = 182.584 GHz in two spectral windows on one sideband, and 
used two additional spectral windows in another sideband.
In band 5, the atmospheric transmission of Earth is very low at 
the observed frequency $\nu_{\rm obs}$ = 182--184.5 GHz by 
H$_{2}$O absorption centered at $\nu_{\rm obs}$ $\sim$ 183.3 GHz.
Although the targeted molecular lines of our sources were confirmed to be 
not strongly affected by the Earth's atmospheric absorption, the
other sideband in some sources fell into the frequency of low Earth's
atmospheric transmission. 
These data were completely flagged by ALMA pipeline processes and 
not used for our analysis.

We started our analysis from pipeline-processed data by ALMA, 
using CASA \citep{mcm07} 
\footnote{https://casa.nrao.edu}.
We determined the continuum level using channels that did not contain 
discernible emission lines, and subtracted it using the 
CASA task ``uvcontsub''.
We then applied the CASA task ``clean'' 
(Briggs weighting; robust $=$ 0.5 and gain $=$ 0.1) to create maps of 
the continuum-subtracted molecular line data, by binning 20 channels 
for all ULIRGs (velocity resolution $\sim$ 35 km s$^{-1}$).
For the LIRG NGC 1614, we applied 10 channels binning 
($\sim$20 km s$^{-1}$), 
because molecular emission lines were much narrower than those of other ULIRGs.
In addition, we created cleaned continuum maps by integrating the 
line-free-channels' data.

The pixel scales were set as 0$\farcs$025--0$\farcs$1 pixel$^{-1}$, 
which were less than one-fifth of the synthesized beam size of each data.
We adopted $\lesssim$10\% absolute flux calibration uncertainty 
for our band 4--5 data, based on the ALMA Cycle 5 Proposer's Guide
\footnote{https://almascience.eso.org/documents-and-tools/cycle5/alma-proposers-guide}
 
\section{Results} 
 
Continuum emission at $\sim$2 mm is detected in all sources and is
displayed in Figure 1 as contours.
Table 3 summarizes detailed continuum emission properties. 
The achieved beam size is $\lesssim$1$''$ (i.e., sub-arcsecond) 
for all the obtained data.
According to the ALMA Cycle 5 Technical Handbook (equation 7.6)
\footnote{https://almascience.eso.org/documents-and-tools/cycle5/alma-technical-handbook},  
the maximum recoverable scale (MRS) is $>$10$''$ at $\sim$2 mm for
the minimum baseline of 15 m (Table 2), which corresponds to $>$8 kpc for all
the ULIRGs and $>$3 kpc for the LIRG NGC 1614.
Thus, our targeting dense molecular line emission in compact nuclei 
($\lesssim$a few kpc) should be fully recovered. 

\begin{figure*}
\begin{center}
\hspace*{-4.7cm}
\includegraphics[angle=0,scale=.195]{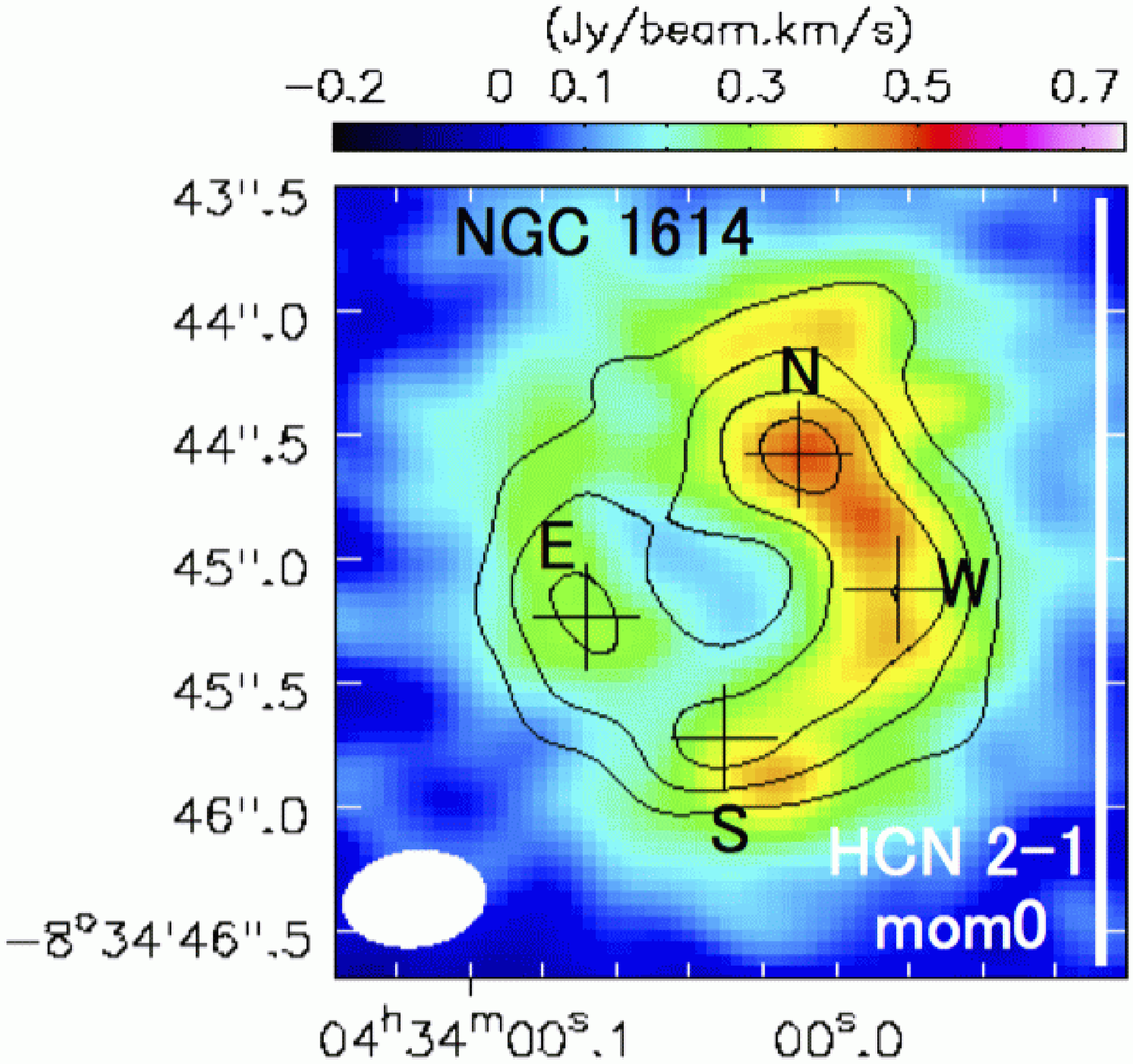} 
\includegraphics[angle=0,scale=.195]{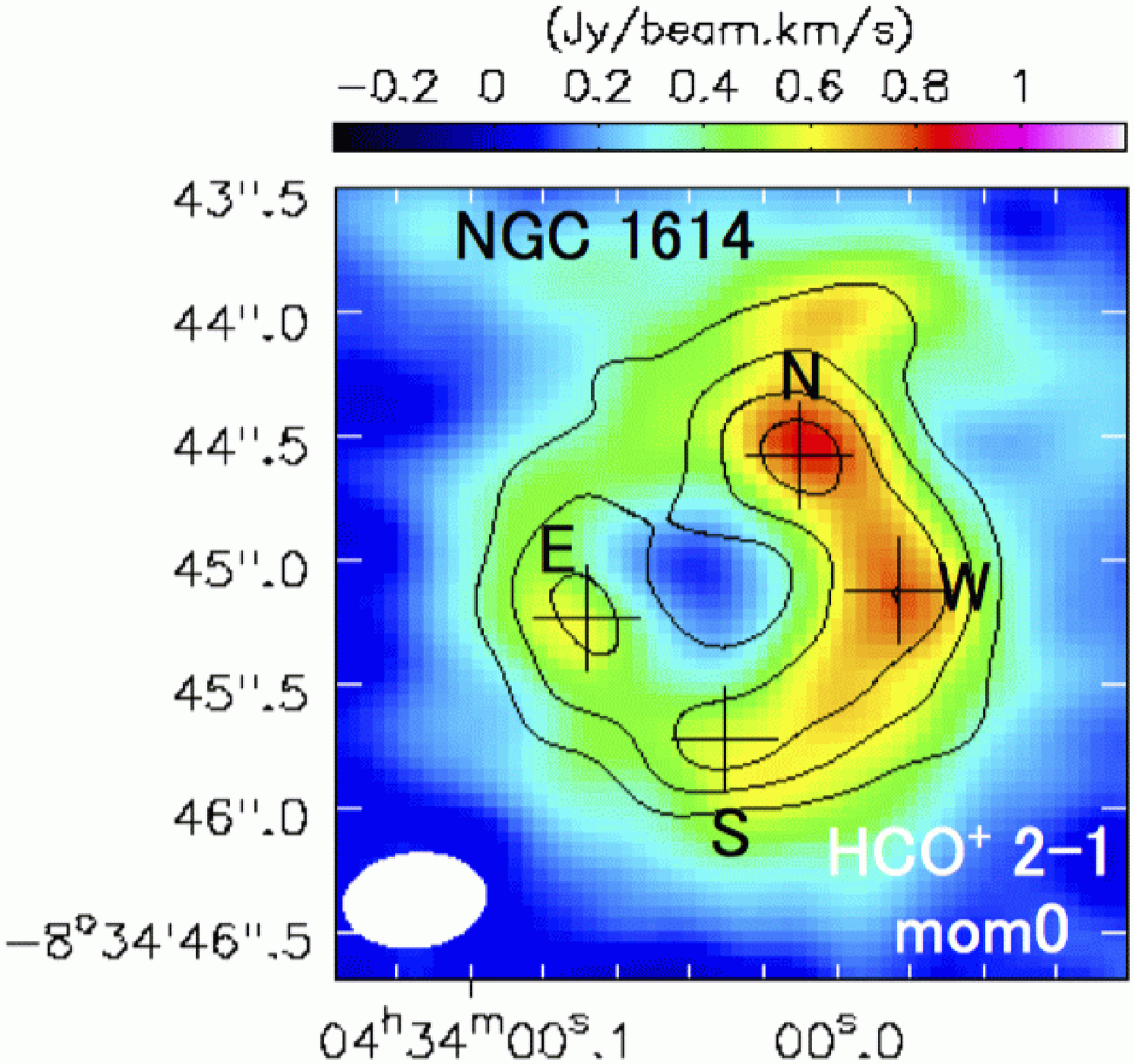} 
\includegraphics[angle=0,scale=.195]{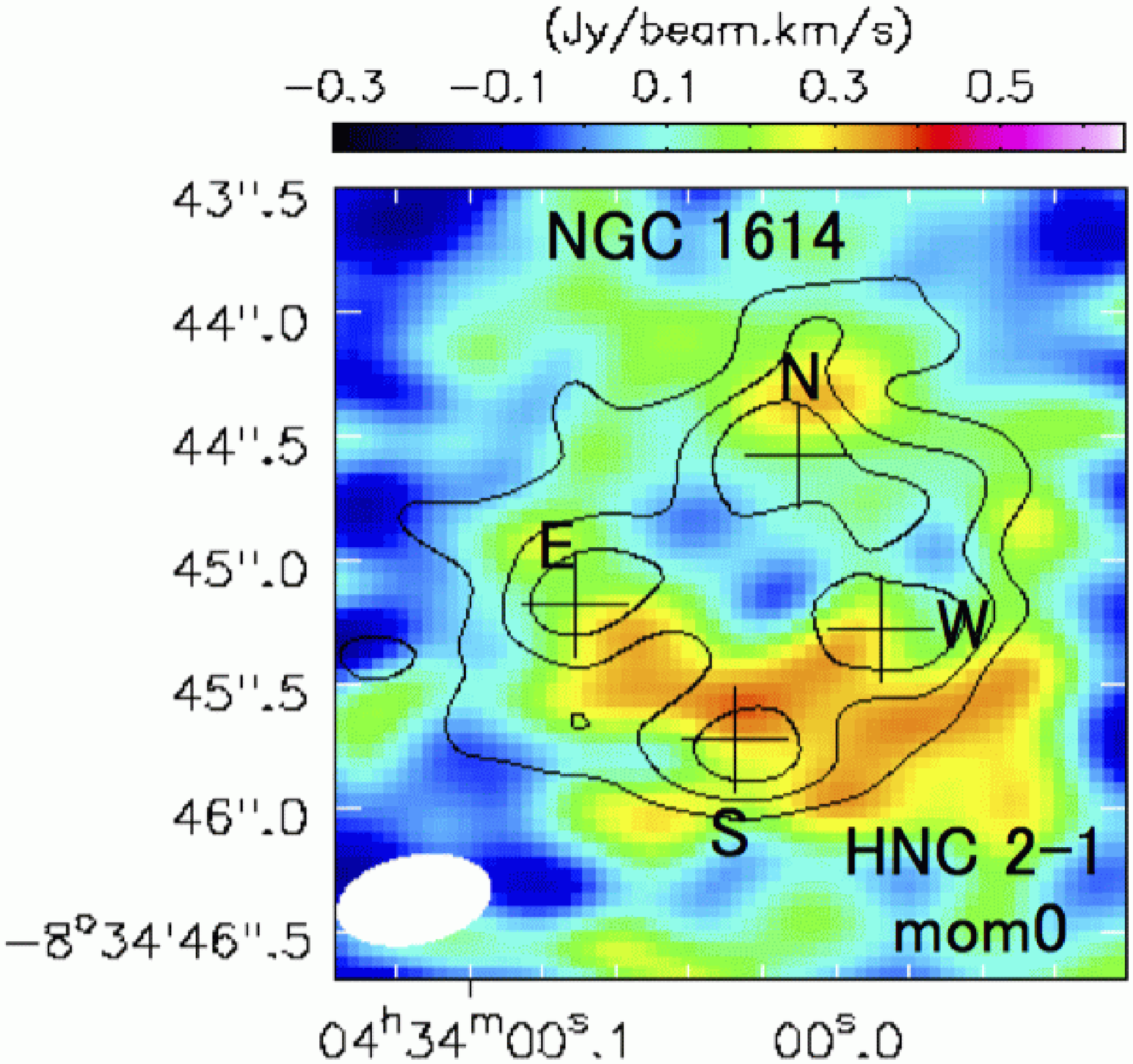} \\
\includegraphics[angle=0,scale=.195]{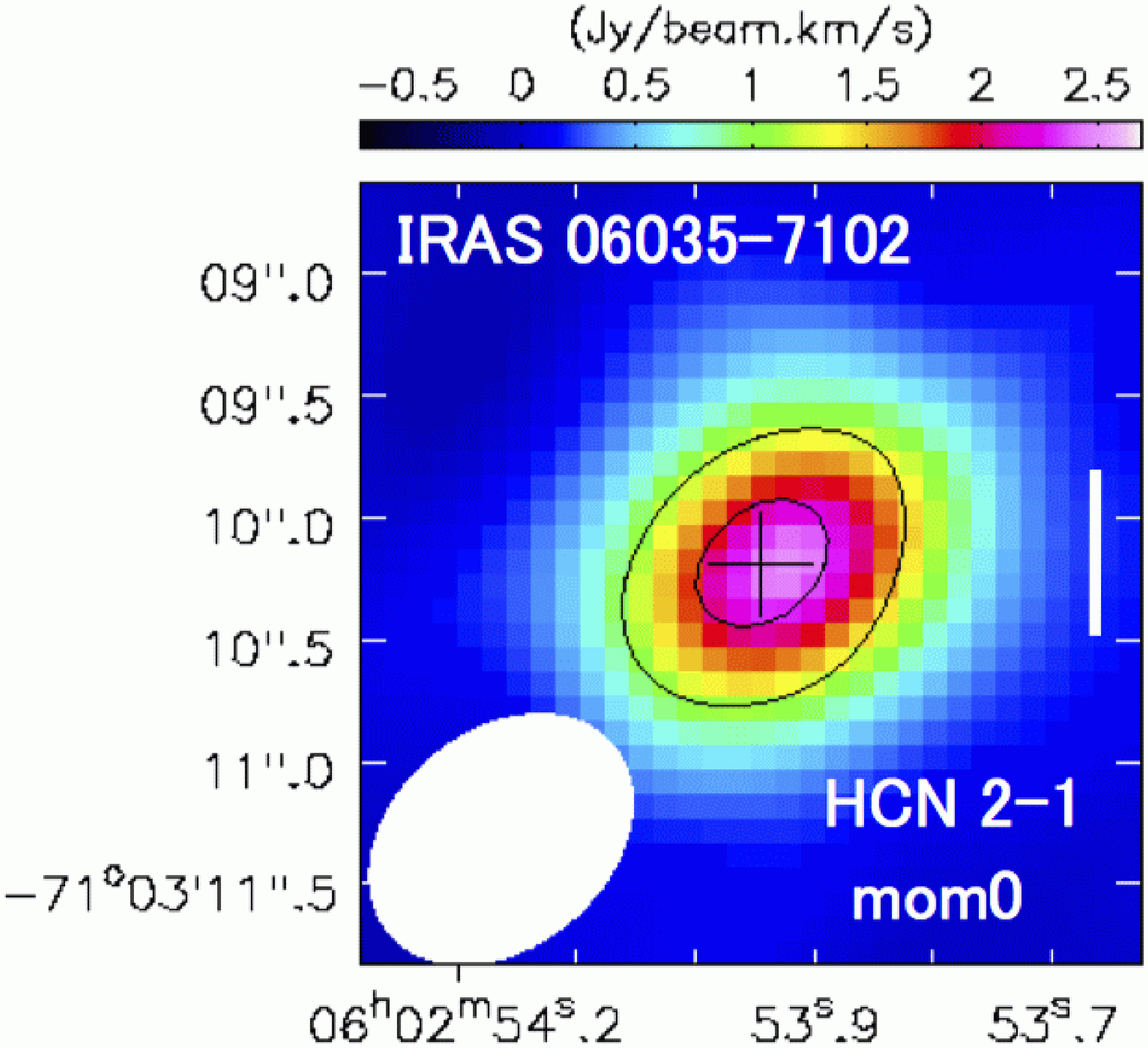} 
\includegraphics[angle=0,scale=.195]{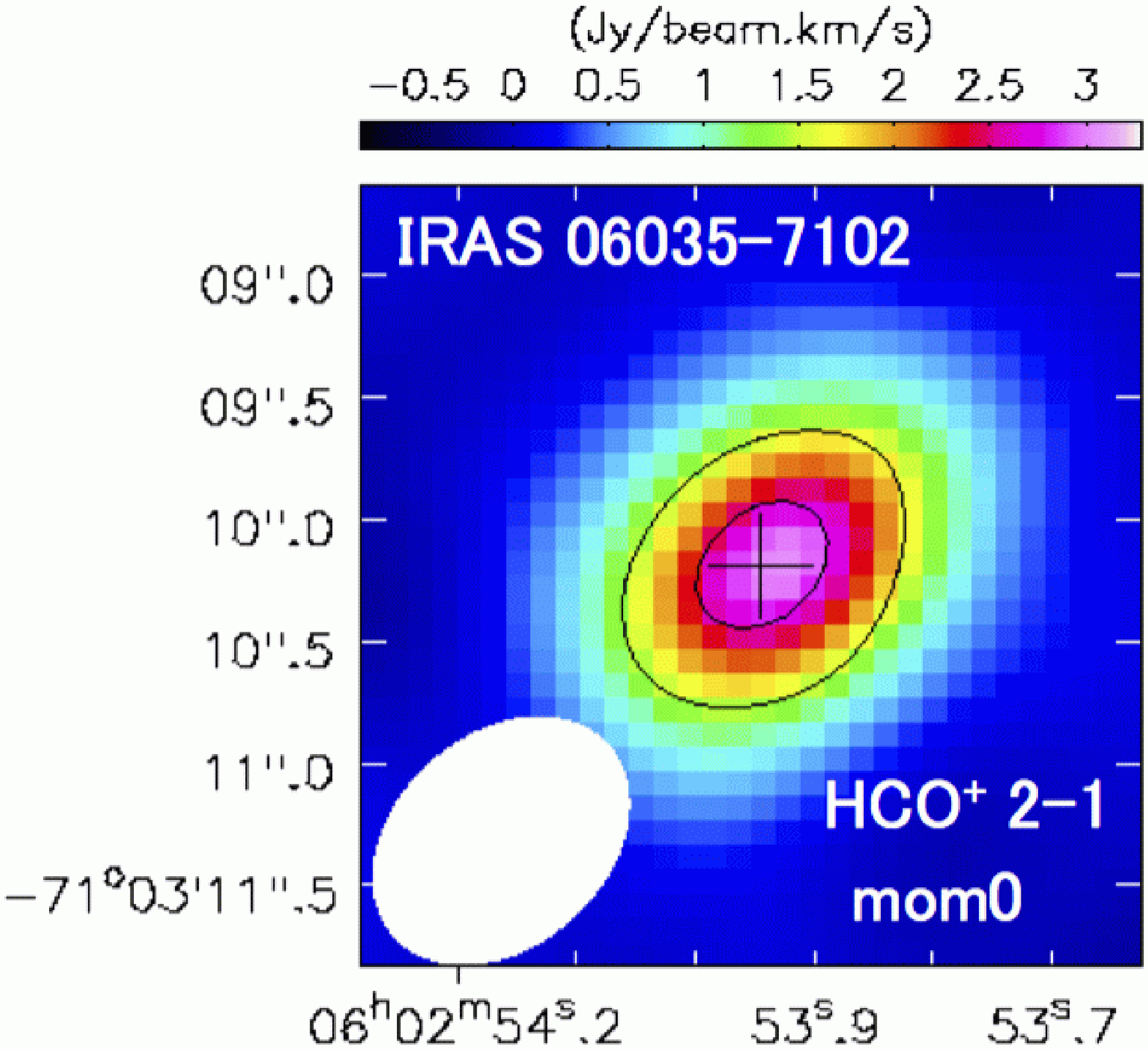} 
\includegraphics[angle=0,scale=.195]{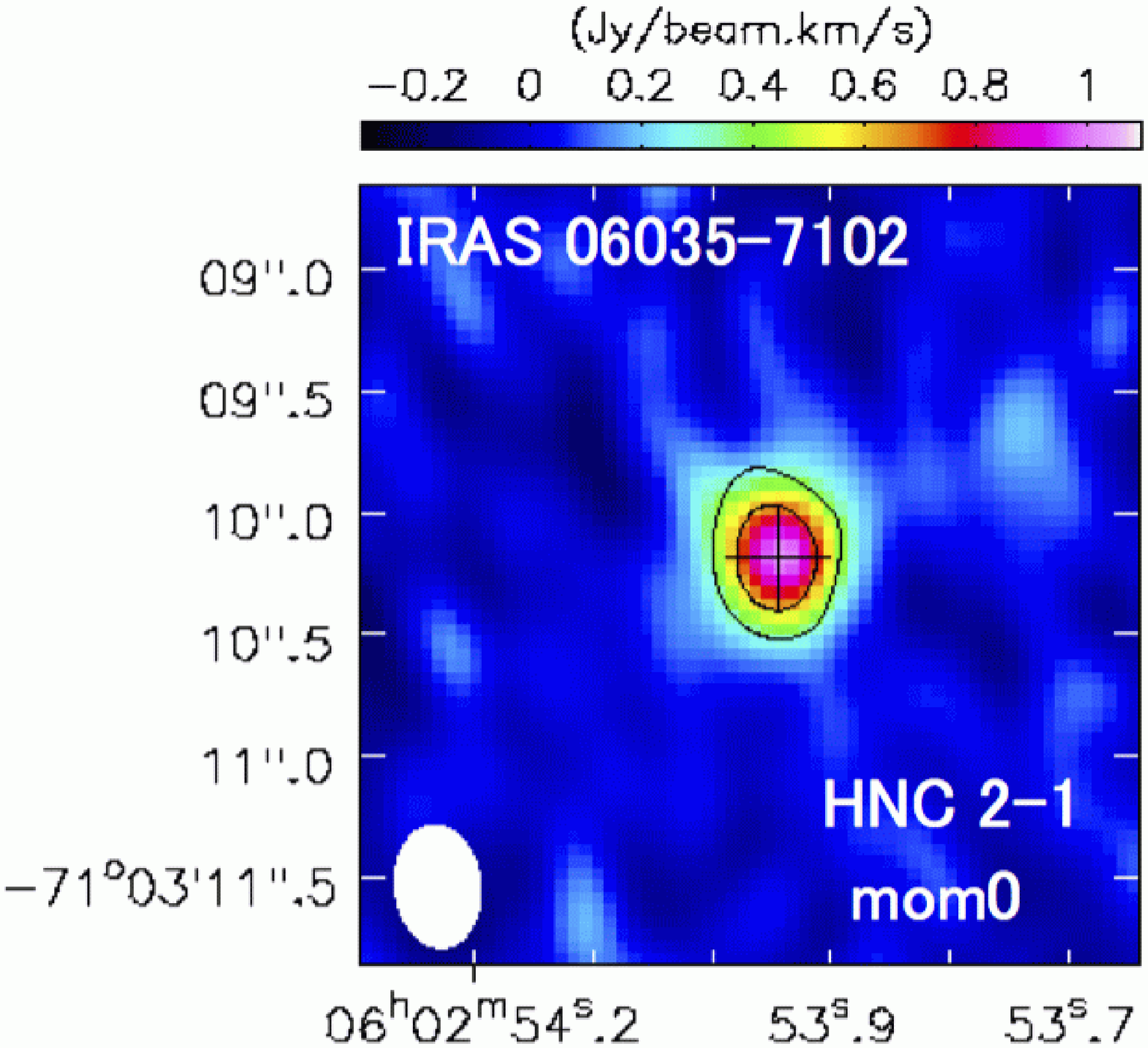} 
\includegraphics[angle=0,scale=.195]{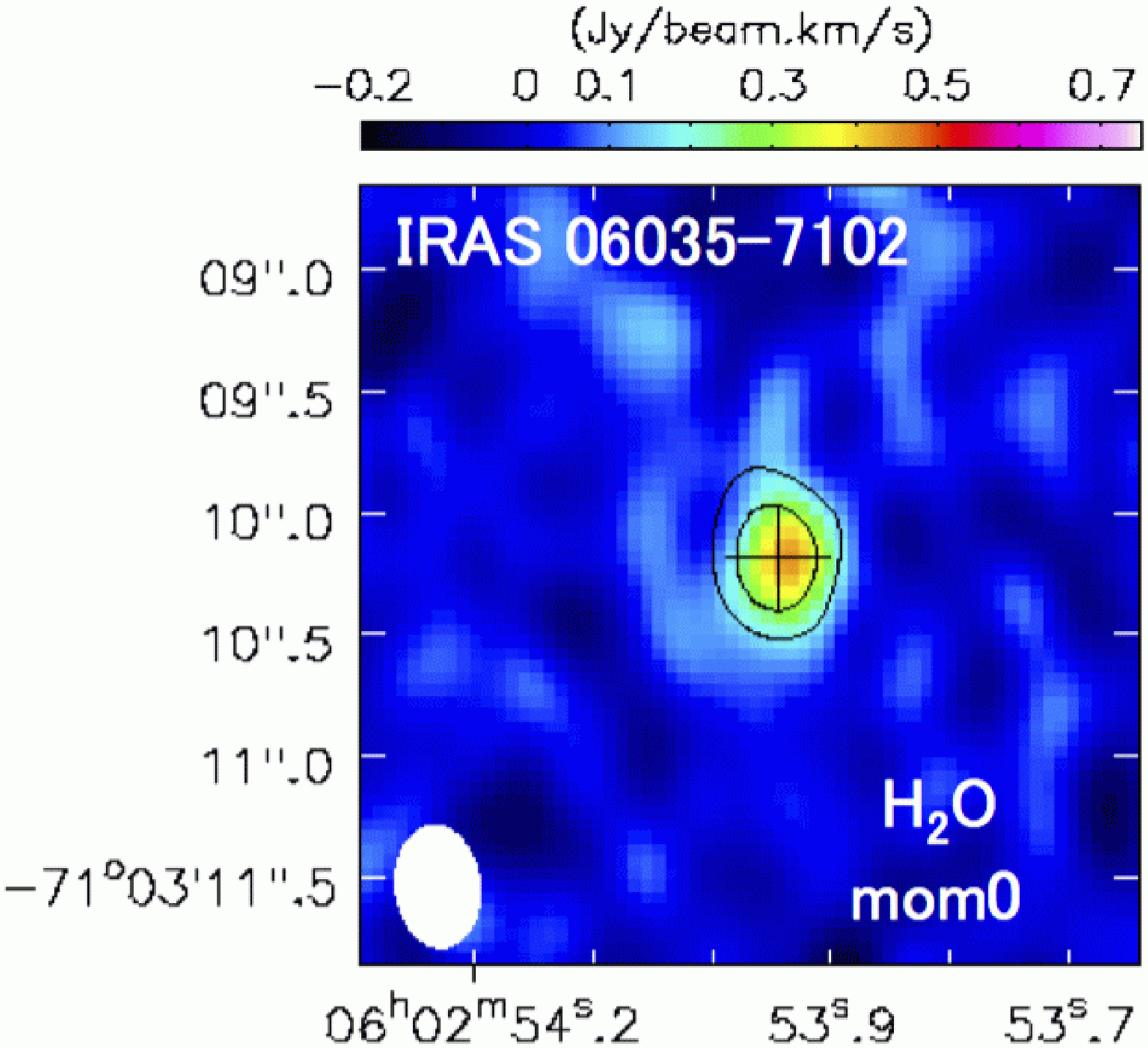} \\
\includegraphics[angle=0,scale=.19]{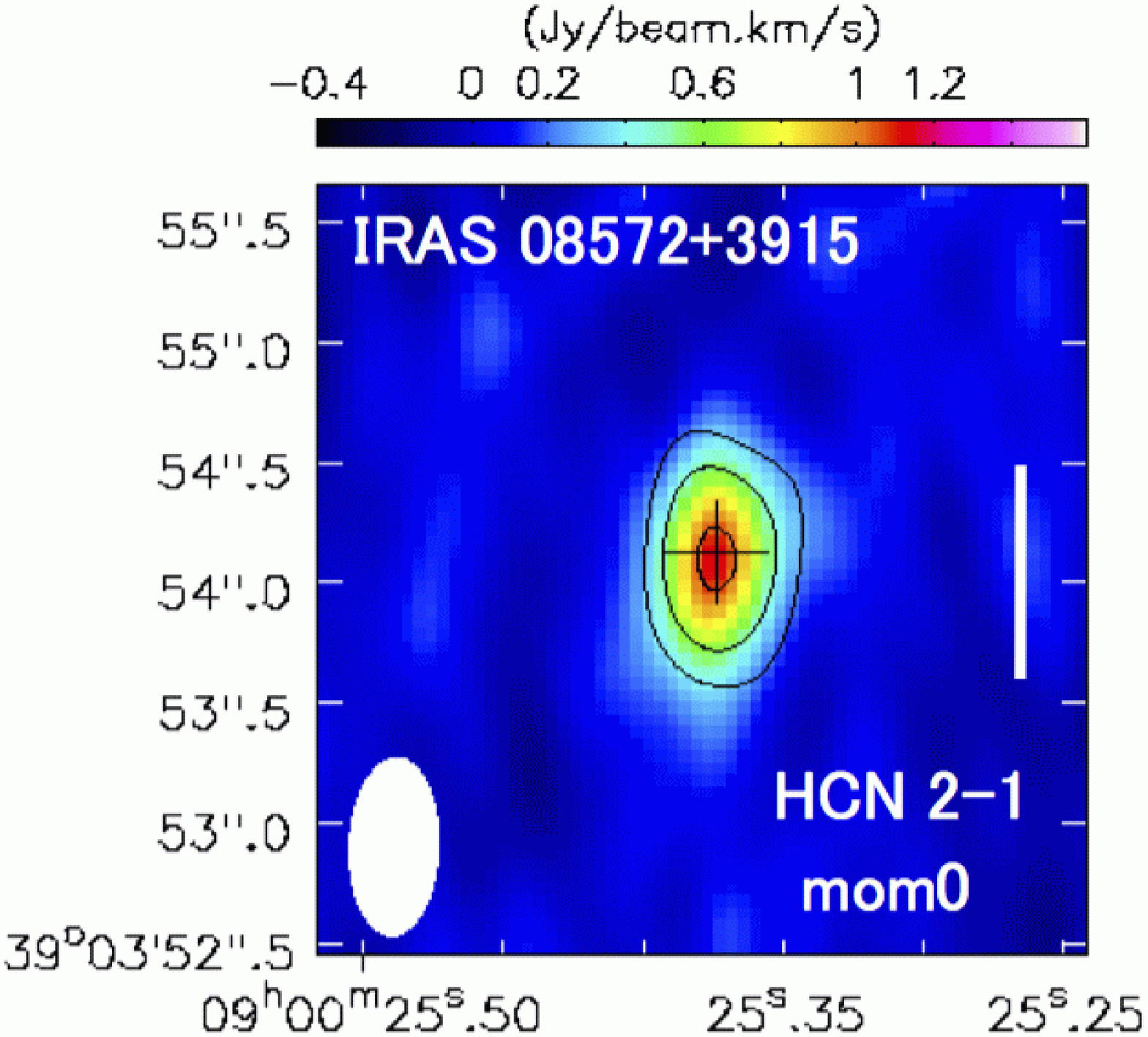} 
\includegraphics[angle=0,scale=.19]{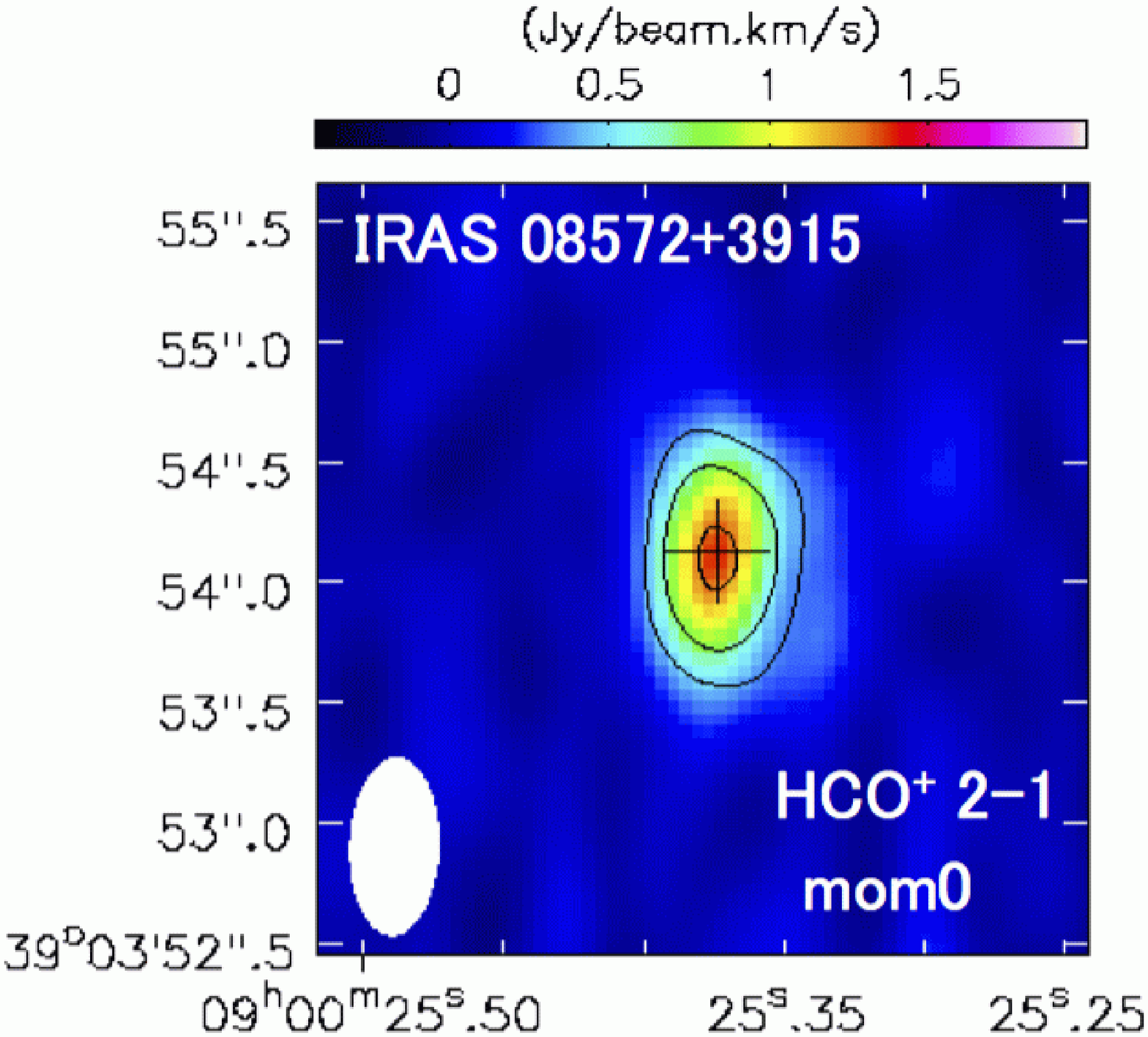} 
\includegraphics[angle=0,scale=.19]{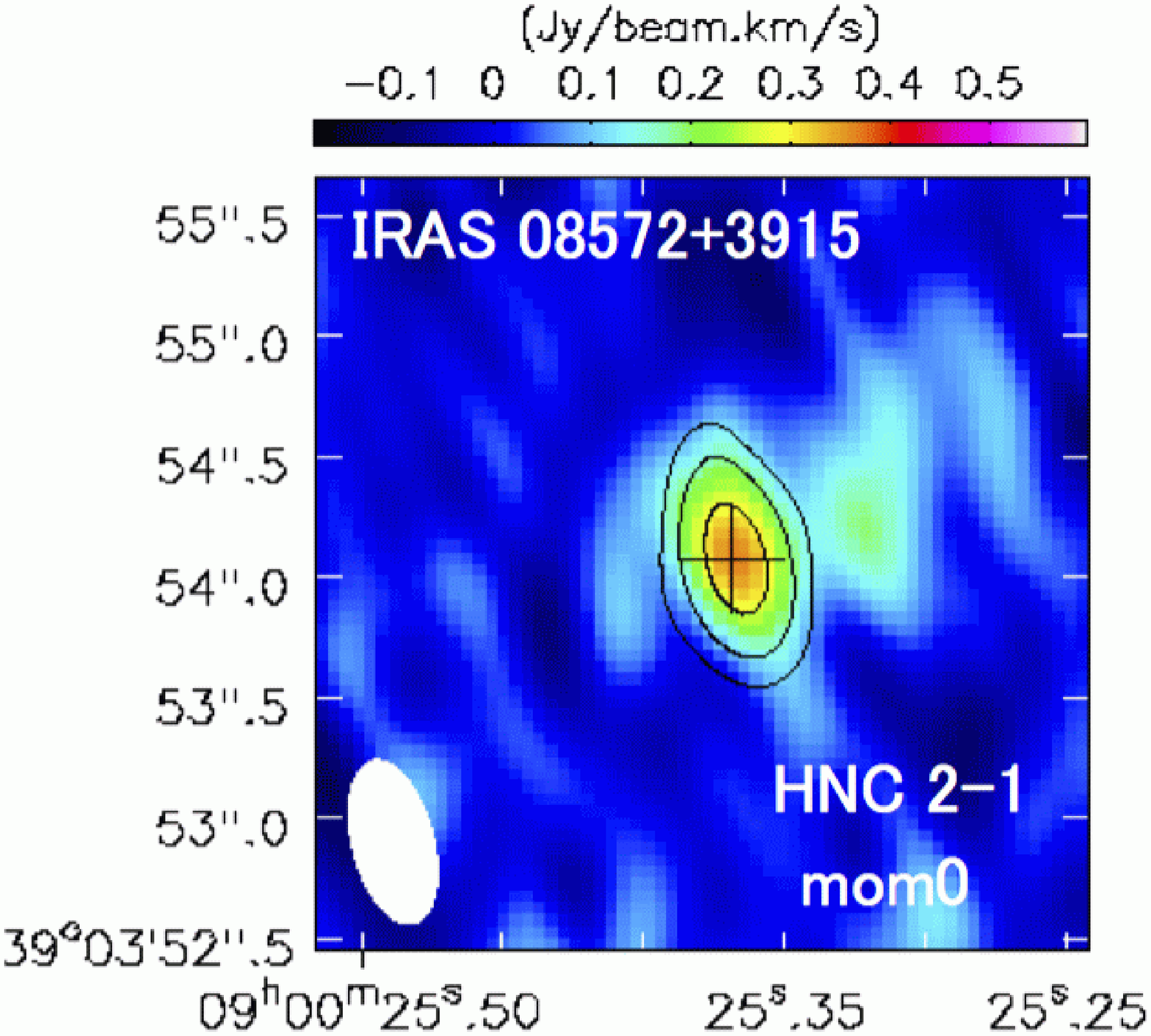} 
\includegraphics[angle=0,scale=.19]{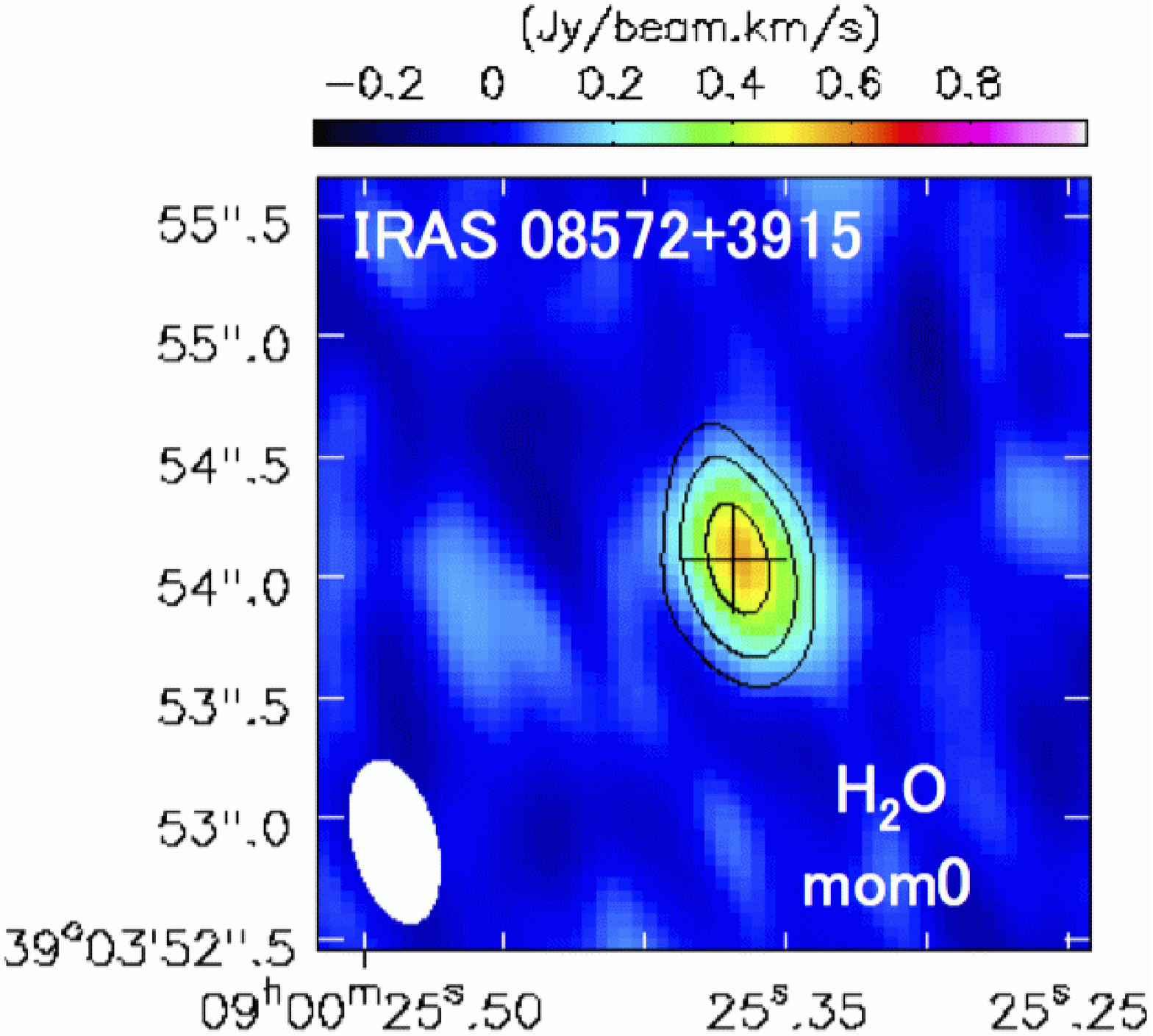} \\
\includegraphics[angle=0,scale=.195]{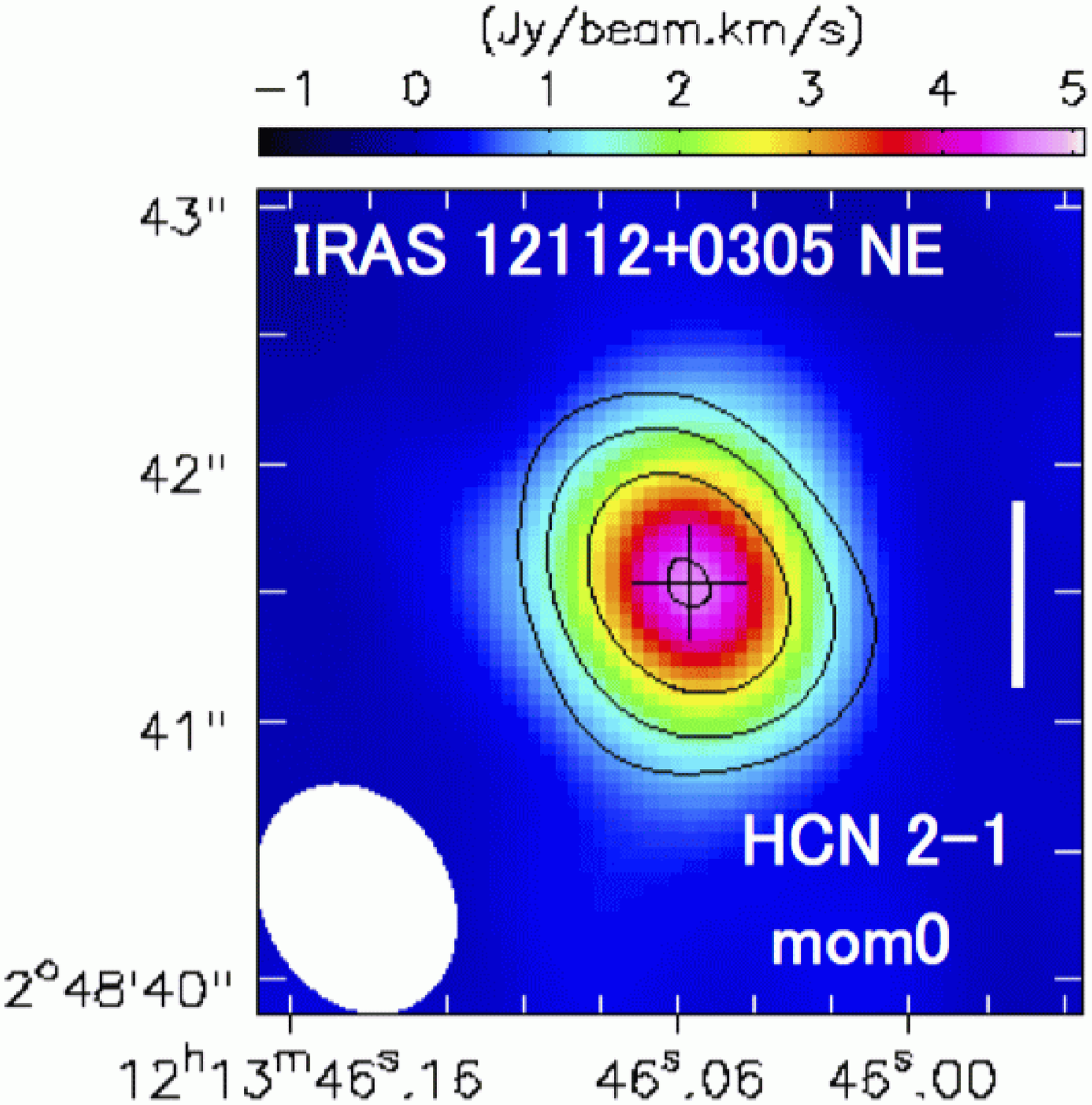} 
\includegraphics[angle=0,scale=.195]{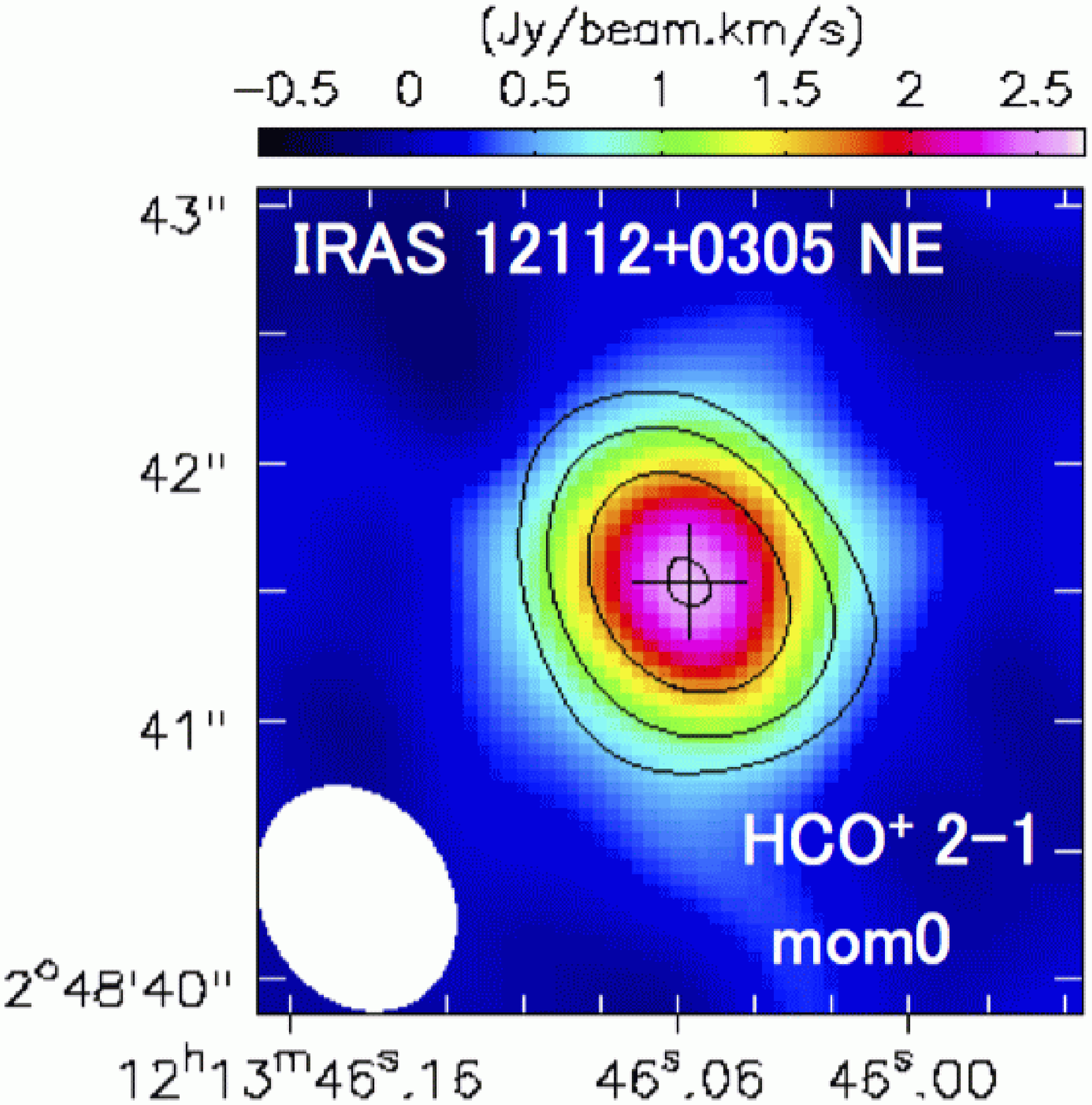} 
\includegraphics[angle=0,scale=.195]{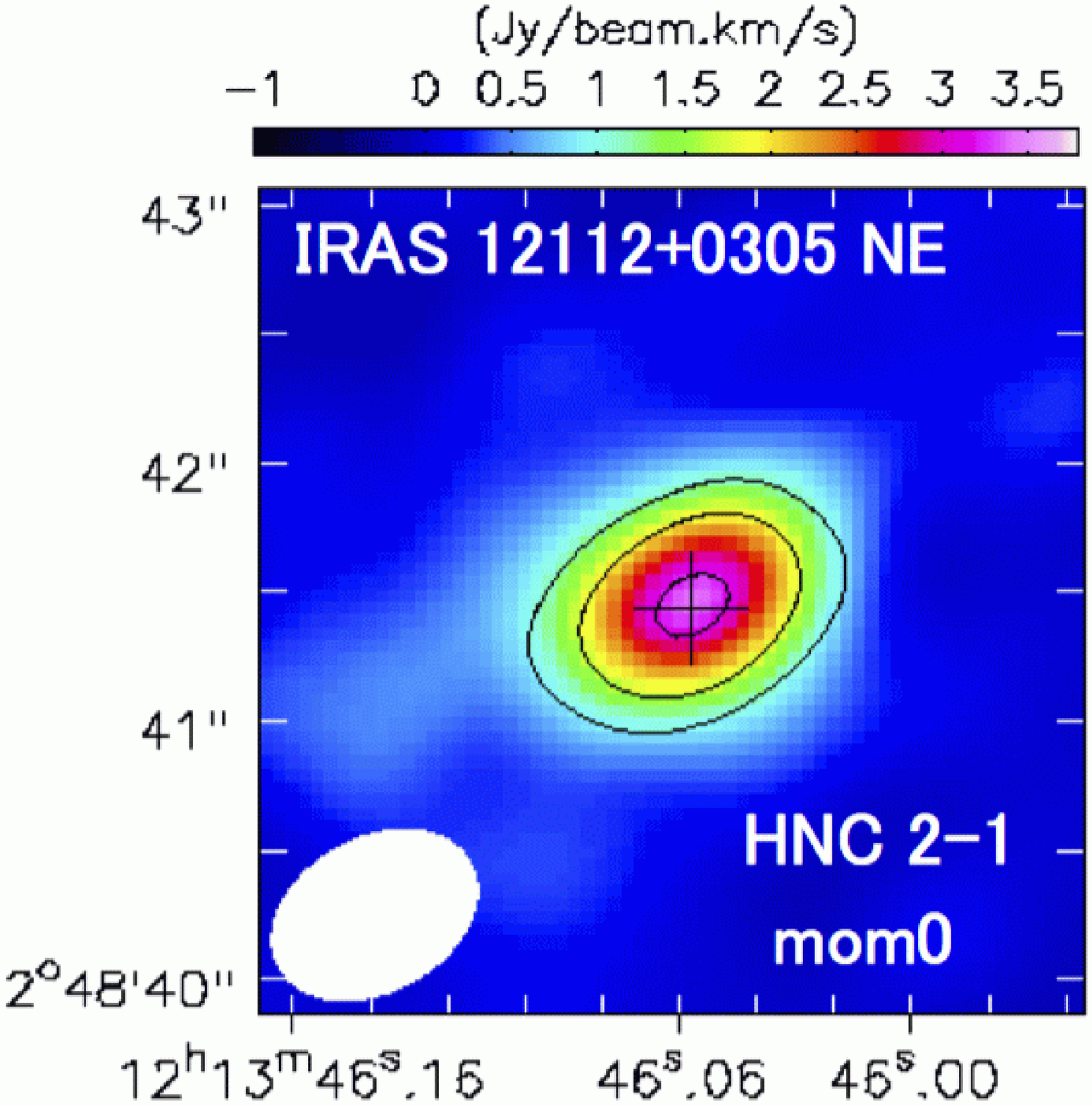} 
\includegraphics[angle=0,scale=.195]{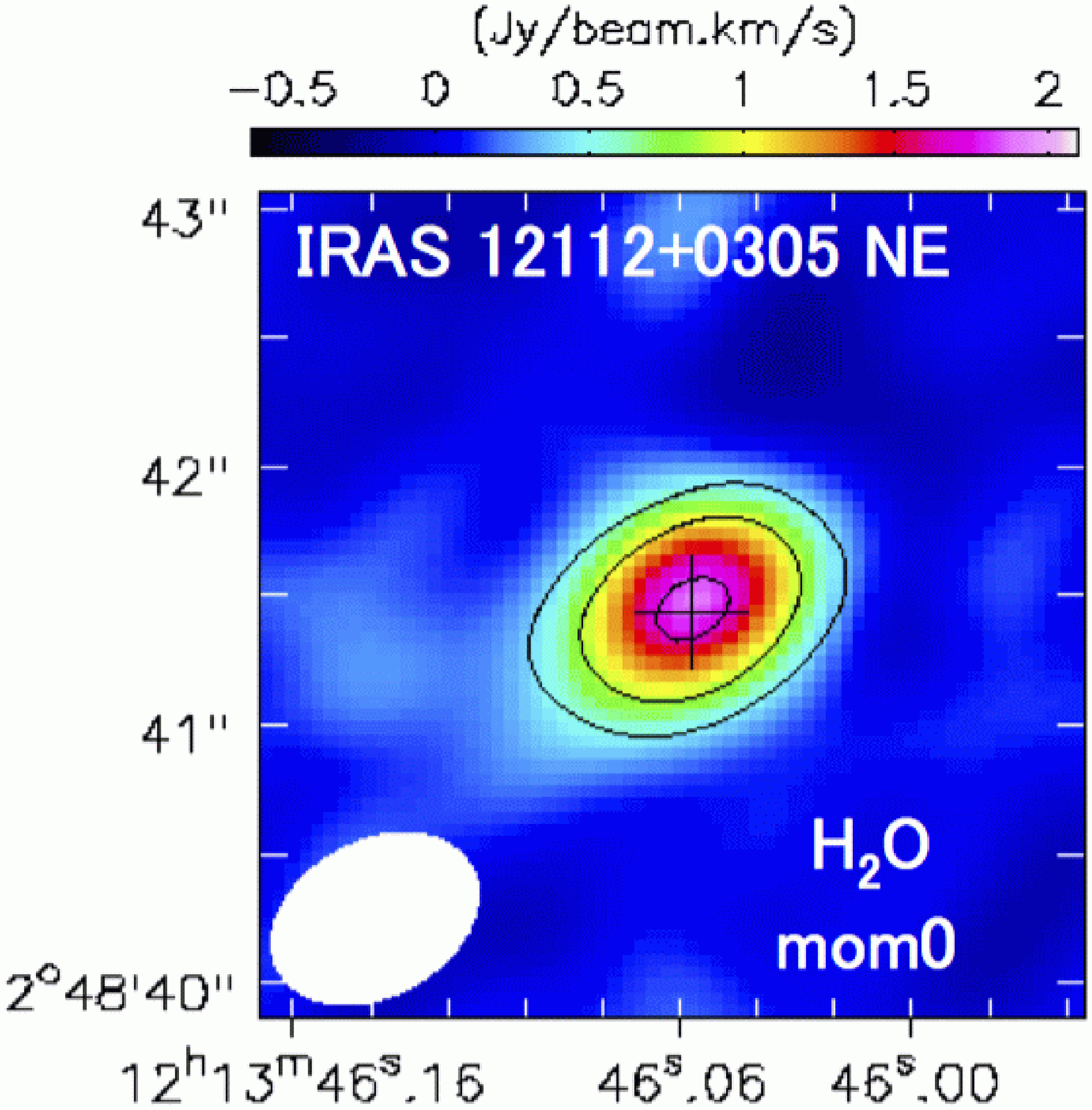} \\
\hspace*{-9.0cm}
\includegraphics[angle=0,scale=.195]{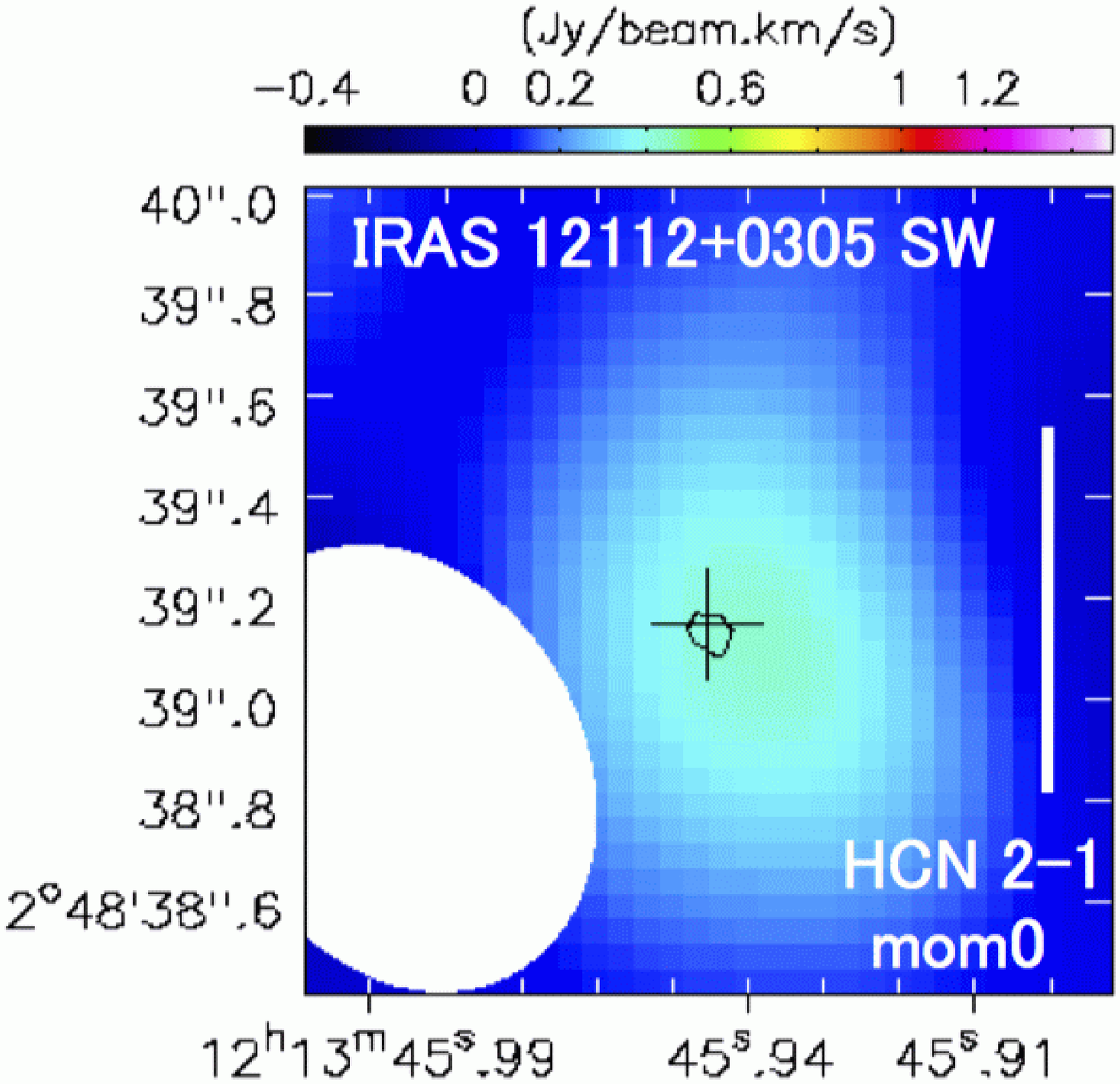} 
\includegraphics[angle=0,scale=.195]{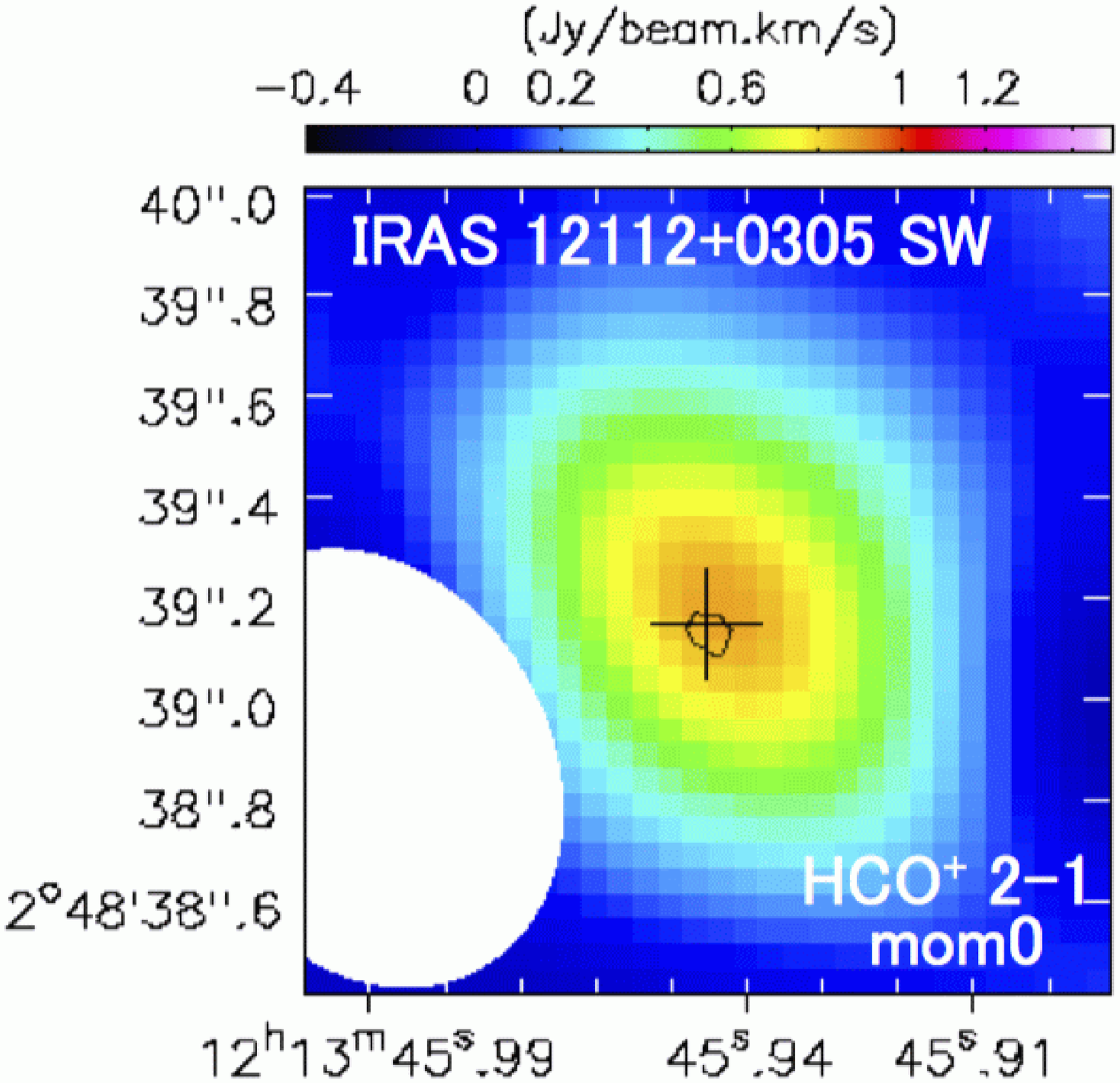} 
\\
\includegraphics[angle=0,scale=.195]{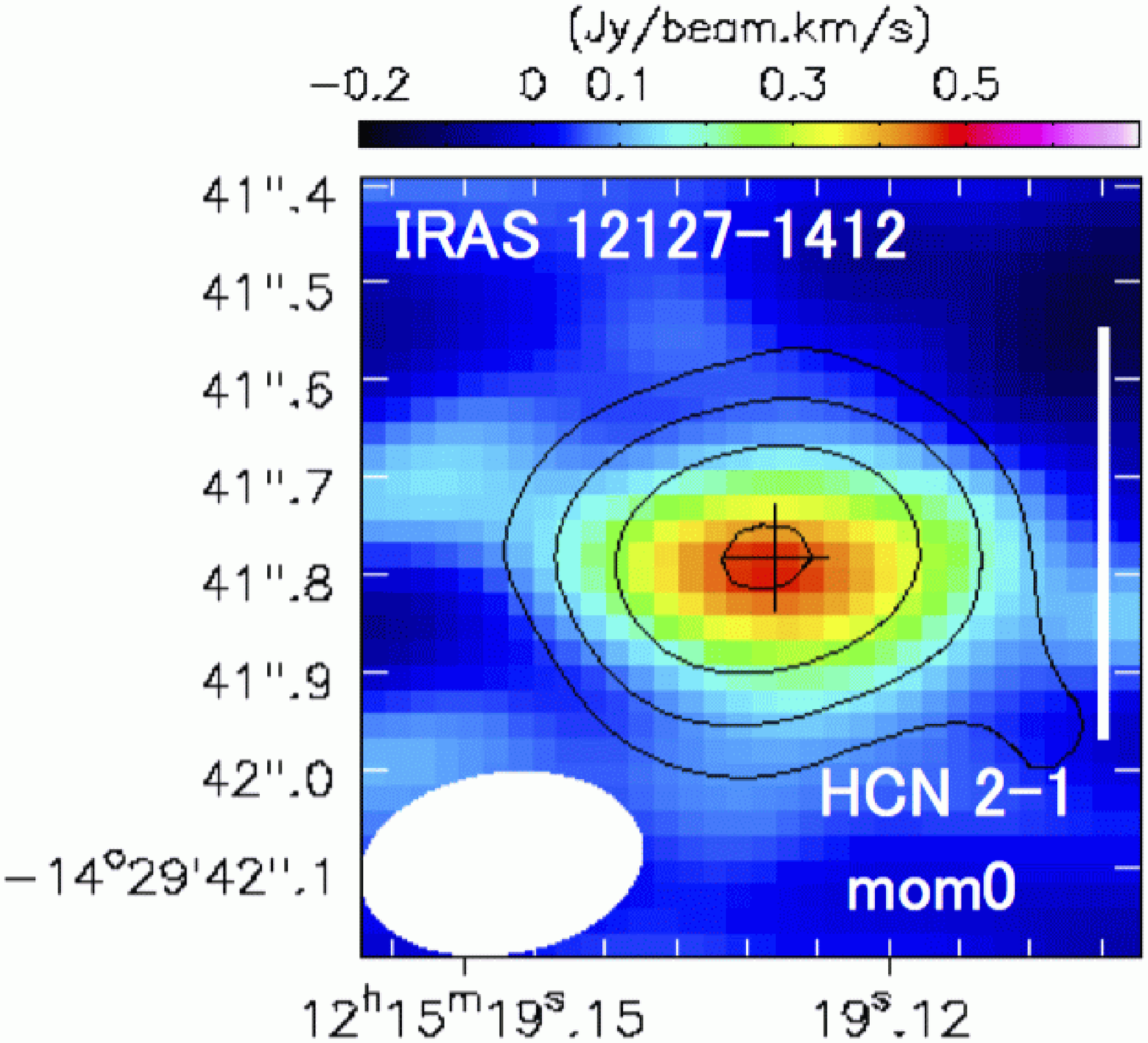} 
\includegraphics[angle=0,scale=.195]{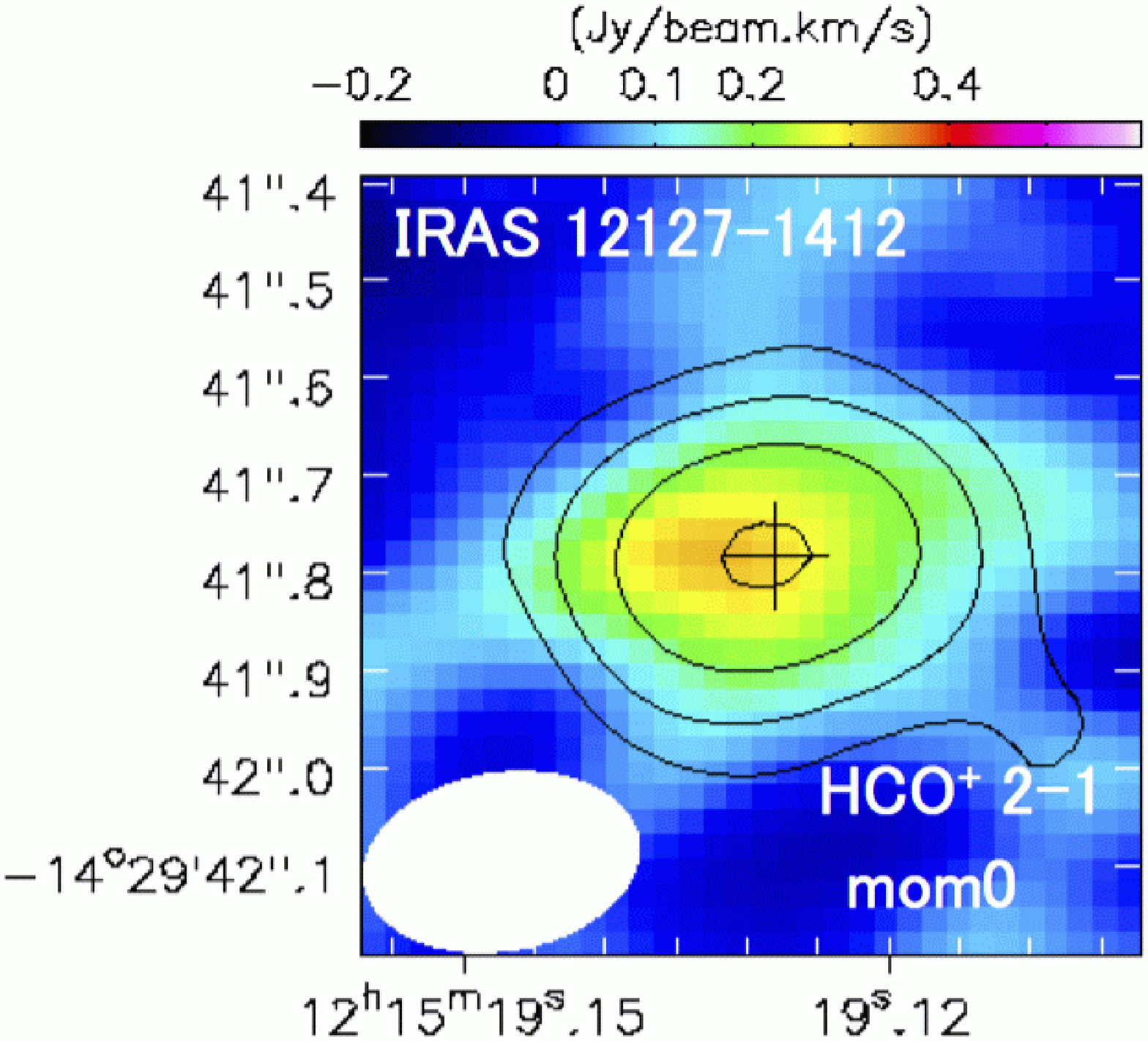} 
\includegraphics[angle=0,scale=.195]{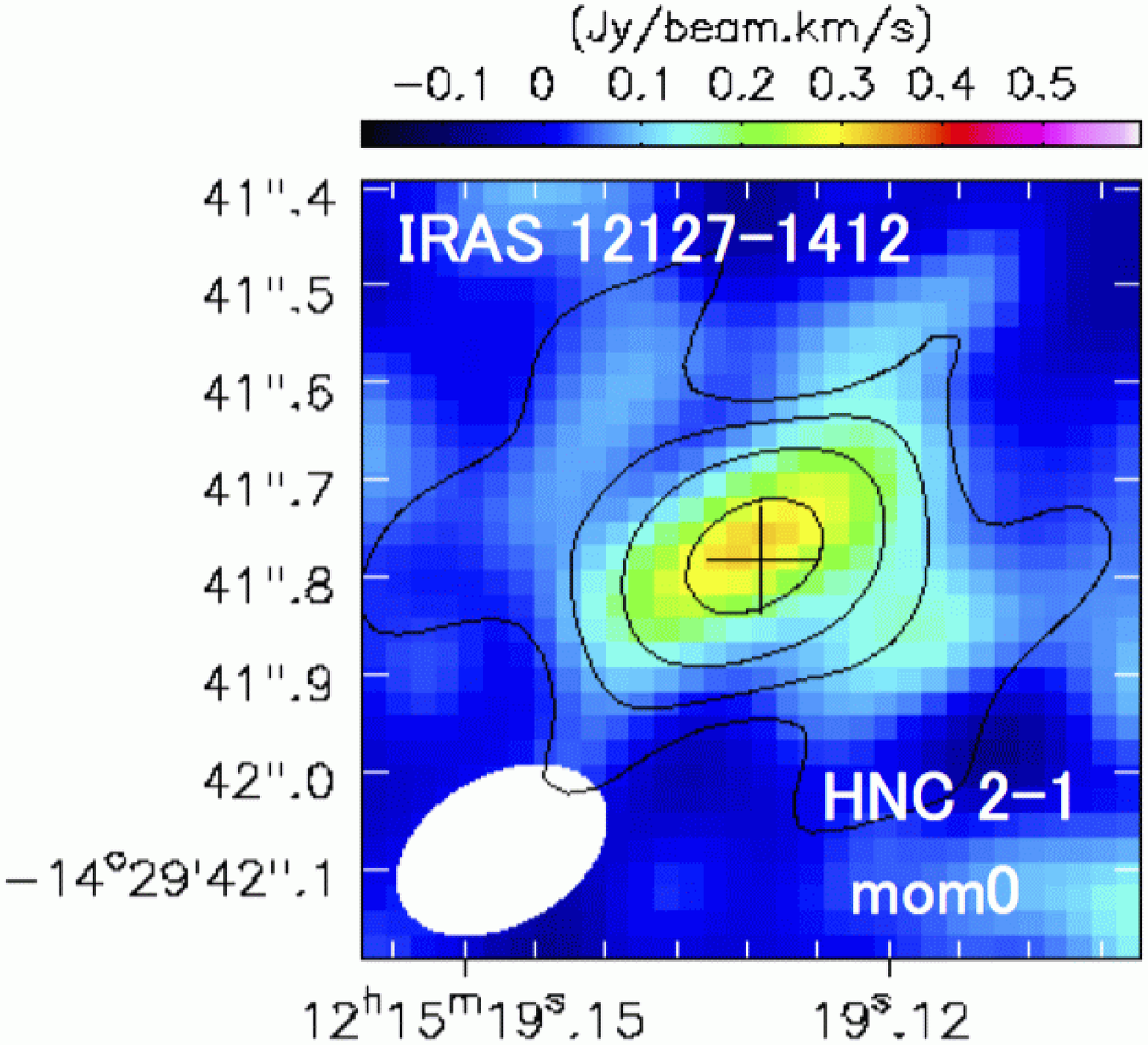} 
\includegraphics[angle=0,scale=.195]{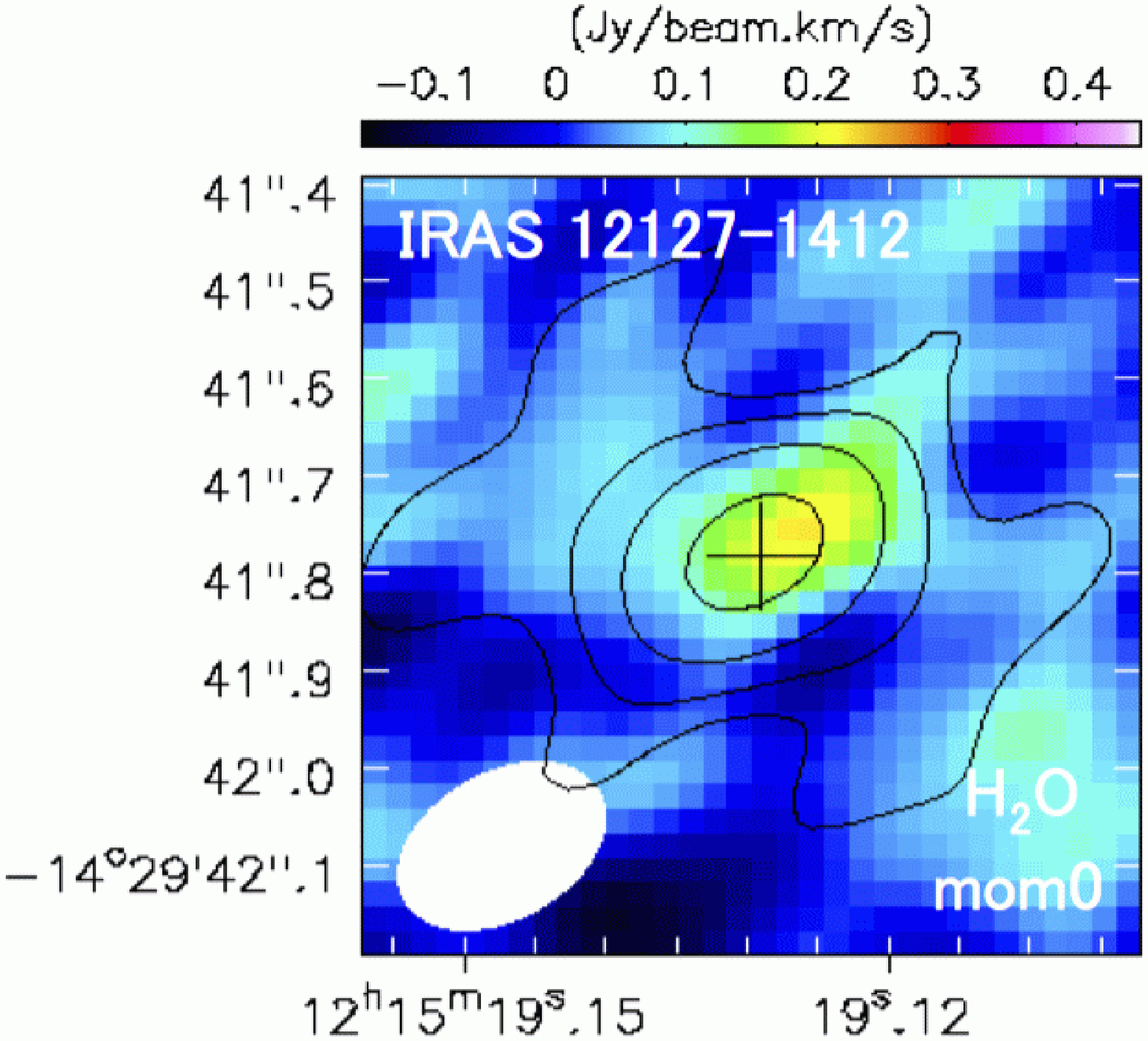} \\
\end{center}
\end{figure*}


\begin{figure*}
\begin{center}
\includegraphics[angle=0,scale=.195]{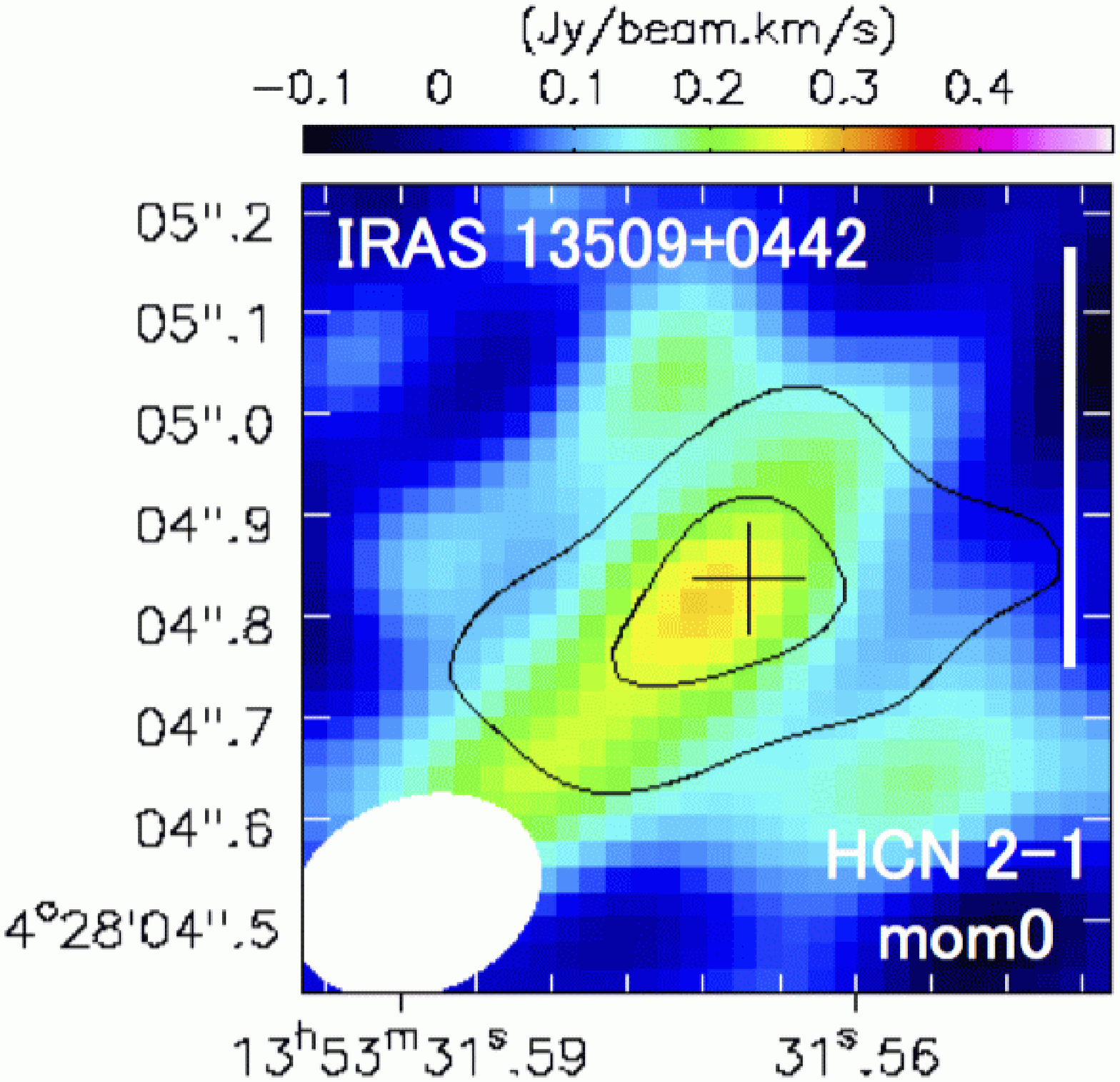} 
\includegraphics[angle=0,scale=.195]{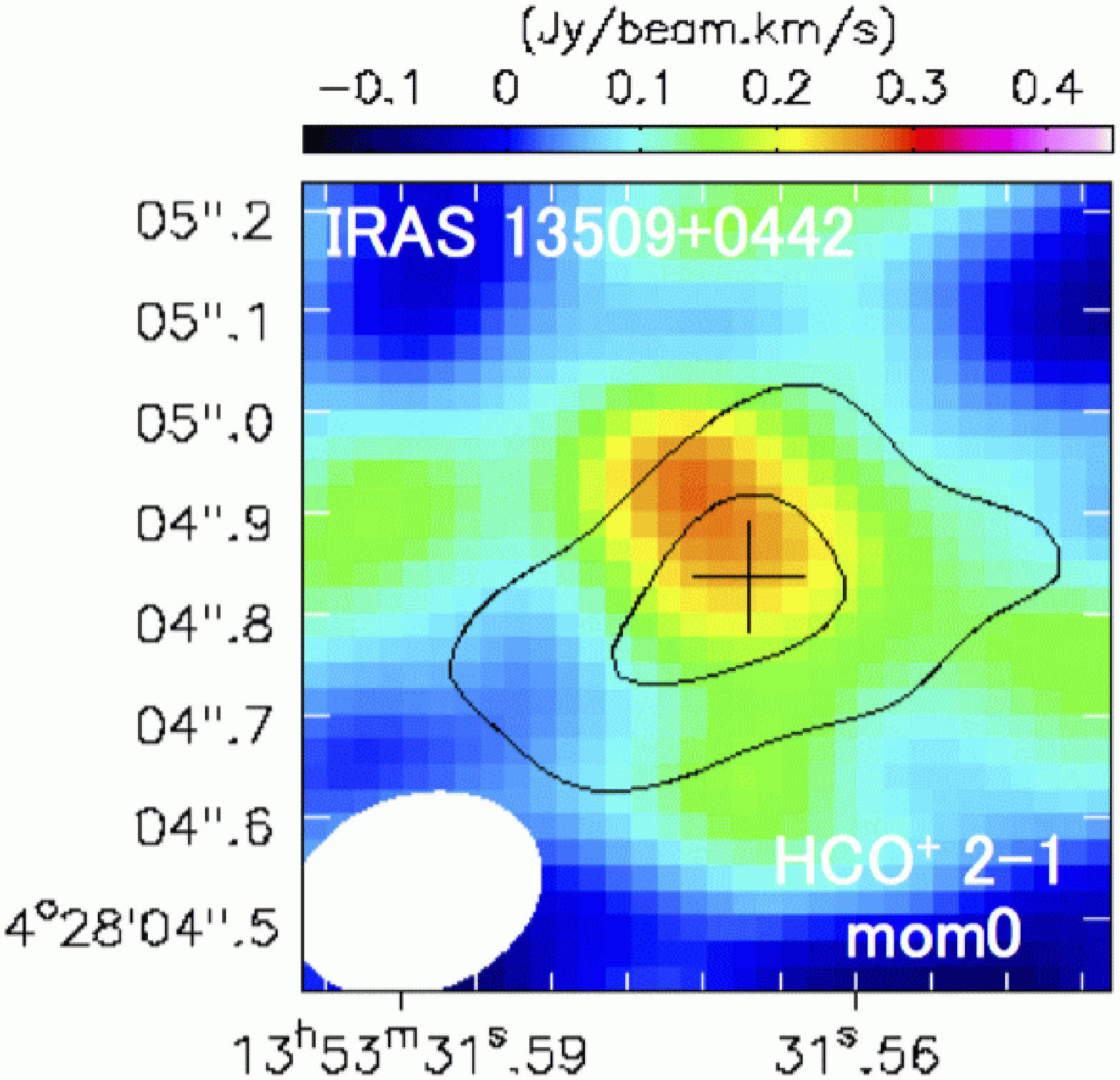} 
\includegraphics[angle=0,scale=.195]{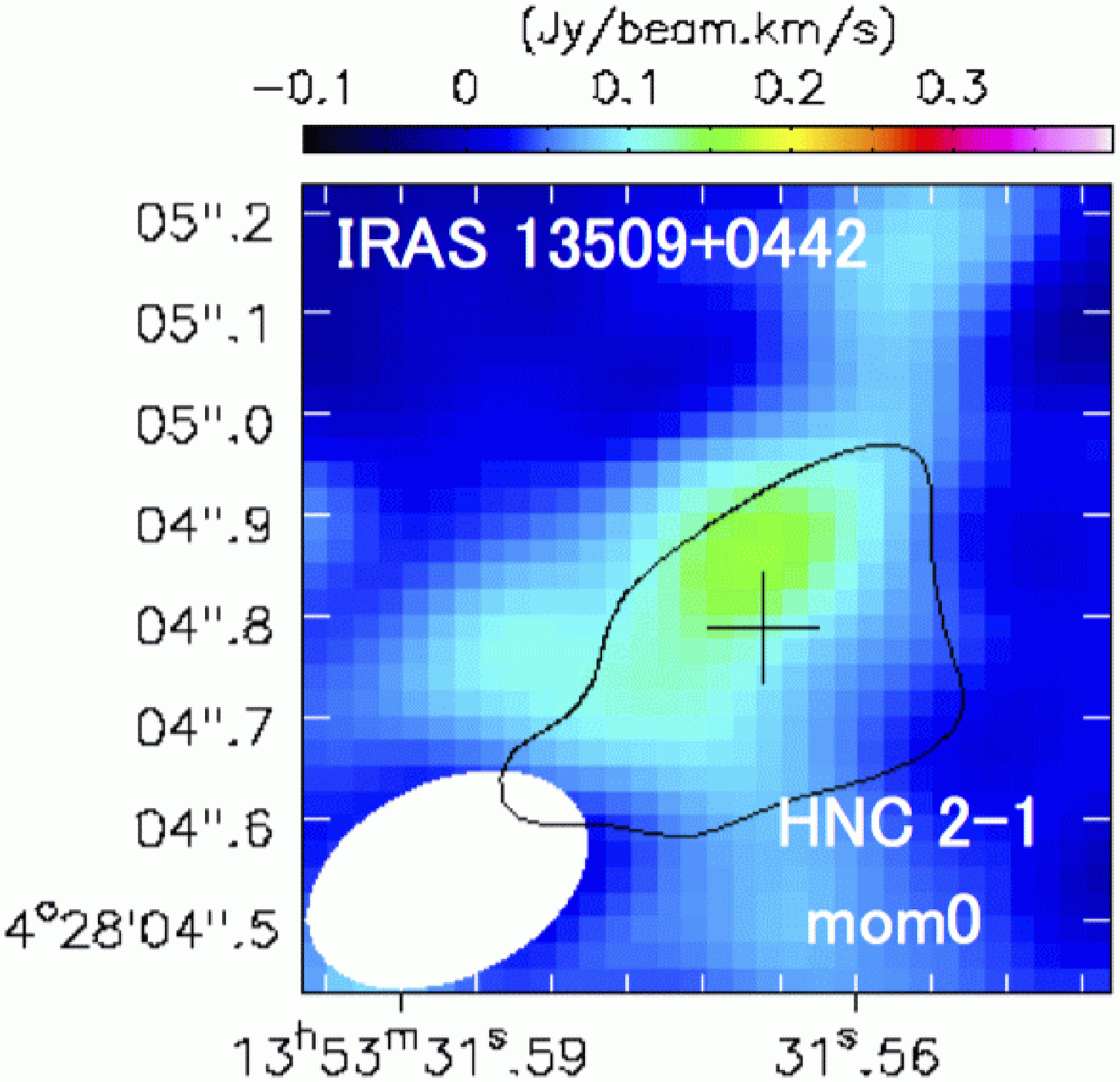} 
\includegraphics[angle=0,scale=.195]{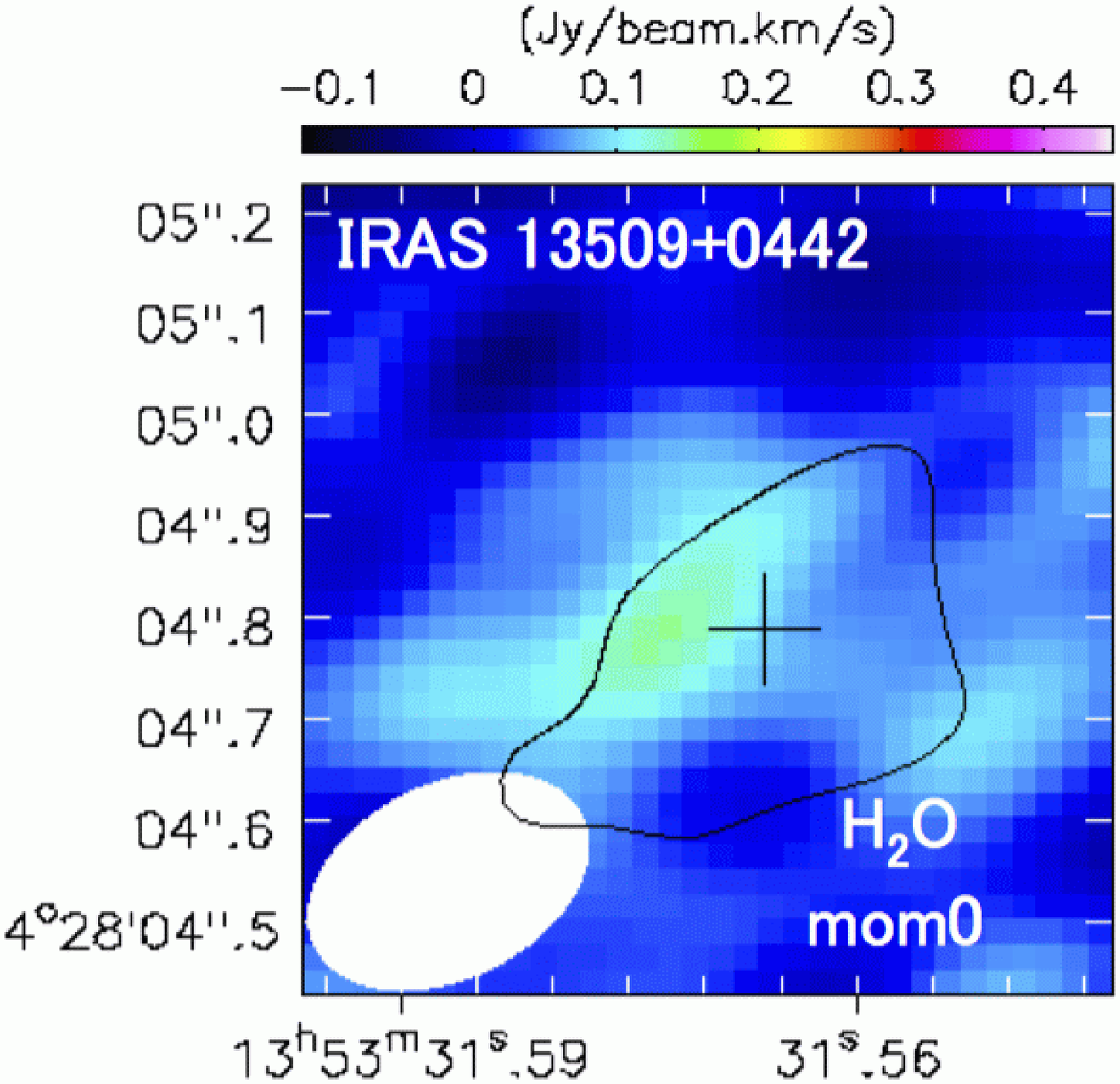} \\
\includegraphics[angle=0,scale=.195]{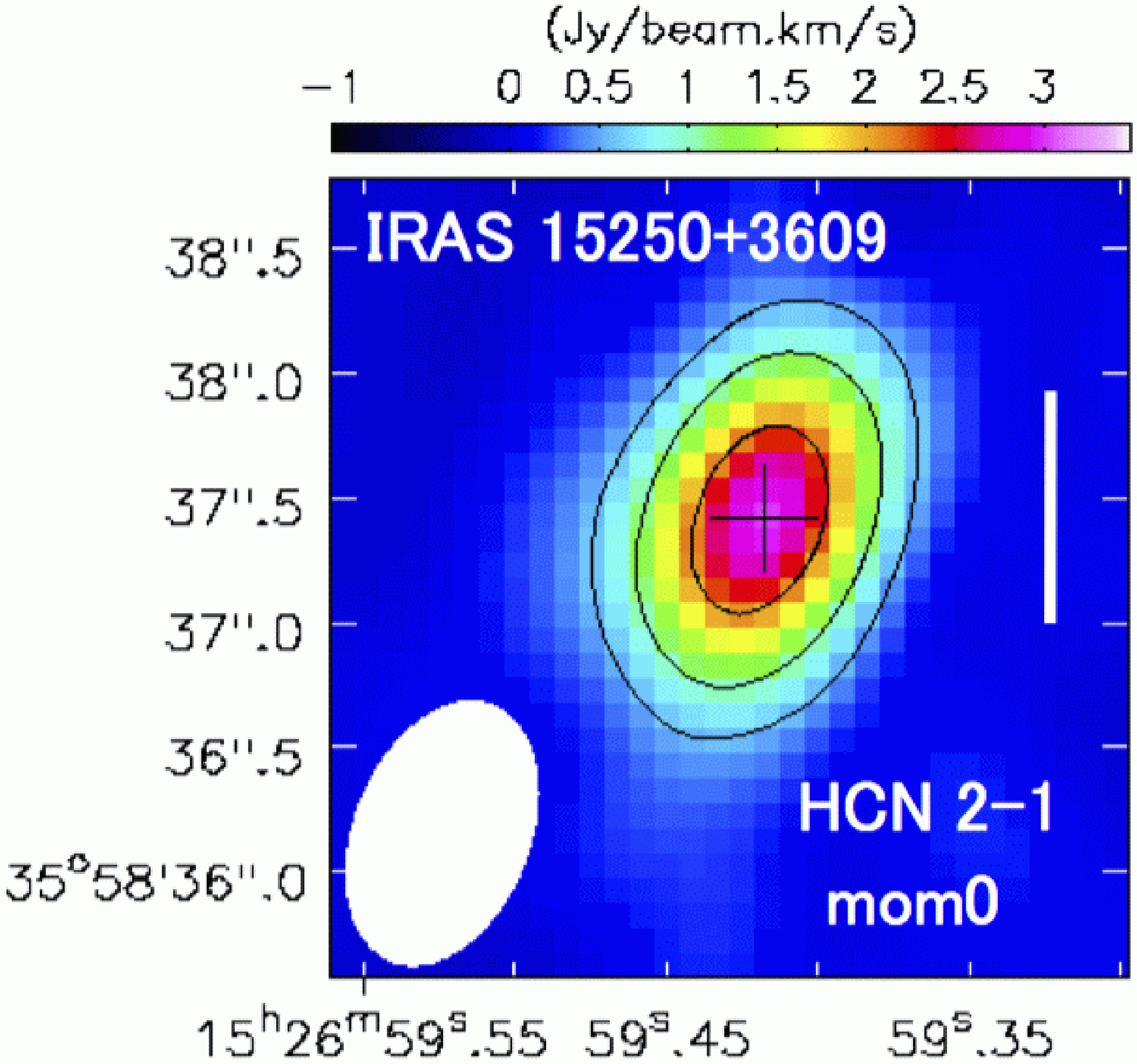} 
\includegraphics[angle=0,scale=.195]{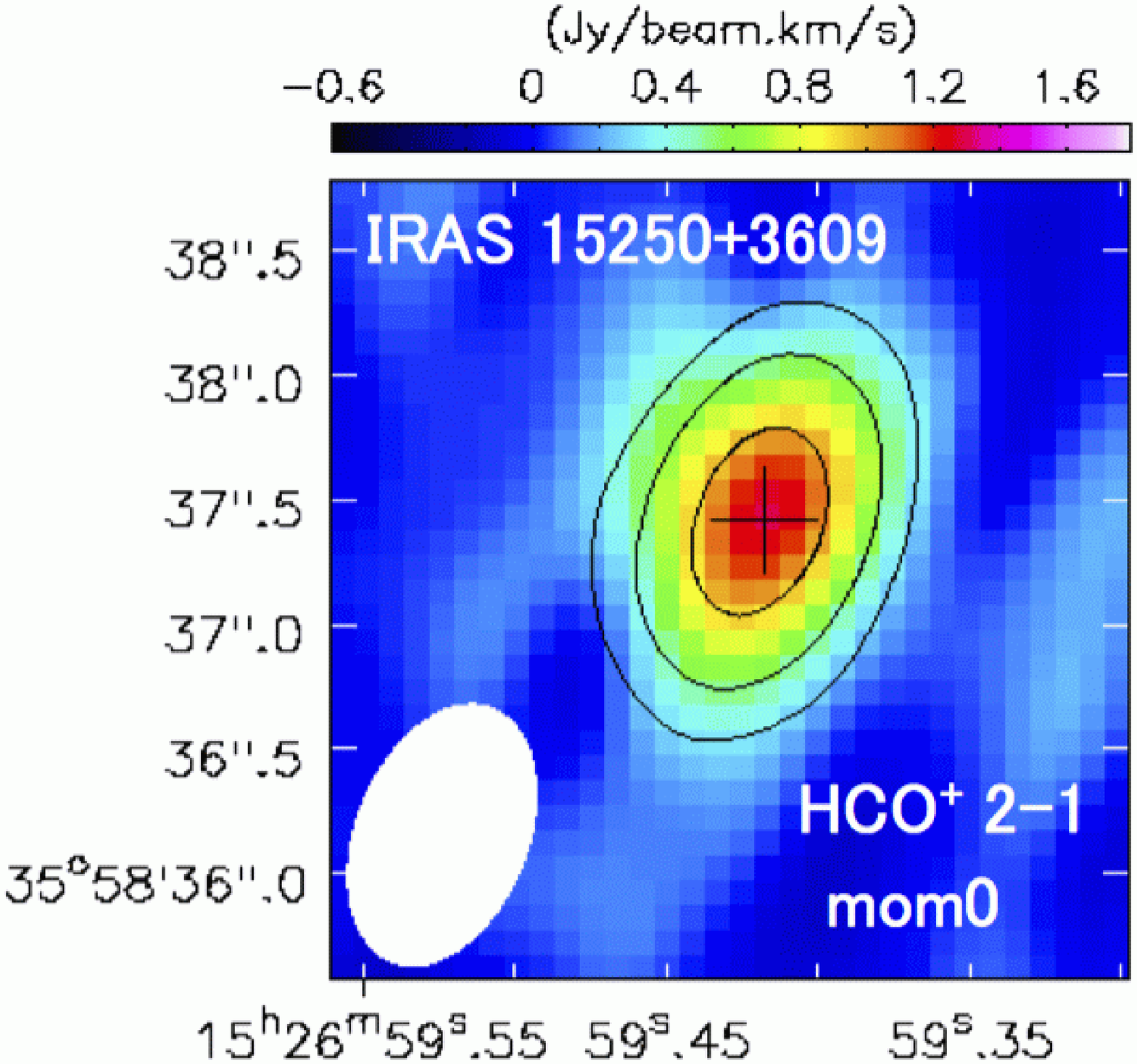} 
\includegraphics[angle=0,scale=.195]{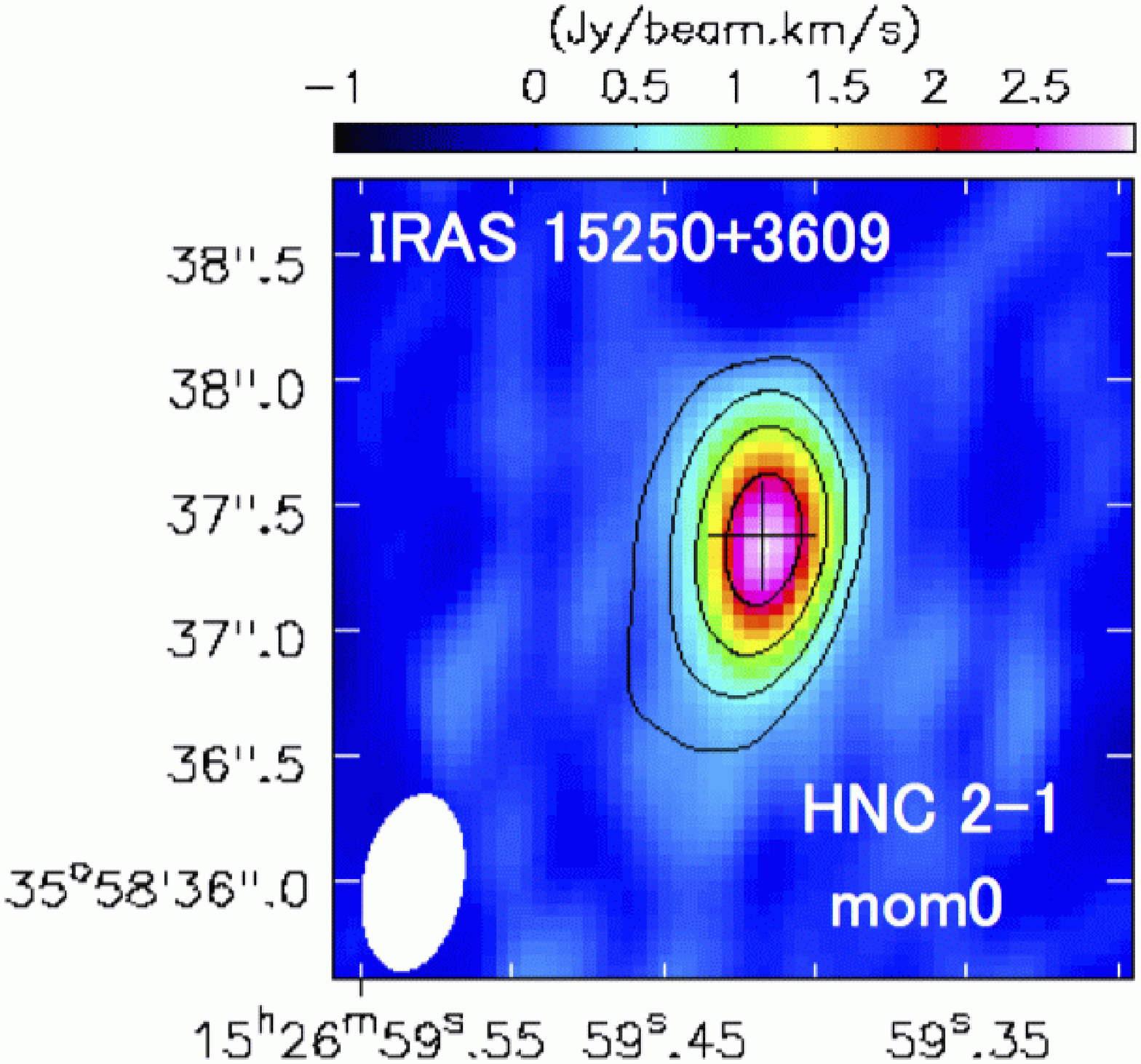} 
\includegraphics[angle=0,scale=.195]{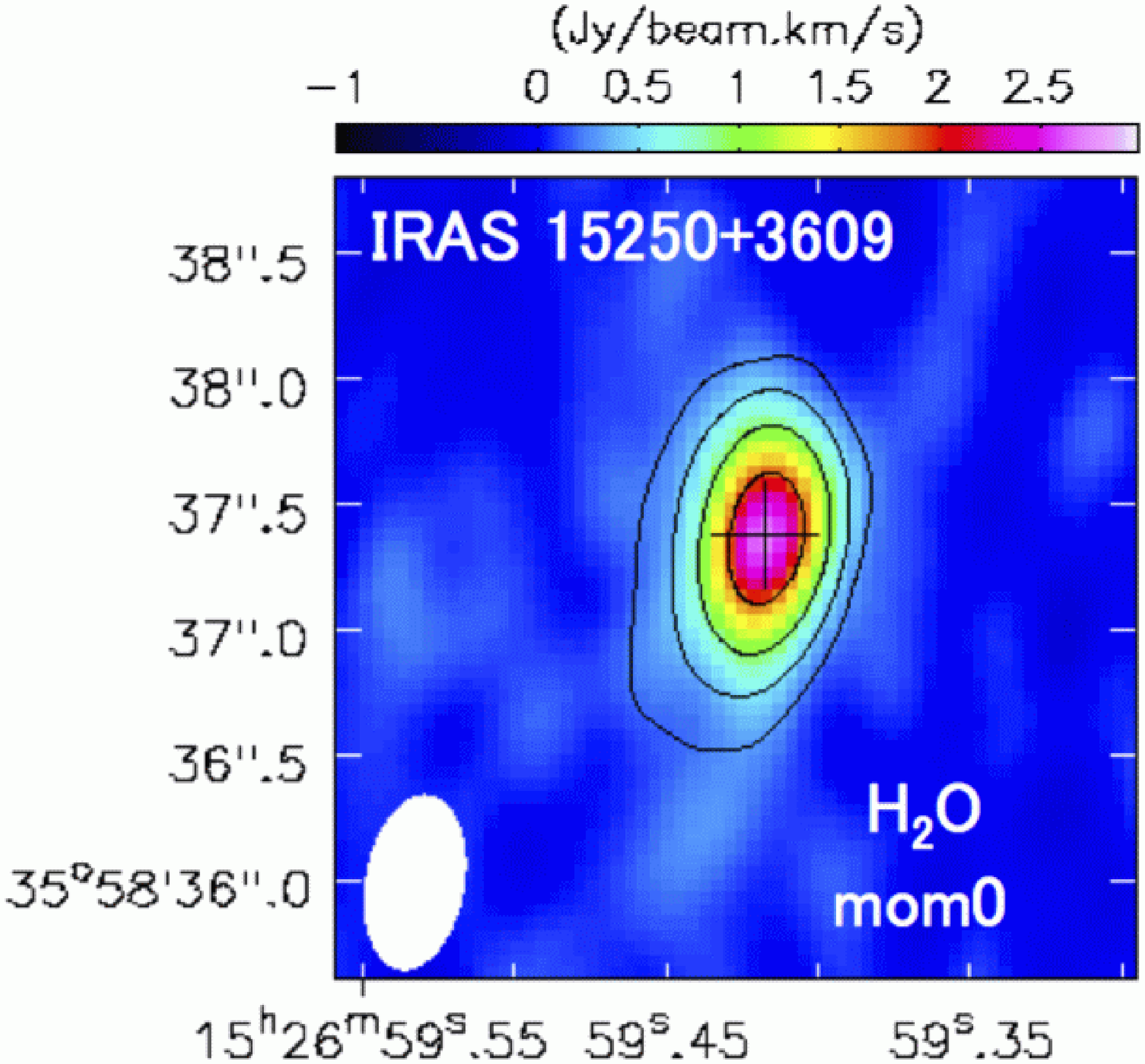} \\
\includegraphics[angle=0,scale=.195]{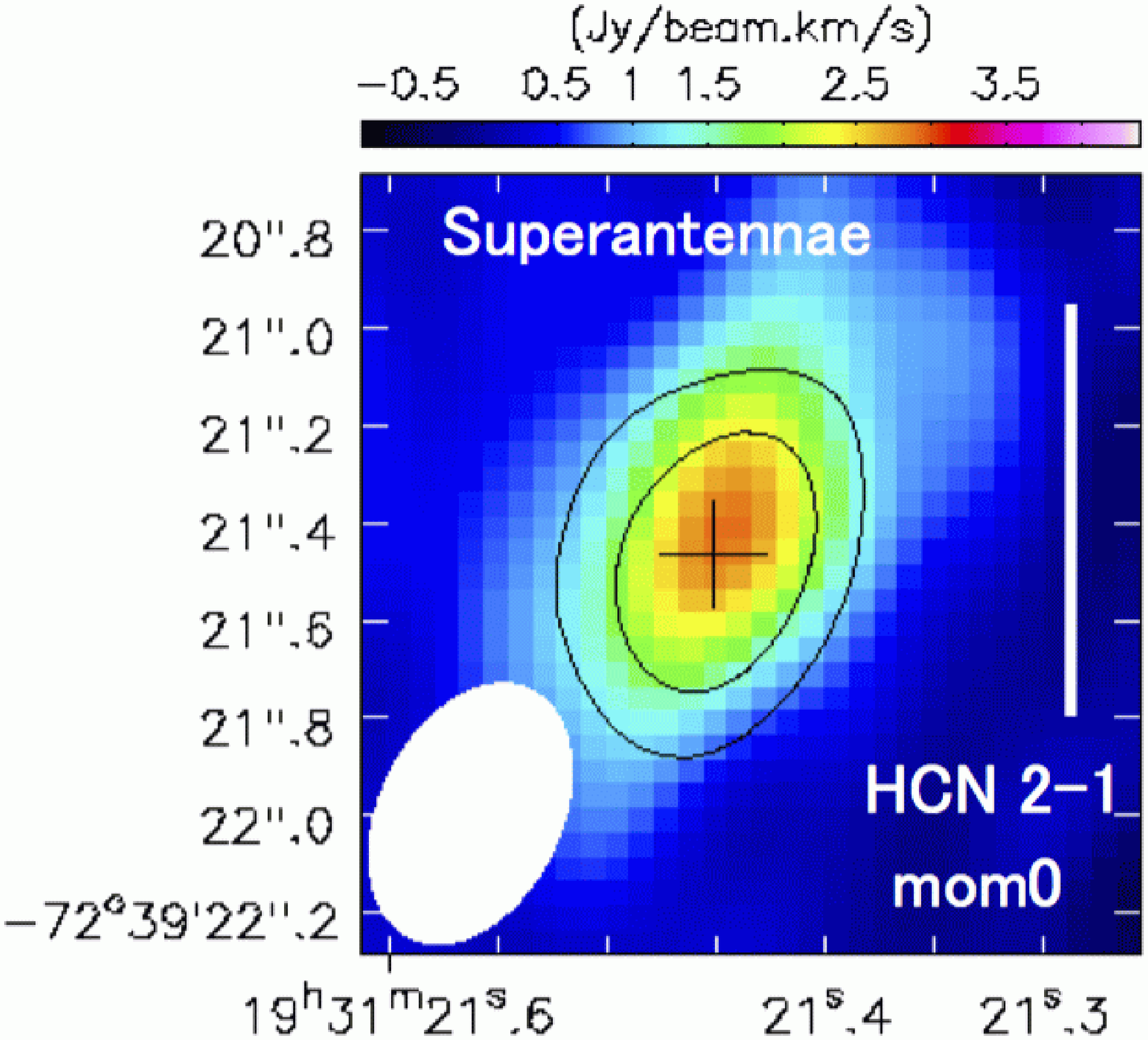} 
\includegraphics[angle=0,scale=.195]{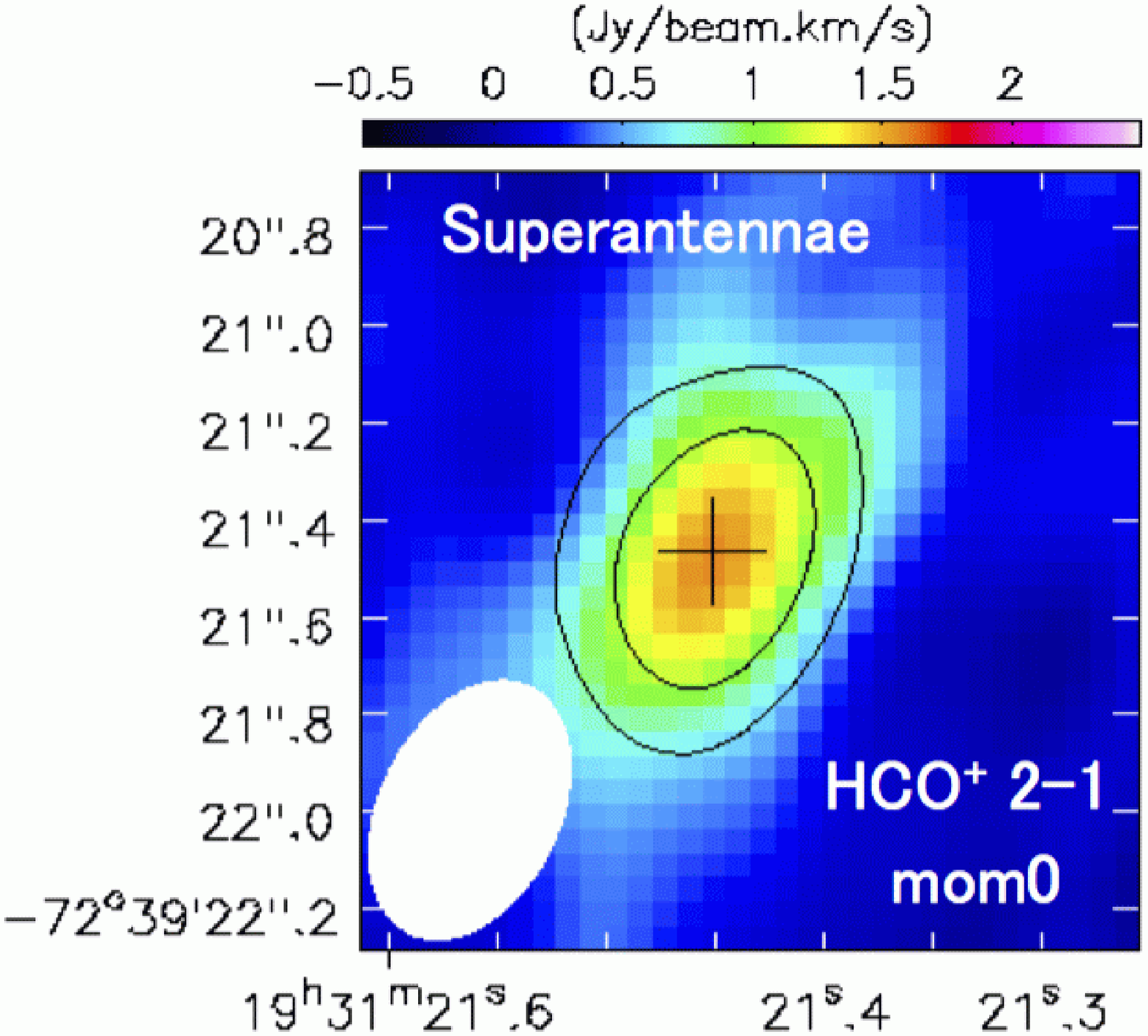} 
\includegraphics[angle=0,scale=.195]{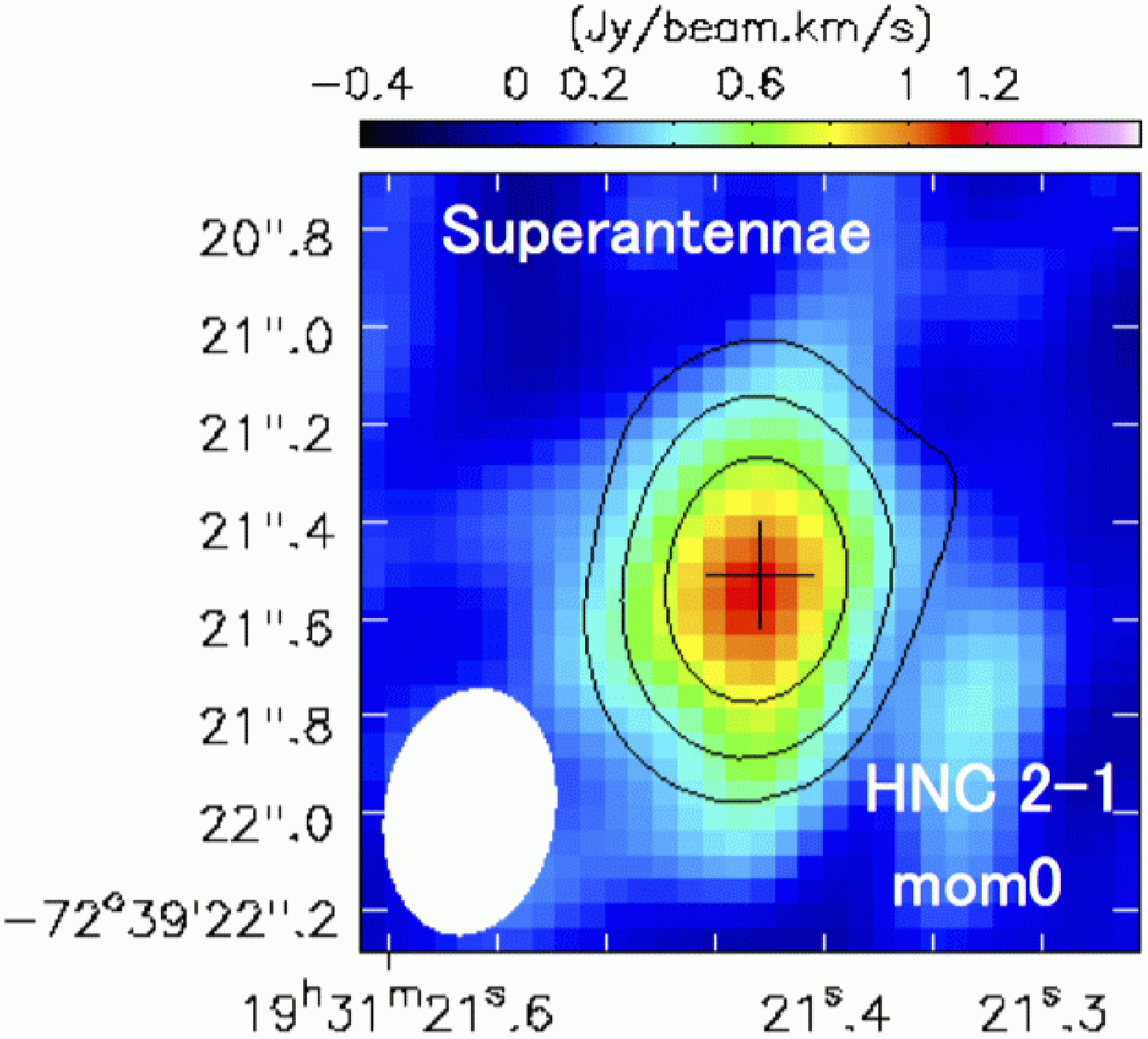} 
\includegraphics[angle=0,scale=.195]{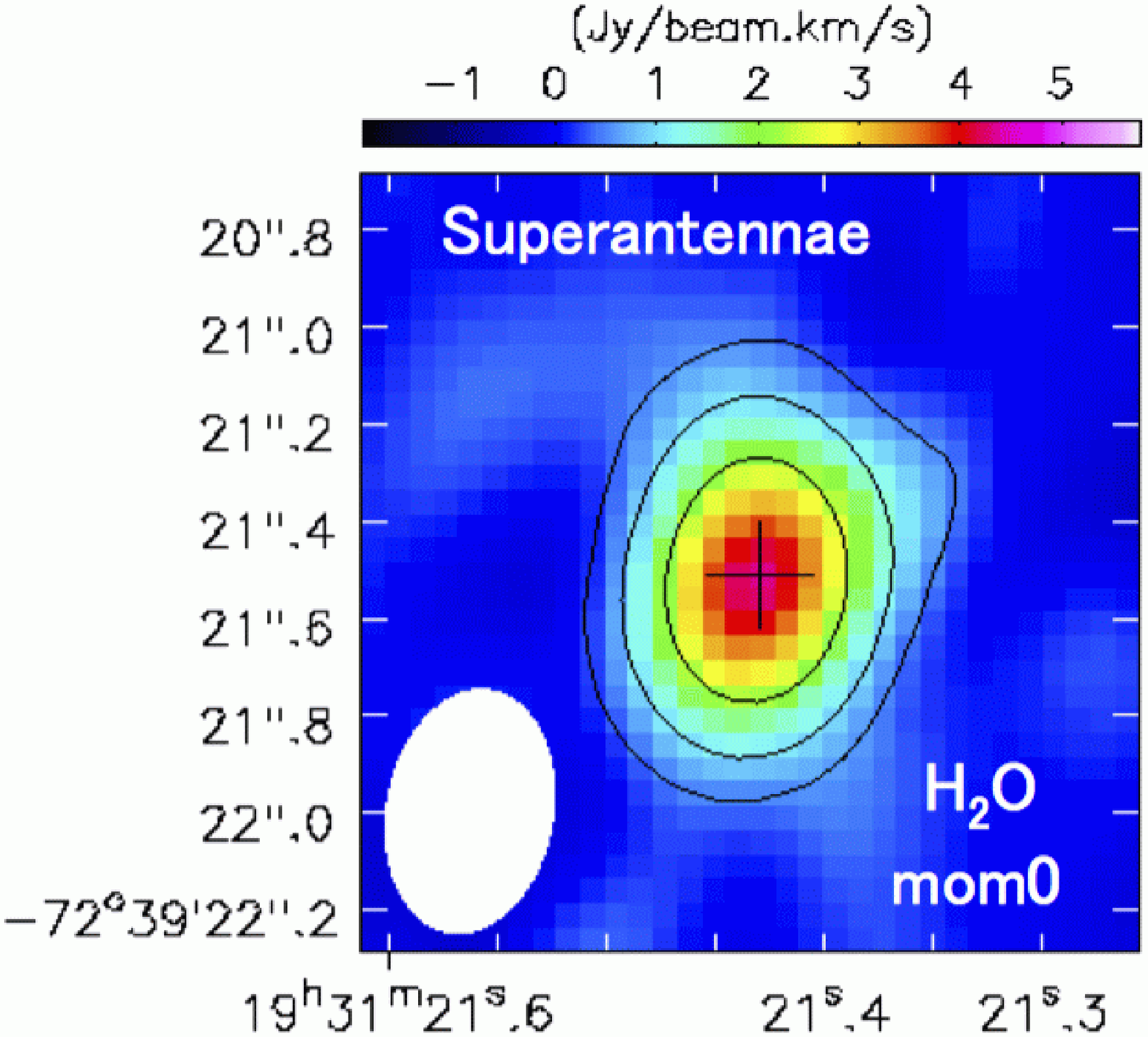} \\
\includegraphics[angle=0,scale=.195]{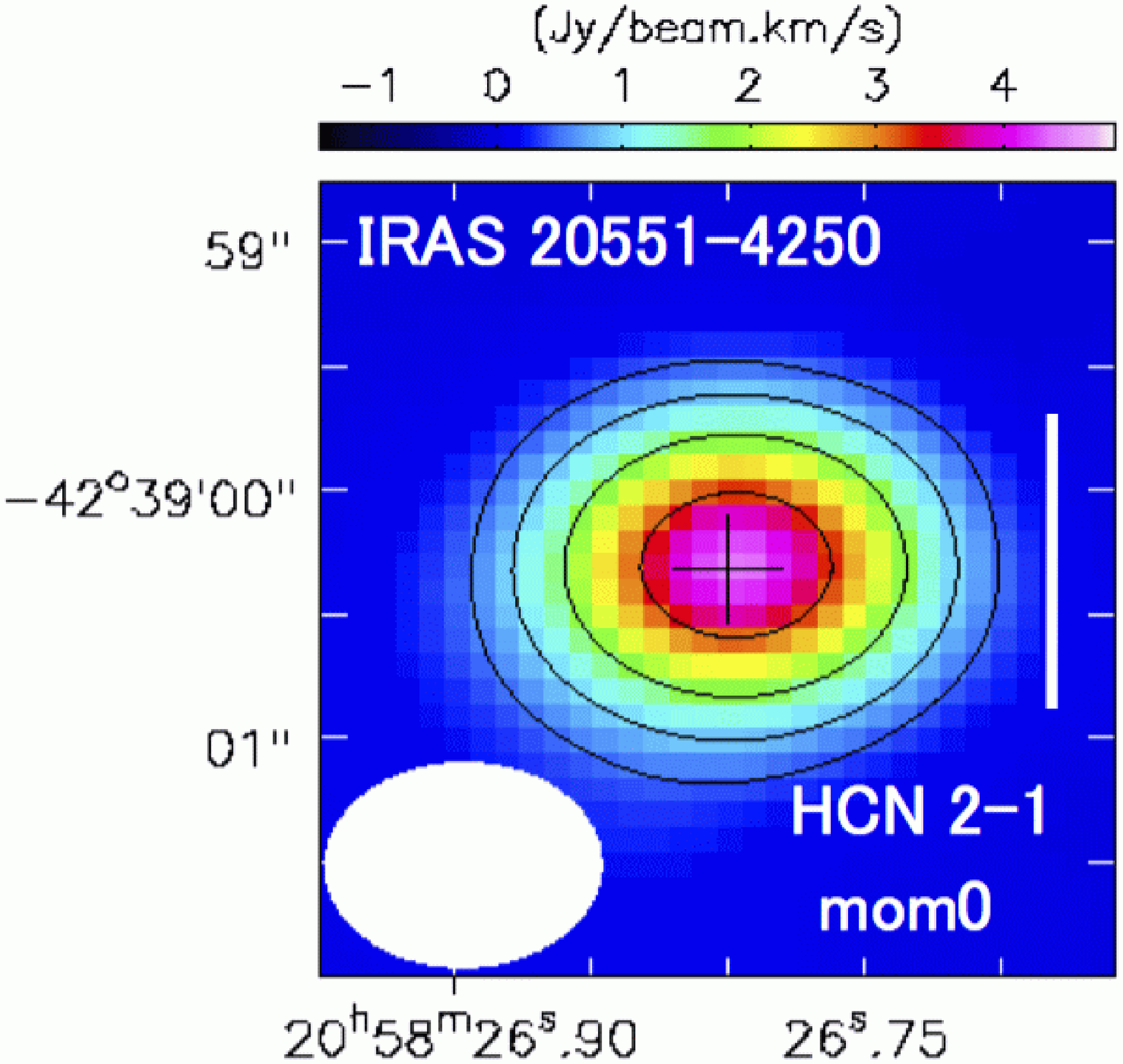} 
\includegraphics[angle=0,scale=.195]{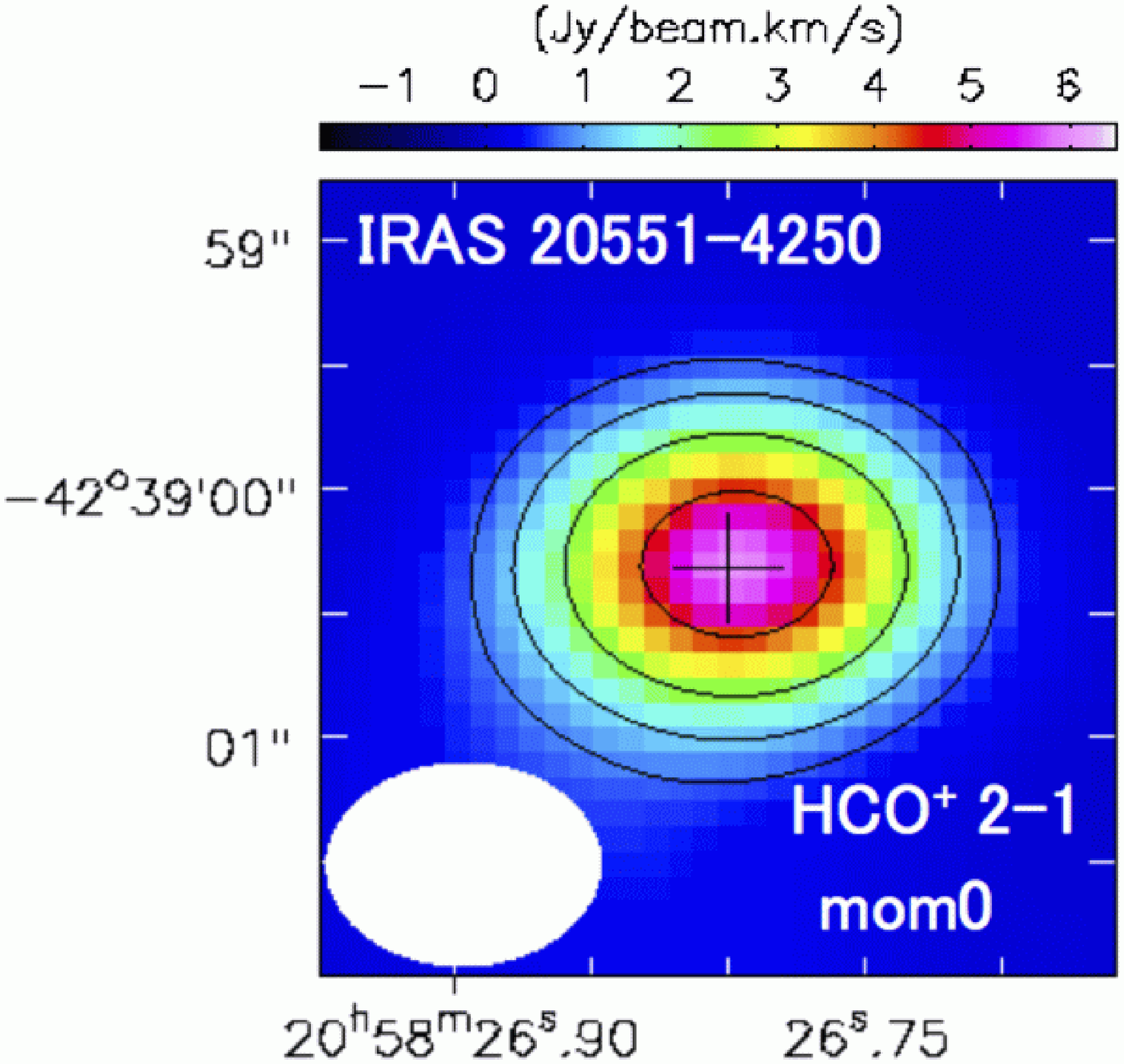} 
\includegraphics[angle=0,scale=.195]{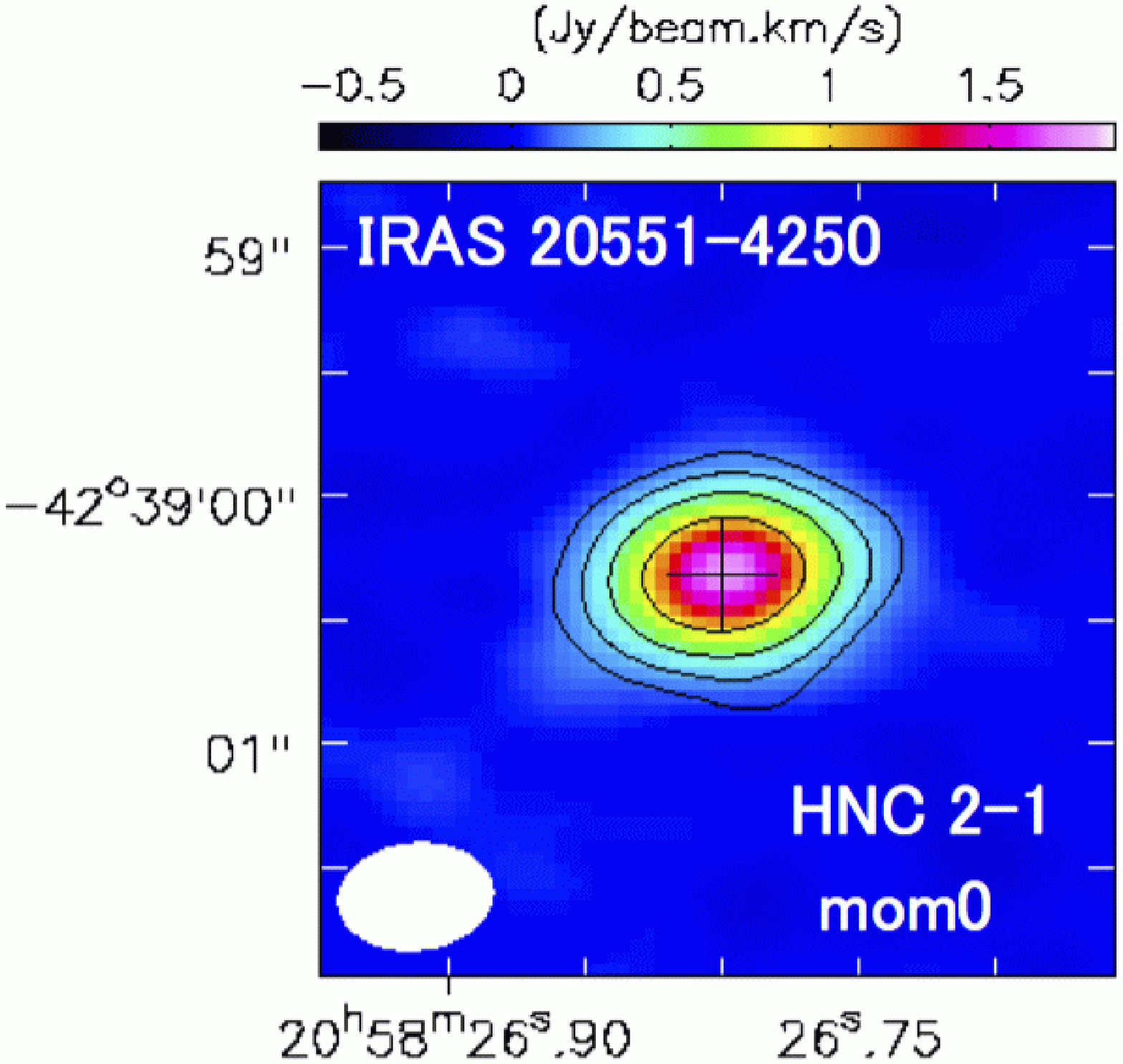} 
\includegraphics[angle=0,scale=.195]{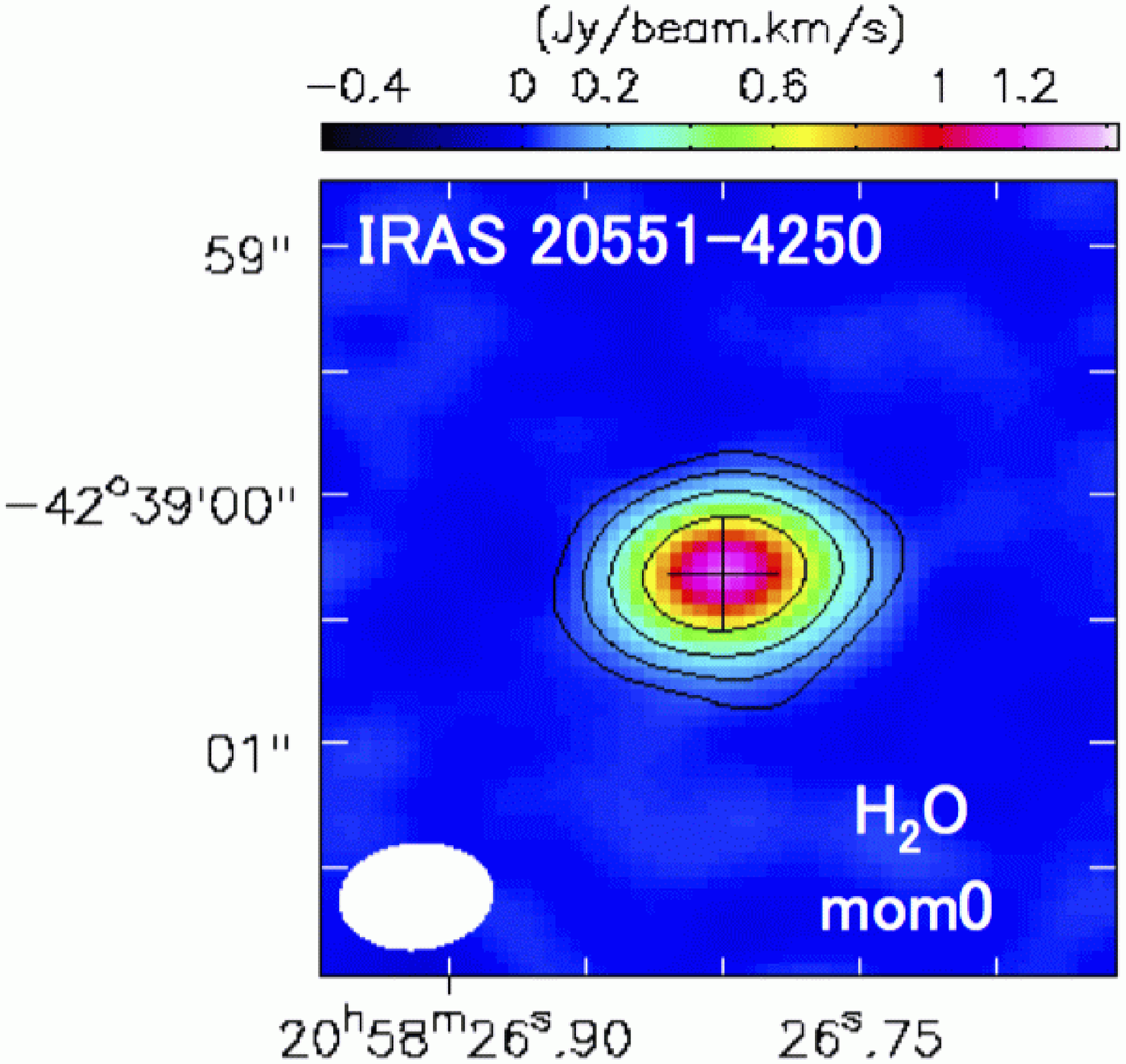} \\
\includegraphics[angle=0,scale=.195]{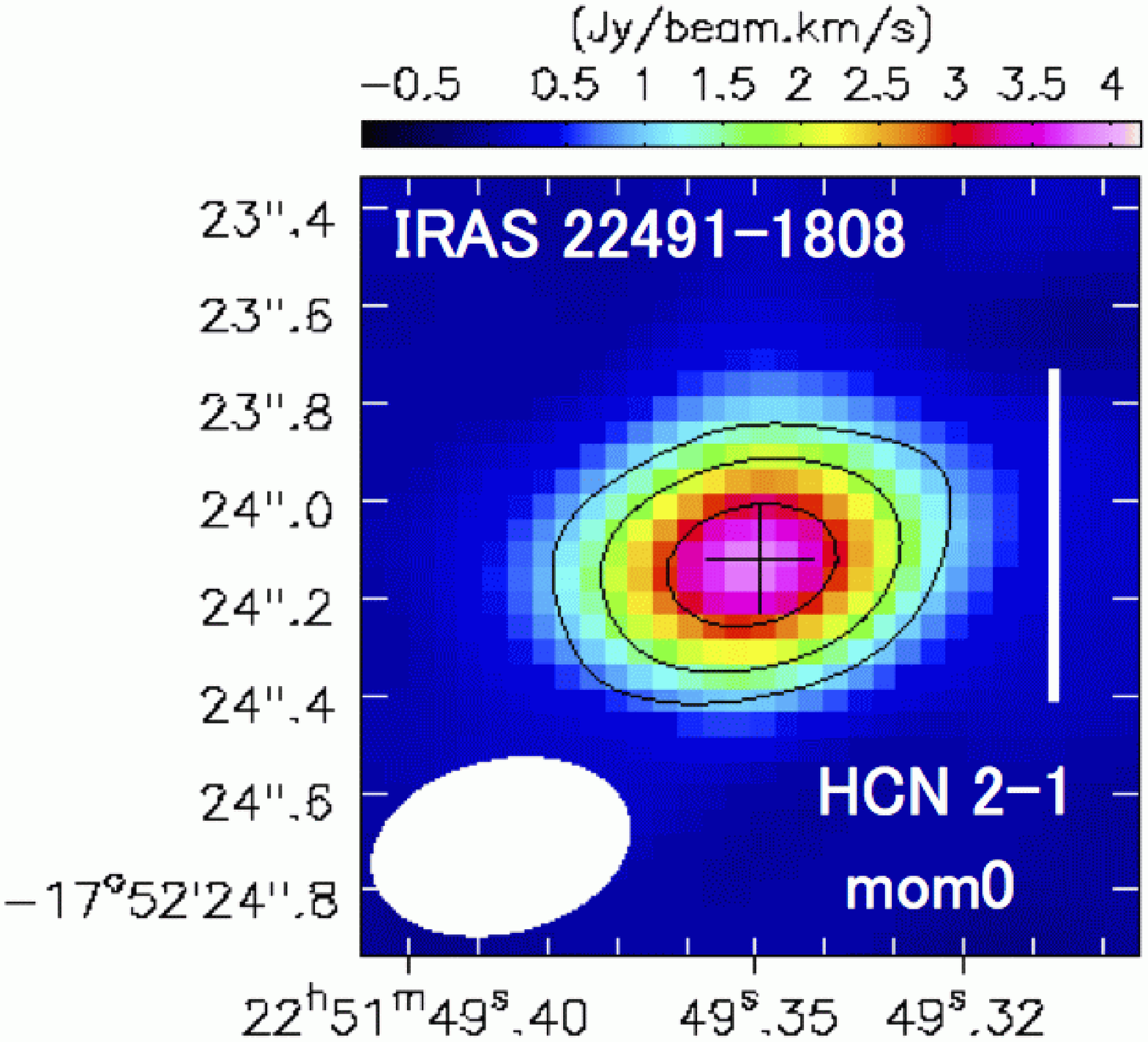} 
\includegraphics[angle=0,scale=.195]{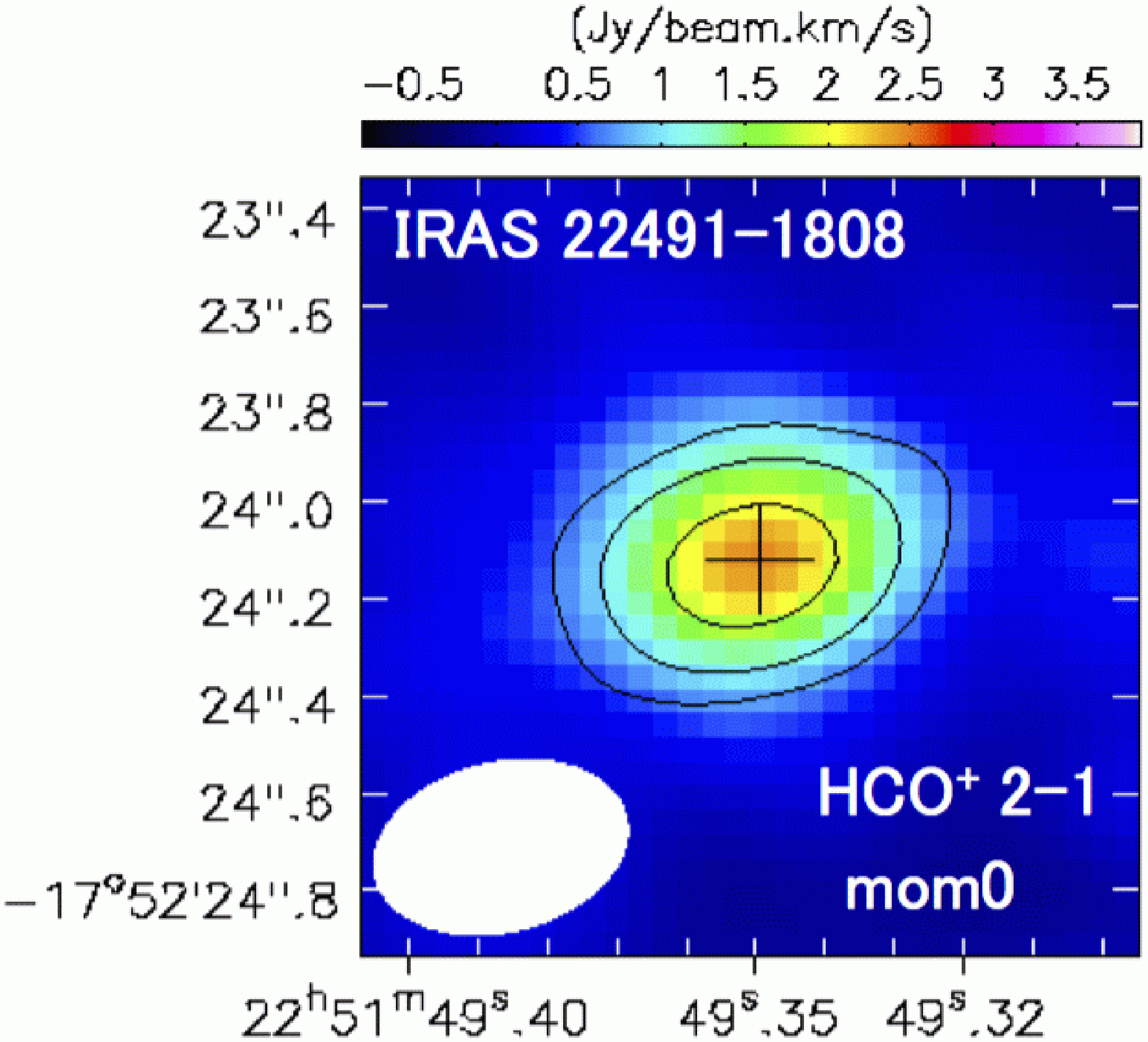} 
\includegraphics[angle=0,scale=.195]{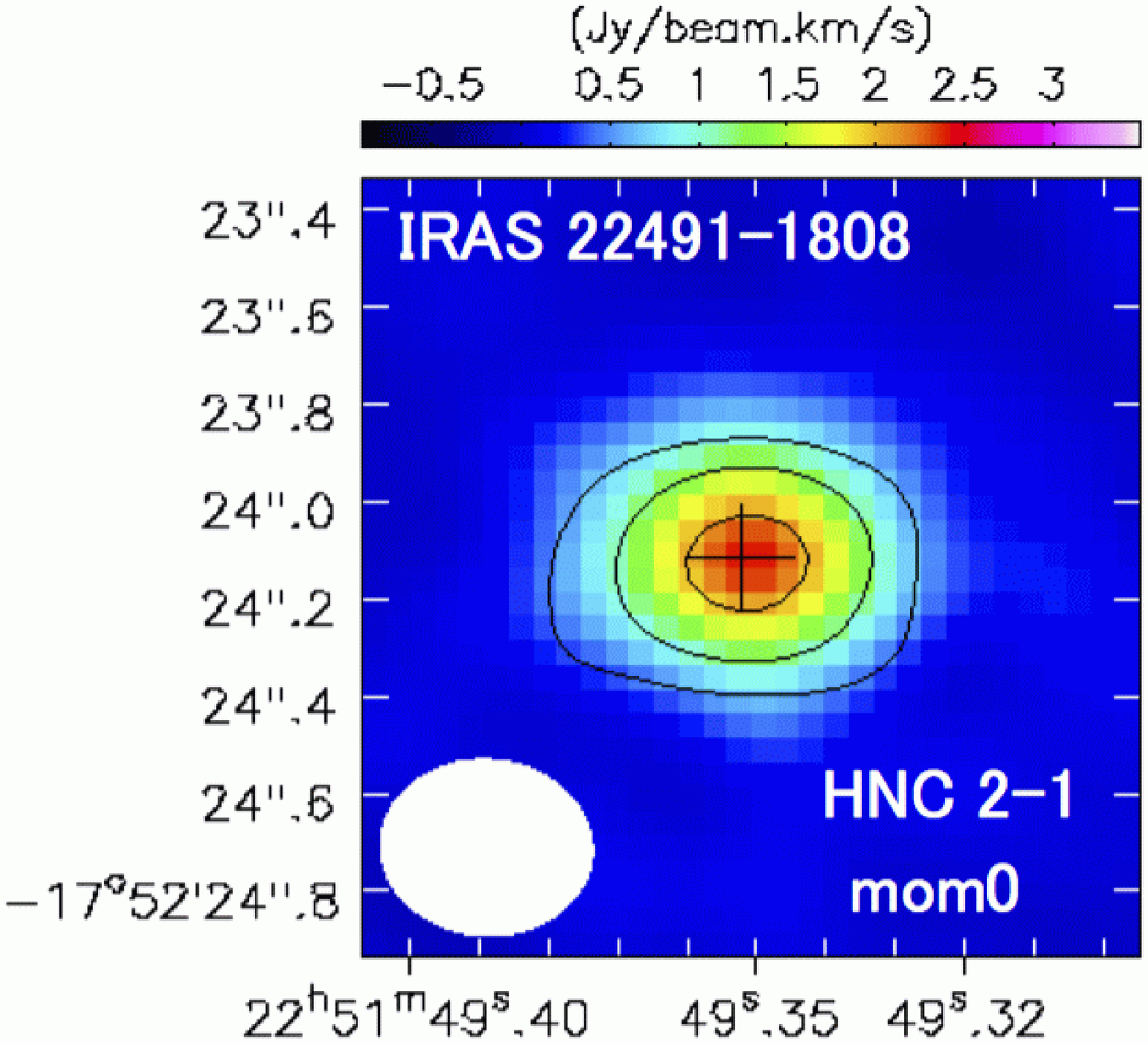} 
\includegraphics[angle=0,scale=.195]{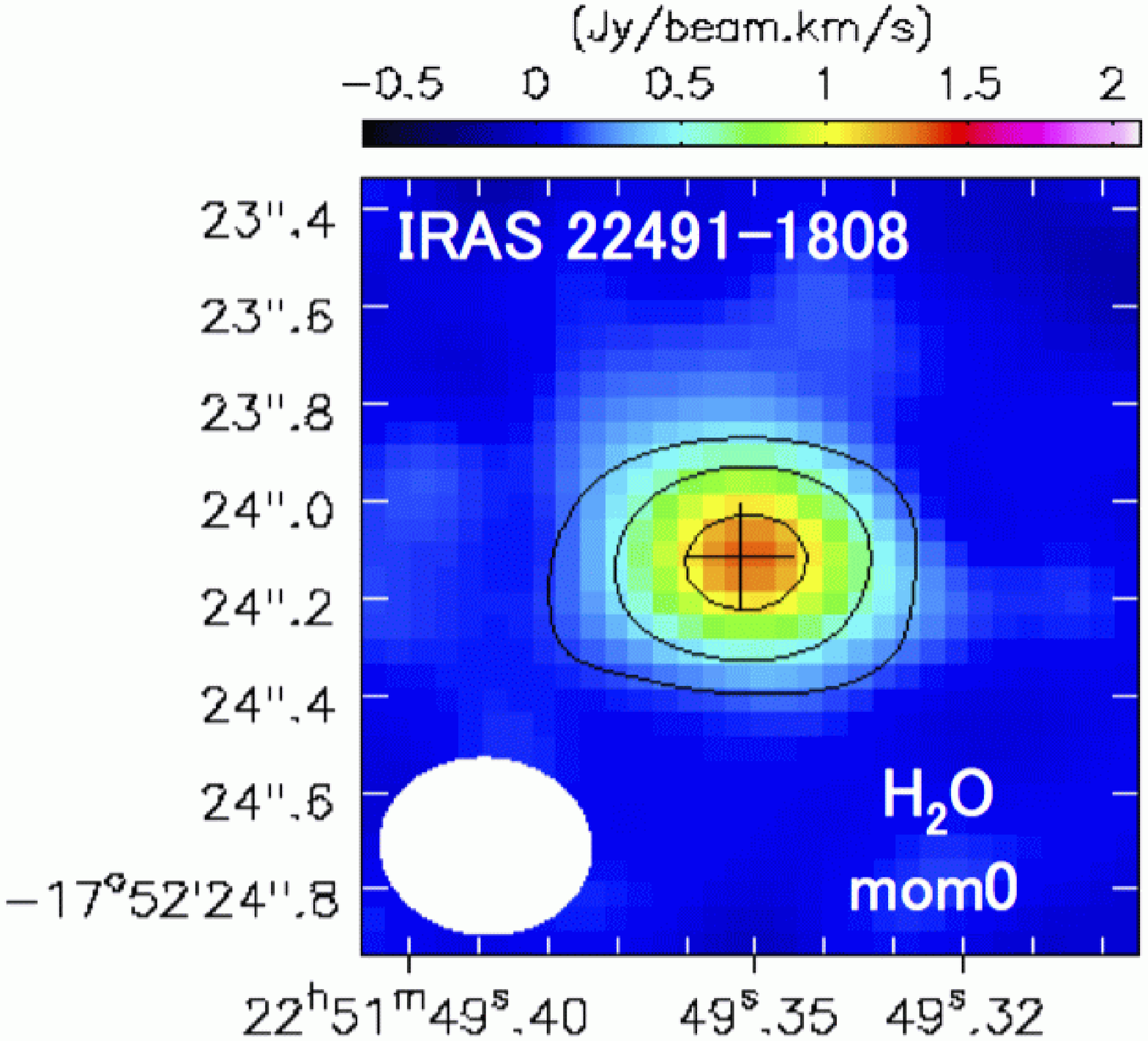} \\
\end{center}
\vspace{0.3cm}
\caption{Integrated intensity (moment 0) map of HCN J=2--1 {\it (Left)}, 
HCO$^{+}$ J=2--1 {\it (Second left)}, 
HNC J=2--1 {\it (Second right)}, and 183 GHz H$_{2}$O {\it (Right)} 
emission line.
Continuum emission that is simultaneously obtained with individual lines 
is shown as contours.
The contours start from 4$\sigma$ and increase by a factor of 2 
(i.e., 8$\sigma$, 16$\sigma$, and 32$\sigma$) for all sources except NGC 1614, 
for which contours are 4$\sigma$, 6$\sigma$, 8$\sigma$, and 10$\sigma$.
Continuum peak position is shown as a cross.
The length of the vertical white solid bar at the right side of 
HCN J=2--1 data {\it (Left)} corresponds to 1 kpc.
Beam size for each moment 0 map is shown as a filled circle 
in the lower-left region.
The beam size of HCN and HCO$^{+}$ (HNC and H$_{2}$O) is
comparable to that of continuum-J21a (continuum-J21b) data (Table 3,
column 6).
}
\end{figure*}

\begin{deluxetable}{llrlcl}
\tabletypesize{\scriptsize}
\tablecaption{Continuum Emission in the Original Beam Data \label{tbl-3}}
\tablewidth{0pt}
\tablehead{
\colhead{Object} & \colhead{Frequency} & \colhead{Flux} & 
\colhead{Peak Coordinate} & \colhead{rms} & \colhead{Synthesized beam} \\
\colhead{} & \colhead{[GHz]} & \colhead{[mJy/beam]} & 
\colhead{(RA,DEC)ICRS} & \colhead{[mJy/beam]} & 
\colhead{[$''$ $\times$ $''$] ($^{\circ}$)} \\  
\colhead{(1)} & \colhead{(2)} & \colhead{(3)}  & \colhead{(4)}  &
\colhead{(5)} & \colhead{(6)} 
}
\startdata 
NGC 1614 W & J21a (173.2--176.9, 185.4--189.1) & 0.88 (10$\sigma$) & 
(04$^{h}$33$^{m}$59.98$^{s}$, $-$08$^{\circ}$34$'$45.1$''$) & 0.088 & 0.55$\times$0.37 ($-$83$^{\circ}$)\\
& J21b (177.4--181.2, 189.6--193.3) & 0.76 (8.1$\sigma$) & 
(04$^{h}$33$^{m}$59.99$^{s}$, $-$08$^{\circ}$34$'$45.3$''$) & 0.093 & 0.58$\times$0.33 ($-$79$^{\circ}$) \\  
NGC 1614 N & J21a (173.2--176.9, 185.4--189.1) & 0.96 (11$\sigma$) & 
(04$^{h}$34$^{m}$00.01$^{s}$, $-$08$^{\circ}$34$'$44.6$''$) & 0.088 & 0.55$\times$0.37 ($-$83$^{\circ}$)\\
& J21b (177.4--181.2, 189.6--193.3) & 0.86 (9.3$\sigma$) & 
(04$^{h}$34$^{m}$00.01$^{s}$, $-$08$^{\circ}$34$'$44.6$''$) & 0.093 & 0.58$\times$0.33 ($-$79$^{\circ}$) \\
NGC 1614 S & J21a (173.2--176.9, 185.4--189.1) & 0.80 (9.1$\sigma$) & 
(04$^{h}$34$^{m}$00.03$^{s}$, $-$08$^{\circ}$34$'$45.7$''$) & 0.088 & 0.55$\times$0.37 ($-$83$^{\circ}$)\\
& J21b (177.4--181.2, 189.6--193.3) & 0.73 (7.8$\sigma$) & 
(04$^{h}$34$^{m}$00.03$^{s}$, $-$08$^{\circ}$34$'$45.7$''$) & 0.093 & 0.58$\times$0.33 ($-$79$^{\circ}$) \\  
NGC 1614 E & J21a (173.2--176.9, 185.4--189.1) & 0.75 (8.5$\sigma$) & 
(04$^{h}$34$^{m}$00.07$^{s}$, $-$08$^{\circ}$34$'$45.2$''$) & 0.088 & 0.55$\times$0.37 ($-$83$^{\circ}$)\\
& J21b (177.4--181.2, 189.6--193.3) & 0.73 (7.9$\sigma$) & 
(04$^{h}$34$^{m}$00.07$^{s}$, $-$08$^{\circ}$34$'$45.2$''$) & 0.093 & 0.58$\times$0.33 ($-$79$^{\circ}$) \\  
IRAS 06035$-$7102 & J21a (163.1--166.7, 175.0--178.7) & 1.0 (9.8$\sigma$) & 
(06$^{h}$02$^{m}$53.94$^{s}$, $-$71$^{\circ}$03$'$10.2$''$) & 0.10 & 1.1$\times$0.80 ($-$50$^{\circ}$)\\
& J21b (166.9--170.7, 178.9--182.6) & 0.83 (13$\sigma$) & 
(06$^{h}$02$^{m}$53.94$^{s}$, $-$71$^{\circ}$03$'$10.2$''$) & 0.064 & 0.48$\times$0.34 (6$^{\circ}$) \\
IRAS 08572$+$3915 & J21a (166.2--169.9) & 1.4 (18$\sigma$) & 
(09$^{h}$00$^{m}$25.37$^{s}$, $+$39$^{\circ}$03$'$54.1$''$) & 0.080 & 0.71$\times$0.35 ($-$1$^{\circ}$) \\
& J21b (170.3--174.1) & 1.6 (22$\sigma$) & 
(09$^{h}$00$^{m}$25.37$^{s}$, $+$39$^{\circ}$03$'$54.1$''$) & 0.073 & 0.68$\times$0.33 (14$^{\circ}$) \\
IRAS 12112$+$0305 NE & J21a (163.9--167.5, 175.7--179.4) & 3.4 (33$\sigma$) & 
(12$^{h}$13$^{m}$46.06$^{s}$, $+$02$^{\circ}$48$'$41.5$''$) & 0.10 & 0.87$\times$0.68 (29$^{\circ}$)\\
& J21b (167.9--171.5) & 3.6 (18$\sigma$) & 
(12$^{h}$13$^{m}$46.06$^{s}$, $+$02$^{\circ}$48$'$41.4$''$) & 0.20 & 0.83$\times$0.57 ($-$61$^{\circ}$) \\
IRAS 12112$+$0305 SW & J21a (163.9--167.5, 175.7--179.4) & 0.39 (3.9$\sigma$) & 
(12$^{h}$13$^{m}$45.95$^{s}$, $+$02$^{\circ}$48$'$39.1$''$) & 0.10 & 0.87$\times$0.68 (29$^{\circ}$)\\
& J21b (167.9--171.7) & 0.39 (1.9$\sigma$) & 
(12$^{h}$13$^{m}$45.93$^{s}$, $+$02$^{\circ}$48$'$39.0$''$) & 0.20 & 0.83$\times$0.57 ($-$61$^{\circ}$) \\
IRAS 12127$-$1412 \tablenotemark{A} & J21a (142.9--146.4, 155.1--158.7) 
\tablenotemark{A} & 0.84 (34$\sigma$) & 
(12$^{h}$15$^{m}$19.13$^{s}$, $-$14$^{\circ}$29$'$41.8$''$) & 0.024 & 0.29$\times$0.19 ($-$81$^{\circ}$) \\
& J21b (159.0--162.7) \tablenotemark{A} & 0.85 (42$\sigma$) & 
(12$^{h}$15$^{m}$19.13$^{s}$, $-$14$^{\circ}$29$'$41.8$''$) & 0.020 & 0.24$\times$0.16 ($-$59$^{\circ}$) \\ 
IRAS 13509$+$0442 \tablenotemark{A} & J21a (142.5--146.0,
154.7--158.2) \tablenotemark{A} & 0.27 (10$\sigma$) &
(13$^{h}$53$^{m}$31.57$^{s}$, $+$04$^{\circ}$28$'$04.8$''$) & 0.027 & 0.27$\times$0.19 ($-$68$^{\circ}$) \\
& J21b (158.5--162.2) \tablenotemark{A} & 0.21 (7.2$\sigma$) & 
(13$^{h}$53$^{m}$31.57$^{s}$, $+$04$^{\circ}$28$'$04.8$''$) & 0.029 & 0.29$\times$0.18 ($-$63$^{\circ}$) \\
IRAS 15250$+$3609 & J21a (166.7--170.3) & 5.1 (22$\sigma$) &
(15$^{h}$26$^{m}$59.42$^{s}$, $+$35$^{\circ}$58$'$37.4$''$) & 0.23 & 1.1$\times$0.68 ($-$22$^{\circ}$) \\
& J21b (170.8--174.6) & 5.6 (47$\sigma$) & 
(15$^{h}$26$^{m}$59.42$^{s}$, $+$35$^{\circ}$38$'$37.4$''$) 
& 0.12 & 0.70$\times$0.39 ($-$10$^{\circ}$) \\
Superantennae & J21a (165.6--169.3) & 3.1 (15$\sigma$) &
(19$^{h}$31$^{m}$21.45$^{s}$, $-$72$^{\circ}$39$'$21.5$''$) & 0.21 & 0.54$\times$0.36 ($-$26$^{\circ}$) \\
& J21b (169.7--173.5) & 3.4 (30$\sigma$) & (19$^{h}$31$^{m}$21.43$^{s}$, $-$72$^{\circ}$39$'$21.5$''$) 
& 0.11 & 0.49$\times$0.34 ($-$10$^{\circ}$) \\
IRAS 20551$-$4250 & J21a (168.6--172.3) & 2.5 (43$\sigma$) & 
(20$^{h}$58$^{m}$26.80$^{s}$, $-$42$^{\circ}$39$'$00.3$''$) & 0.058 
& 1.1$\times$0.82 ($-$89$^{\circ}$) \\
& J21b (172.8--176.5, 185.0--188.7) & 2.3 (63$\sigma$) & 
(20$^{h}$58$^{m}$26.80$^{s}$, $-$42$^{\circ}$39$'$00.3$''$) & 0.036 &  
0.61$\times$0.41 ($-$84$^{\circ}$)  \\
IRAS 22491$-$1808 & J21a (163.2--166.8, 175.0--178.6) & 2.0 (23$\sigma$) &
(22$^{h}$51$^{m}$49.35$^{s}$, $-$17$^{\circ}$52$'$24.1$''$) & 0.088 & 0.50$\times$0.33 ($-$76$^{\circ}$) \\
& J21b (167.2--171.0) & 2.2 (20$\sigma$) 
& (22$^{h}$51$^{m}$49.35$^{s}$, $-$17$^{\circ}$52$'$24.1$''$) & 0.11 
& 0.41$\times$0.34 ($-$89$^{\circ}$) \\
\enddata

\tablenotetext{A}{Band 4 observation because the targeted molecular
lines are redshifted into Band 4 (125--163 GHz).}

\tablecomments{
Col.(1): Object name.
Col.(2): Frequency range in GHz used for continuum extraction.
Frequencies of obvious emission lines were removed.
When only one frequency range is shown, it means that data in another 
sideband were noisy owing to poor Earth's atmospheric transmission, 
and were not used. 
Col.(3): Flux density (in mJy beam$^{-1}$) at the emission peak.
Value at the highest flux pixel (0$\farcs$025--0$\farcs$1 pixel$^{-1}$)
is extracted. 
The detection significance relative to the root mean square (rms) noise 
is shown in parentheses. 
Col.(4): Coordinate of the continuum emission peak in ICRS.
Col.(5): The rms noise level (1$\sigma$) (in mJy beam$^{-1}$), derived
from the standard deviation of sky signals in each continuum map.
Col.(6): Synthesized beam (in arcsec $\times$ arcsec) and position angle
(in degrees). 
The position angle is 0$^{\circ}$ along the north--south direction
and increases in the counterclockwise direction.
}

\end{deluxetable}

In the LIRG NGC 1614, there are four bright continuum emission 
peaks, denoted as NGC 1614 W, N, S, and E (Figure 1). 
In IRAS 12112$+$0305, continuum emission is significantly 
($>$3$\sigma$) 
detected not only in the north-eastern (NE) primary nucleus, but also 
in the south-western (SW) secondary nucleus (Figure 1).
Continuum emission properties of these multiple positions are 
tabulated in Table 3. 
For other ULIRGs, continuum emission is clearly detected only in the 
primary nucleus.
In the fields of IRAS 13509$+$0442 and the Superantennae, a bright continuum 
emitting source is serendipitously detected at $\sim$8$''$ north and 
$\sim$8$''$ south side of the primary ULIRG's nucleus, respectively. 
These continuum sources are unlikely to be physically related to 
each ULIRG, and thus their observed properties are described in Appendix A. 
In short, the former and latter sources are likely to be an infrared 
luminous dusty galaxy at $z >$ 1 and a blazar, respectively. 

Integrated intensity (moment 0) maps of the HCN J=2--1, HCO$^{+}$ J=2--1, 
HNC J=2--1 and 183 GHz H$_{2}$O lines are created by integrating
signals in channels with significant line detection, and the peak emission 
flux as well as detection significance is tabulated in Table 4. 
Figure 1 displays the moment 0 maps of these emission lines with 
significant ($\gtrsim$3$\sigma$) detection.  
The peak positions of these detected molecular emission lines spatially 
agree with the simultaneously obtained continuum data. 

\begin{deluxetable}{l|cccc}
\tabletypesize{\scriptsize}
\tablecaption{Peak Flux of Molecular Emission Line in Integrated Intensity 
(Moment 0) Map with Original Beam Size \label{tbl-4}} 
\tablewidth{0pt}
\tablehead{
\colhead{} & \multicolumn{4}{c}{Peak [Jy beam$^{-1}$ km s$^{-1}$]} \\ 
\colhead{Object} & \colhead{HCN J=2--1} &
\colhead{HCO$^{+}$ J=2--1} & \colhead{HNC J=2--1} & \colhead{183 GHz H$_{2}$O} \\
\colhead{(1)} & \colhead{(2)} & \colhead{(3)} & \colhead{(4)} & 
\colhead{(5)} 
}
\startdata 
NGC 1614 W & 0.40 (6.6$\sigma$) \tablenotemark{A} & 
0.79 (8.3$\sigma$) \tablenotemark{A} & $<$0.29 ($<$3$\sigma$) \tablenotemark{A} 
& $<$0.31 ($<$3$\sigma$) \tablenotemark{A} \\
NGC 1614 N & 0.49 (8.1$\sigma$) \tablenotemark{A} & 
0.84 (8.8$\sigma$) \tablenotemark{A} & $<$0.29 ($<$3$\sigma$) \tablenotemark{A} 
& $<$0.31 ($<$3$\sigma$) \tablenotemark{A} \\
NGC 1614 S & 0.34 (5.5$\sigma$) \tablenotemark{A} & 
0.56 (5.9$\sigma$) \tablenotemark{A} & 0.32 (3.0$\sigma$) \tablenotemark{A}
& $<$0.31 ($<$3$\sigma$) \tablenotemark{A} \\
NGC 1614 E & 0.29 (4.8$\sigma$) \tablenotemark{A} & 
0.54 (5.7$\sigma$) \tablenotemark{A} & $<$0.29 ($<$3$\sigma$) \tablenotemark{A}
& $<$0.31 ($<$3$\sigma$) \tablenotemark{A} \\
IRAS 06035$-$7102 & 2.5 (19$\sigma$) & 3.1 (20$\sigma$) & 1.0 (14$\sigma$) 
&  0.45 (9.4$\sigma$) \\
IRAS 08572$+$3915 & 1.1 (13$\sigma$) & 1.4 (15$\sigma$) & 0.37 (8.3$\sigma$) 
& 0.57 (11$\sigma$) \\
IRAS 12112$+$0305 NE & 4.7 (26$\sigma$) & 2.6 (19$\sigma$) & 3.4 (15$\sigma$) 
& 1.9 (14$\sigma$) \\
IRAS 12112$+$0305 SW & 0.43 (4.2$\sigma$) \tablenotemark{B} & 
0.88 (7.9$\sigma$) \tablenotemark{B} & $<$0.34 ($<$3$\sigma$) \tablenotemark{B}
& $<$0.45 ($<$3$\sigma$) \tablenotemark{B} \\
IRAS 12127$-$1412 & 0.48 (8.2$\sigma$) & 0.33 (6.8$\sigma$) & 0.32 (7.6$\sigma$) 
& 0.22 (4.6$\sigma$)\\
IRAS 13509$+$0442 & 0.28 (5.7$\sigma$) & 0.28 (5.3$\sigma$) & 0.16 (5.5$\sigma$) 
& 0.15 (3.8$\sigma$) \\
IRAS 15250$+$3609 & 3.0 (20$\sigma$) & 1.3 (13$\sigma$) & 3.0 (19$\sigma$) 
& 2.7 (20$\sigma$) \\ 
Superantennae & 2.9 (12$\sigma$) & 1.6 (11$\sigma$) & 1.1 (10$\sigma$) 
& 4.3 (21$\sigma$) \\ 
IRAS 20551$-$4250 & 4.4 (53$\sigma$) & 6.0 (59$\sigma$) & 1.8 (30$\sigma$) 
& 1.3 (27$\sigma$) \\ 
IRAS 22491$-$1808 & 3.9 (24$\sigma$) & 2.5 (20$\sigma$) & 2.4 (23$\sigma$) 
& 1.3 (17$\sigma$) \\ 
\enddata

\tablenotetext{A}{For NGC 1614, integrated channels are determined to 
cover all of the nuclear ($\sim$1 kpc) emission components.
Molecular emission at individual continuum peaks is narrower than the whole 
nuclear emission and has different peak velocity (Table 6 and Figure 2). 
Thus, the detection significance is apparently low, owing to increased 
noise originating from the integration of a large number of channels, 
even if narrow emission line signatures are clearly discernible 
in original beam spectra (Figure 2).}

\tablenotetext{B}{Integrated channels are optimized for the SW nucleus, 
because the molecular emission from the SW nucleus is faint and has a 
significantly different velocity profile from the brighter NE nucleus.}

\tablecomments{Col.(1): Object name.
Cols.(2)--(4): Flux (in Jy beam$^{-1}$ km s$^{-1}$)
at the emission peak in the integrated intensity (moment 0) map 
with original synthesized beam.
Detection significance relative to the rms noise (1$\sigma$) in the 
moment 0 map is shown in parentheses. 
These moment 0 maps of the original beam are primarily used for 
(a) verification of significant detection at the peak position of the 
molecular line 
and (b) confirming its spatial coincidence with continuum peak within peak 
determination uncertainty (= beam-size/signal-to-noise [S/N] ratio).
Col.(2): HCN J=2--1 ($\nu_{\rm rest}$=177.261 GHz).
Col.(3): HCO$^{+}$ J=2--1 ($\nu_{\rm rest}$=178.375 GHz).
Col.(4): HNC J=2--1 ($\nu_{\rm rest}$=181.325 GHz).
Col.(5): Para-H$_{2}$O 3$_{\rm 1,3}$--2$_{\rm 2,0}$ ($\nu_{\rm rest}$=183.310 GHz).
In each object, the synthesized beam size of the HCN J=2--1 and 
HCO$^{+}$ J=2--1 (HNC J=2--1 and H$_{2}$O) is virtually identical 
to that of continuum J21a (J21b) shown in Table 3, column 6.}

\end{deluxetable}

For the nearby LIRG NGC 1614 ($z =$ 0.0160), the continuum and molecular 
line emission are spatially extended, and are substantially larger than the 
synthesized beam size (Figure 1). 
We create spectra from original beam data cubes (hereafter
``original beam'') at four bright continuum emission peaks, which are
shown in Figure 2. 
In Figure 3, we display original beam spectra at the 
primary continuum peak in each ULIRG, except for the double nuclei 
ULIRG IRAS 12112$+$0305 
for which the spectra at both nuclei are shown separately.

Figure 4 shows the intensity-weighted mean velocity (moment 1) maps of 
the HCN J=2--1, HCO$^{+}$ J=2--1, HNC J=2--1, and 183 GHz H$_{2}$O emission 
lines whenever they are detected with significantly high S/N ratios 
to obtain meaningful velocity information (Table 4). 
Intensity-weighted velocity dispersion (moment 2) maps are also shown 
in Figure 5.

\begin{figure}
\begin{center}
\includegraphics[angle=-90,scale=.35]{f2a.eps} 
\hspace{0.5cm}
\includegraphics[angle=-90,scale=.35]{f2b.eps} \\ 
\includegraphics[angle=-90,scale=.35]{f2c.eps} 
\hspace{0.5cm}
\includegraphics[angle=-90,scale=.35]{f2d.eps} \\
\includegraphics[angle=-90,scale=.35]{f2e.eps} 
\hspace{0.5cm}
\includegraphics[angle=-90,scale=.35]{f2f.eps} \\
\includegraphics[angle=-90,scale=.35]{f2g.eps} 
\hspace{0.5cm}
\includegraphics[angle=-90,scale=.35]{f2h.eps} \\
\end{center}
\end{figure}

\clearpage

\begin{figure}
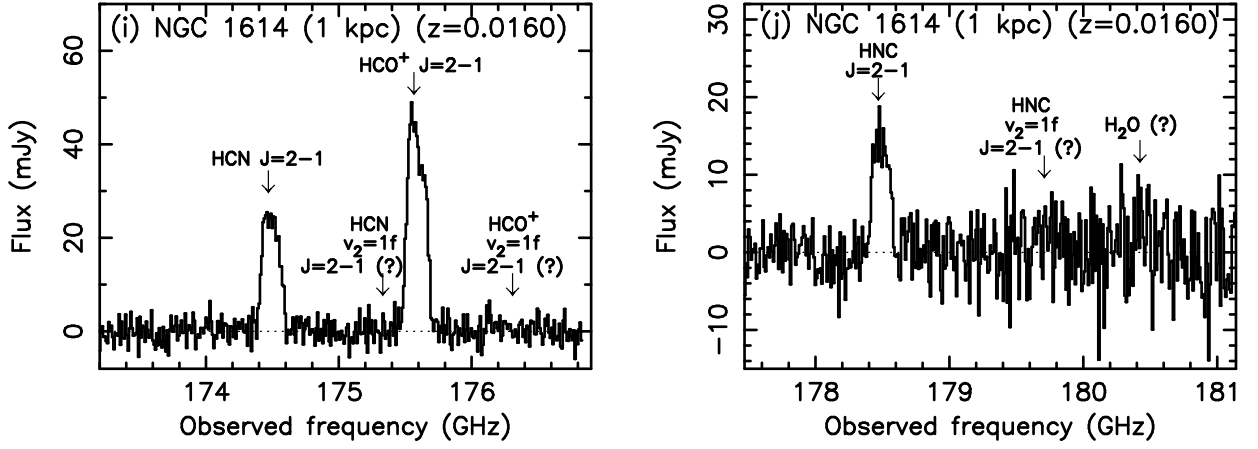

\begin{center}
\includegraphics[angle=-90,scale=.35]{f2i.eps} 
\hspace{0.5cm}
\includegraphics[angle=-90,scale=.35]{f2j.eps} \\
\end{center}
\caption{
{\it (a)--(h)} : Spectra at the W-, N-, S-, and E-continuum peak 
positions (Figure 1 top) within original elliptic beam 
(black solid line) and 200 pc circular beam (red dotted line).
{\it (i)--(j)}: Spectra with 1 kpc circular beam.
The abscissa represents the observed frequency in GHz and the ordinate 
represents flux density in mJy.
Downward arrows are plotted at the redshifted frequency of certain 
emission lines, where $z =$ 0.0160 (Table 1, column 2) is adopted 
for all positions.
When the detection is not significant, the mark ``(?)'' is added.
The horizontal black thin dotted line indicates the zero flux 
level.
}
\end{figure}

\begin{figure}
\begin{center}
\includegraphics[angle=-90,scale=.35]{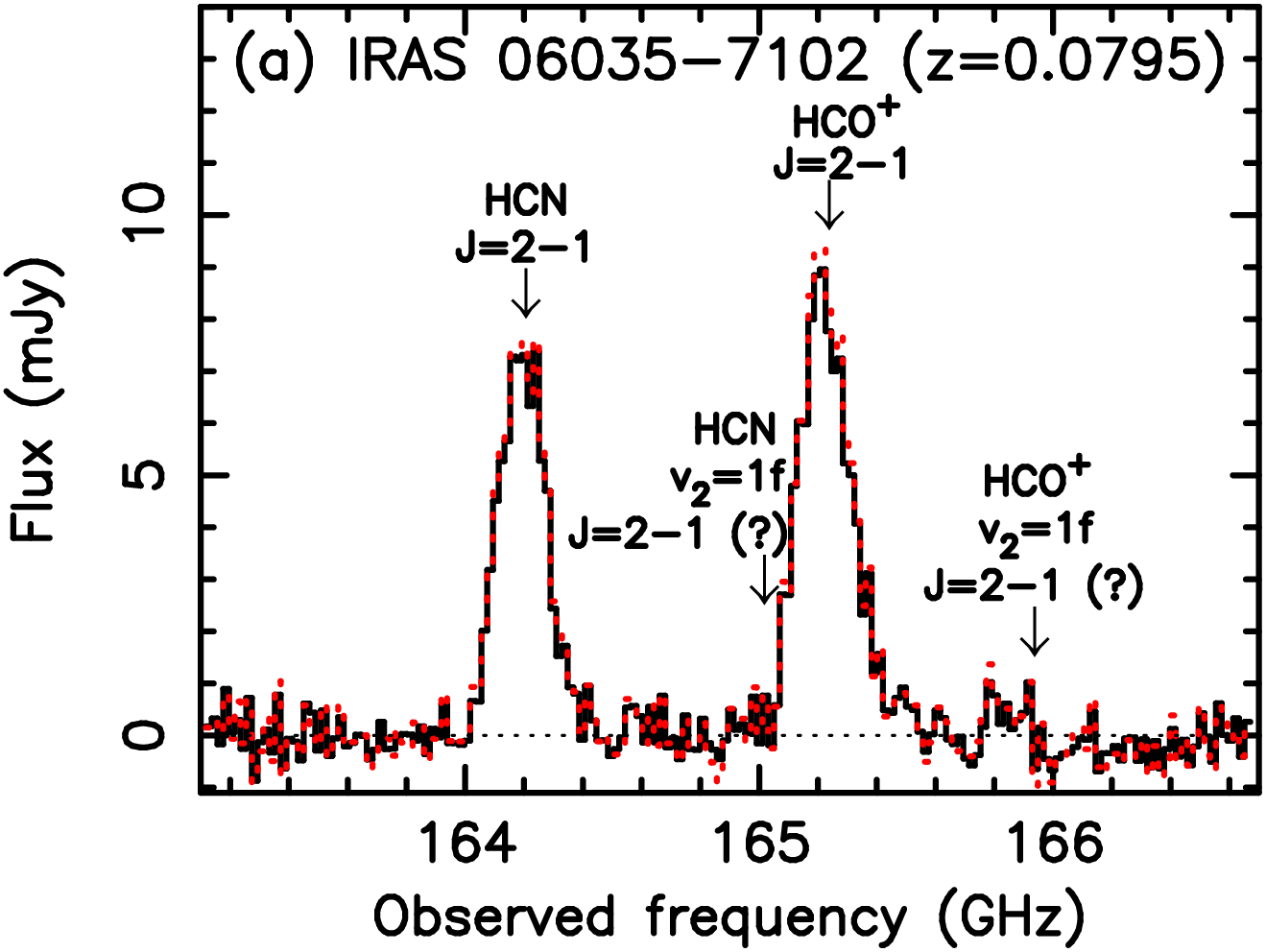} 
\hspace{0.5cm}
\includegraphics[angle=-90,scale=.35]{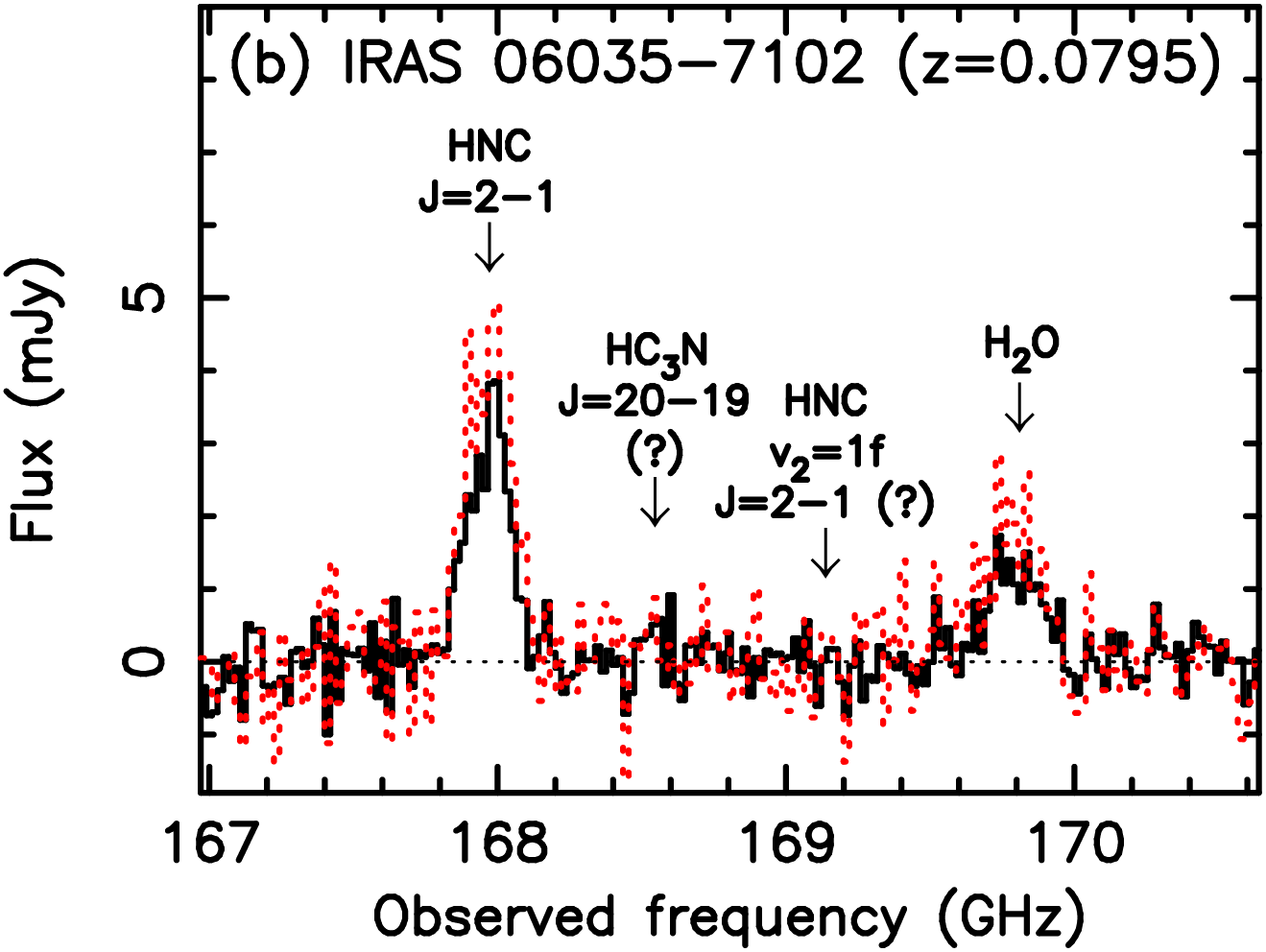} \\
\includegraphics[angle=-90,scale=.35]{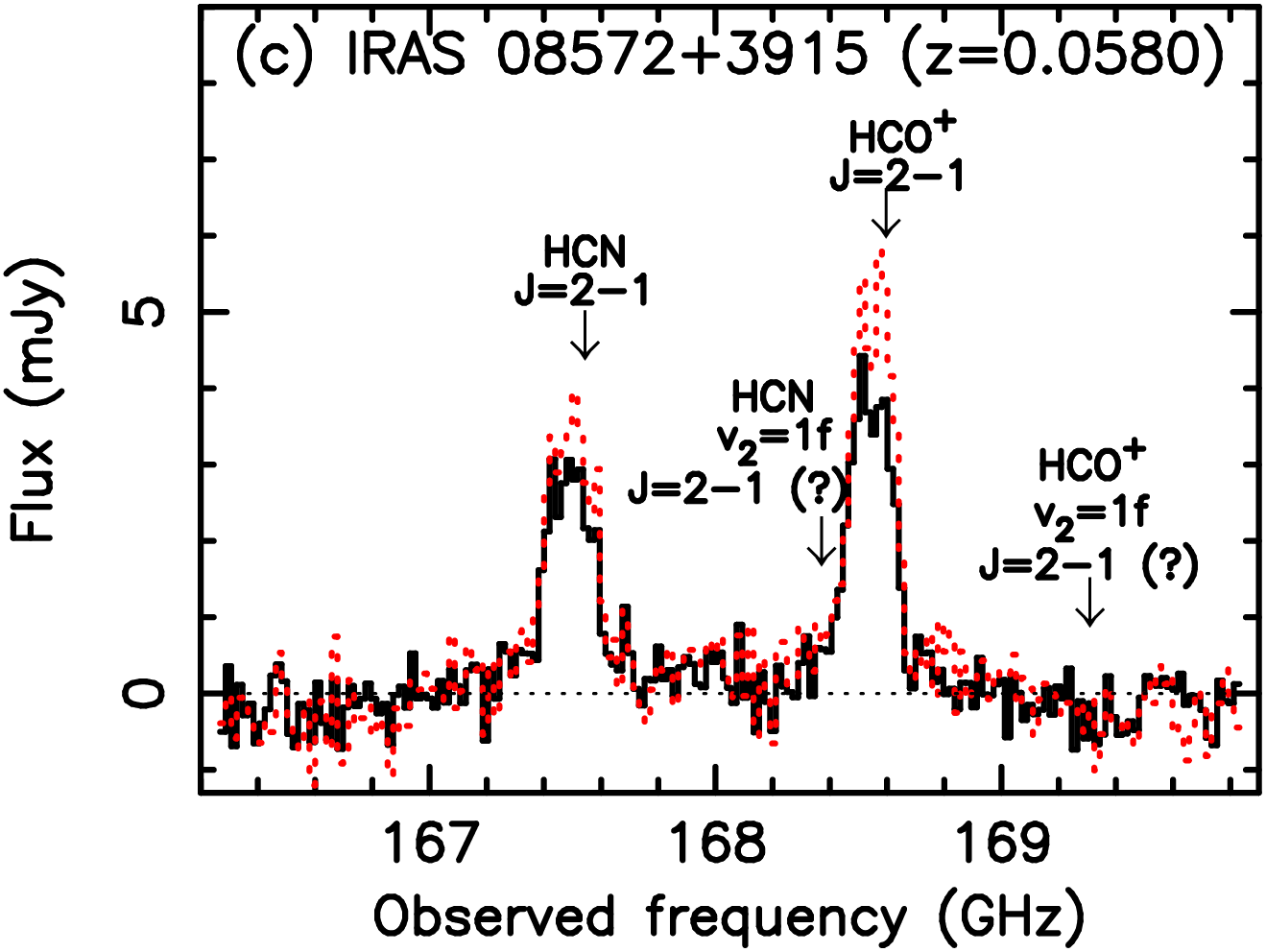} 
\hspace{0.5cm}
\includegraphics[angle=-90,scale=.35]{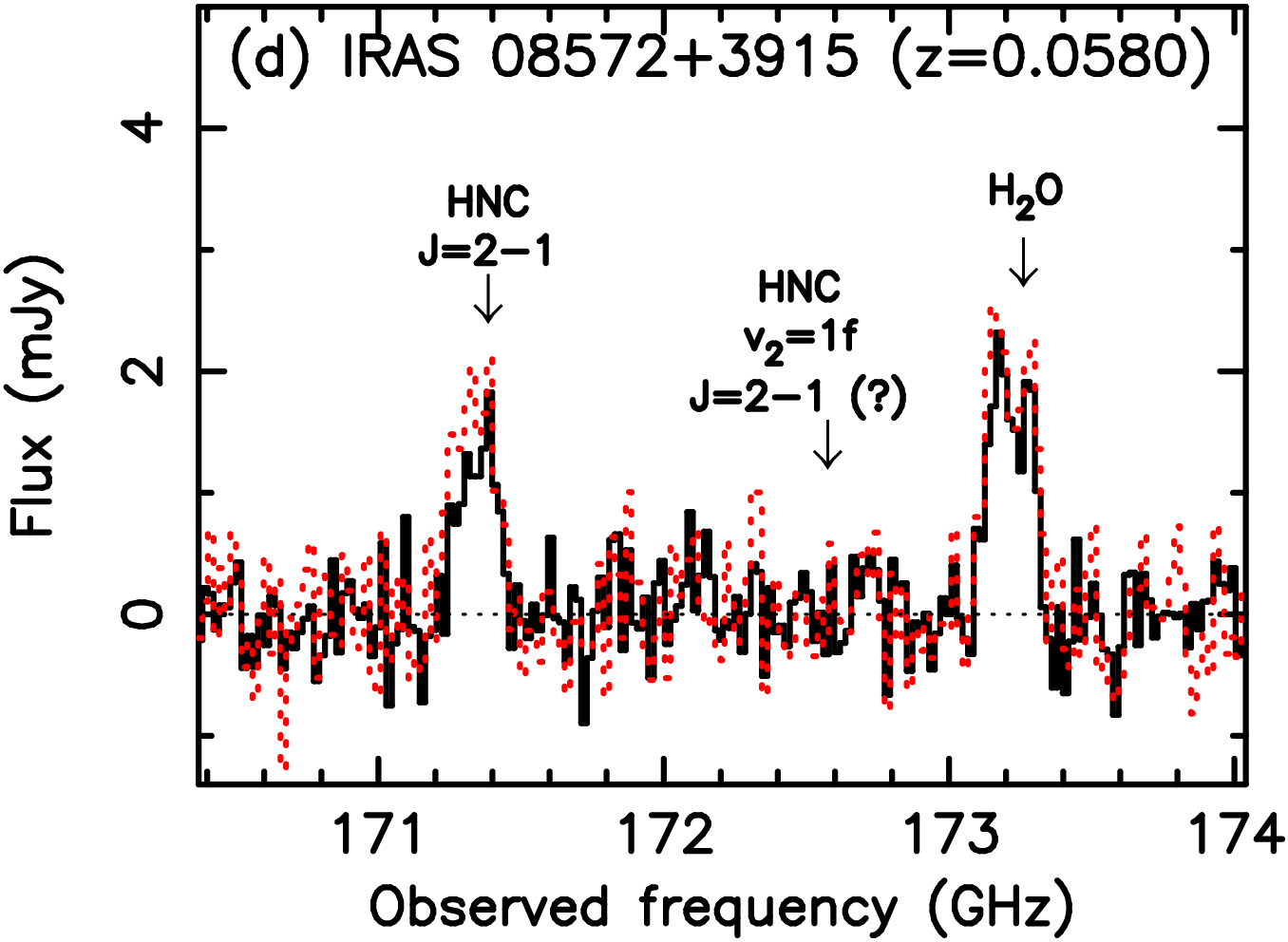} \\
\includegraphics[angle=-90,scale=.35]{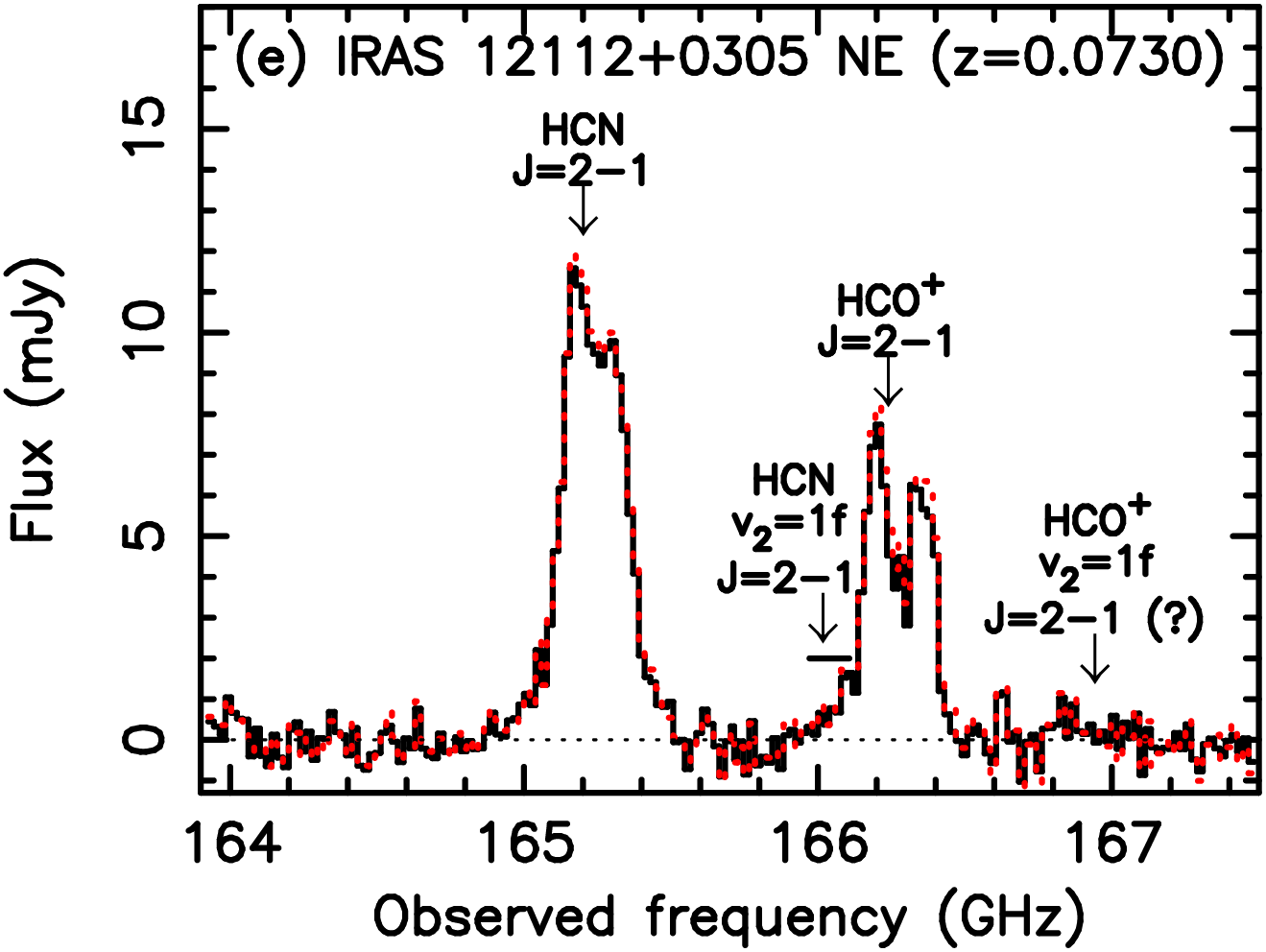} 
\hspace{0.5cm}
\includegraphics[angle=-90,scale=.35]{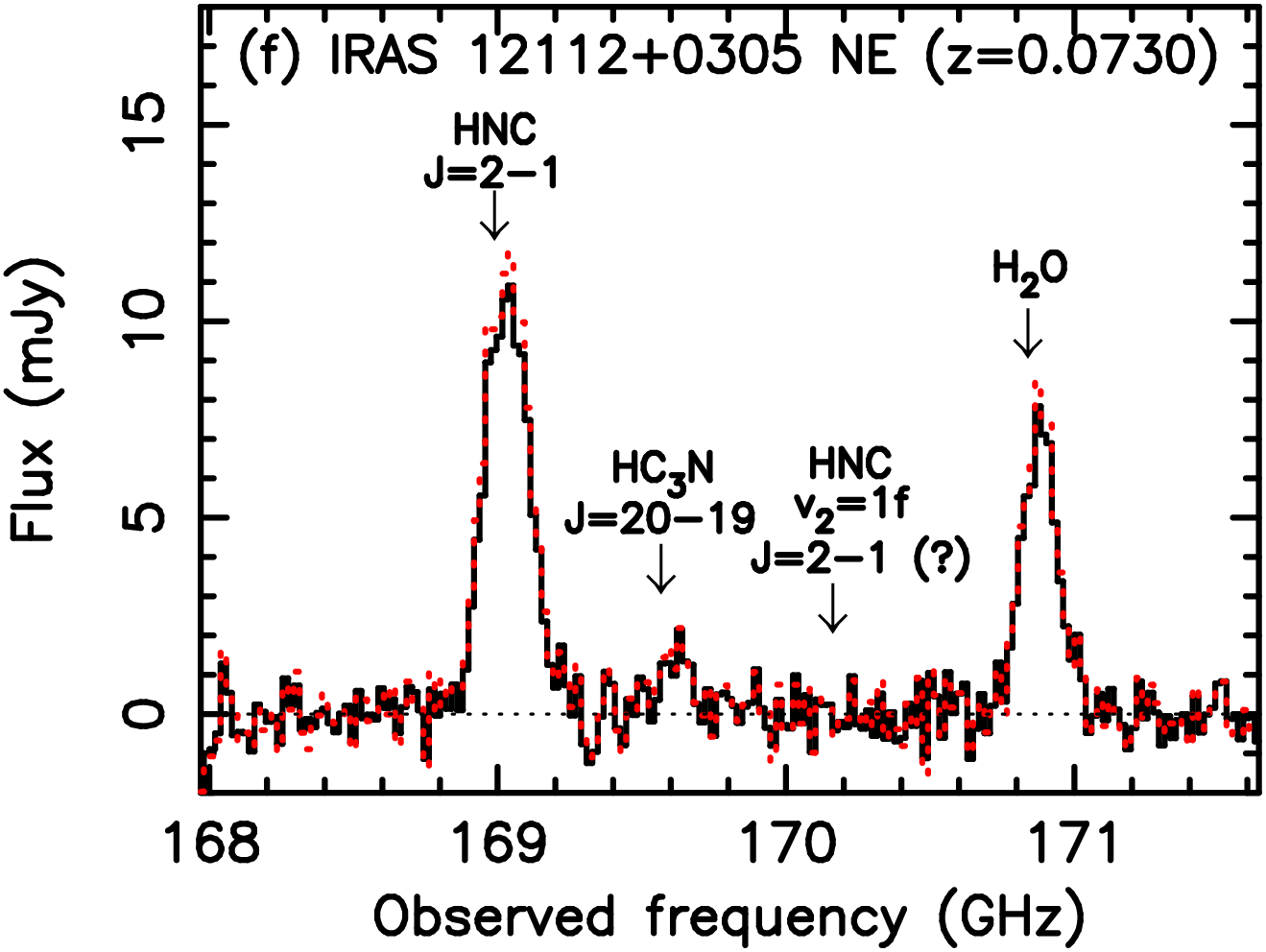} \\
\includegraphics[angle=-90,scale=.35]{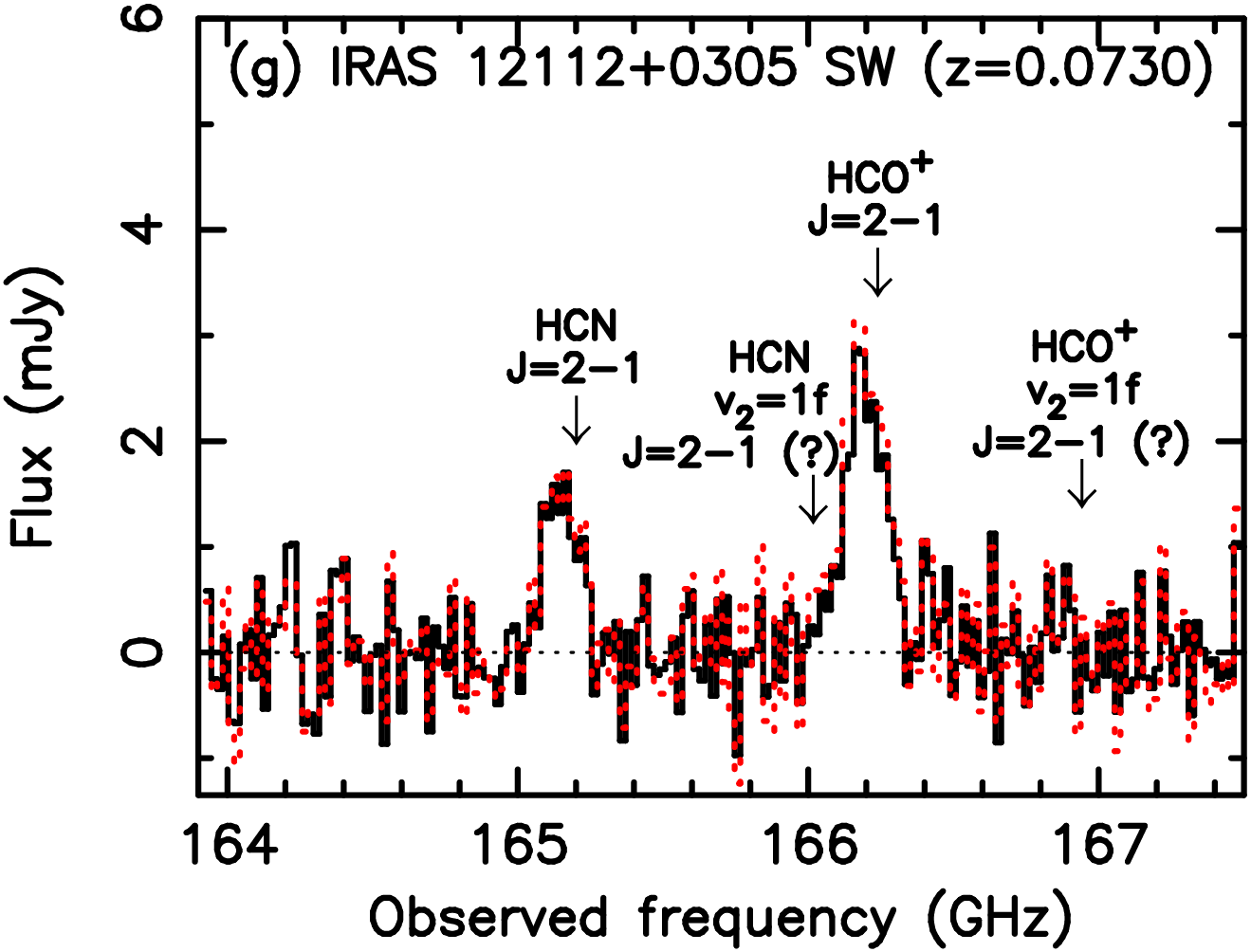} 
\hspace{0.5cm}
\includegraphics[angle=-90,scale=.35]{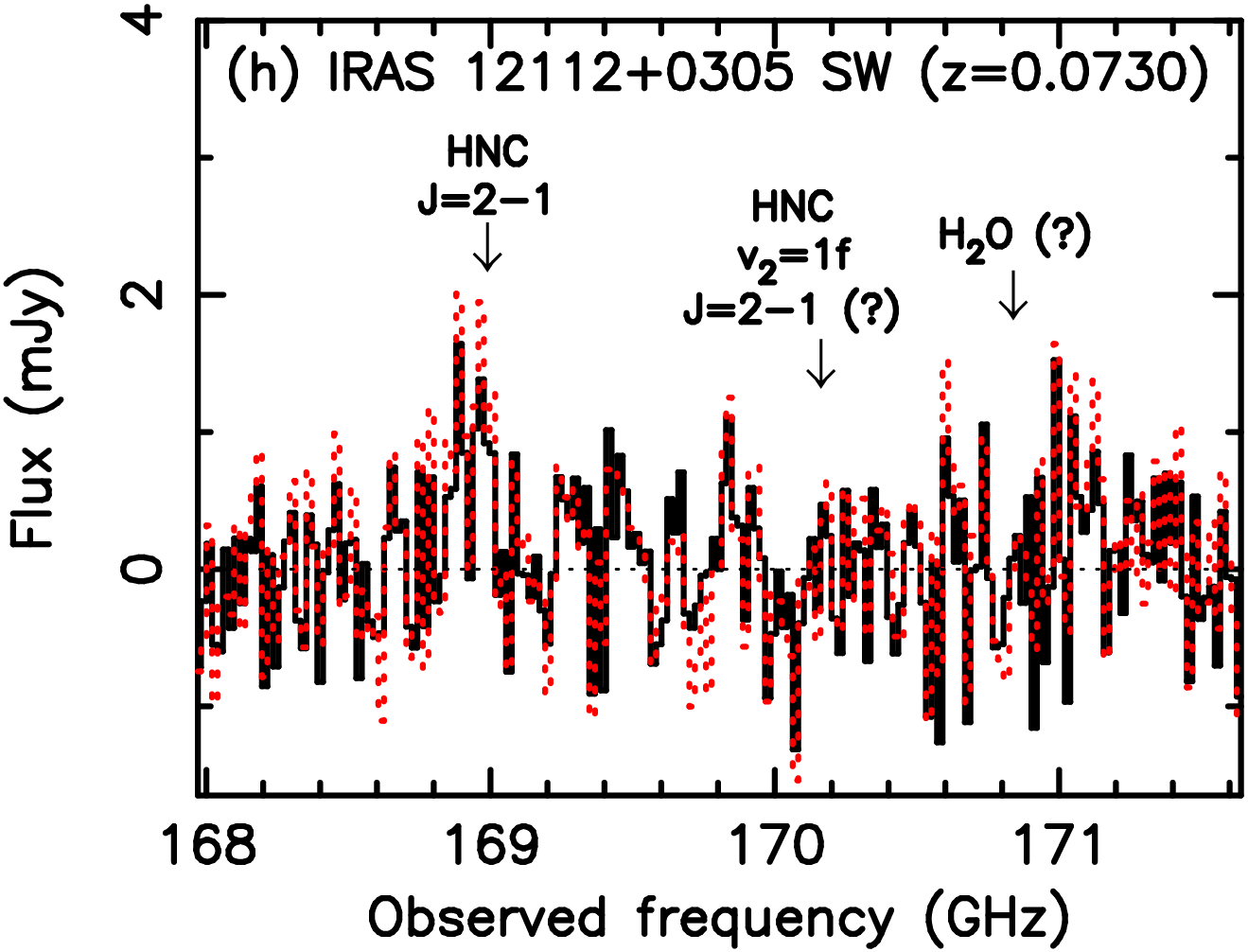} \\
\end{center}
\end{figure}

\clearpage

\begin{figure}
\begin{center}
\includegraphics[angle=-90,scale=.35]{f3i.eps} 
\hspace{0.5cm}
\includegraphics[angle=-90,scale=.35]{f3j.eps} \\
\includegraphics[angle=-90,scale=.35]{f3k.eps}
\hspace{0.5cm}
\includegraphics[angle=-90,scale=.35]{f3l.eps} \\
\includegraphics[angle=-90,scale=.35]{f3m.eps} 
\hspace{0.5cm}
\includegraphics[angle=-90,scale=.35]{f3n.eps} \\
\includegraphics[angle=-90,scale=.35]{f3o.eps} 
\hspace{0.5cm}
\includegraphics[angle=-90,scale=.35]{f3p.eps} \\
\end{center}
\end{figure}

\clearpage

\begin{figure}
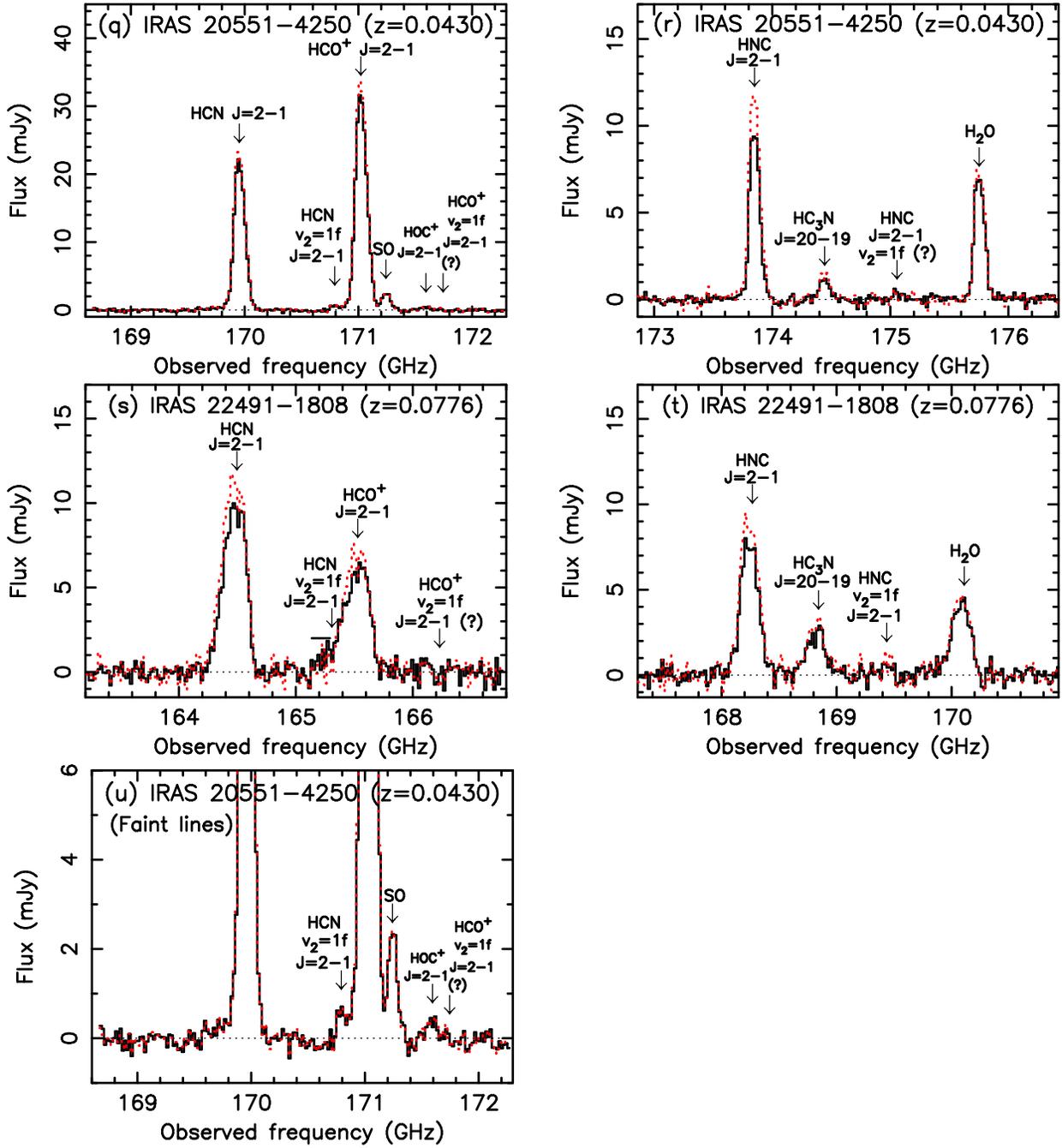

\begin{center}
\includegraphics[angle=-90,scale=.35]{f3q.eps} 
\hspace{0.5cm}
\includegraphics[angle=-90,scale=.35]{f3r.eps} \\
\includegraphics[angle=-90,scale=.35]{f3s.eps} 
\hspace{0.5cm}
\includegraphics[angle=-90,scale=.35]{f3t.eps} \\
\hspace*{-8.5cm}
\includegraphics[angle=-90,scale=.35]{f3u.eps} 
\end{center}
\caption{
Spectra of the HCN J=2--1, HCO$^{+}$ J=2--1, HNC J=2--1, and 183 GHz 
H$_{2}$O lines of ULIRGs.
The abscissa represents the observed frequency in GHz and the ordinate is 
flux density in mJy.
The black solid and red dotted lines represent the original 
elliptic beam and a $\sim$1kpc circular beam spectrum at
the continuum peak, respectively. 
For IRAS 20551$–$4250, a spectrum is added to {\it (u)} to highlight 
faint emission lines.
Other symbols are the same as in Figure 2.
}
\end{figure}

\begin{figure}
\begin{center}
\hspace*{-4.7cm}
\includegraphics[angle=0,scale=.195]{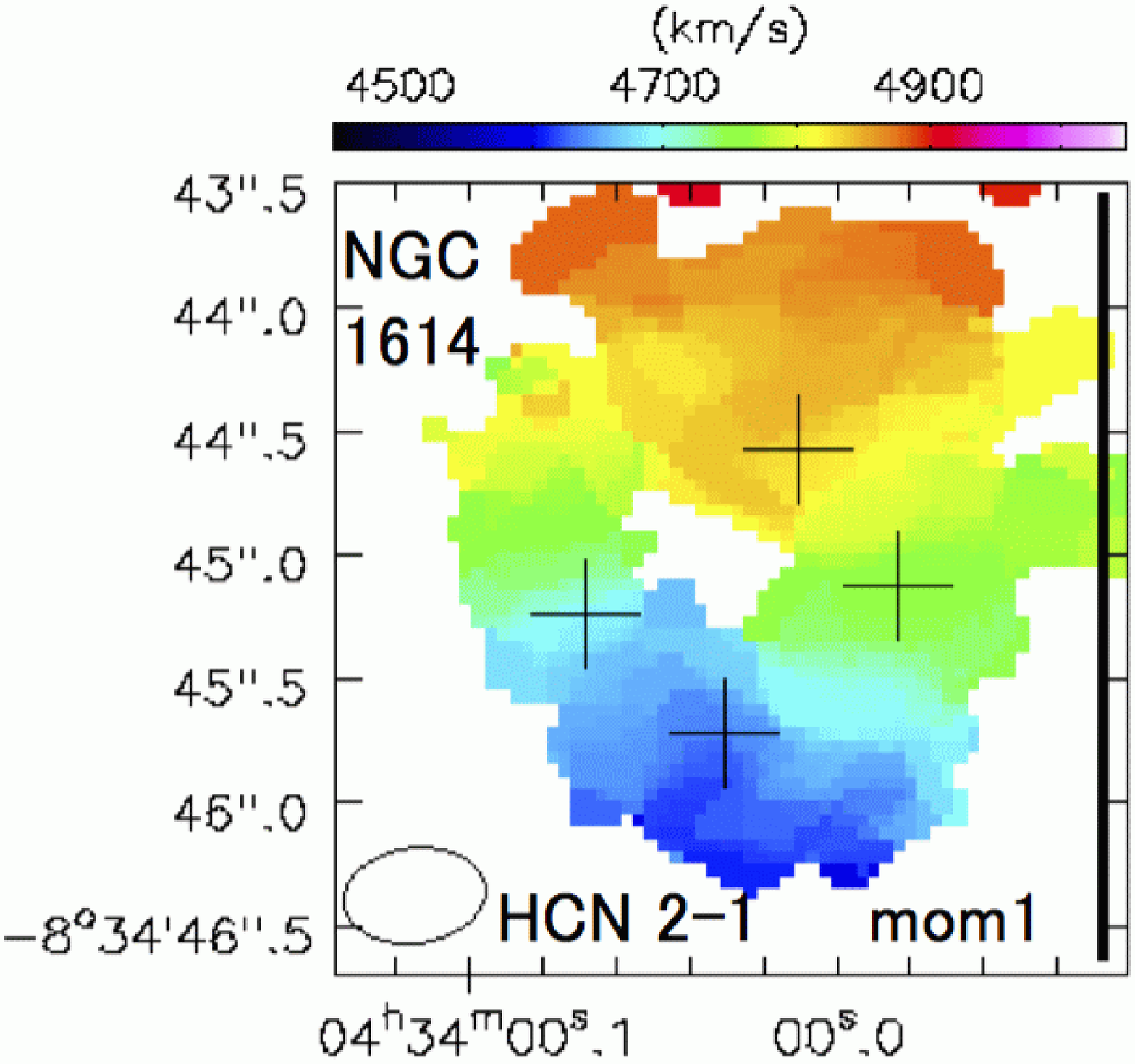} 
\includegraphics[angle=0,scale=.195]{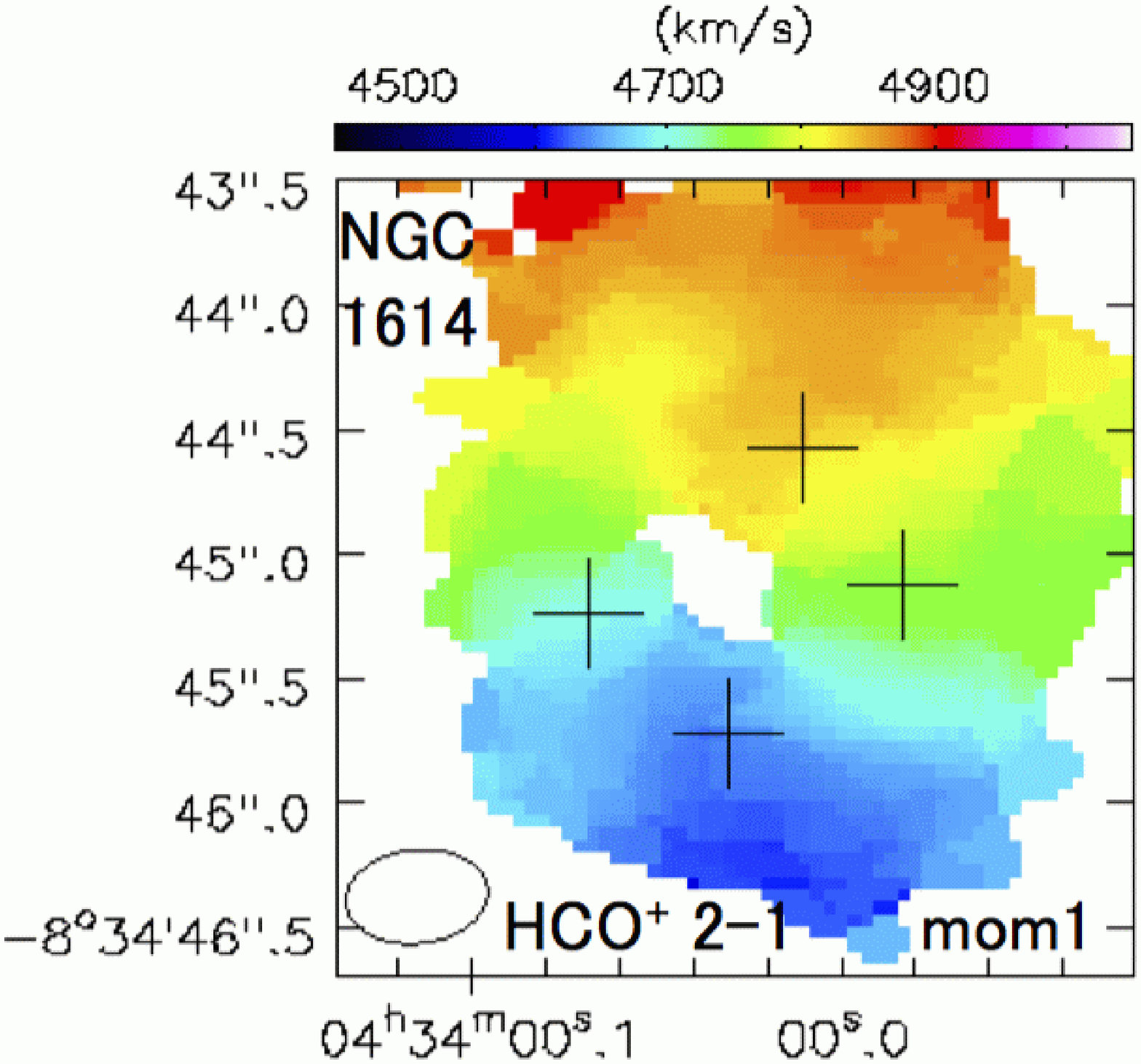} 
\includegraphics[angle=0,scale=.195]{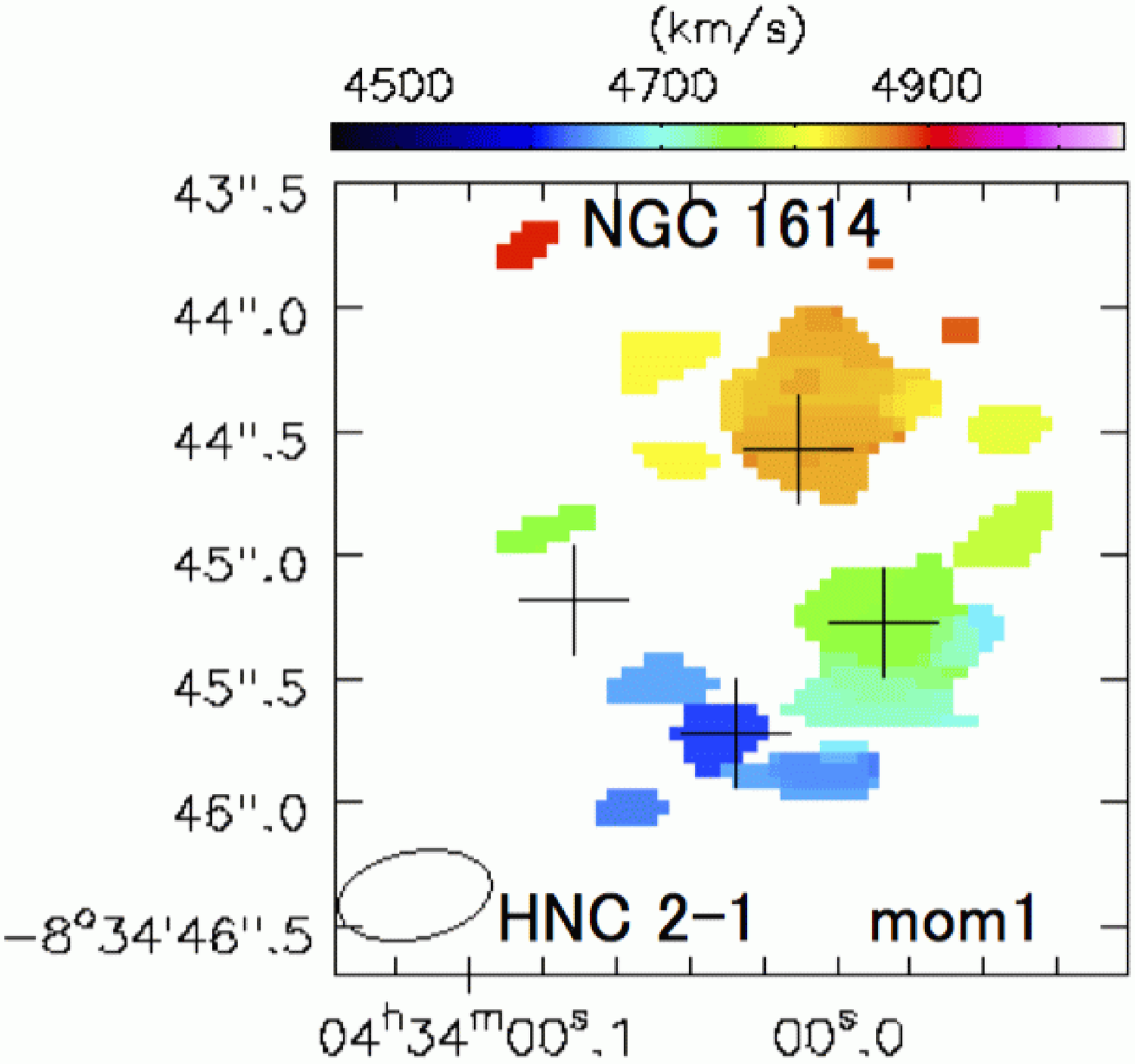} \\
\includegraphics[angle=0,scale=.195]{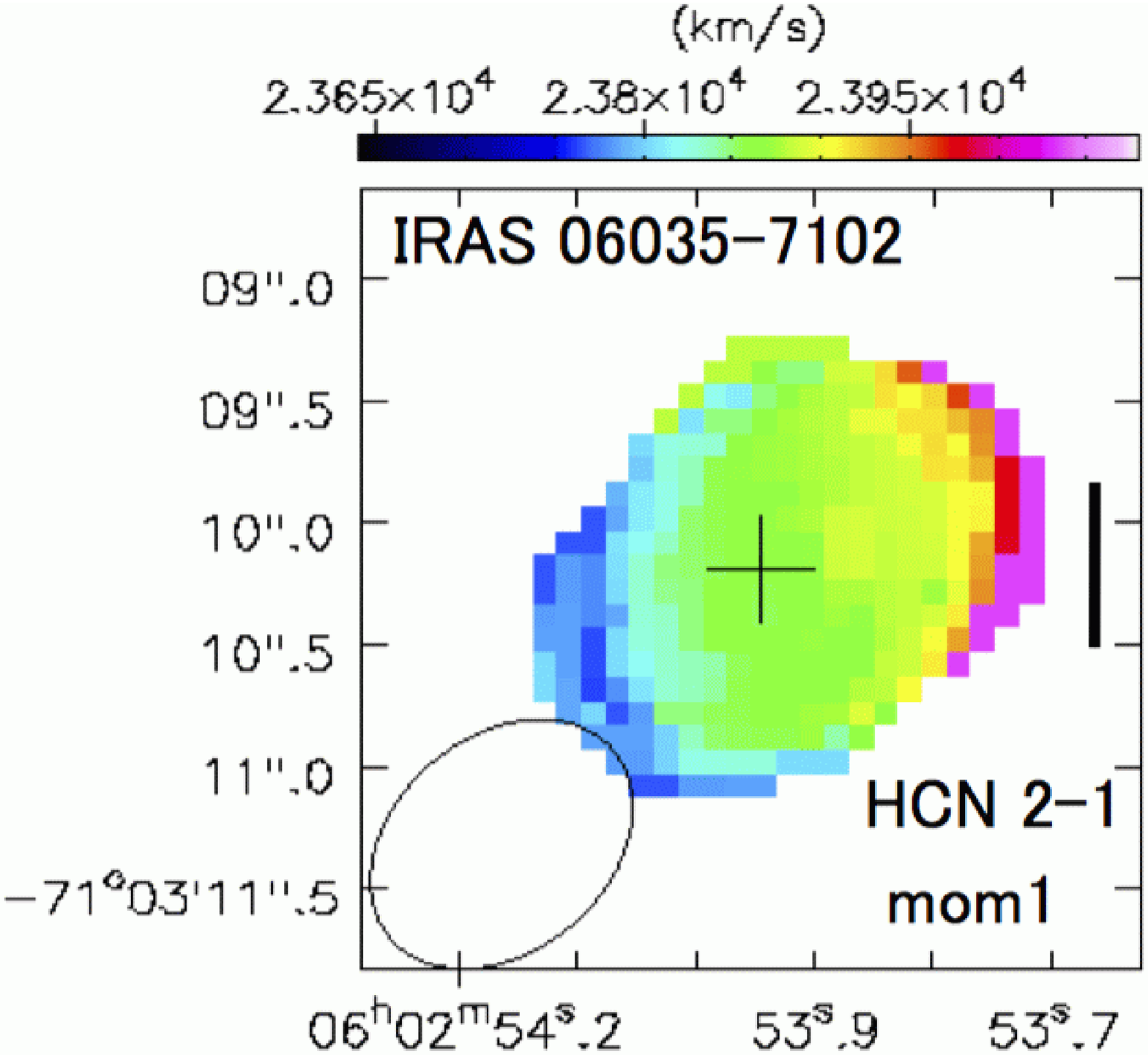} 
\includegraphics[angle=0,scale=.195]{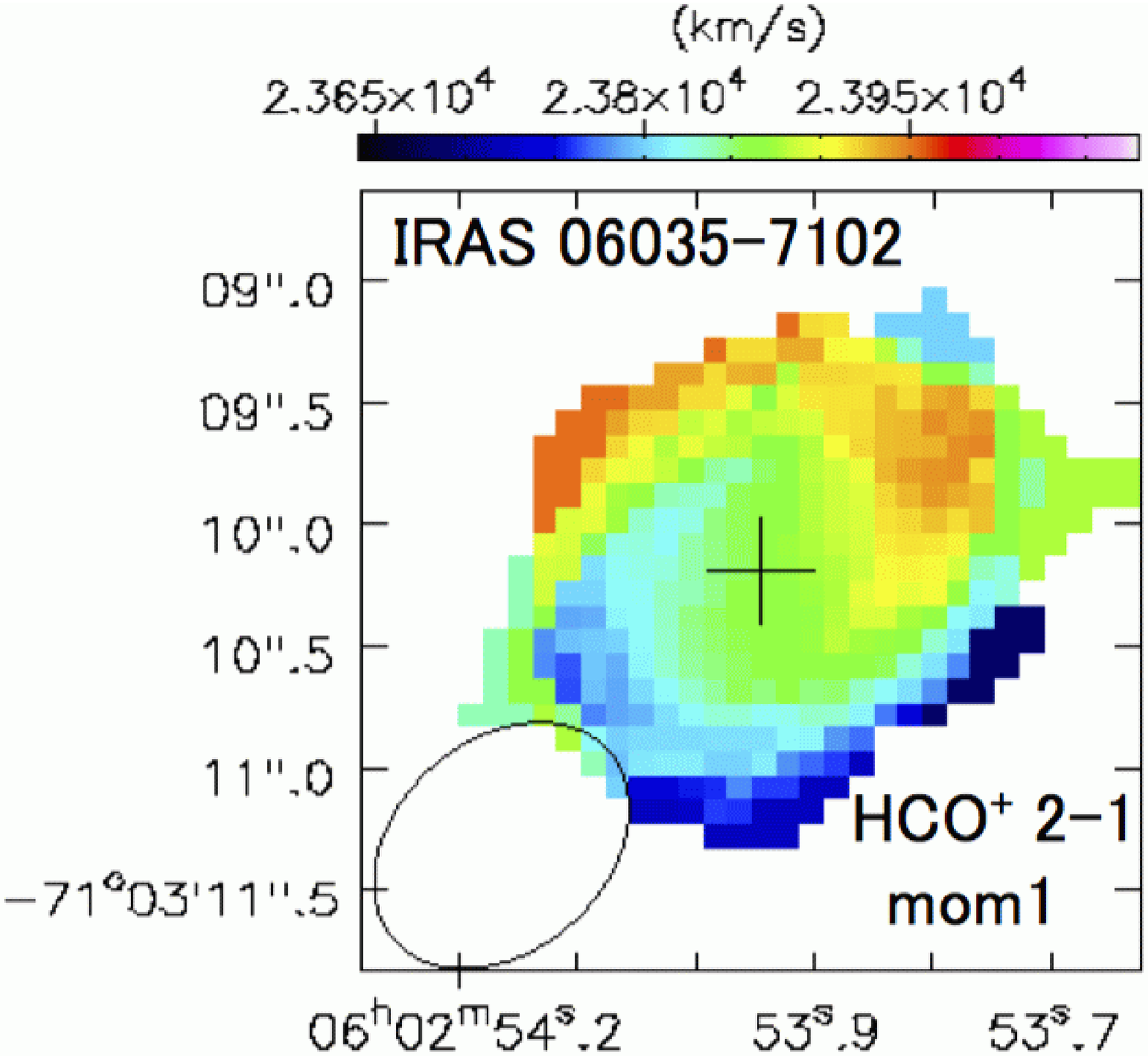} 
\includegraphics[angle=0,scale=.195]{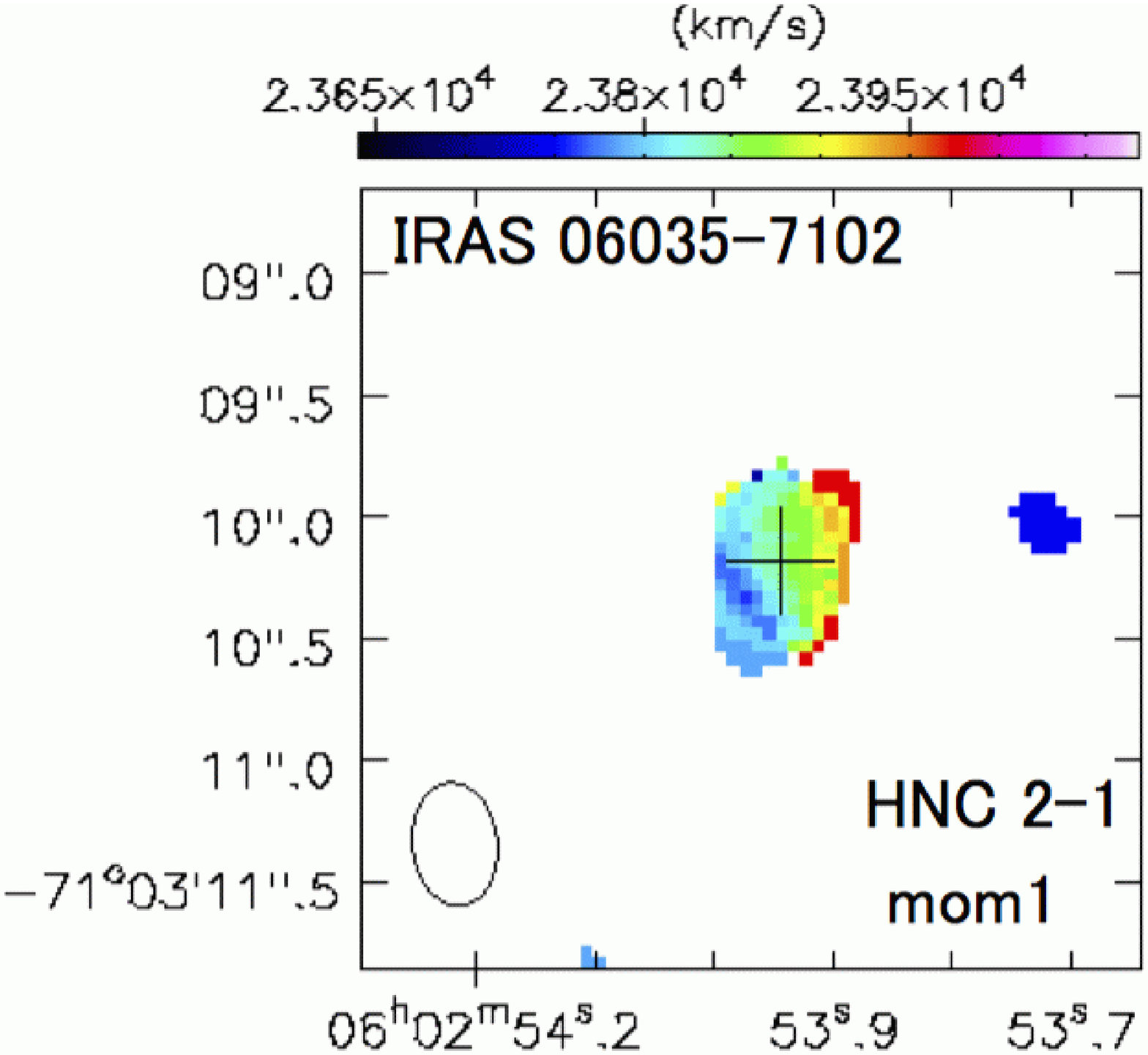} 
\includegraphics[angle=0,scale=.195]{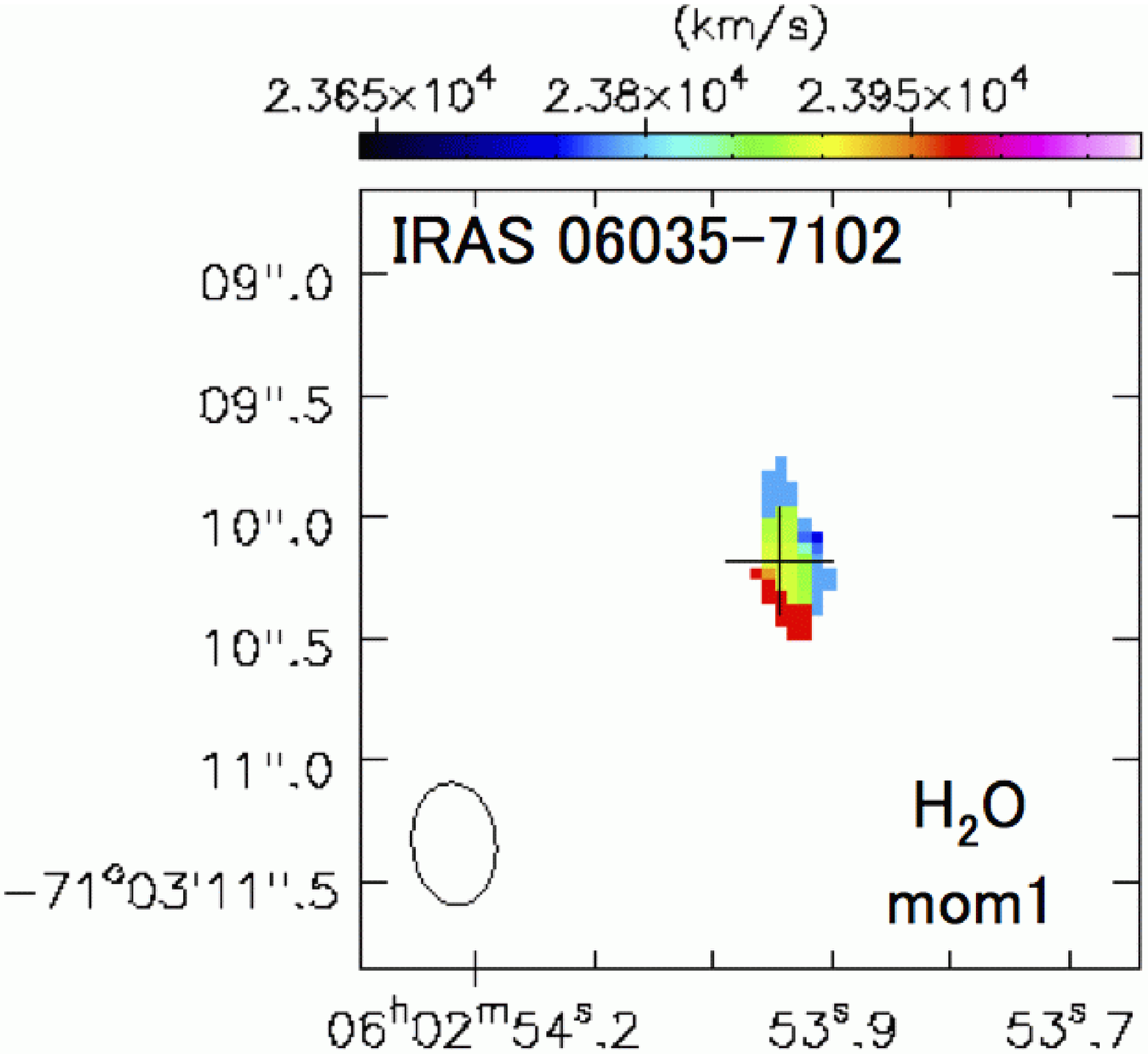} \\
\includegraphics[angle=0,scale=.19]{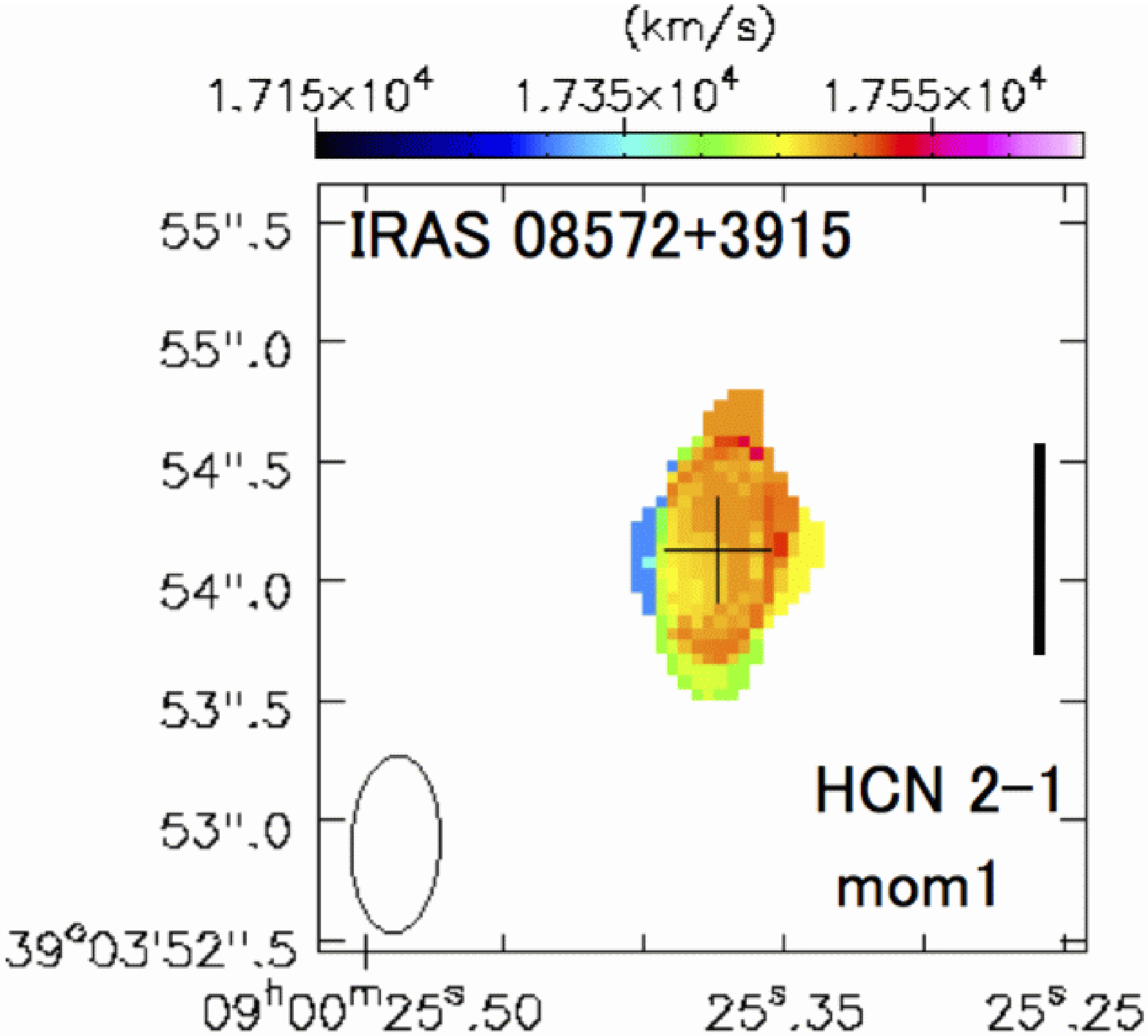} 
\includegraphics[angle=0,scale=.19]{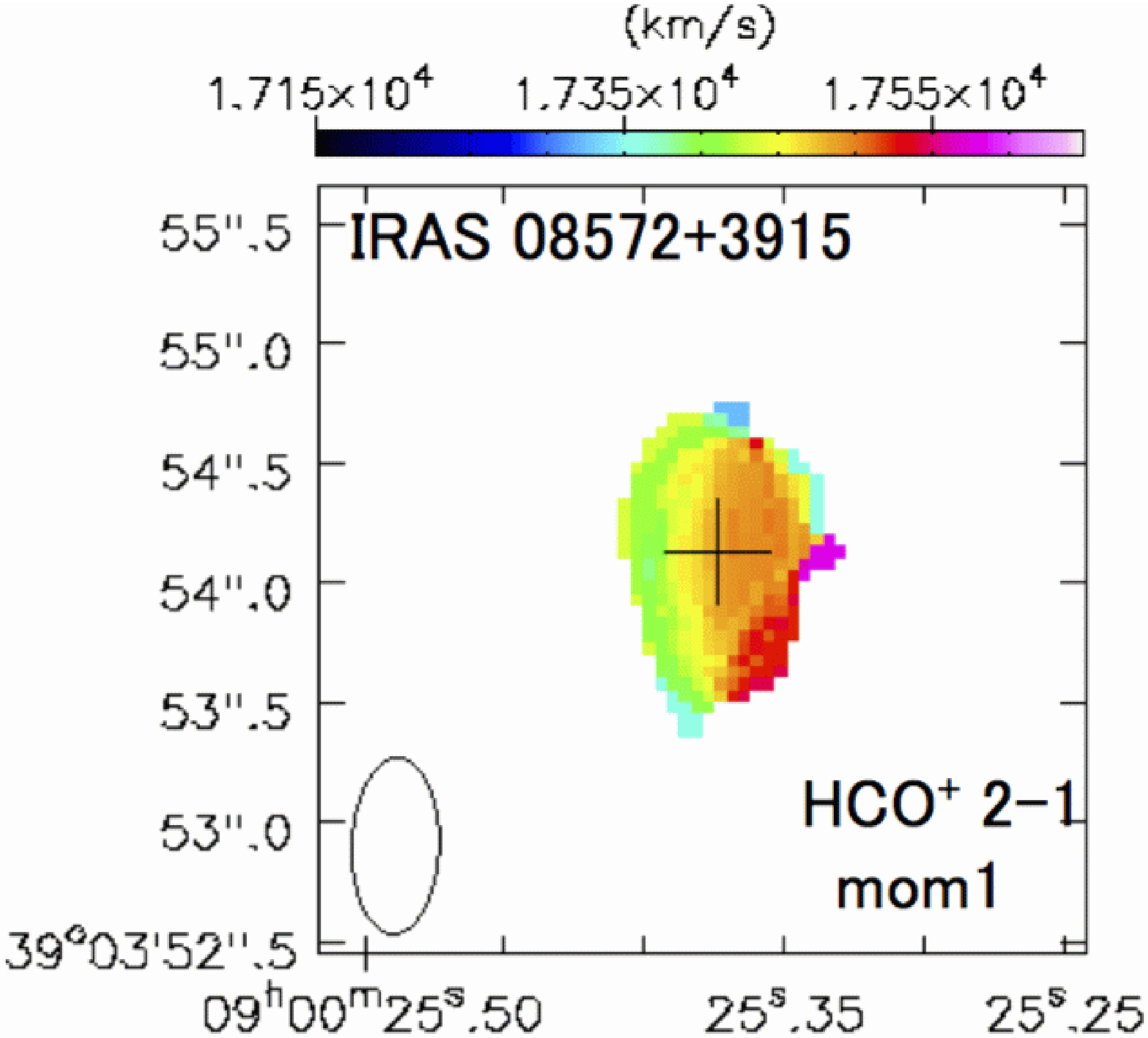} 
\includegraphics[angle=0,scale=.19]{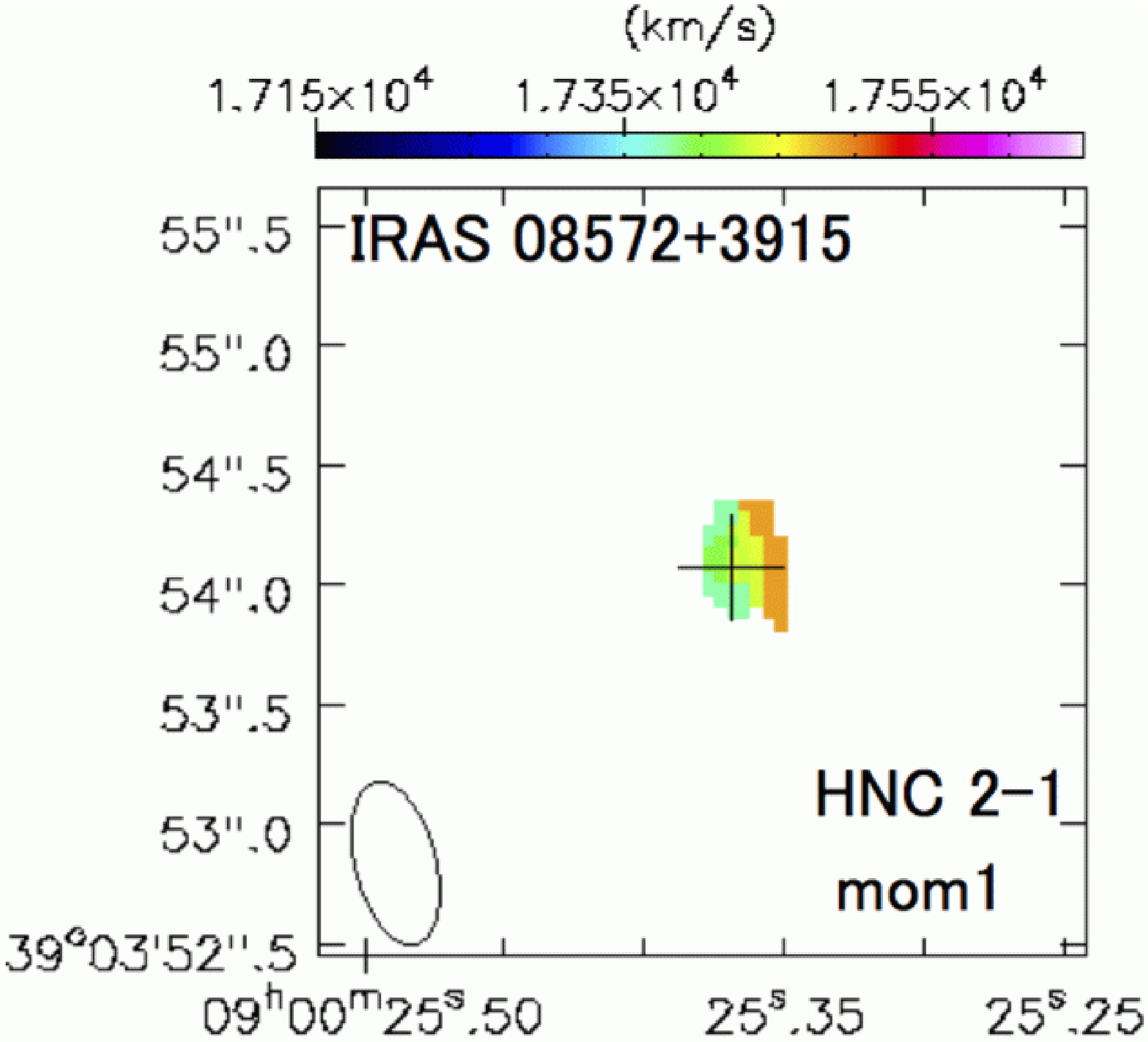} 
\includegraphics[angle=0,scale=.19]{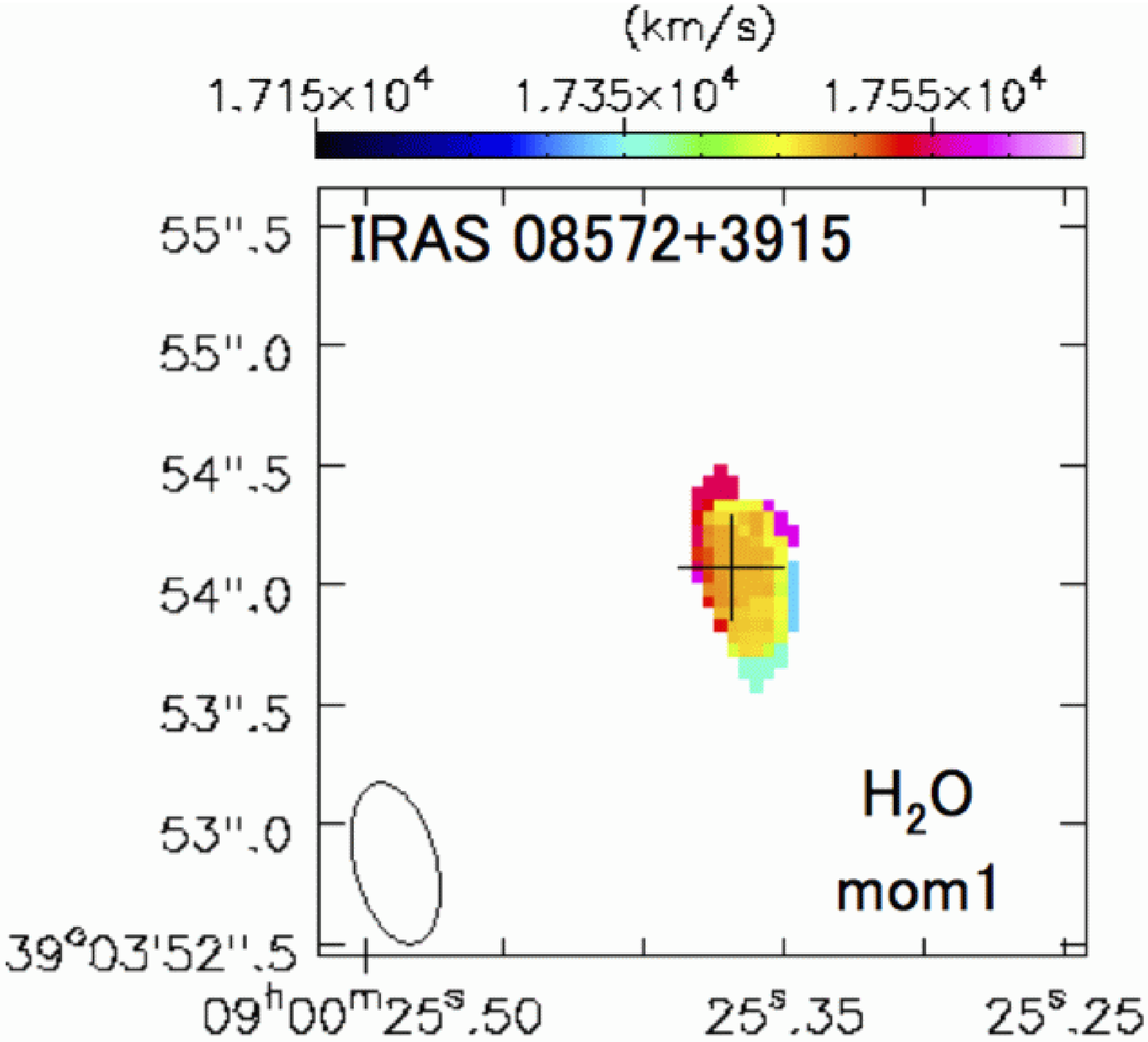} \\
\includegraphics[angle=0,scale=.195]{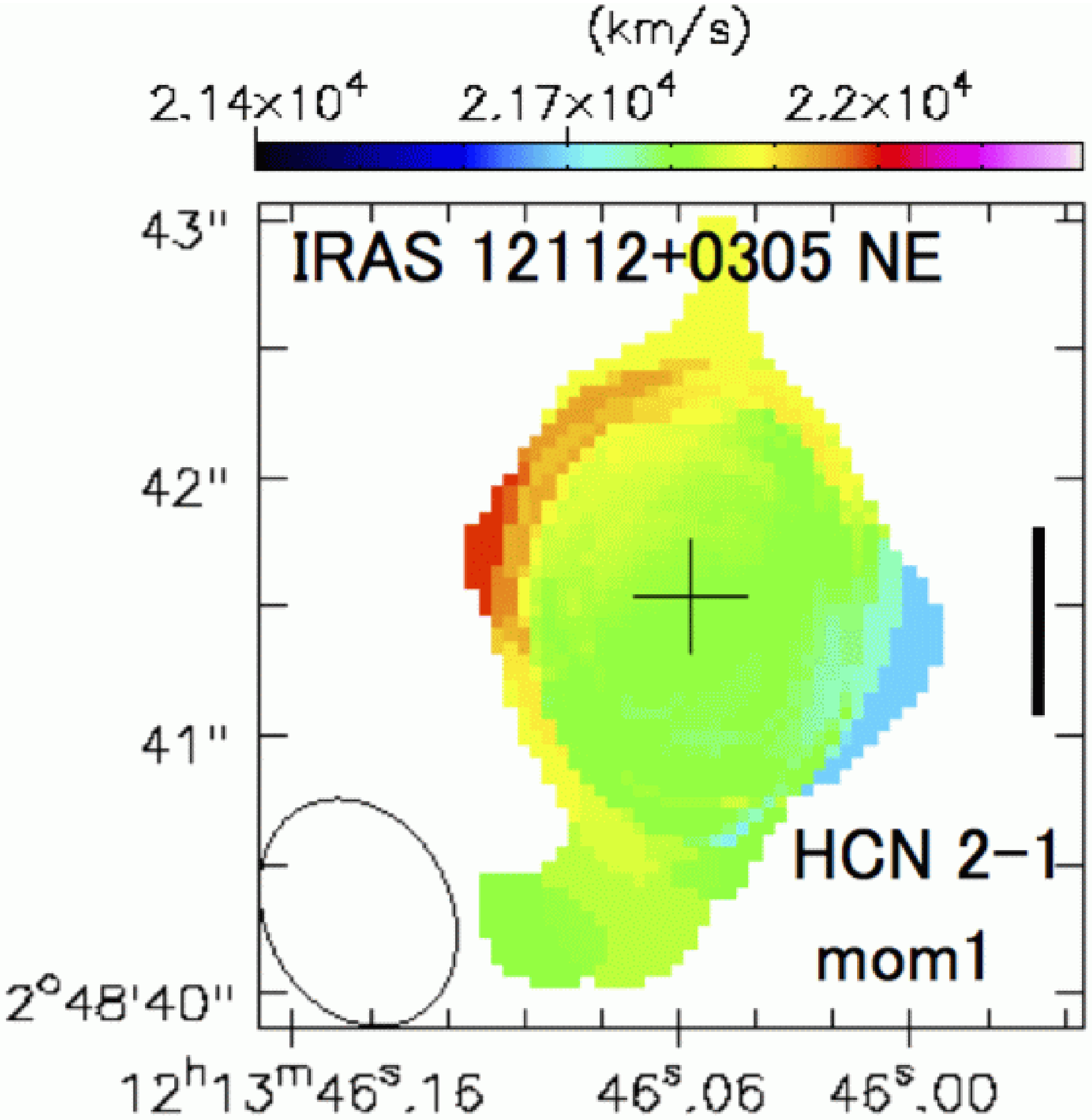} 
\includegraphics[angle=0,scale=.195]{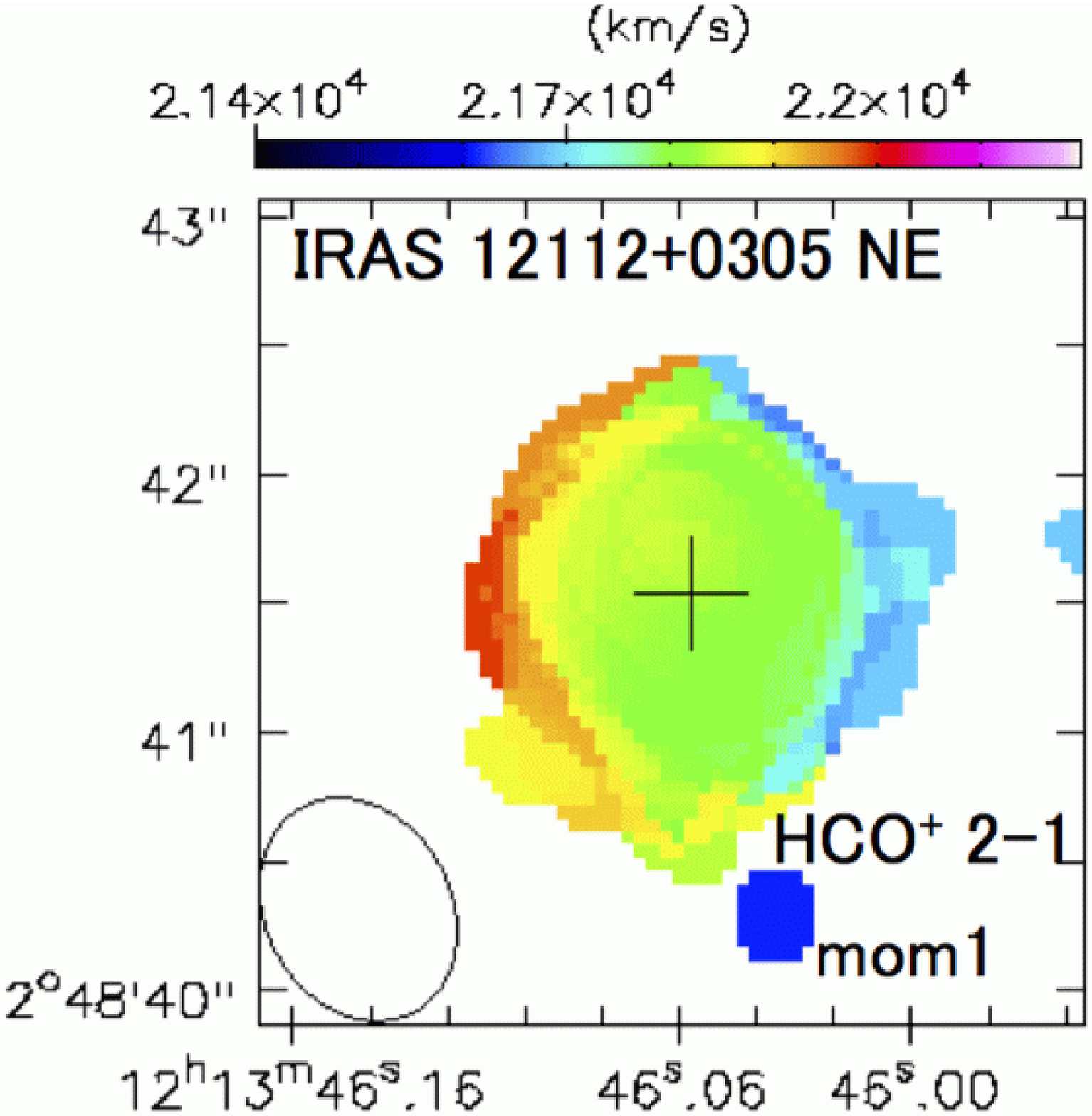} 
\includegraphics[angle=0,scale=.195]{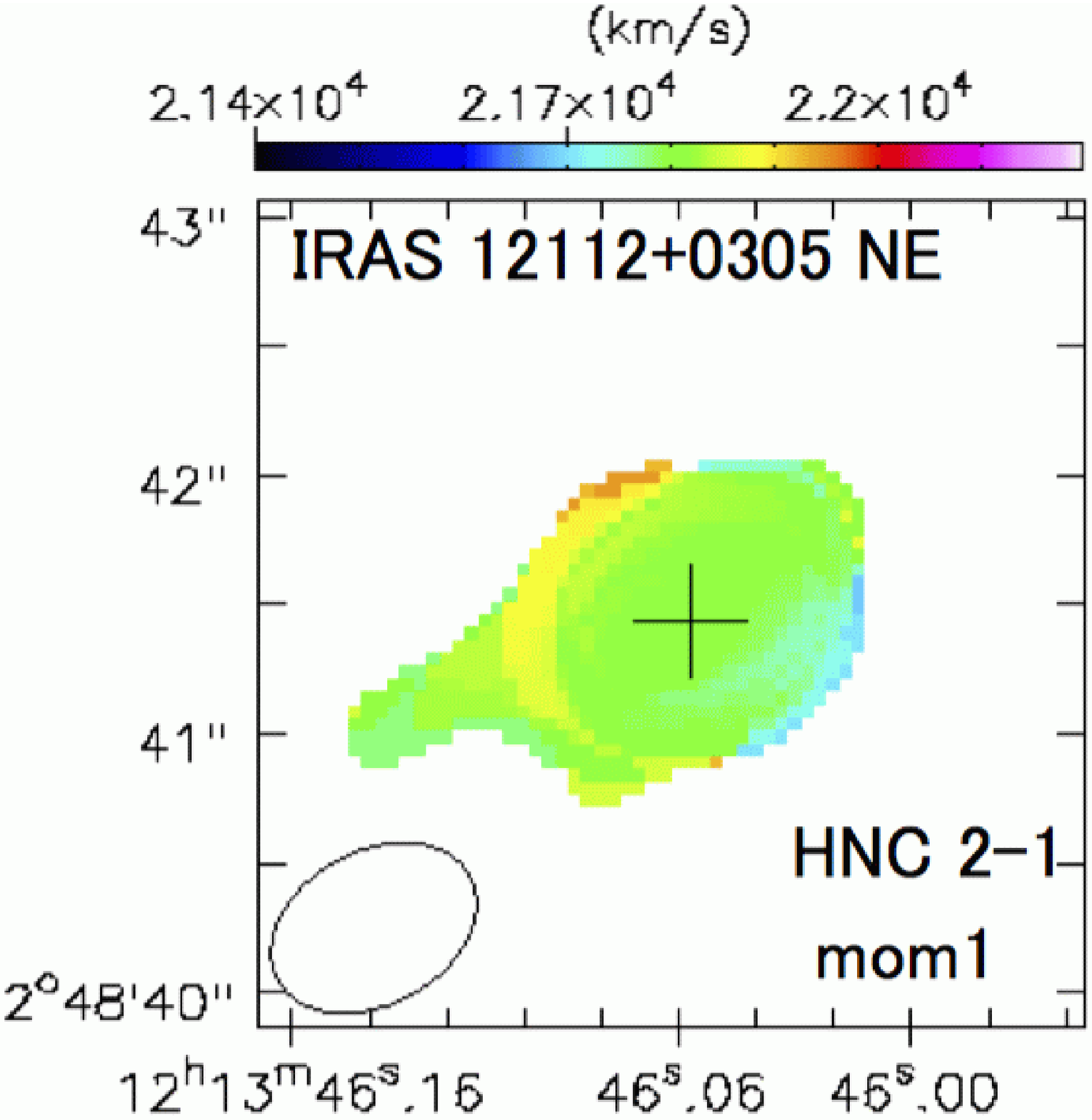} 
\includegraphics[angle=0,scale=.195]{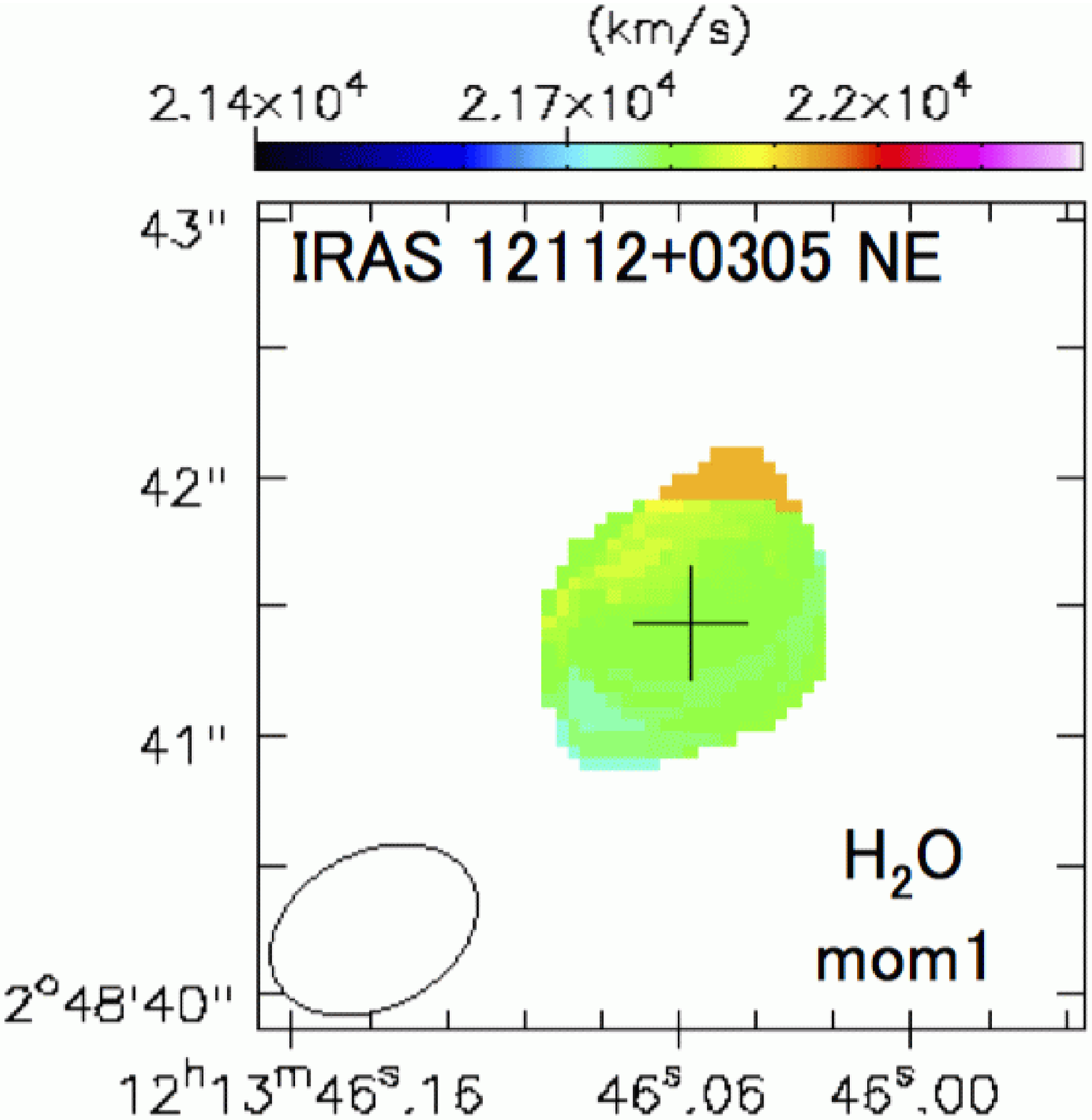} \\
\hspace*{-9.0cm}
\includegraphics[angle=0,scale=.195]{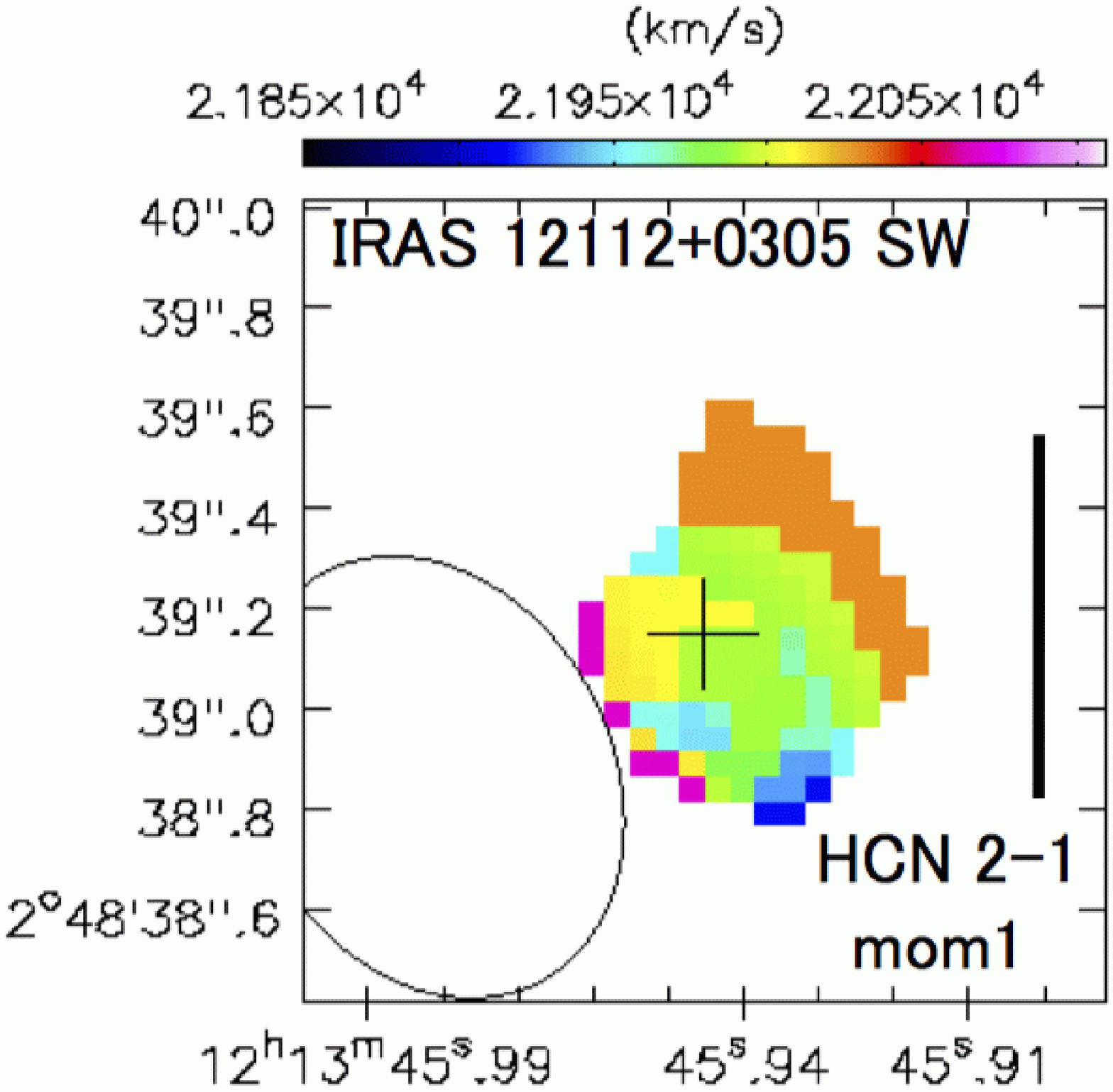} 
\includegraphics[angle=0,scale=.195]{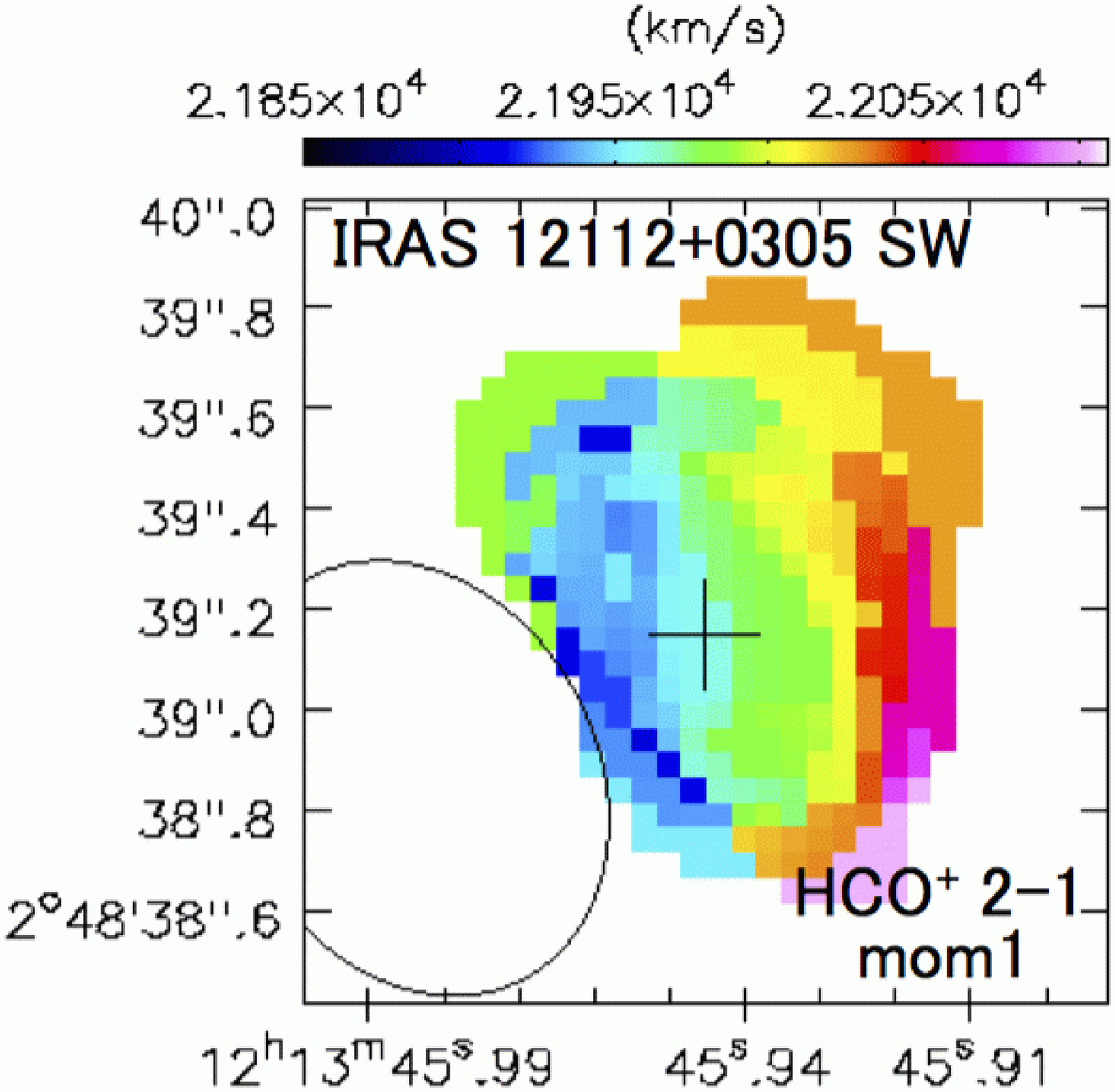} \\
\includegraphics[angle=0,scale=.195]{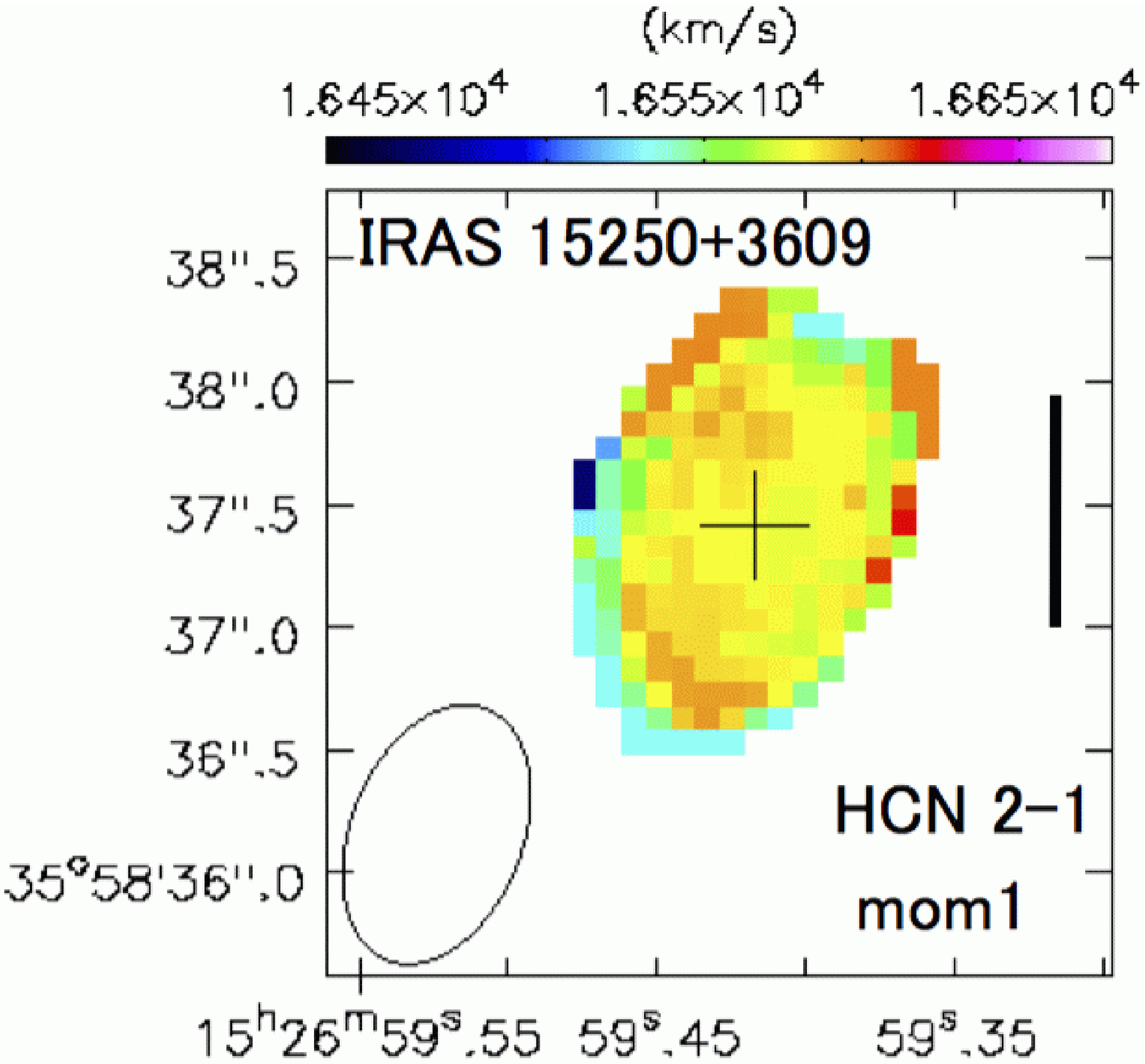} 
\includegraphics[angle=0,scale=.195]{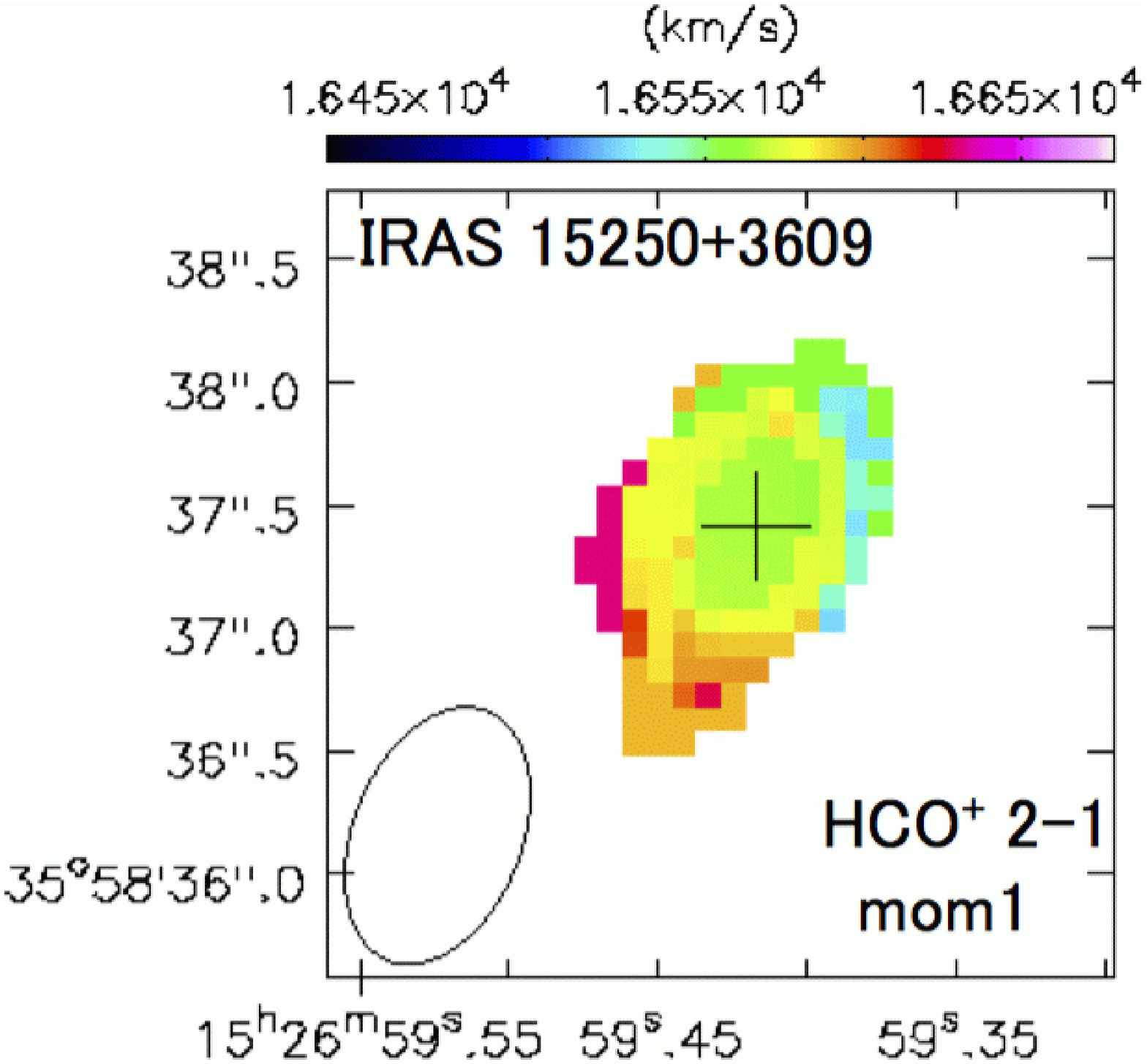} 
\includegraphics[angle=0,scale=.195]{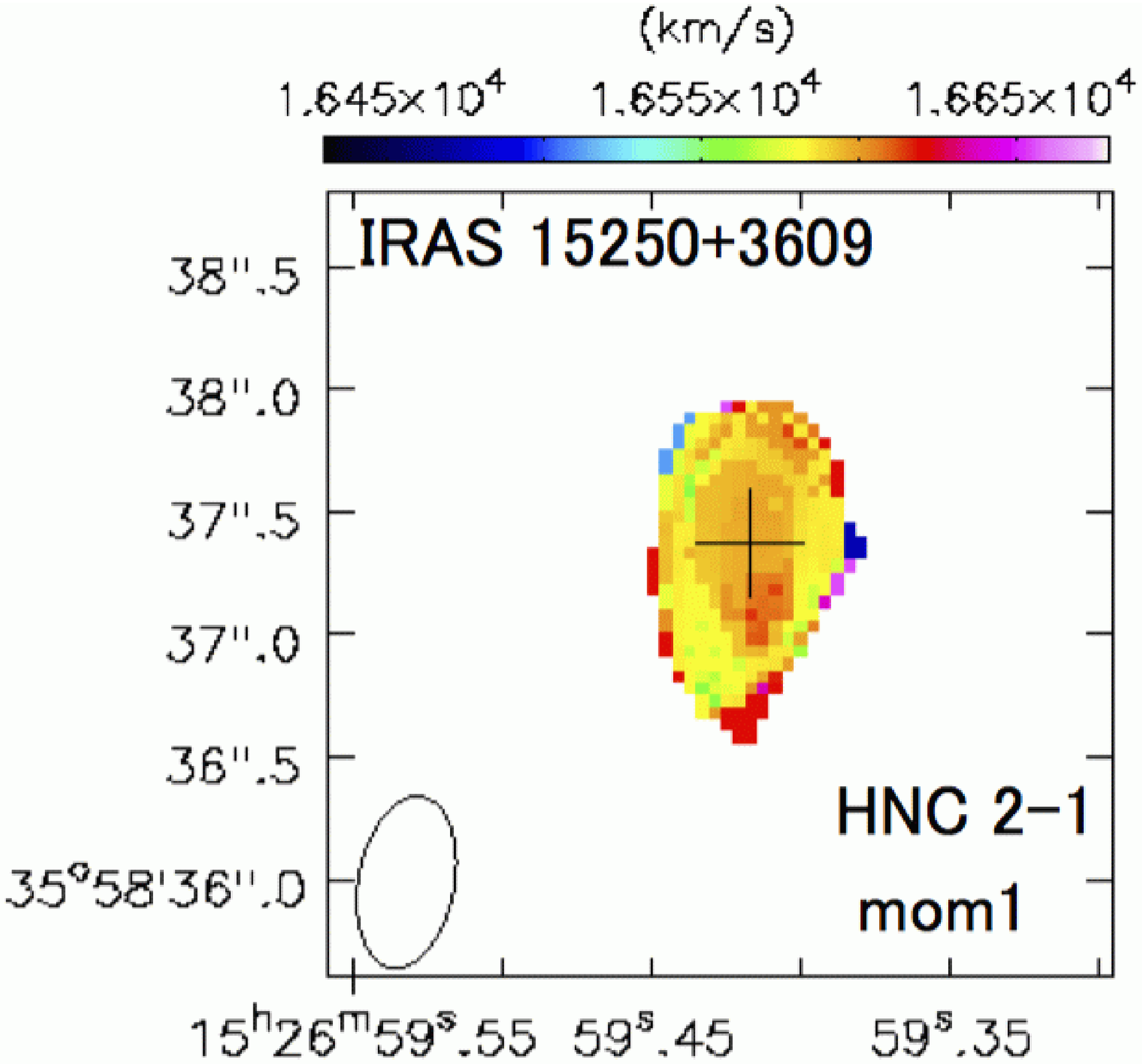} 
\includegraphics[angle=0,scale=.195]{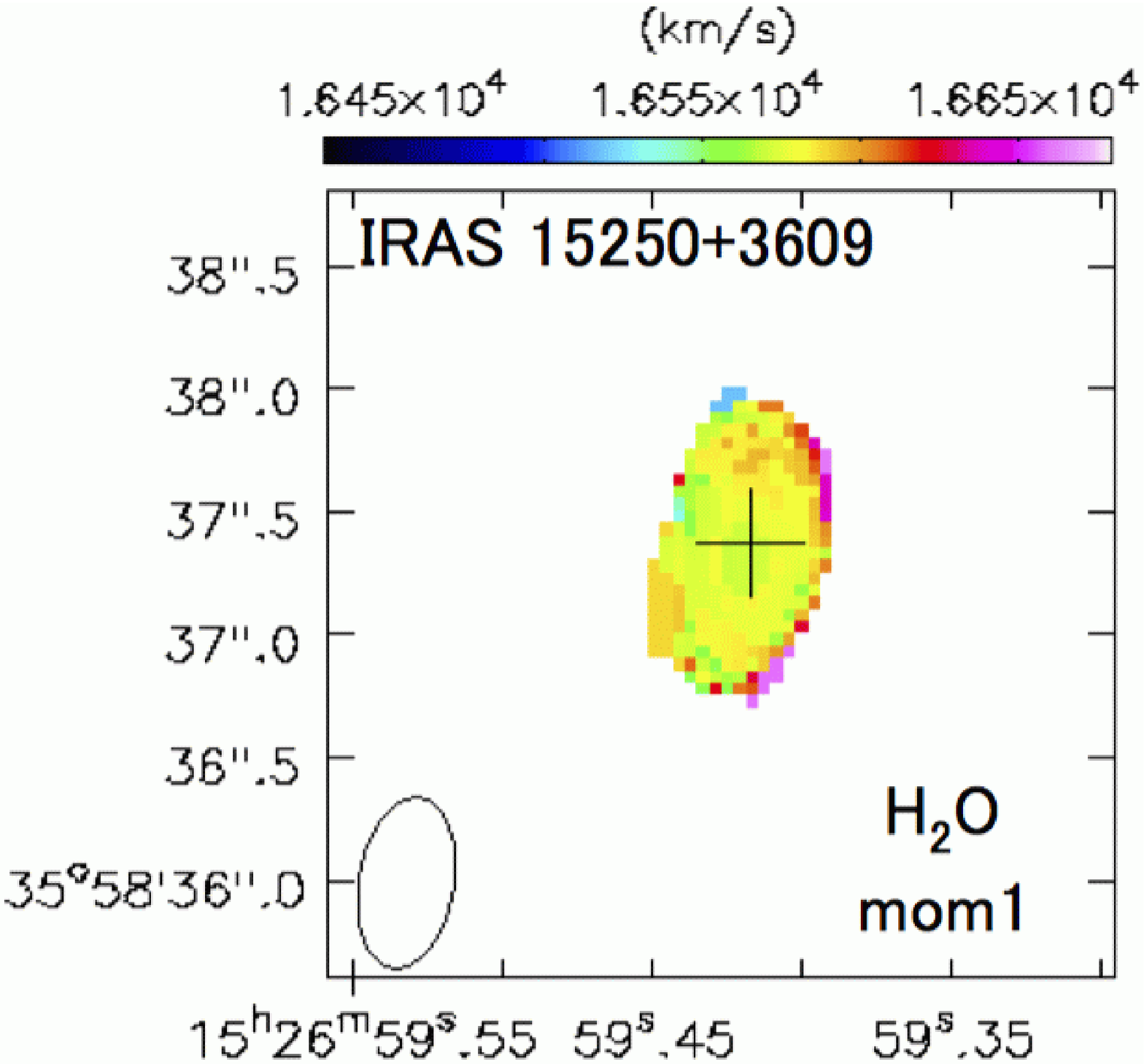} \\
\end{center}
\end{figure}

\clearpage

\begin{figure}
\begin{center}
\includegraphics[angle=0,scale=.195]{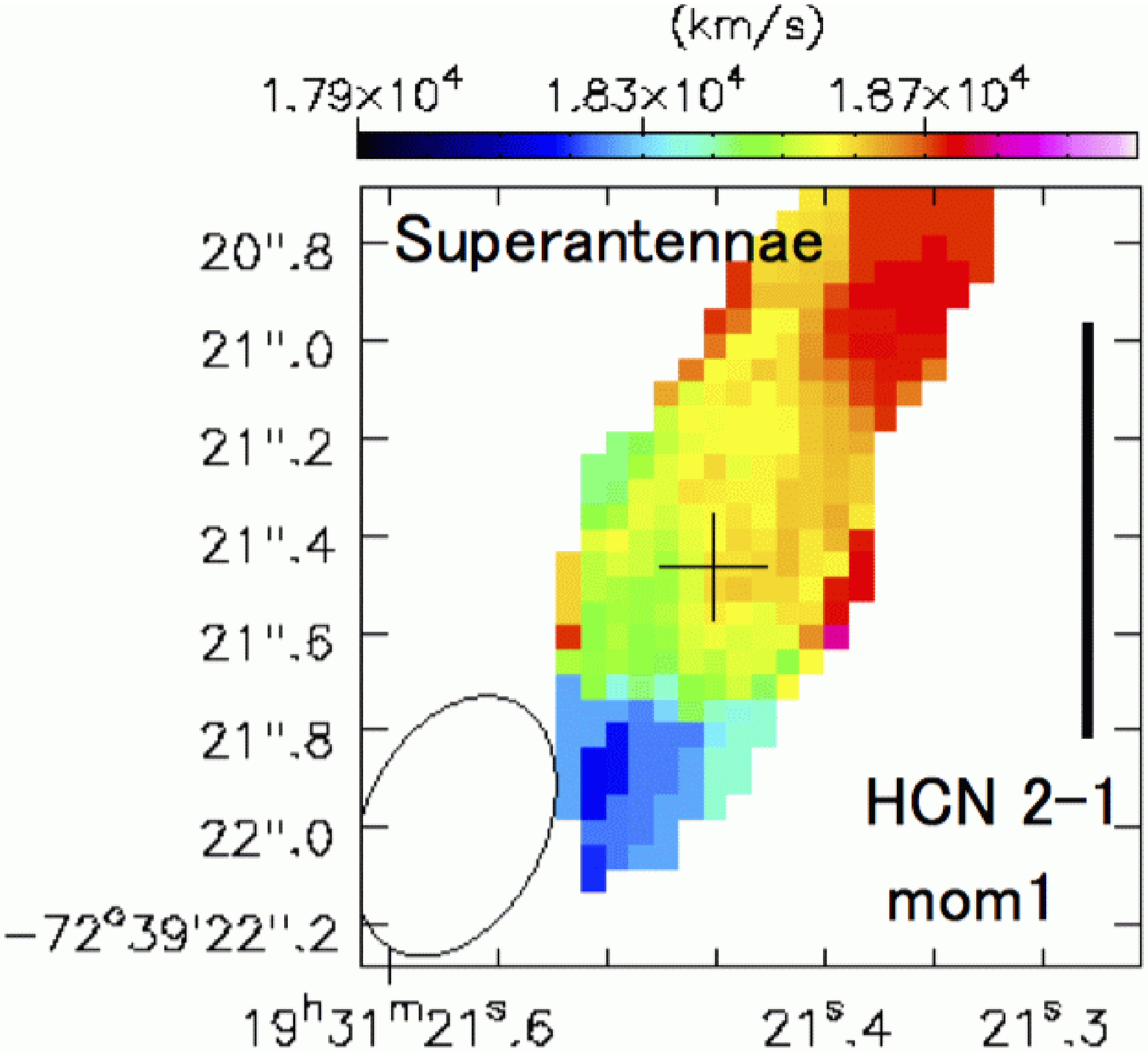} 
\includegraphics[angle=0,scale=.195]{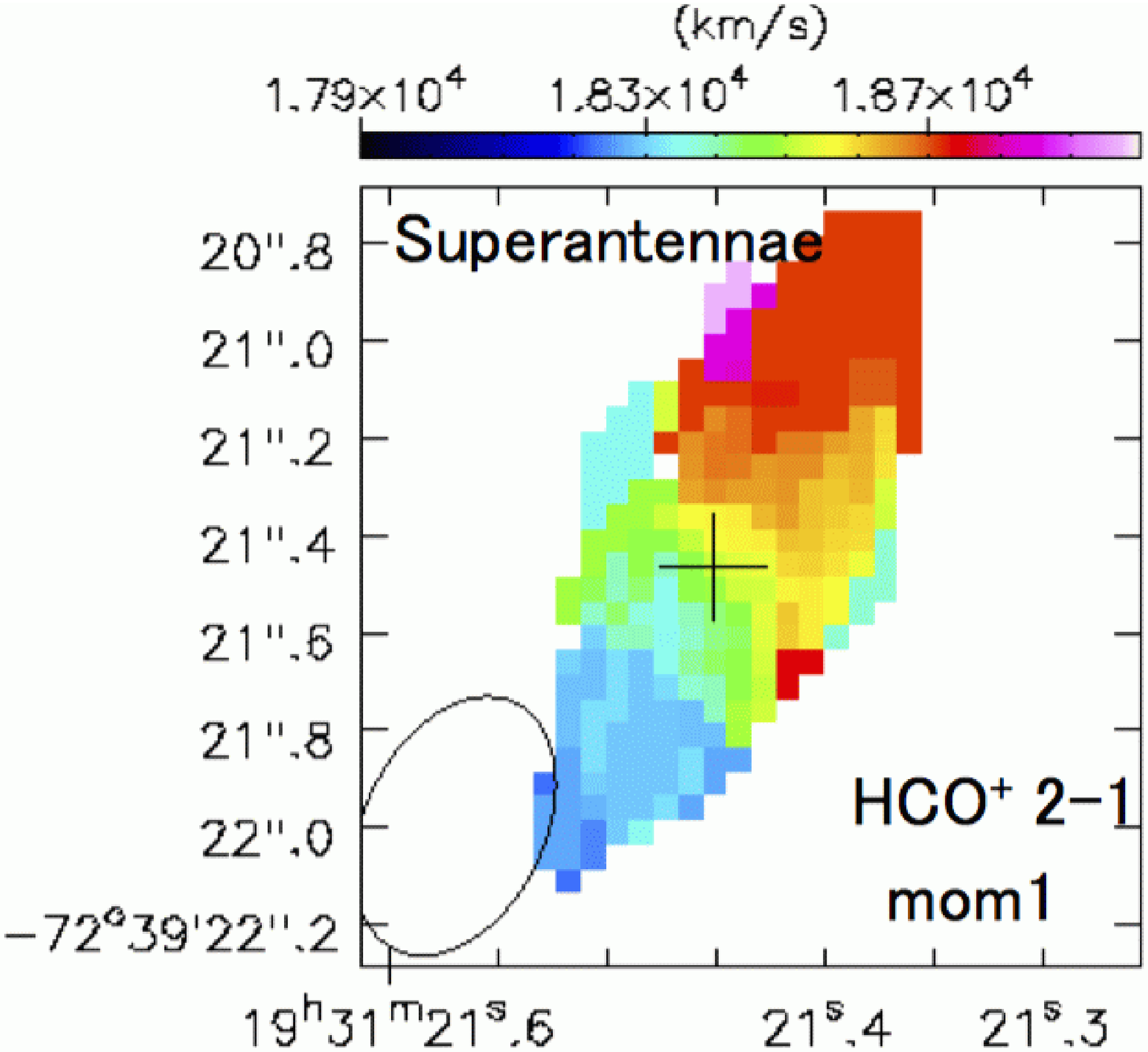} 
\includegraphics[angle=0,scale=.195]{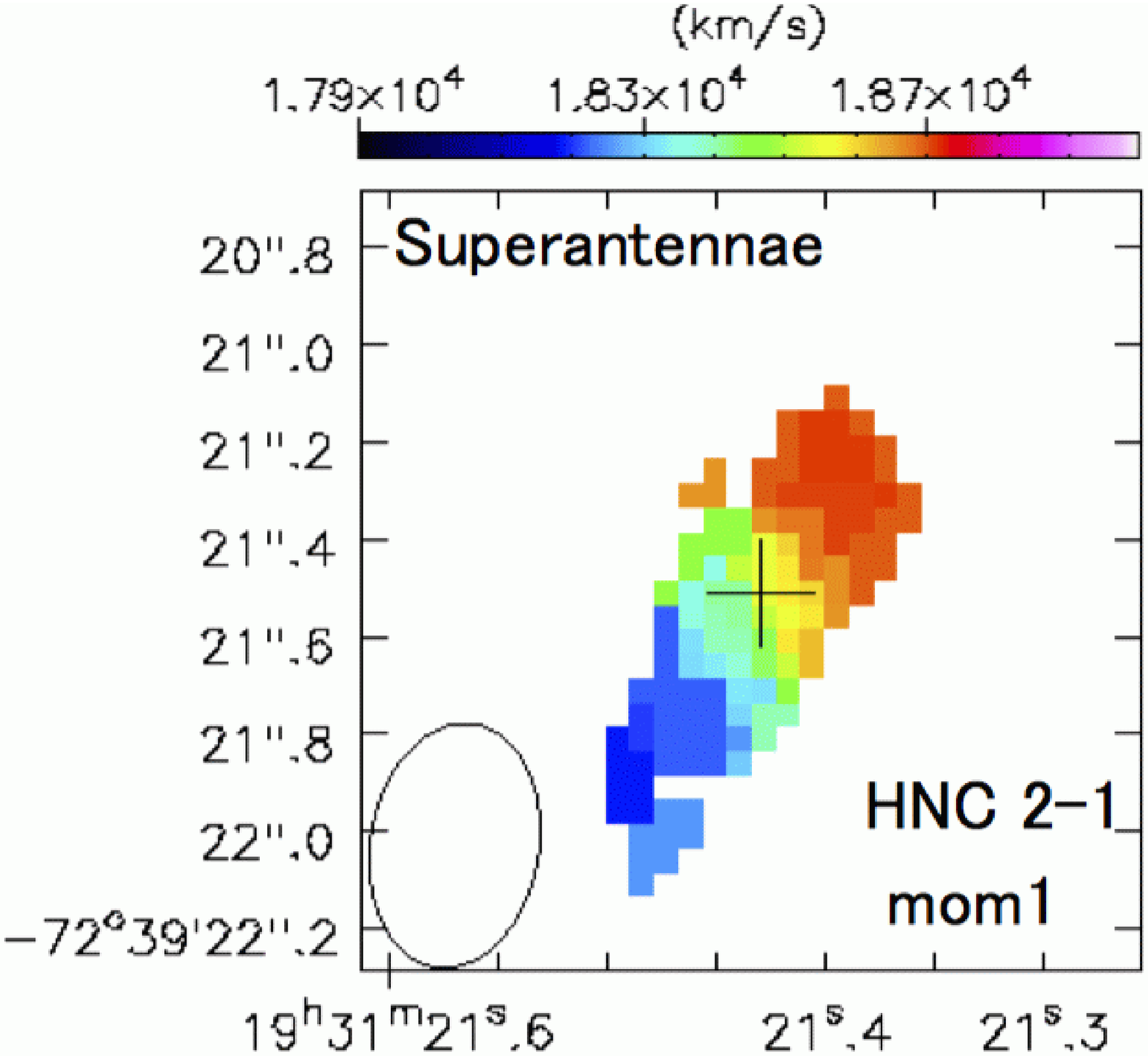} 
\includegraphics[angle=0,scale=.195]{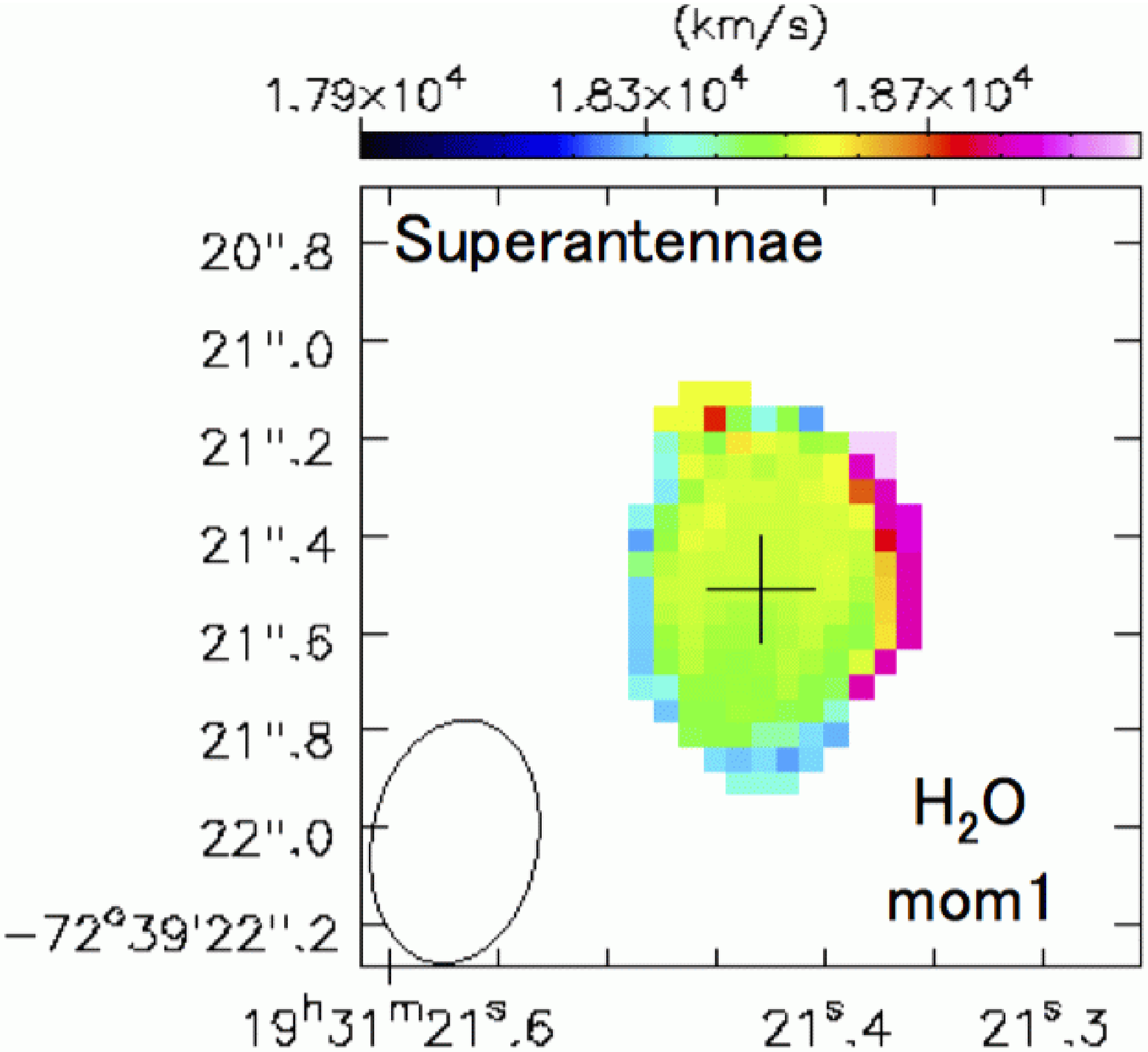} \\
\includegraphics[angle=0,scale=.19]{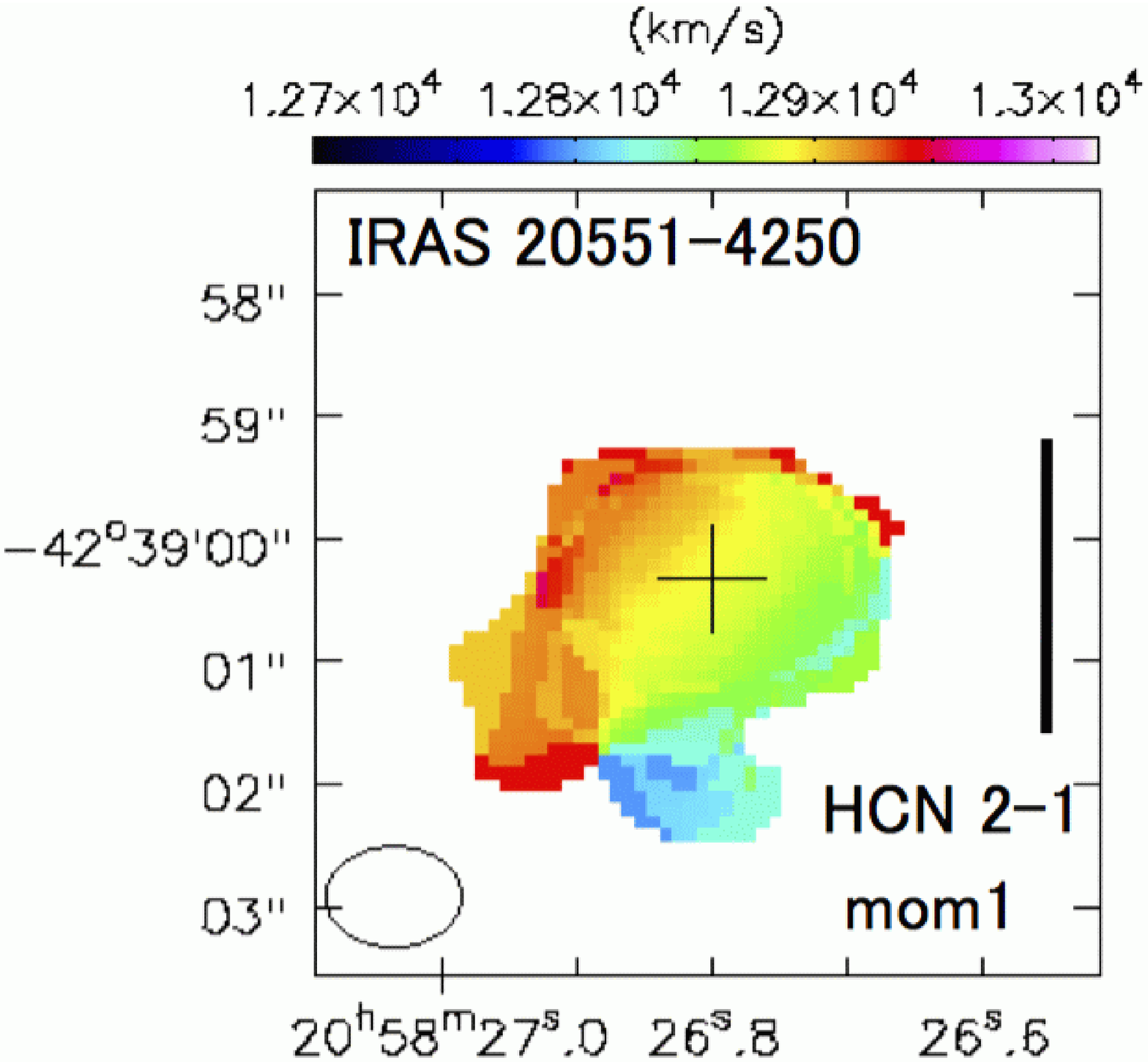} 
\includegraphics[angle=0,scale=.19]{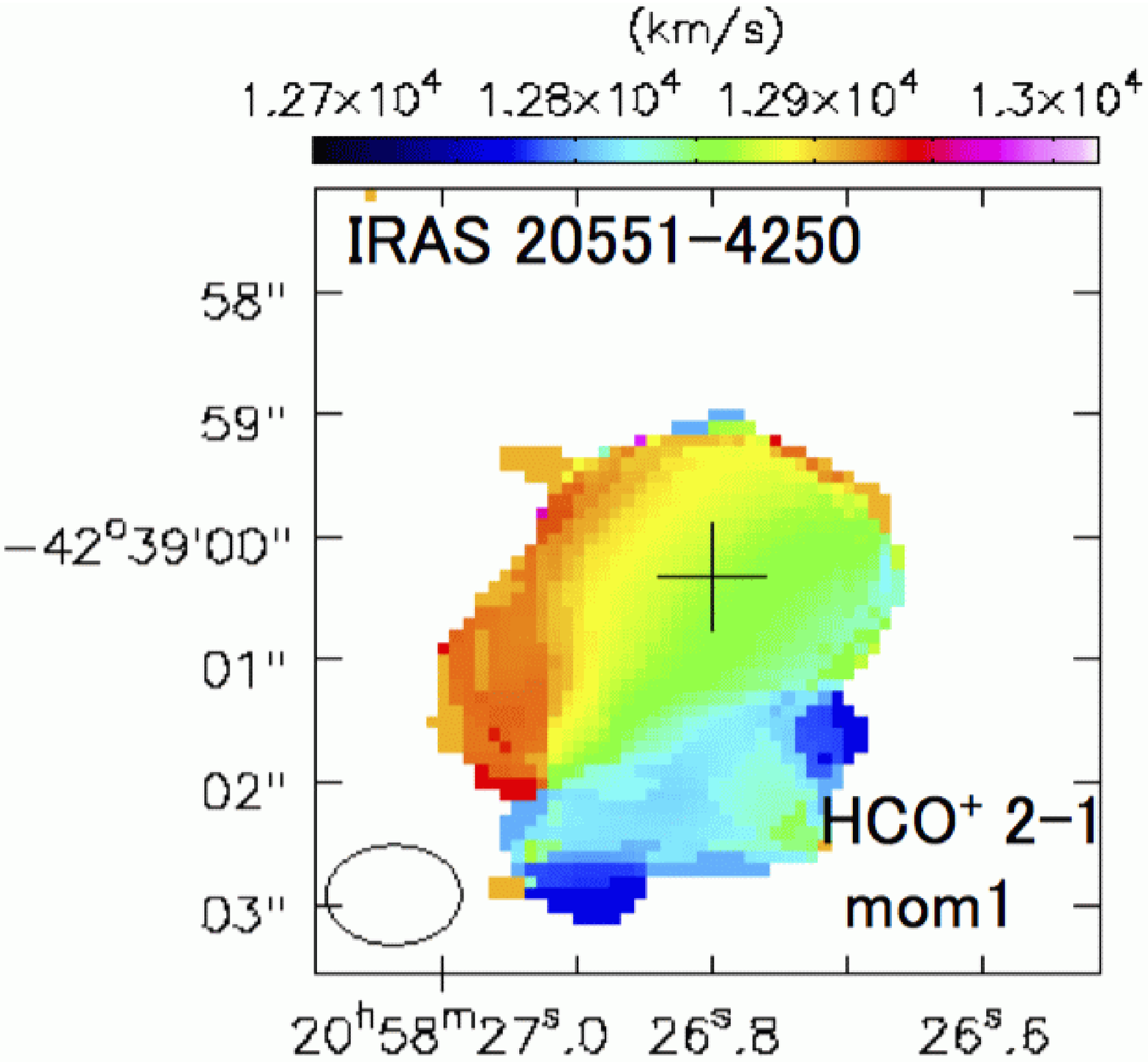} 
\includegraphics[angle=0,scale=.19]{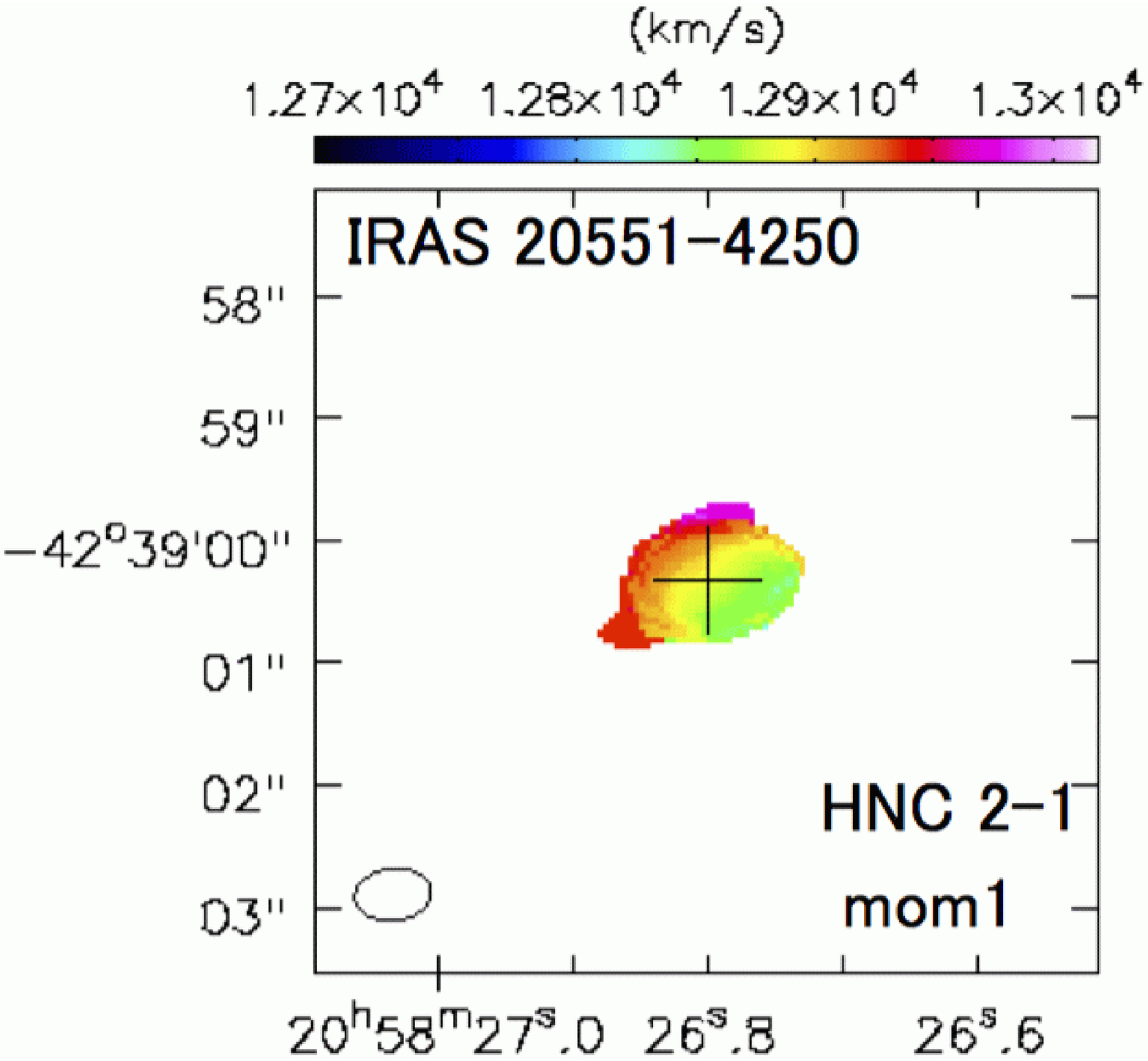} 
\includegraphics[angle=0,scale=.19]{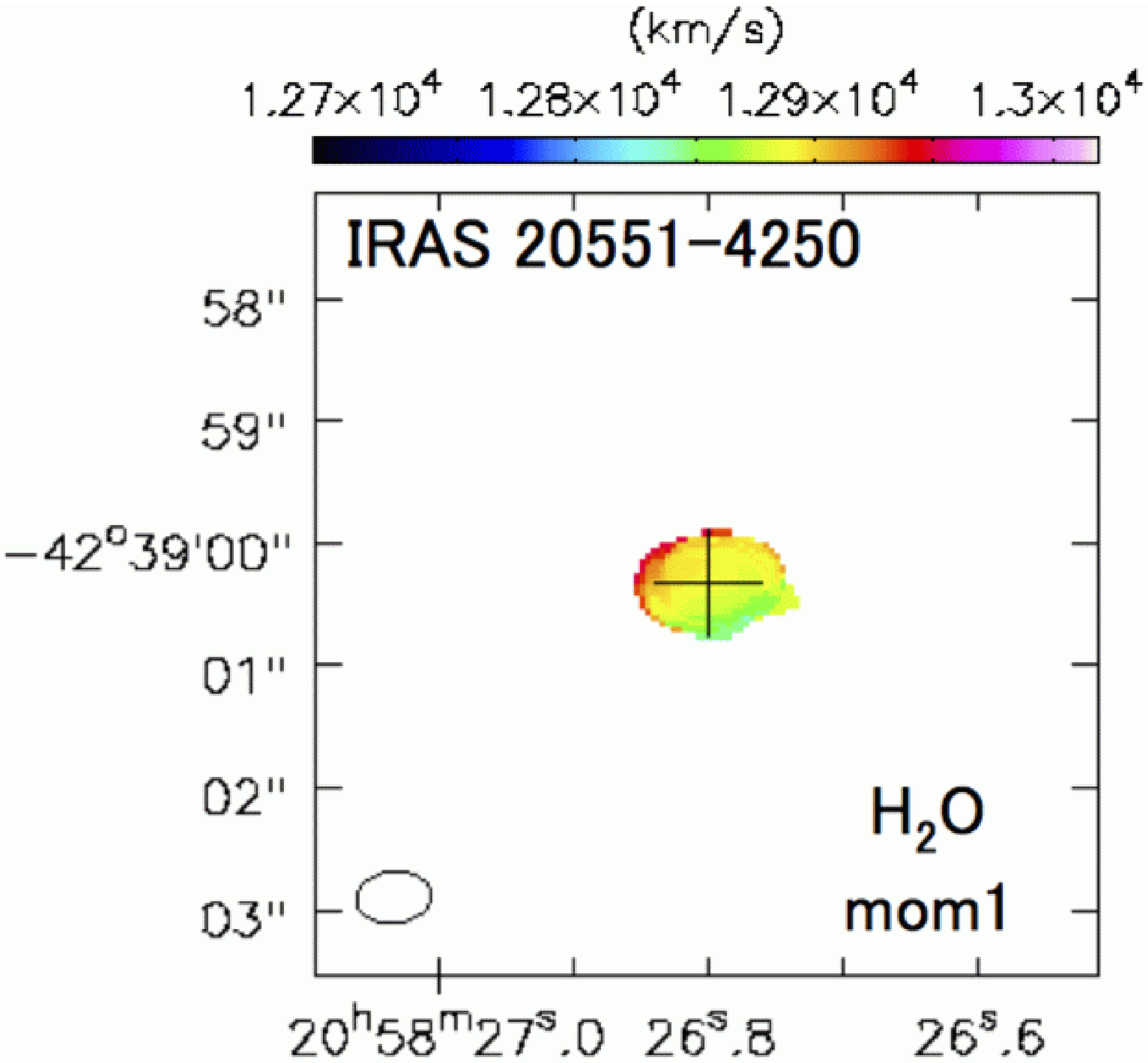} \\
\includegraphics[angle=0,scale=.195]{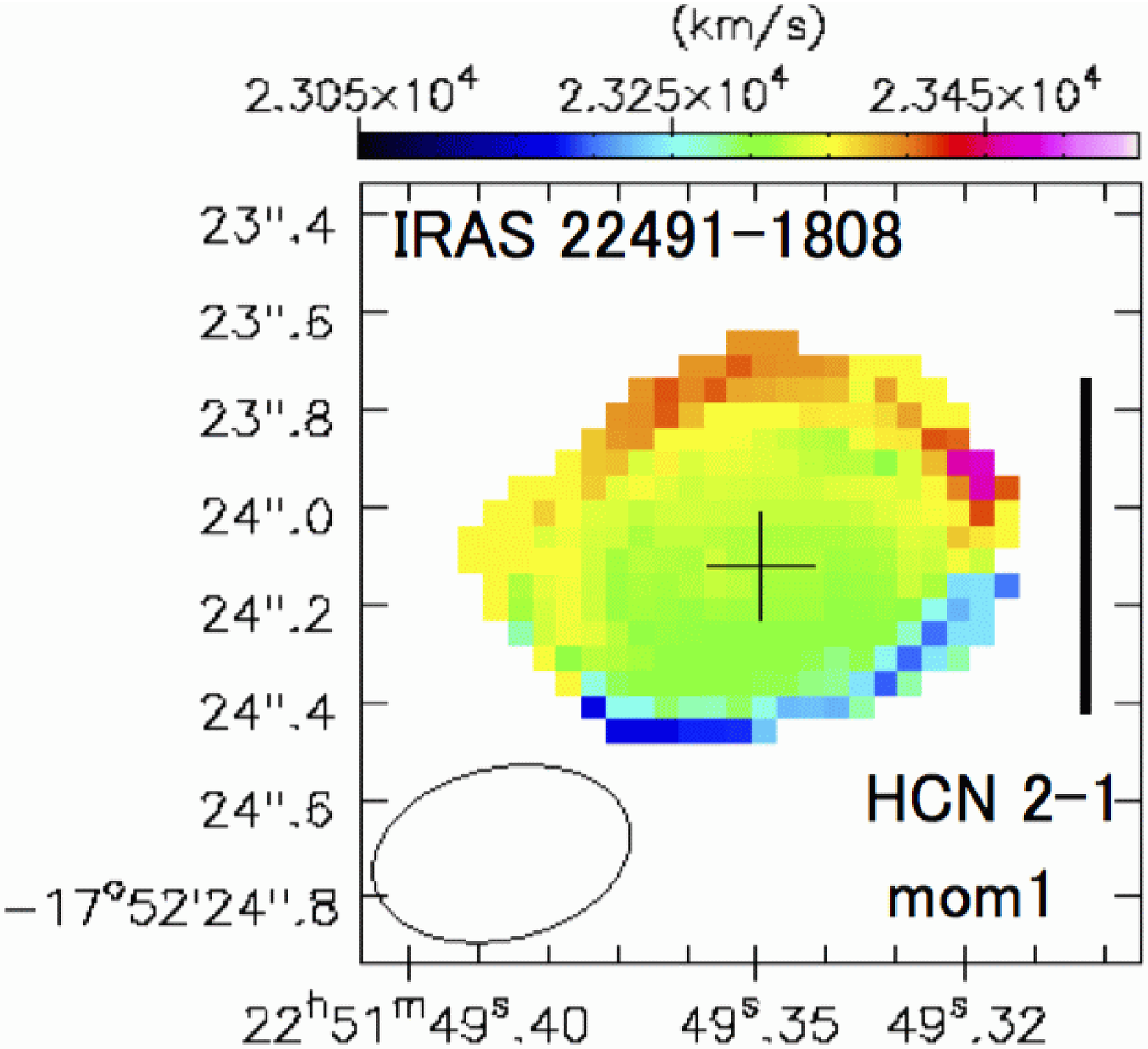} 
\includegraphics[angle=0,scale=.195]{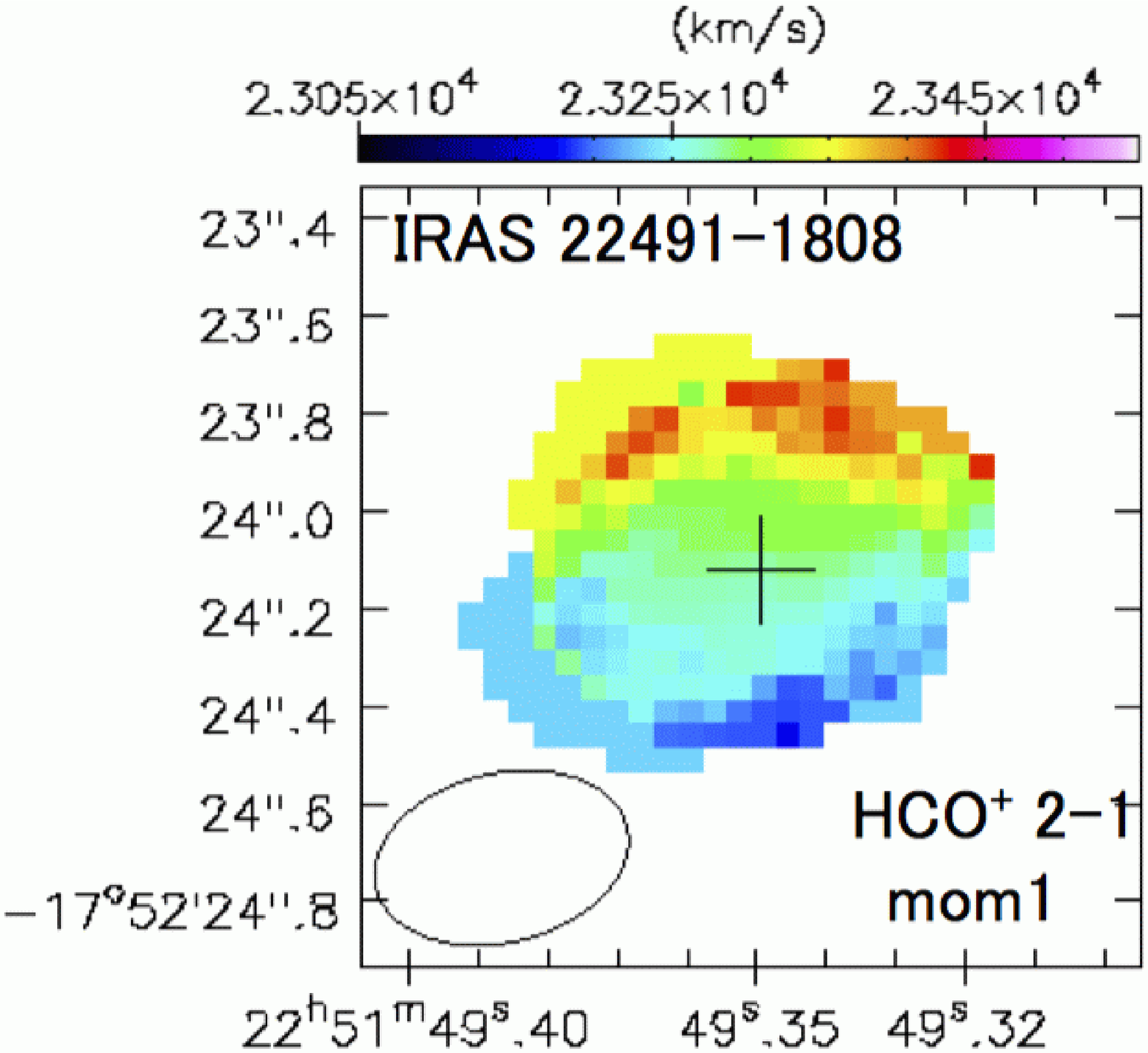} 
\includegraphics[angle=0,scale=.195]{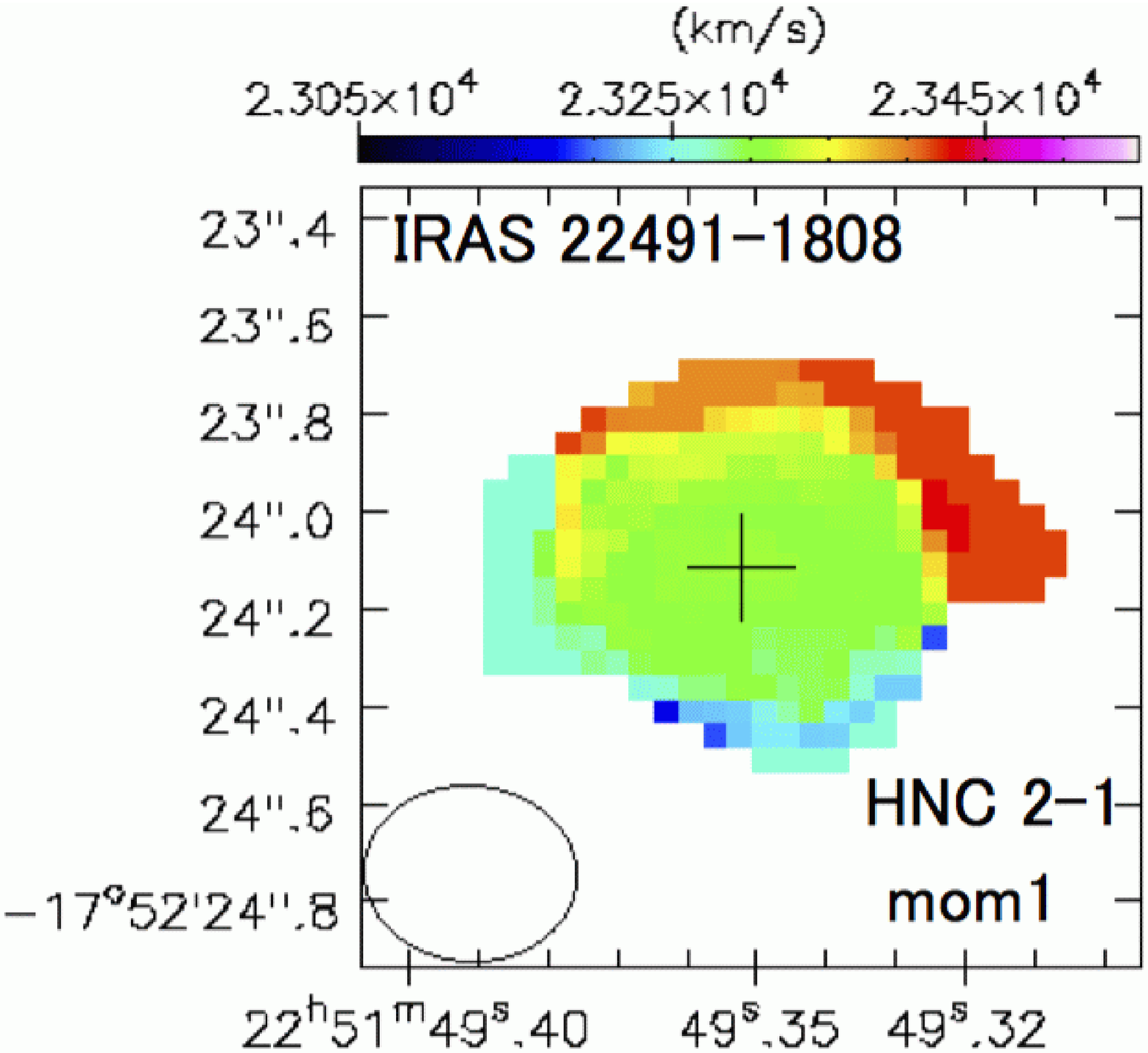} 
\includegraphics[angle=0,scale=.195]{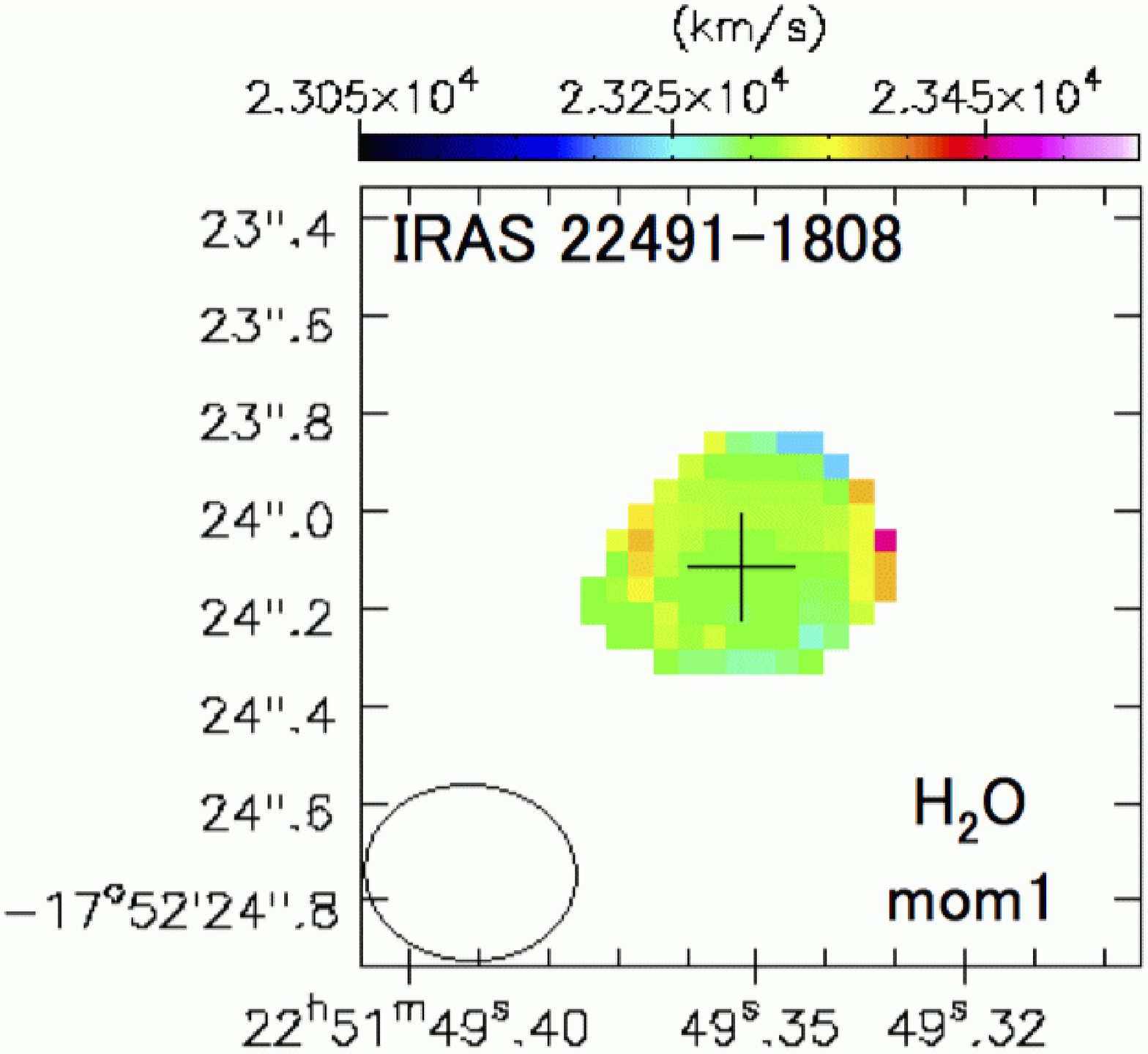} \\
\end{center}
\vspace{0.5cm}
\caption{
Intensity-weighted mean velocity (moment 1) maps of the HCN J=2--1 
{\it (Left)}, 
HCO$^{+}$ J=2--1 {\it (Second left)}, HNC J=2--1 {\it (Second right)}, 
and 183 GHz H$_{2}$O {\it (Right)} lines for sources with sufficiently 
bright emission.
Continuum peak position is shown as a cross.
The length of the vertical black solid bar at the right side of 
HCN J=2--1 data {\it (Left)} corresponds to 1 kpc.
IRAS 12127$-$1412 and IRAS 13509$+$0442 are excluded because all the four 
molecular emission lines are not sufficiently bright to obtain 
meaningful velocity information.
The maps of the 183 GHz H$_{2}$O line of NGC 1614, and HNC J=2--1 and 
183 GHz H$_{2}$O lines of IRAS 12112$+$0305 SW, are not shown 
because these lines are faint.
An appropriate cutoff ($\sim$2$\sigma$) is applied when we make 
these moment 1 maps to prevent them from being dominated by noise. 
}
\end{figure}

\begin{figure}
\begin{center}
\hspace*{-4.7cm}
\includegraphics[angle=0,scale=.195]{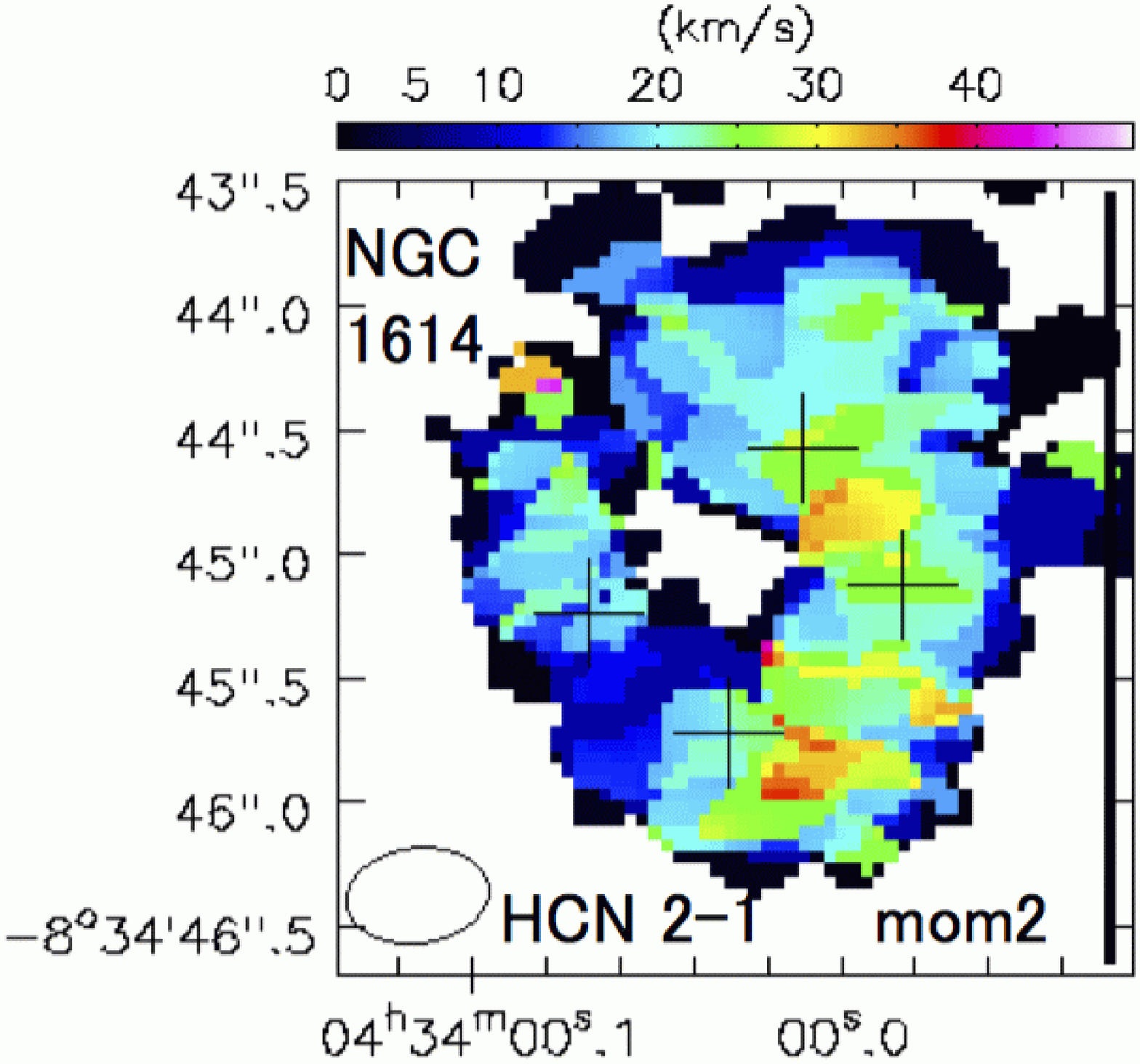} 
\includegraphics[angle=0,scale=.195]{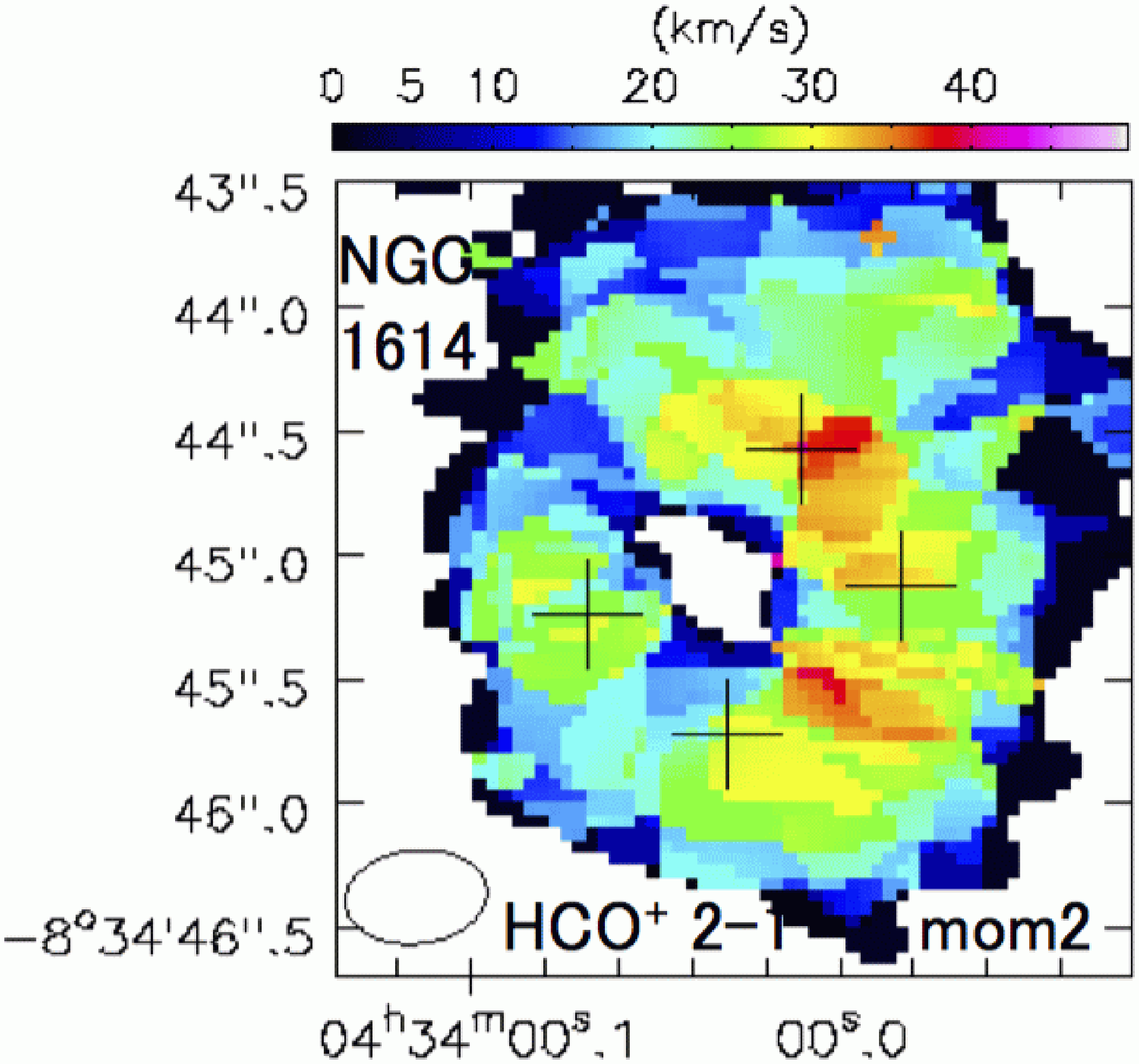} 
\includegraphics[angle=0,scale=.195]{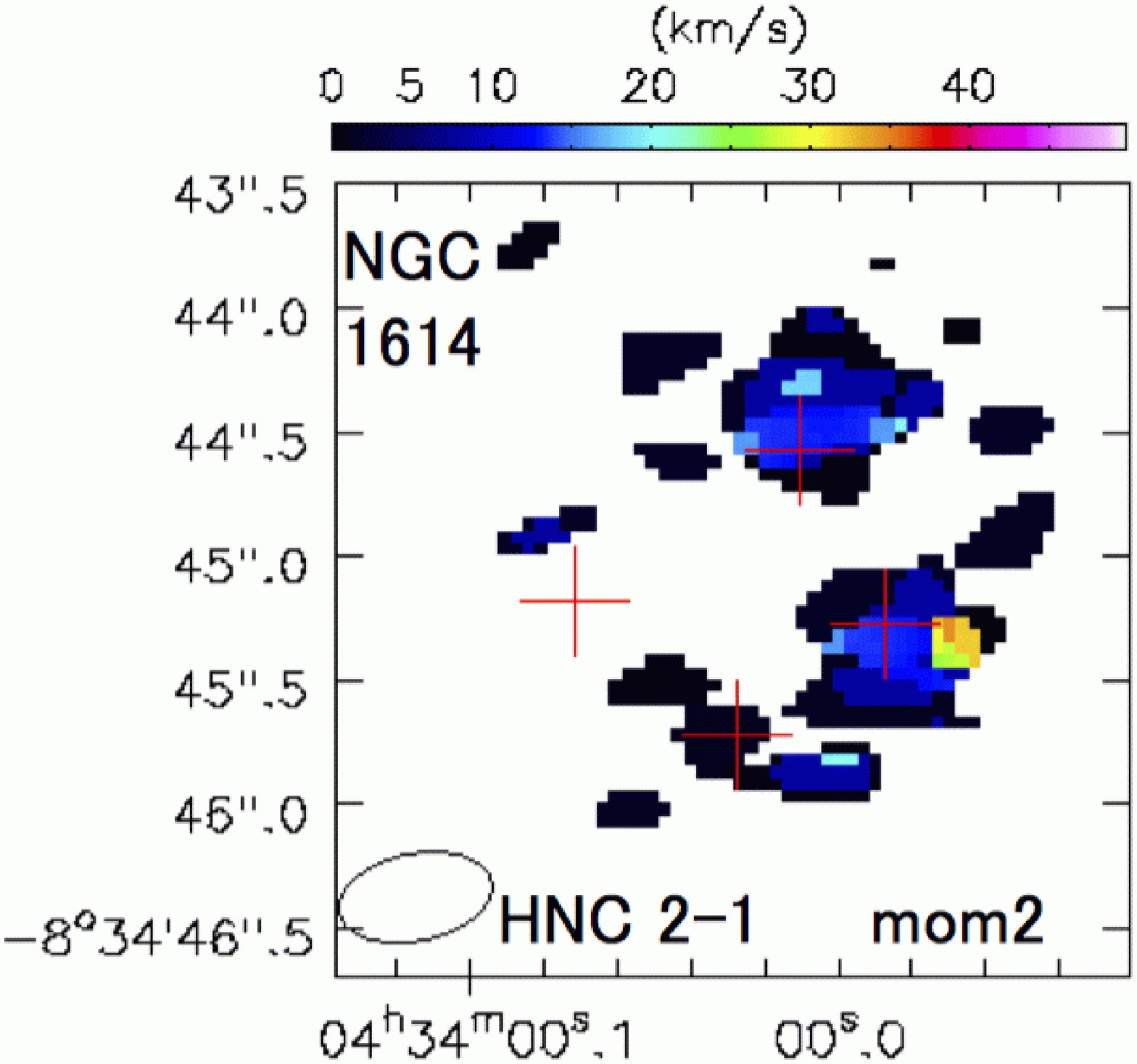} \\
\includegraphics[angle=0,scale=.195]{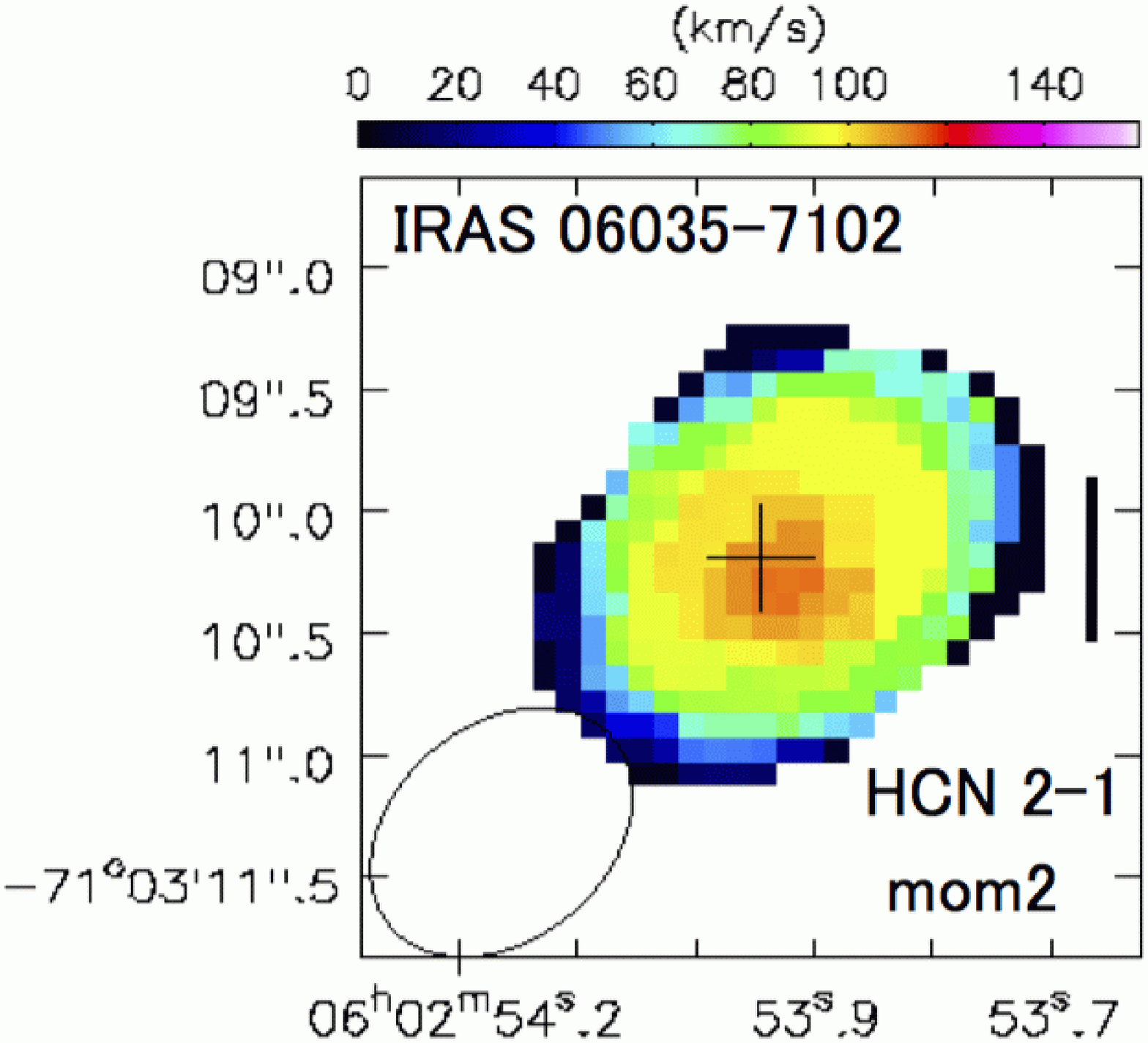} 
\includegraphics[angle=0,scale=.195]{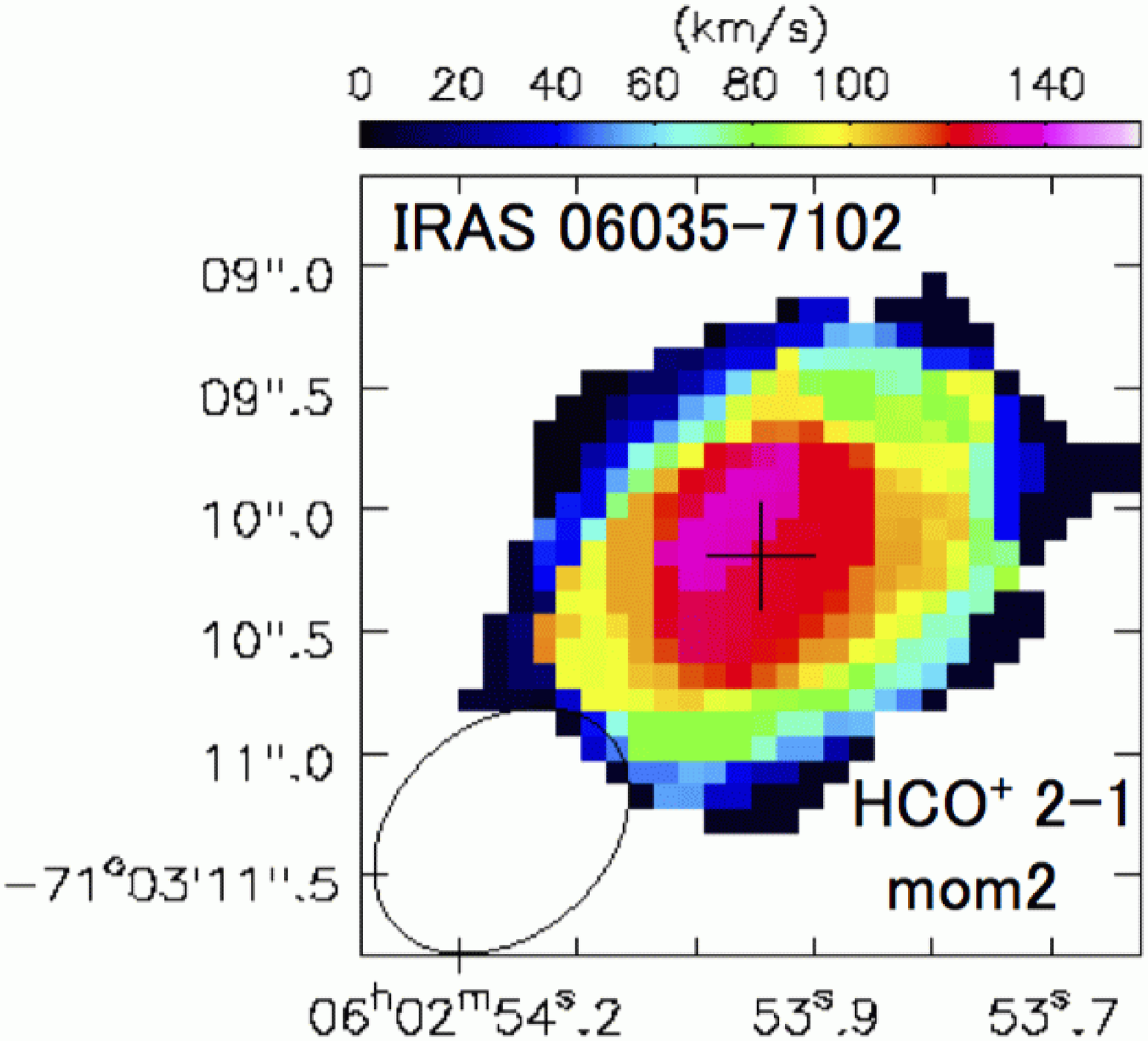} 
\includegraphics[angle=0,scale=.195]{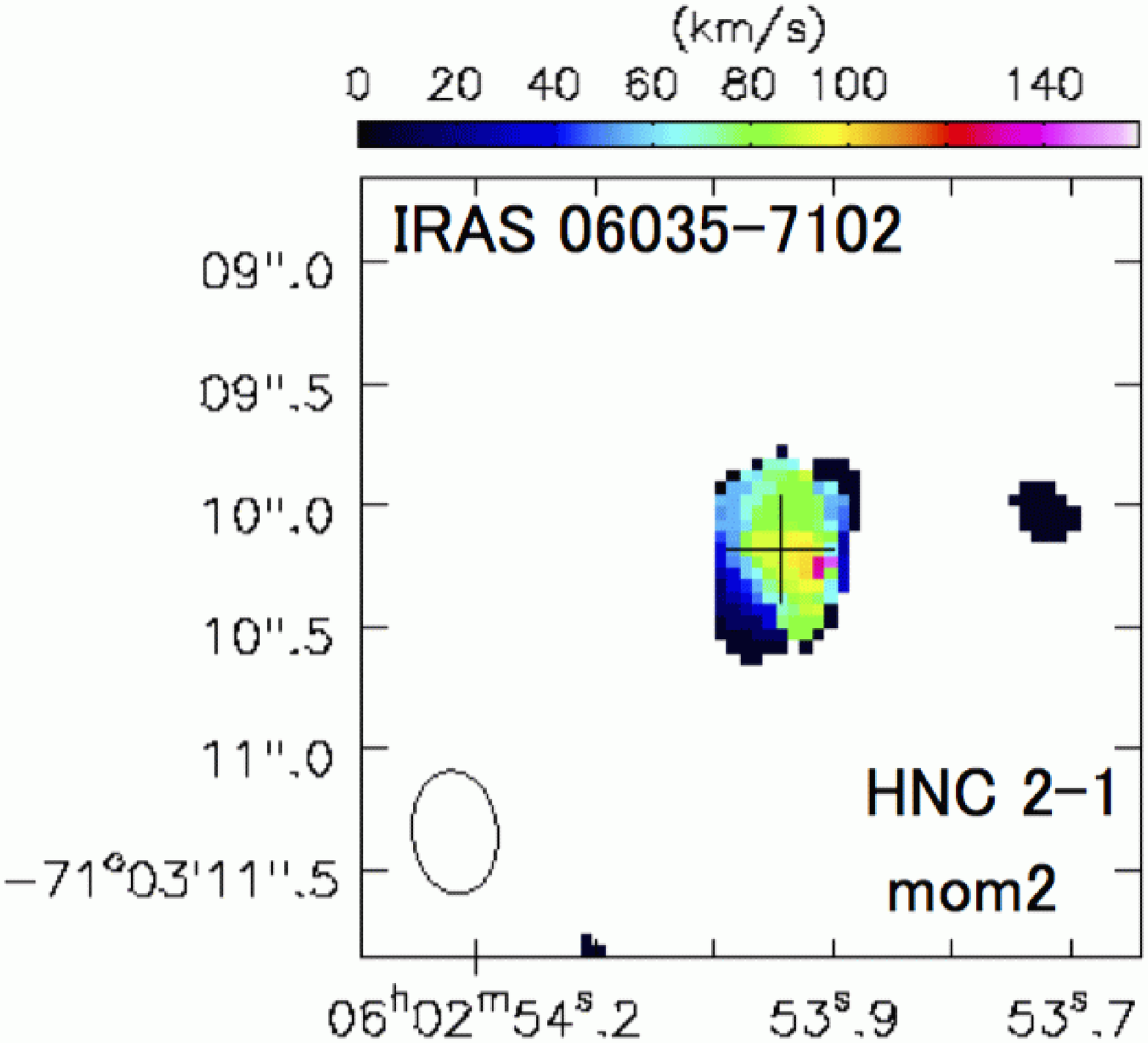} 
\includegraphics[angle=0,scale=.195]{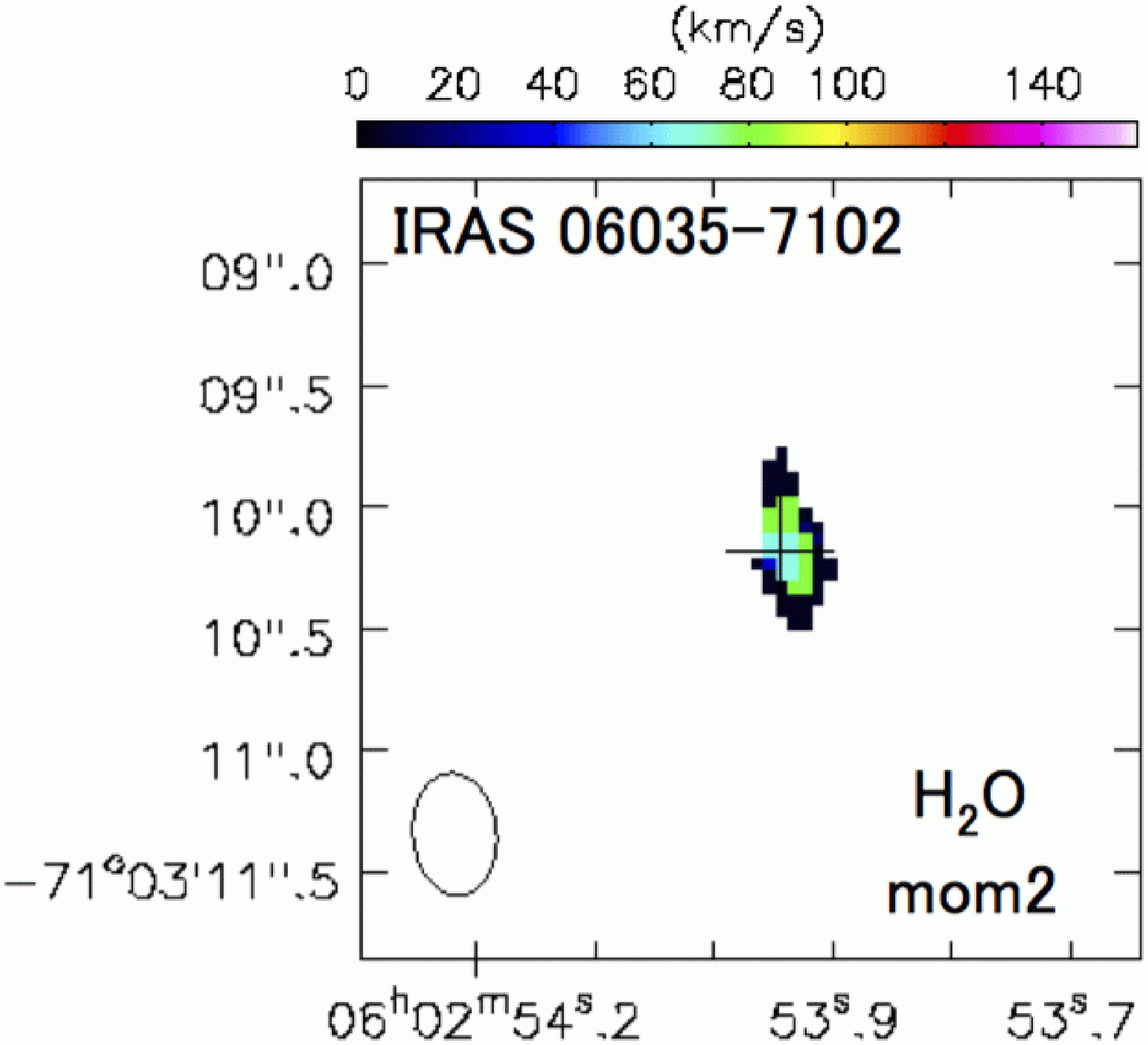} \\
\includegraphics[angle=0,scale=.19]{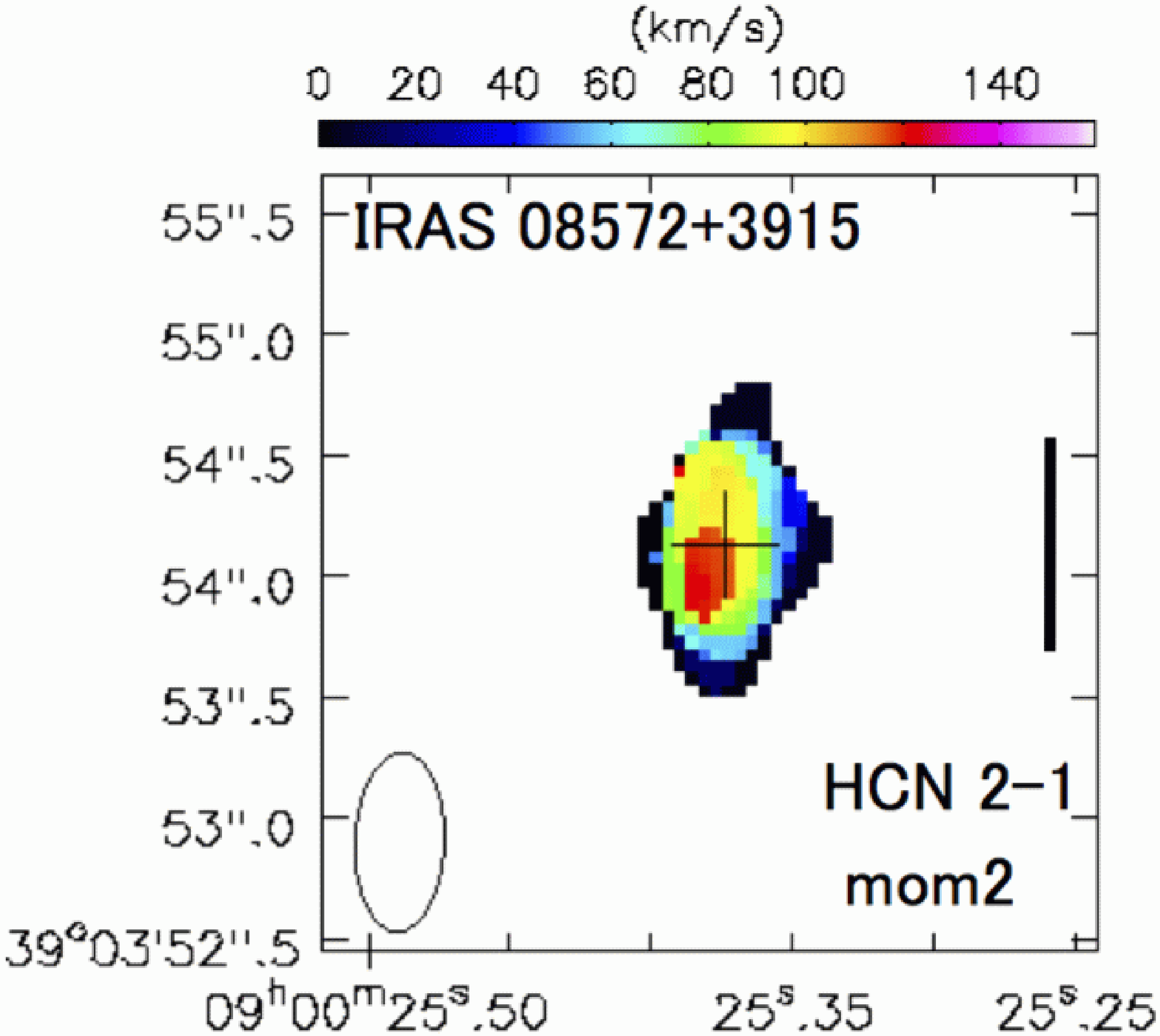} 
\includegraphics[angle=0,scale=.19]{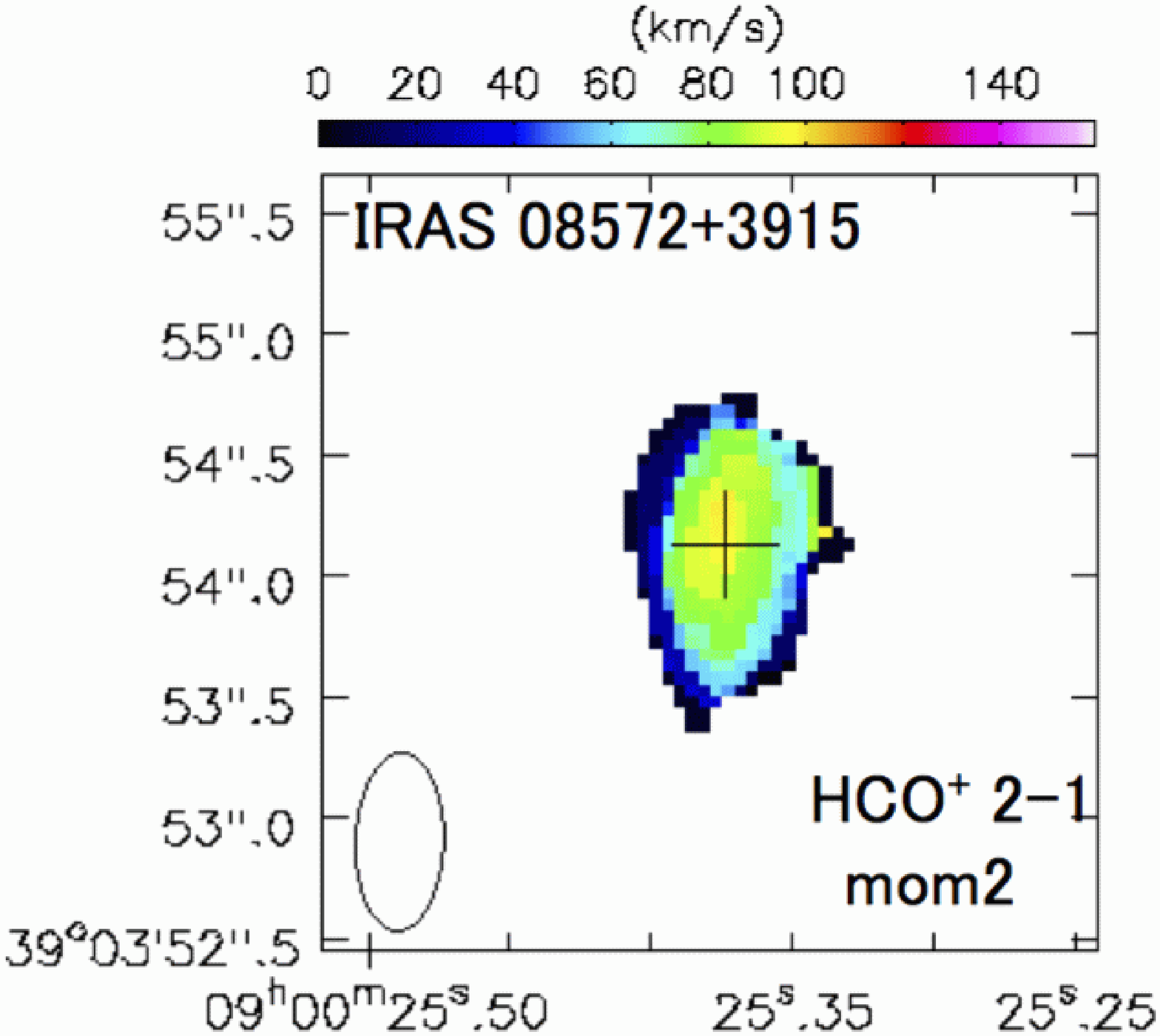} 
\includegraphics[angle=0,scale=.19]{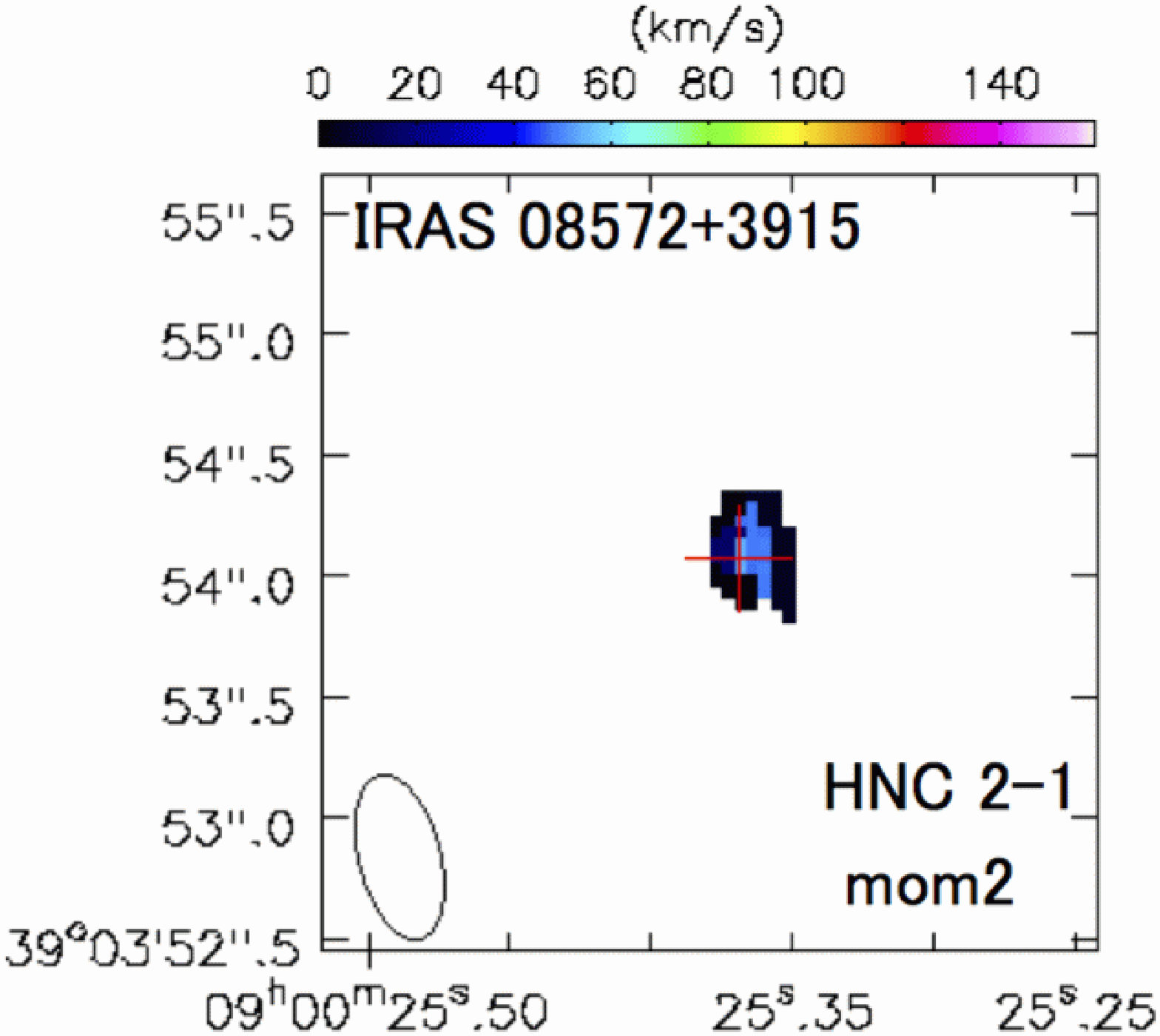} 
\includegraphics[angle=0,scale=.19]{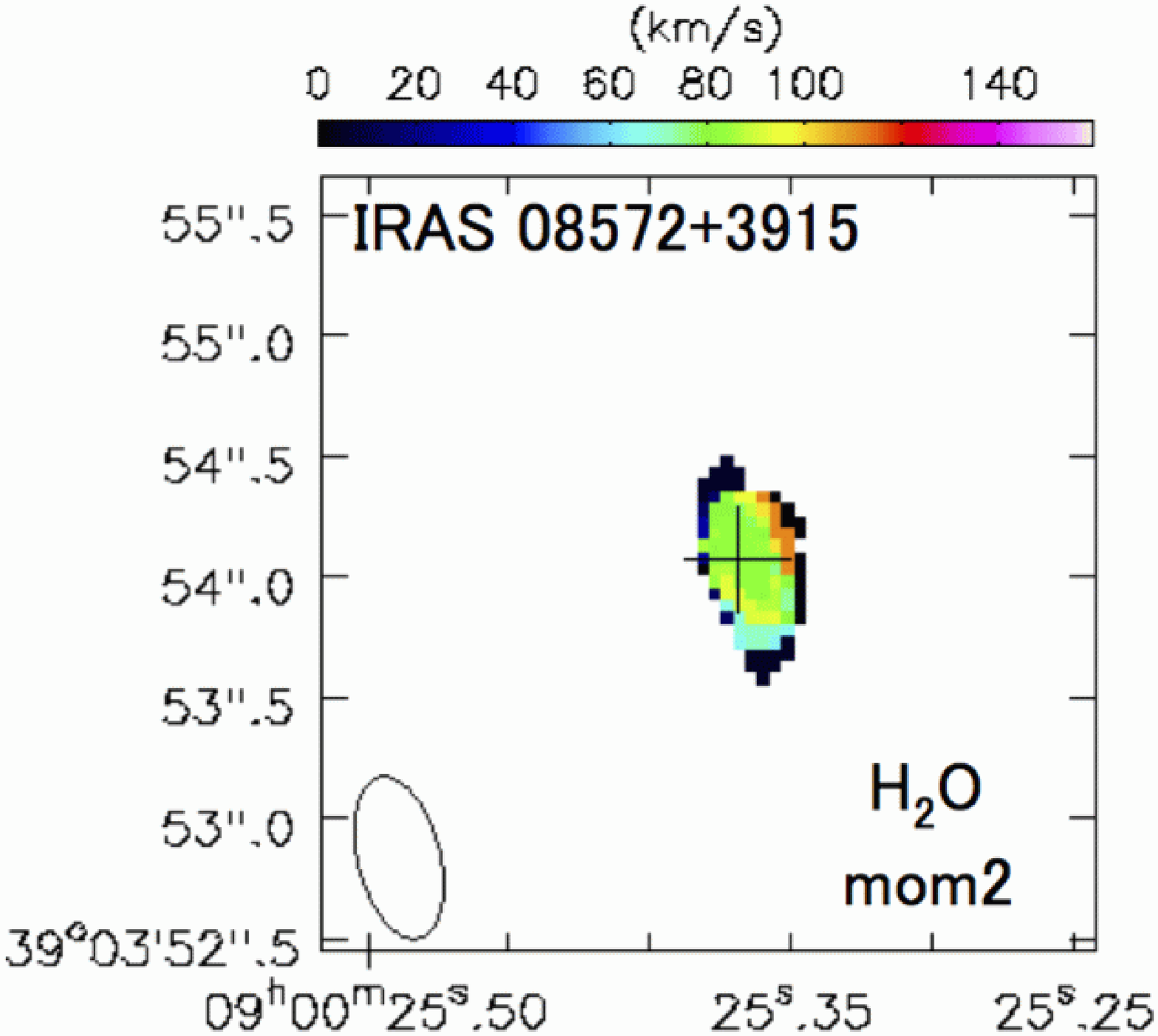} \\
\includegraphics[angle=0,scale=.195]{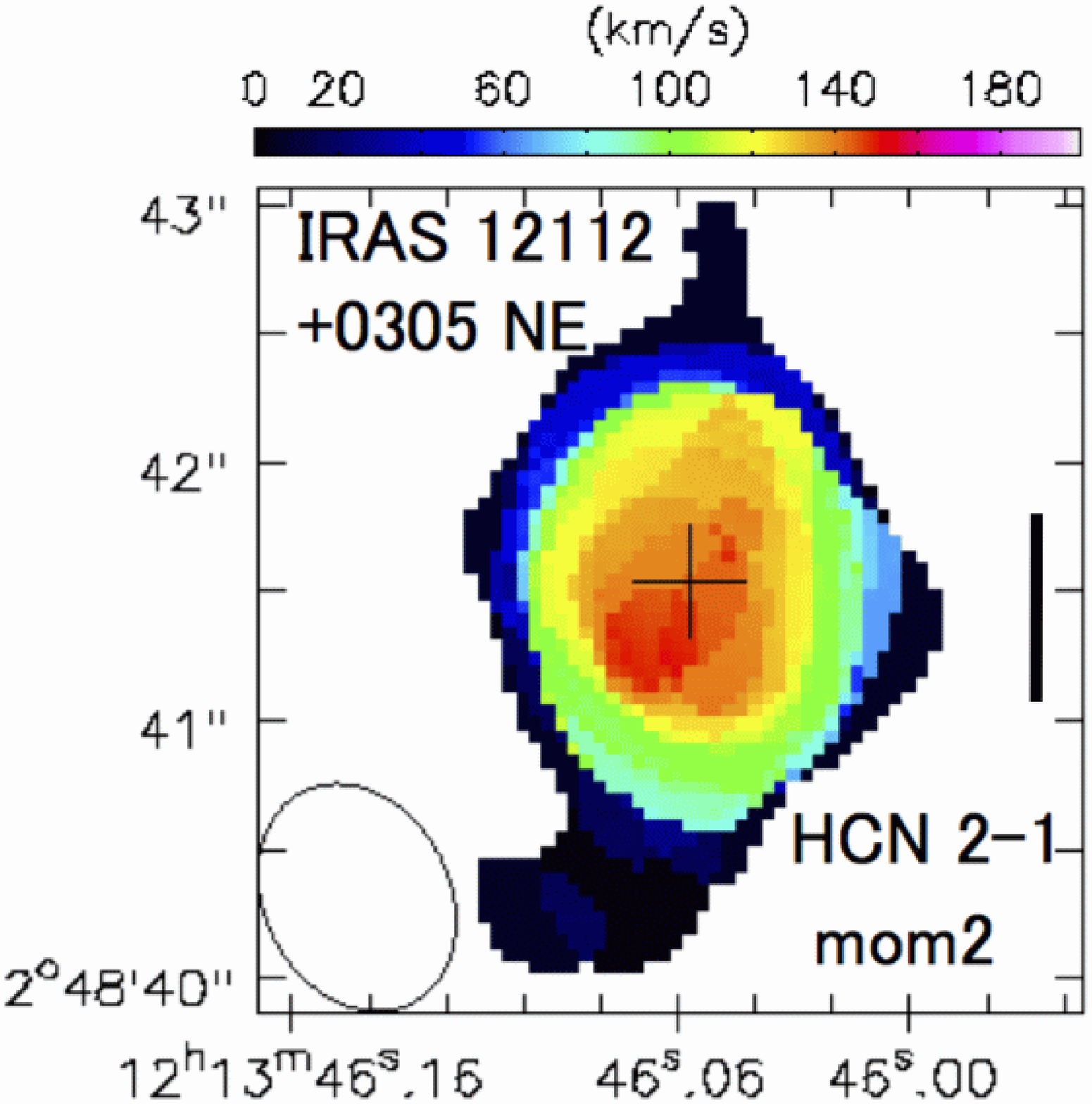} 
\includegraphics[angle=0,scale=.195]{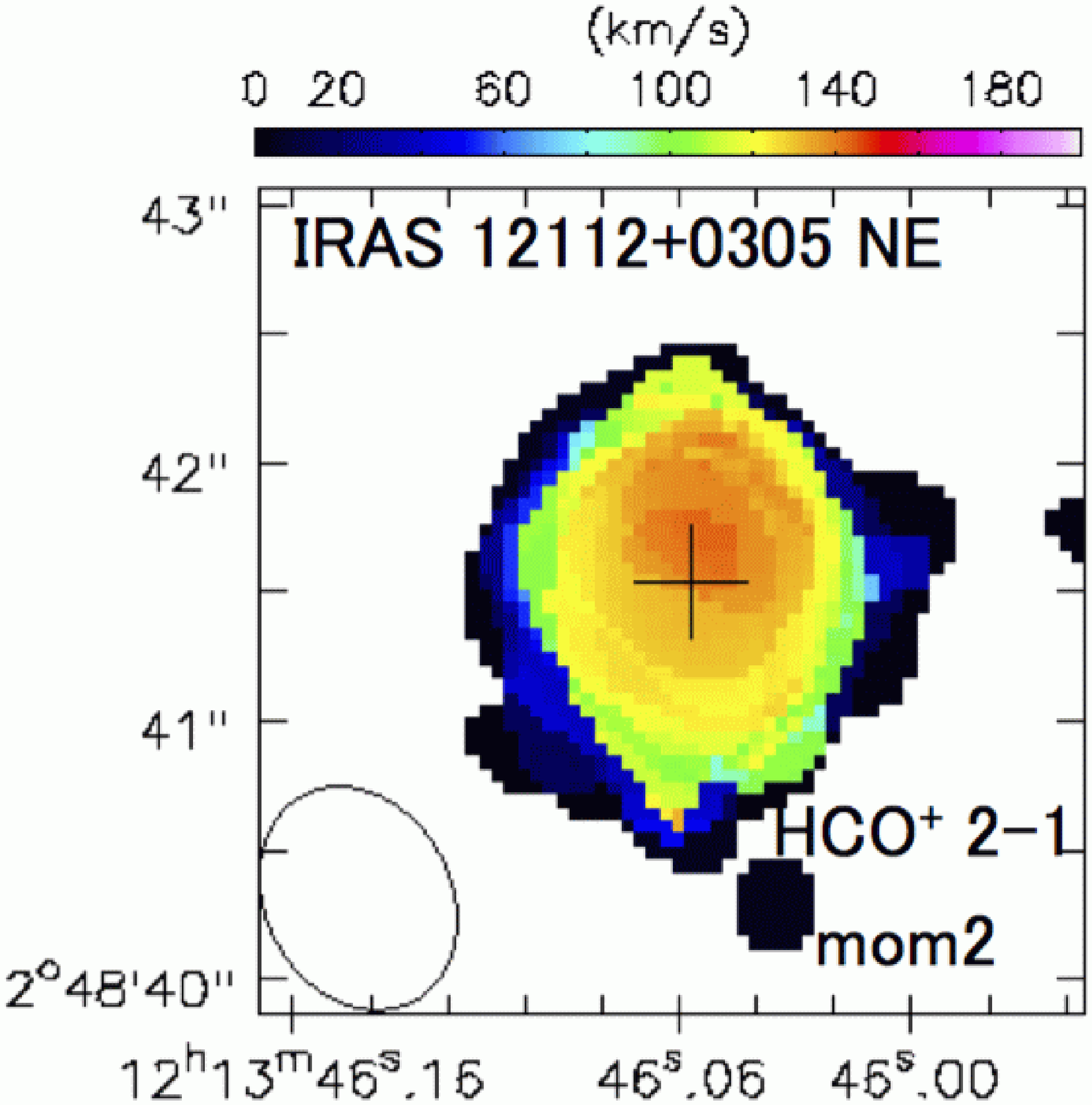} 
\includegraphics[angle=0,scale=.195]{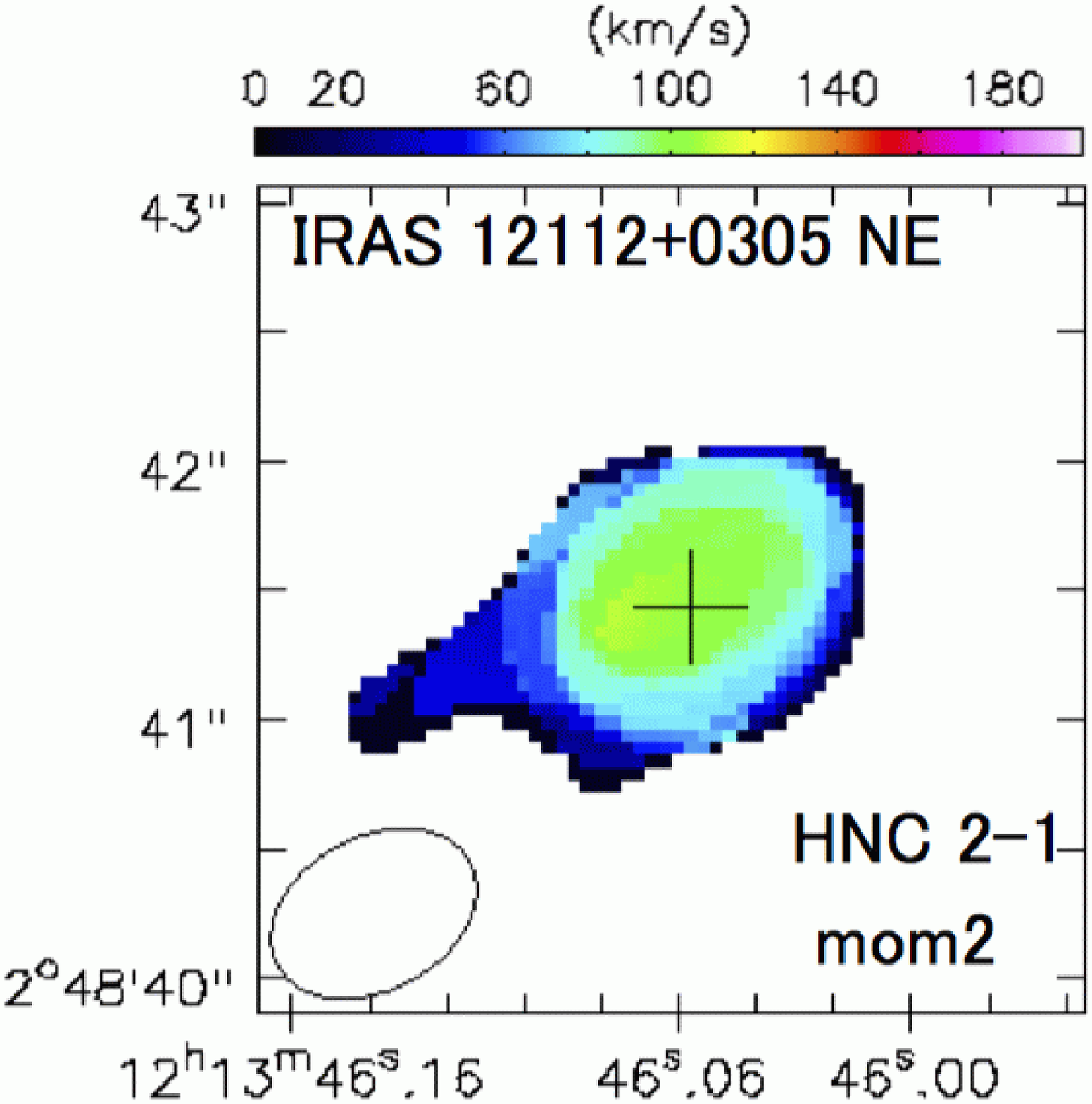} 
\includegraphics[angle=0,scale=.195]{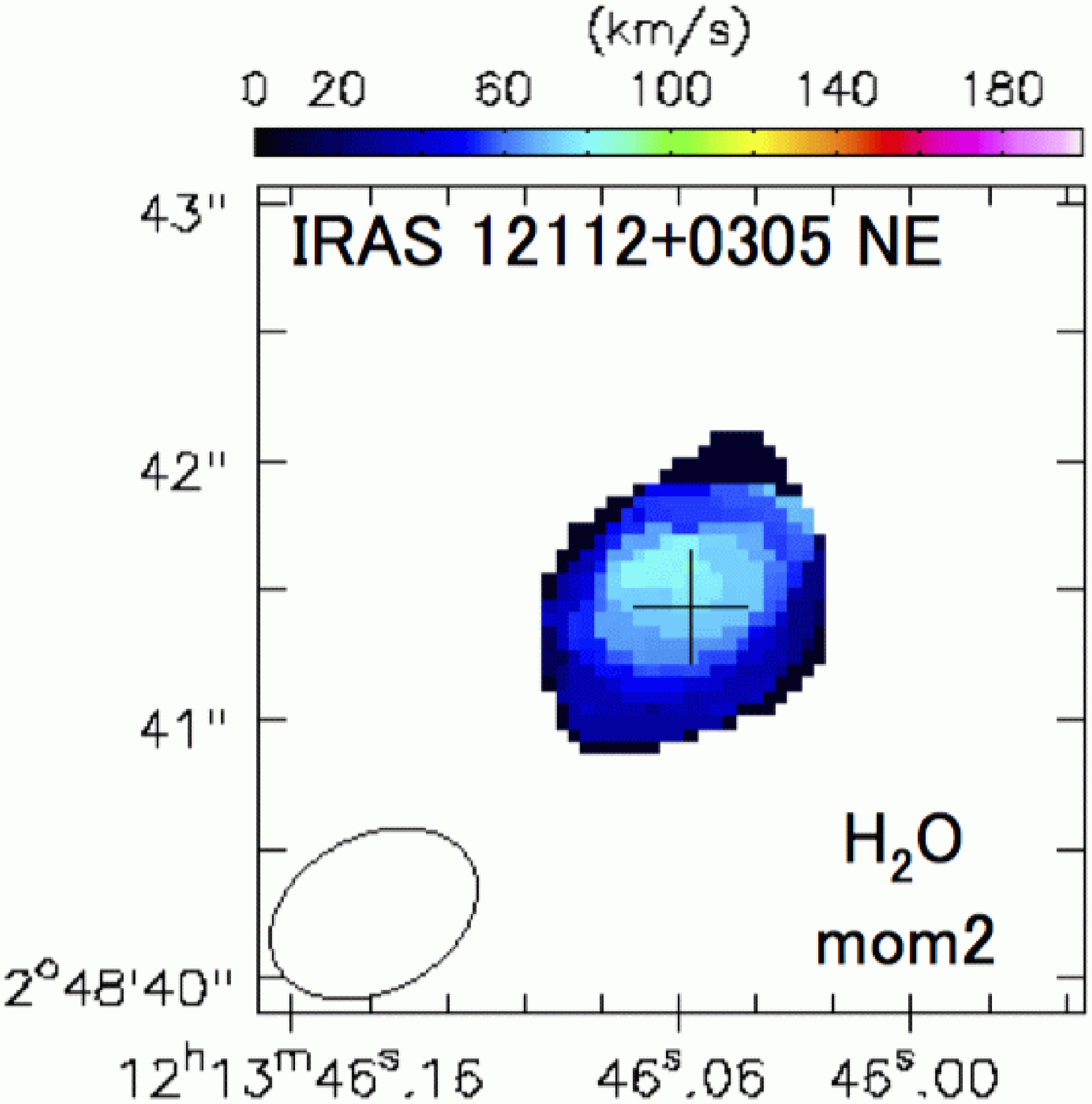} \\
\hspace*{-9.0cm}
\includegraphics[angle=0,scale=.195]{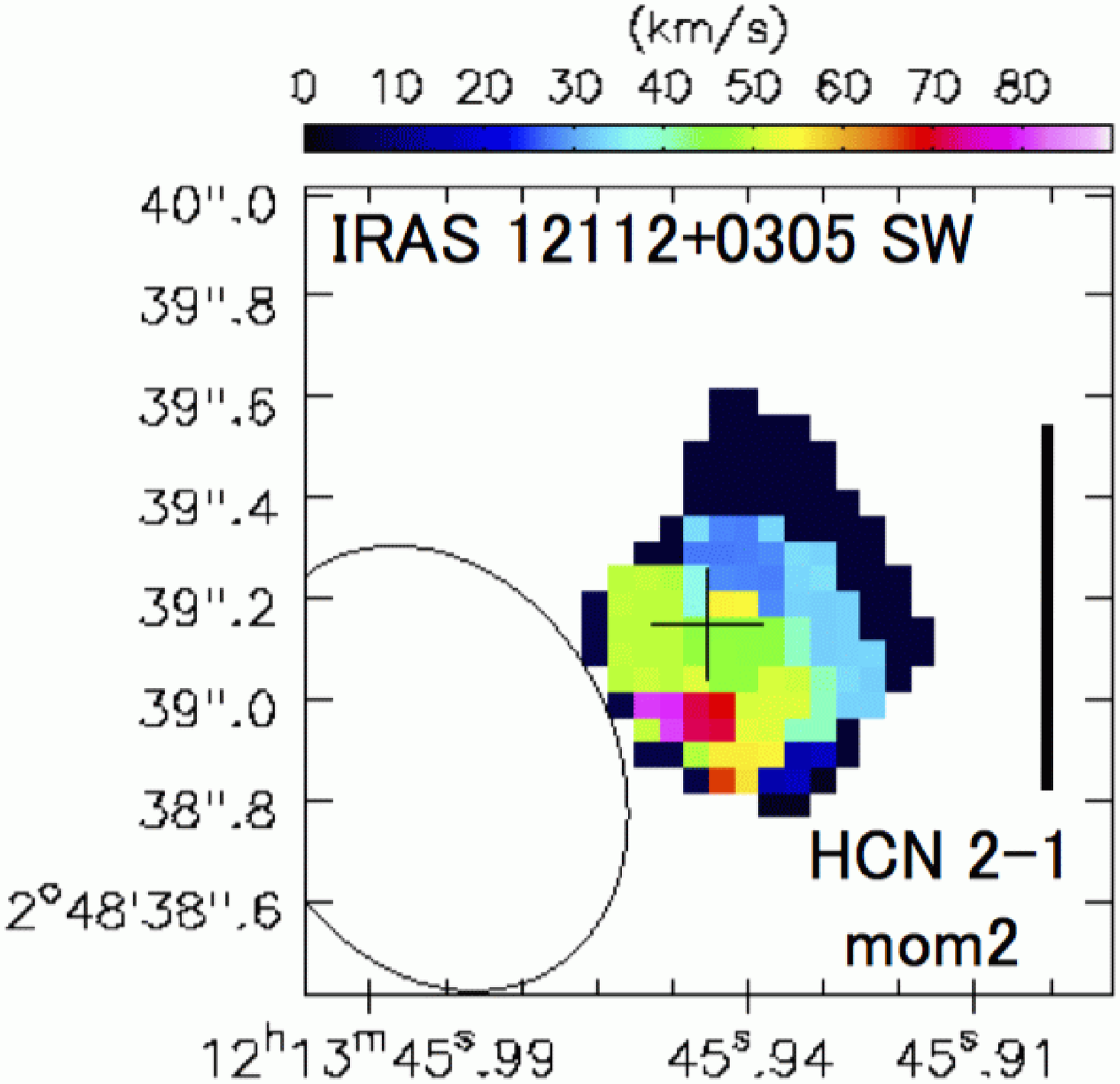} 
\includegraphics[angle=0,scale=.195]{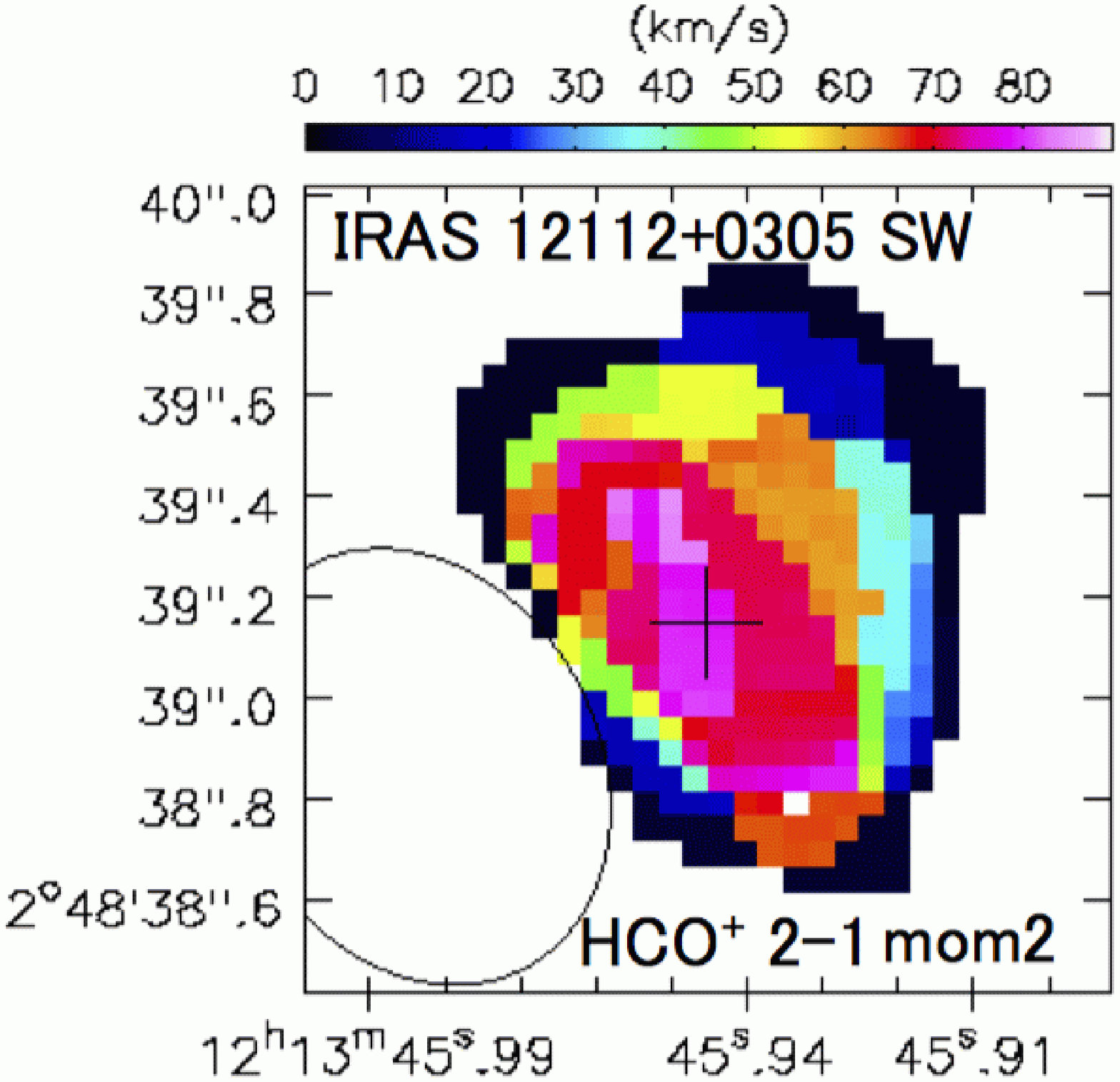} \\ 
\includegraphics[angle=0,scale=.195]{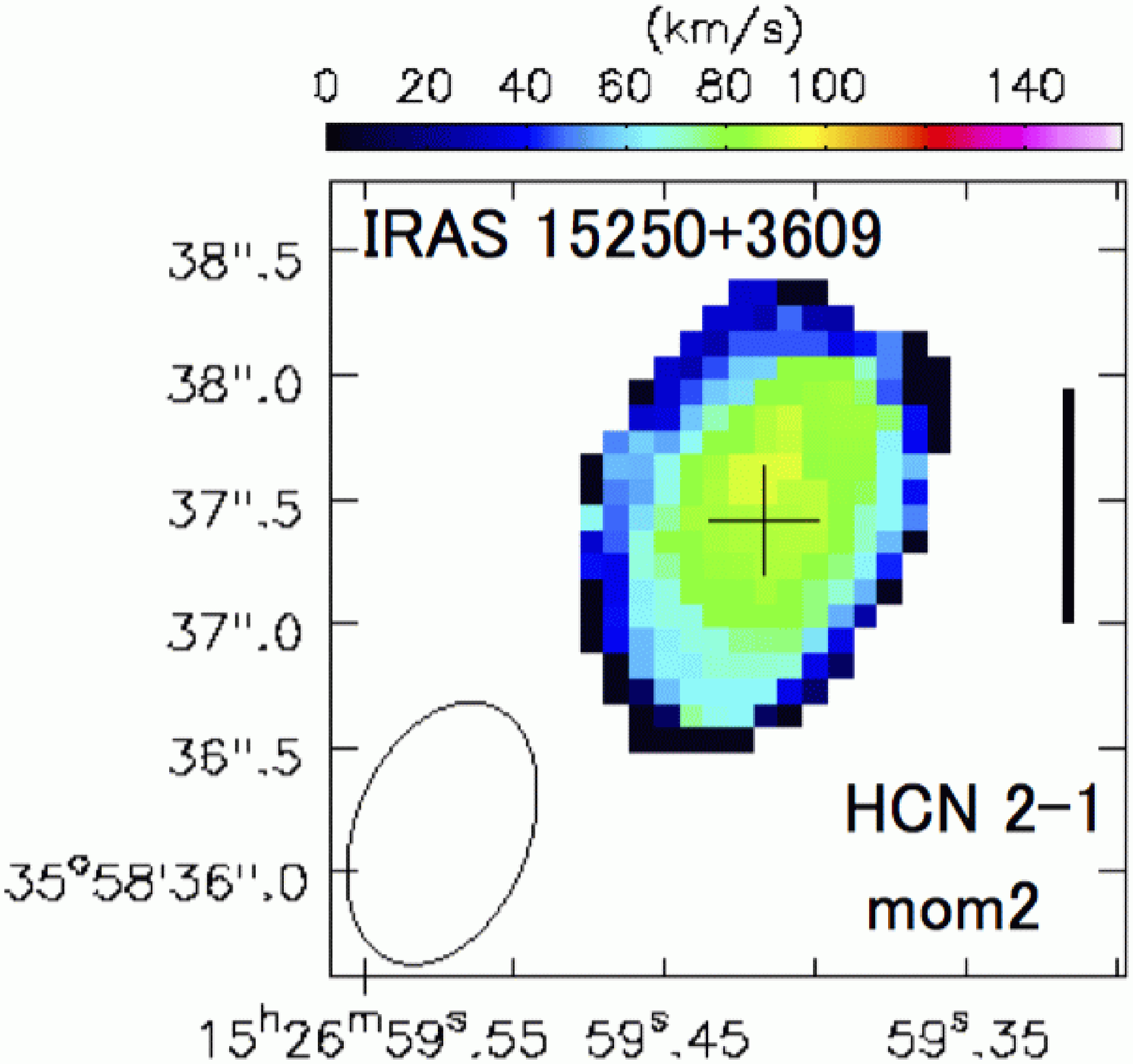} 
\includegraphics[angle=0,scale=.195]{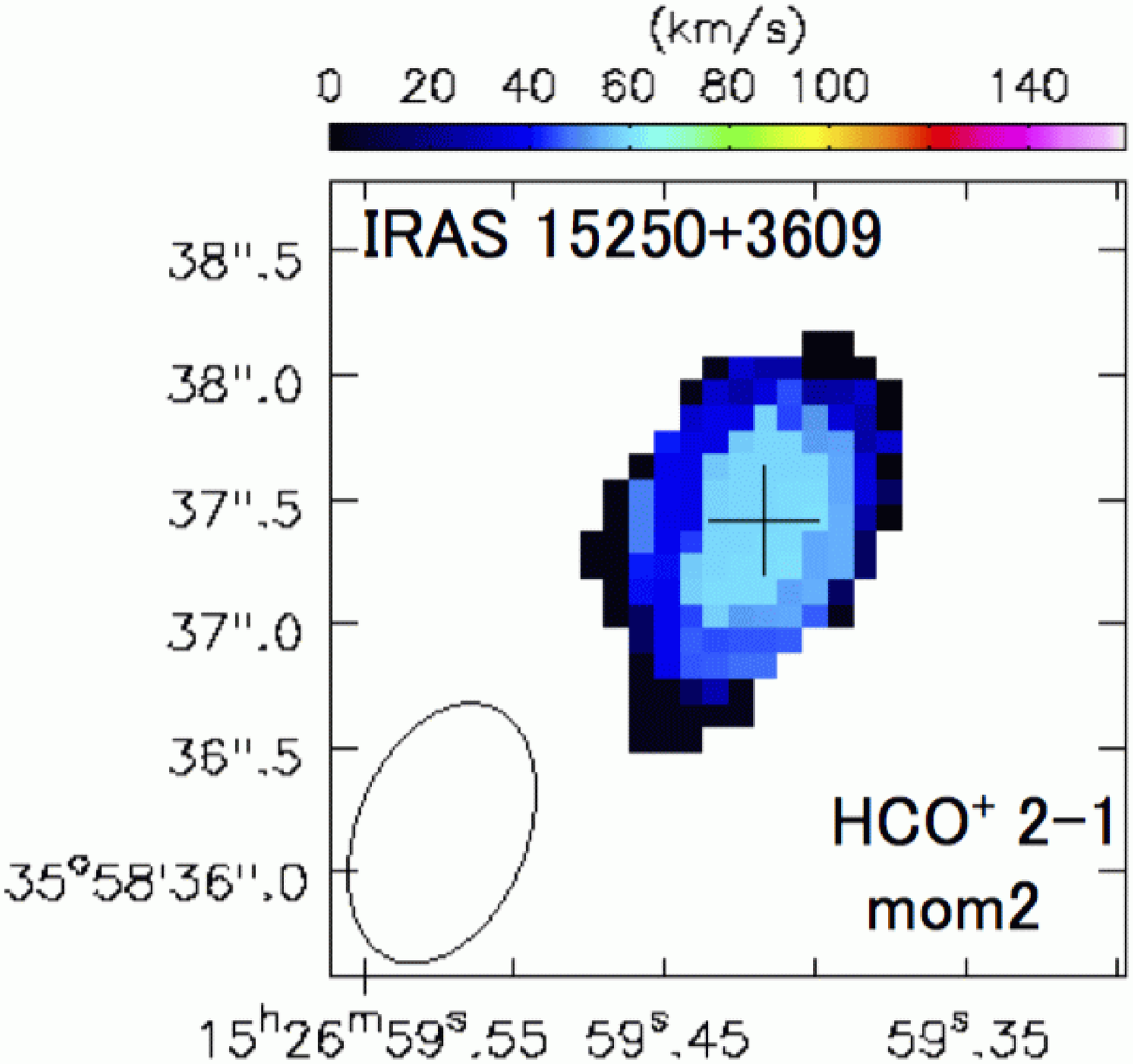} 
\includegraphics[angle=0,scale=.195]{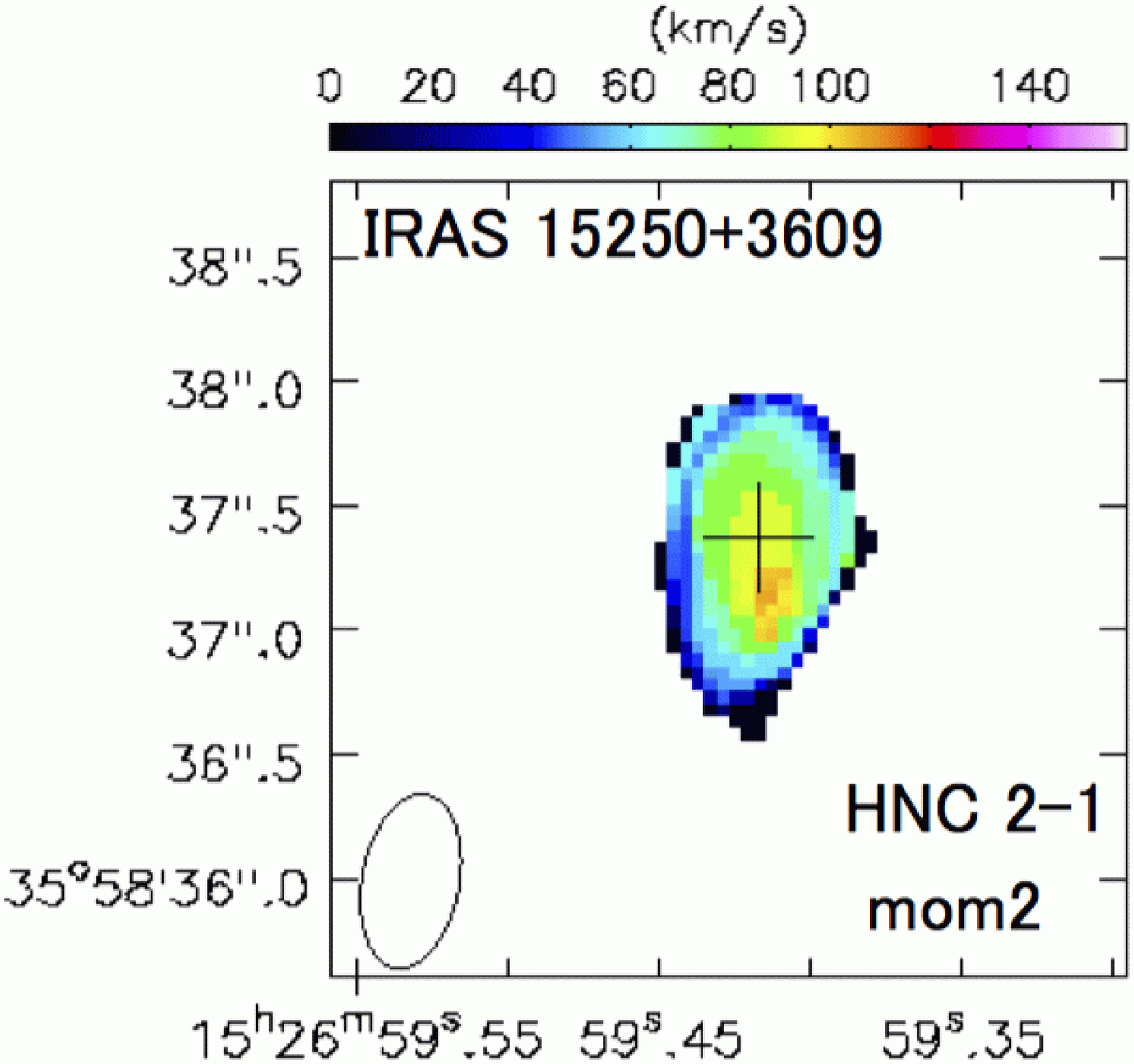} 
\includegraphics[angle=0,scale=.195]{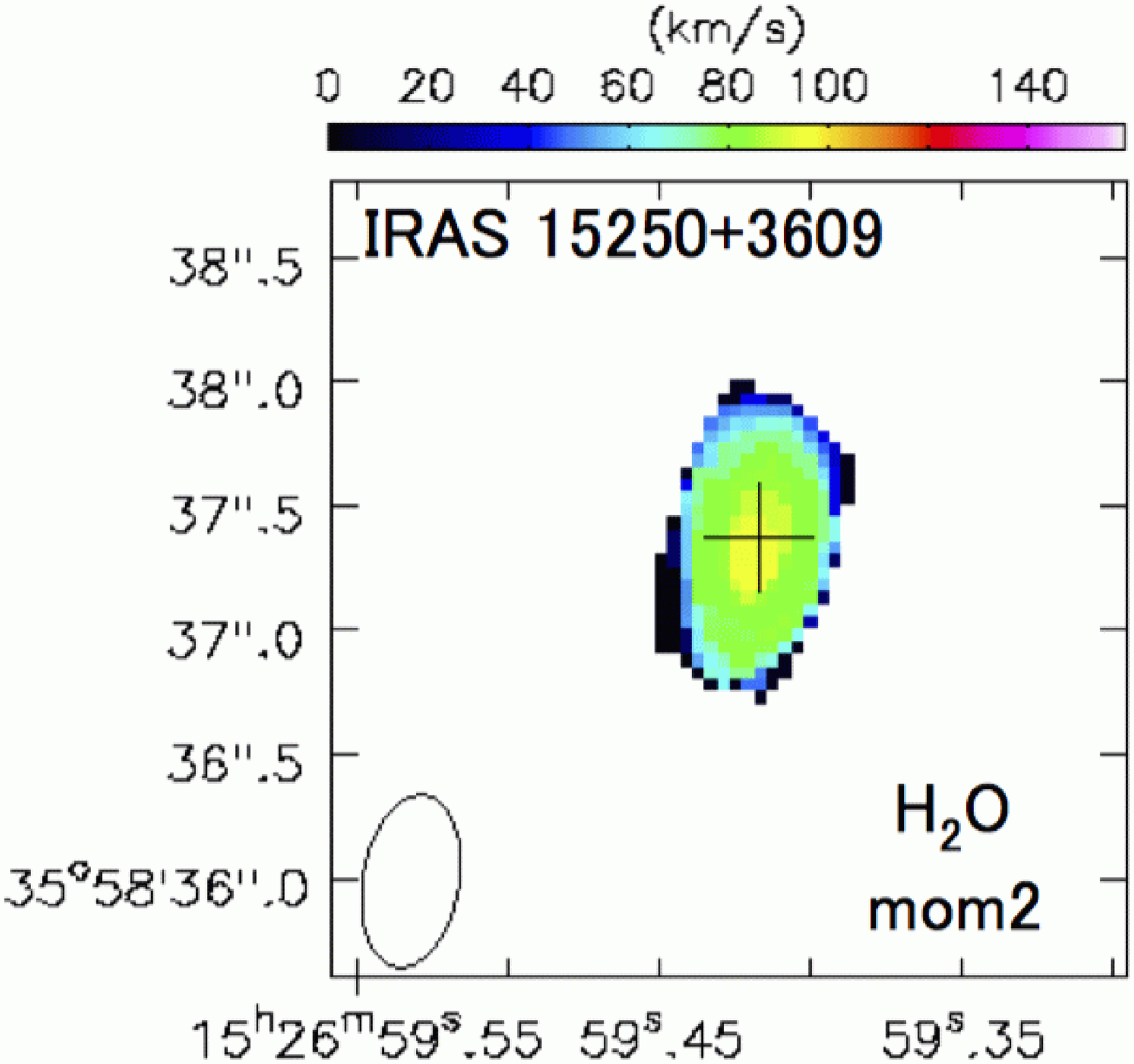} \\
\end{center}
\end{figure}

\clearpage

\begin{figure}
\begin{center}
\includegraphics[angle=0,scale=.195]{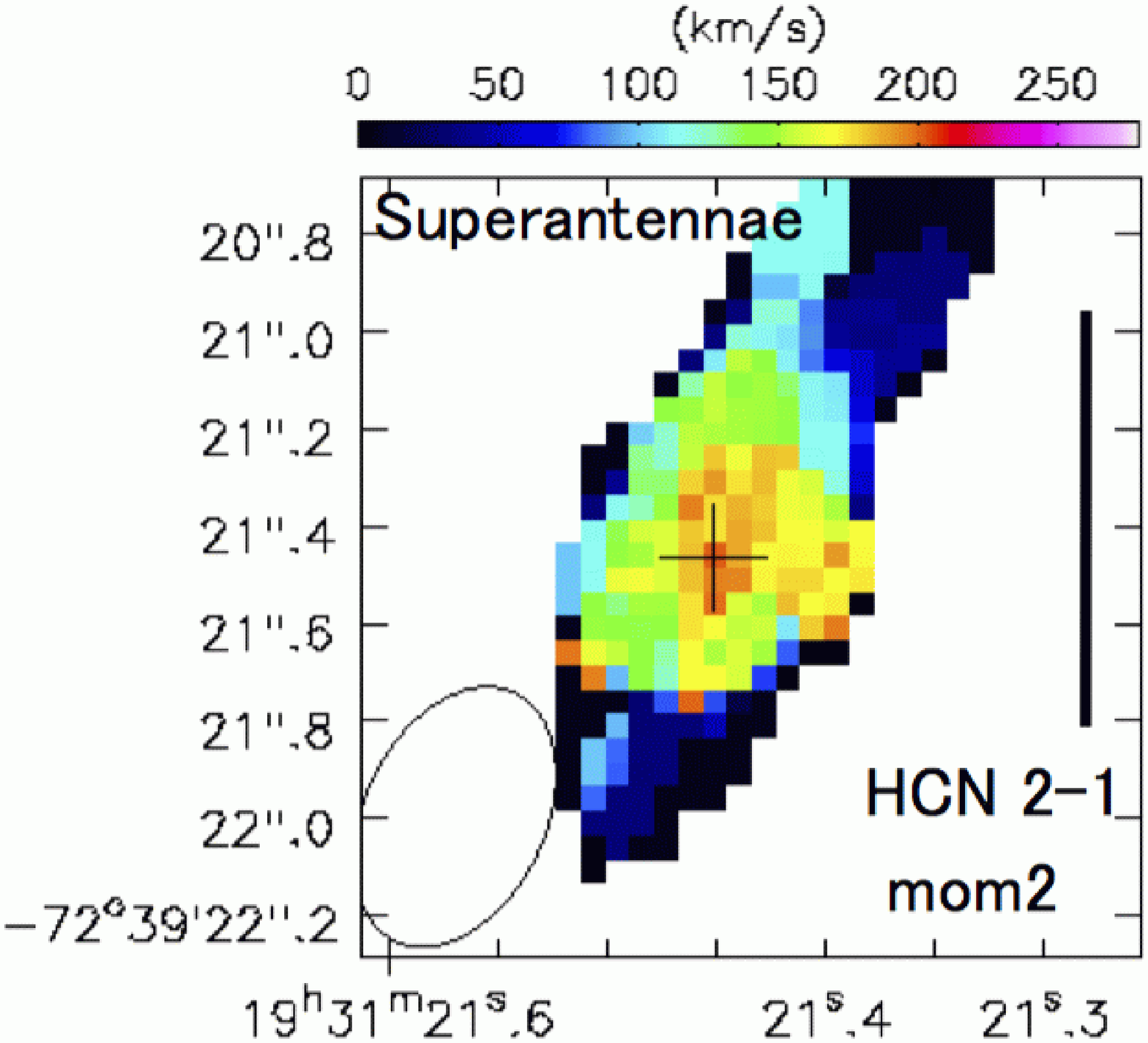} 
\includegraphics[angle=0,scale=.195]{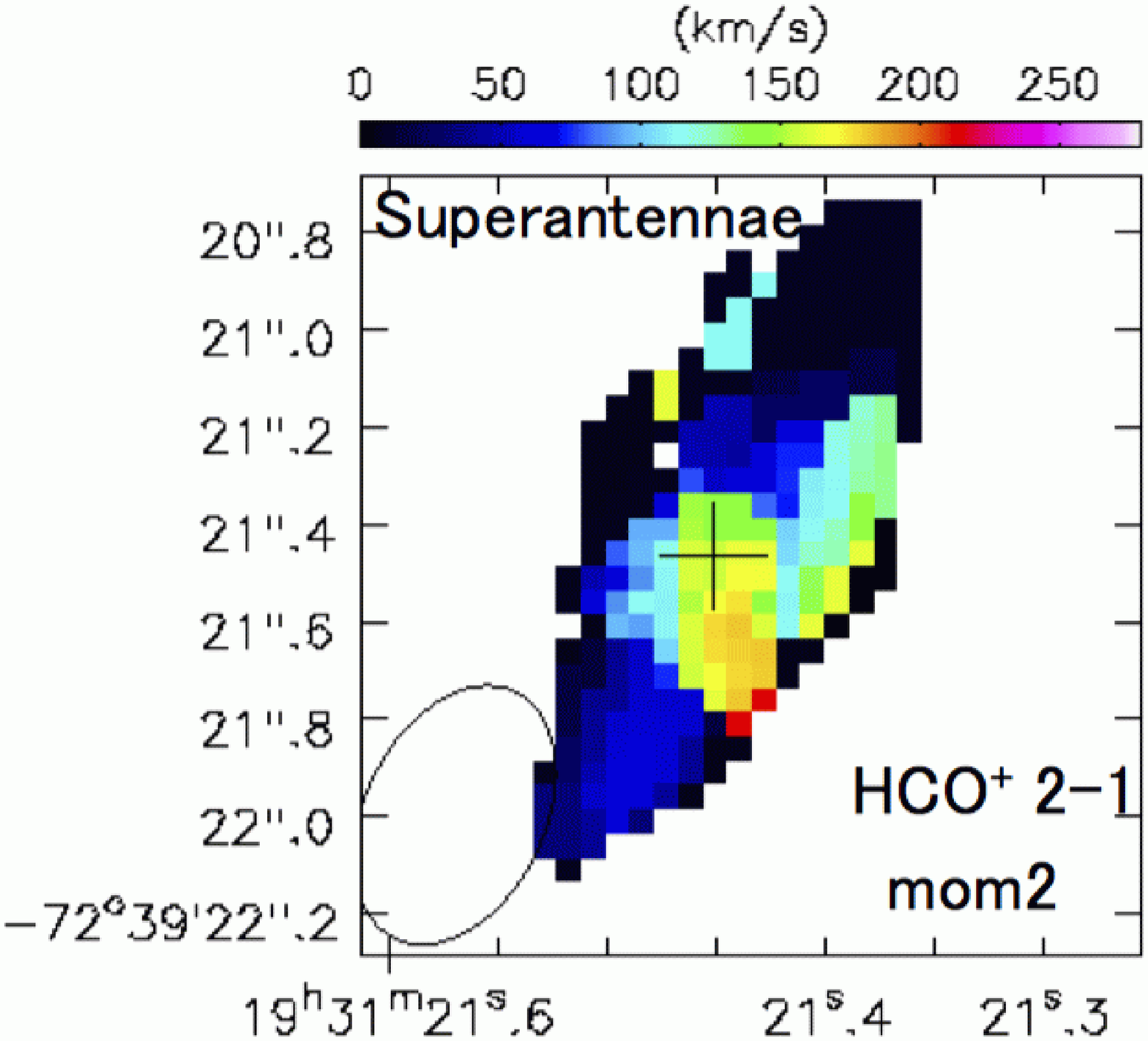} 
\includegraphics[angle=0,scale=.195]{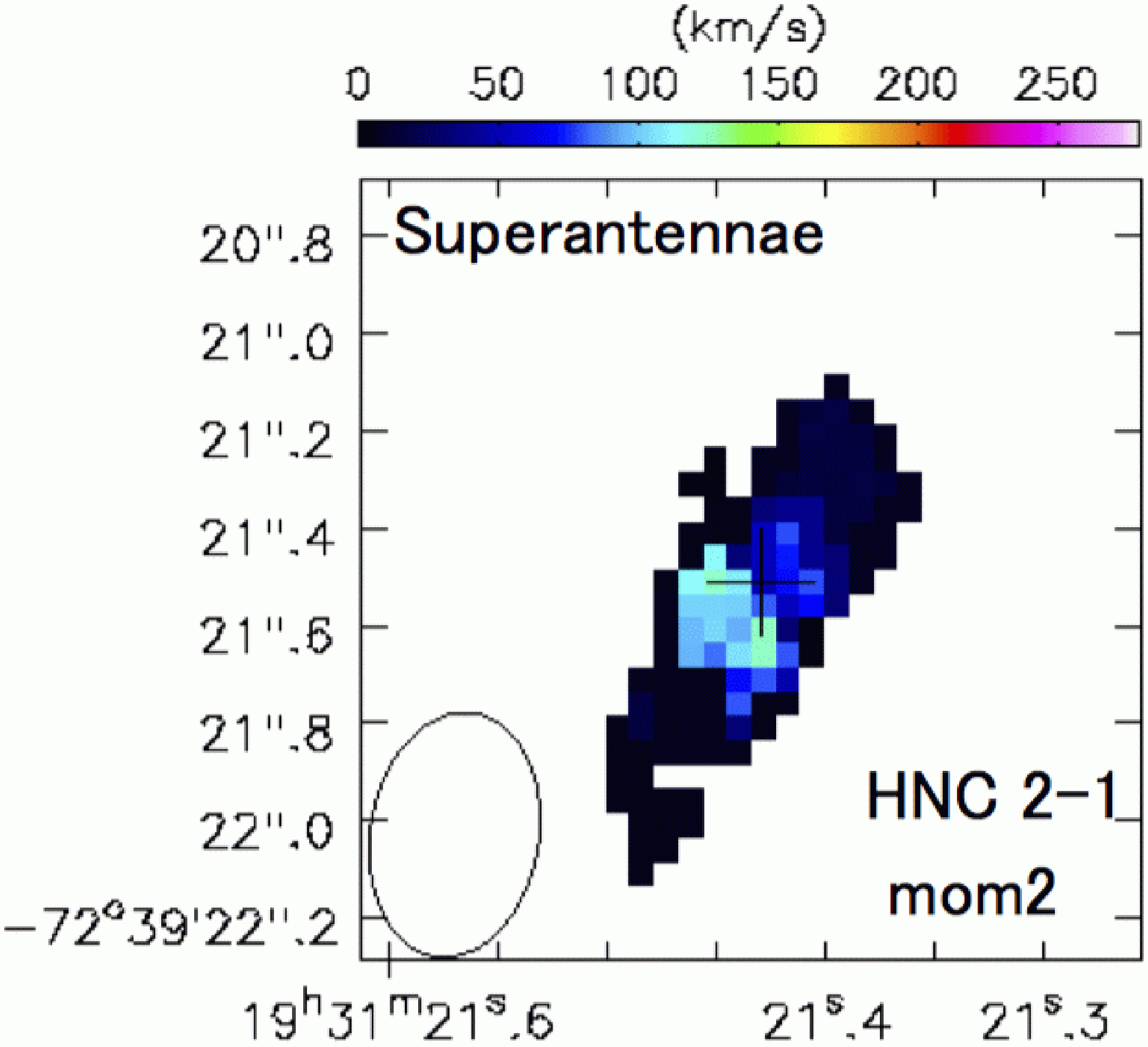} 
\includegraphics[angle=0,scale=.195]{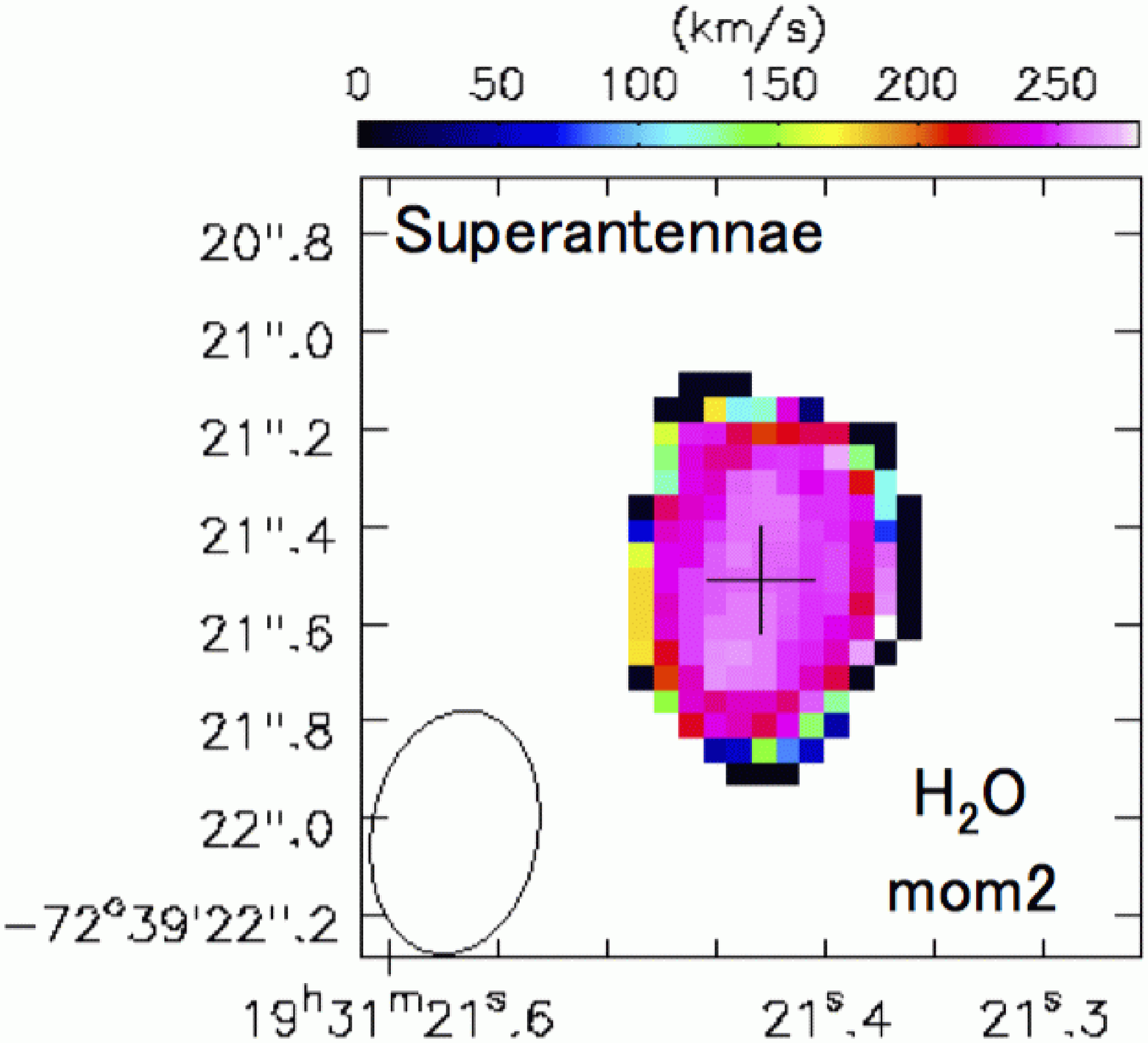} \\
\includegraphics[angle=0,scale=.195]{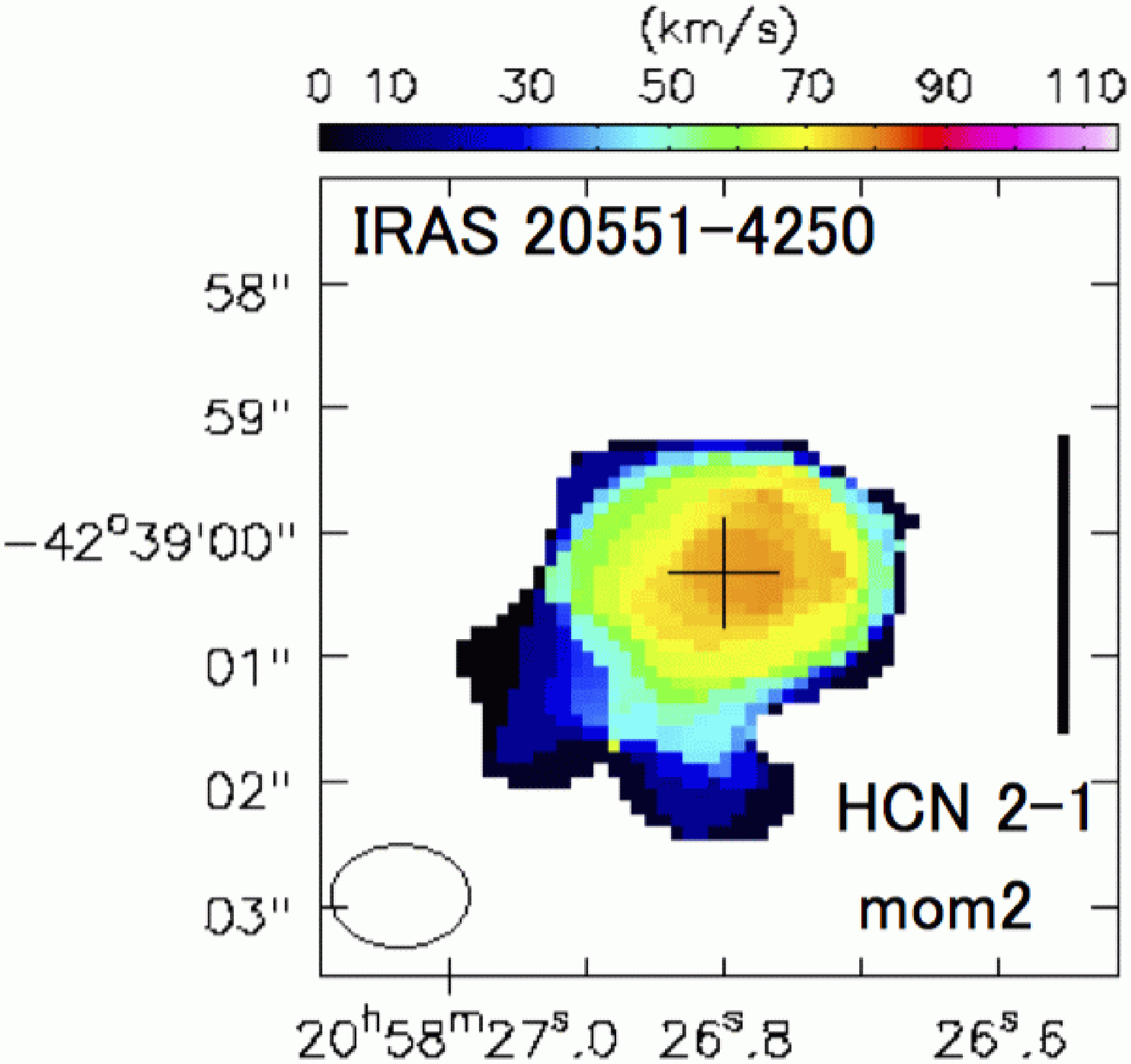} 
\includegraphics[angle=0,scale=.195]{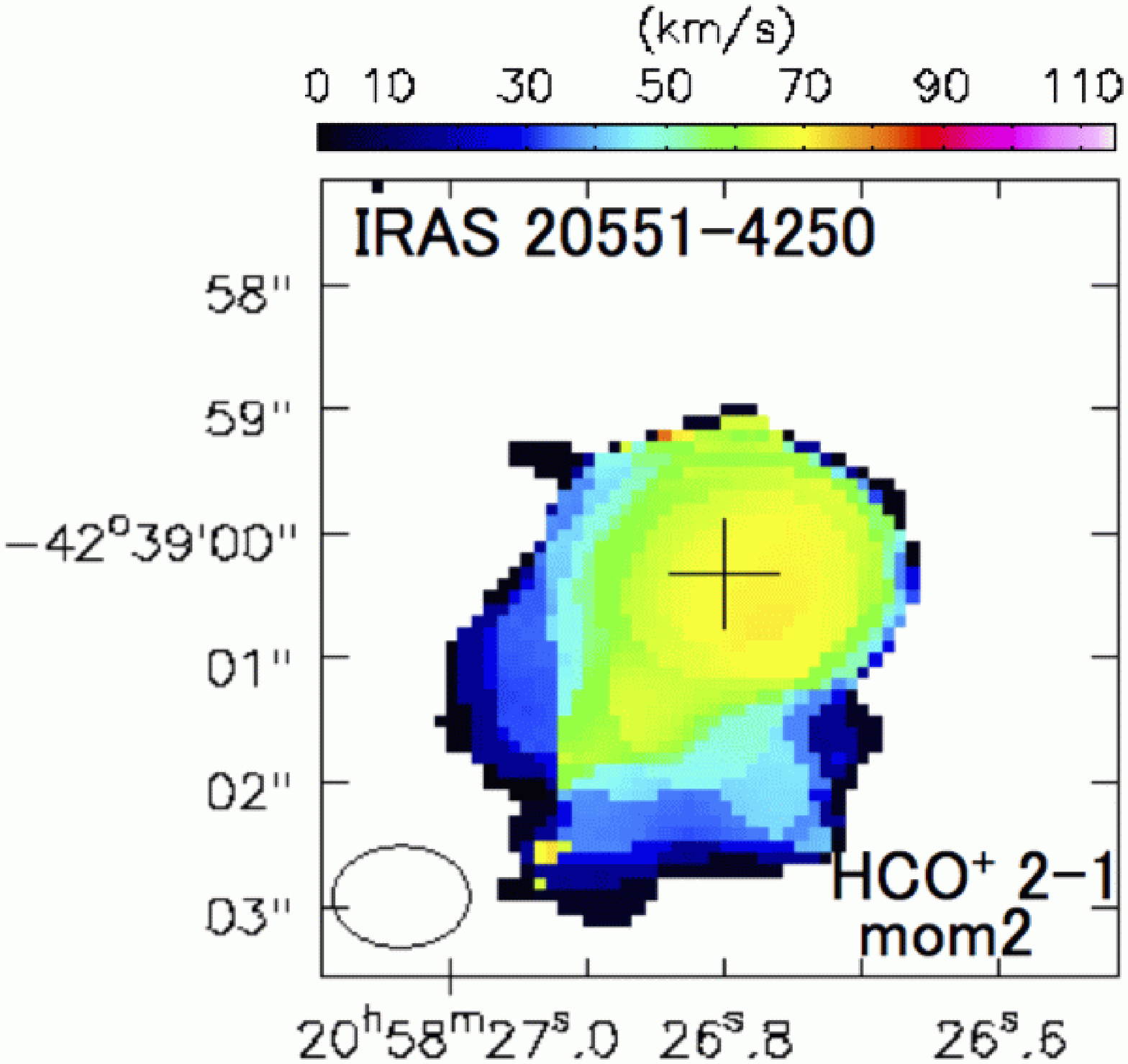} 
\includegraphics[angle=0,scale=.195]{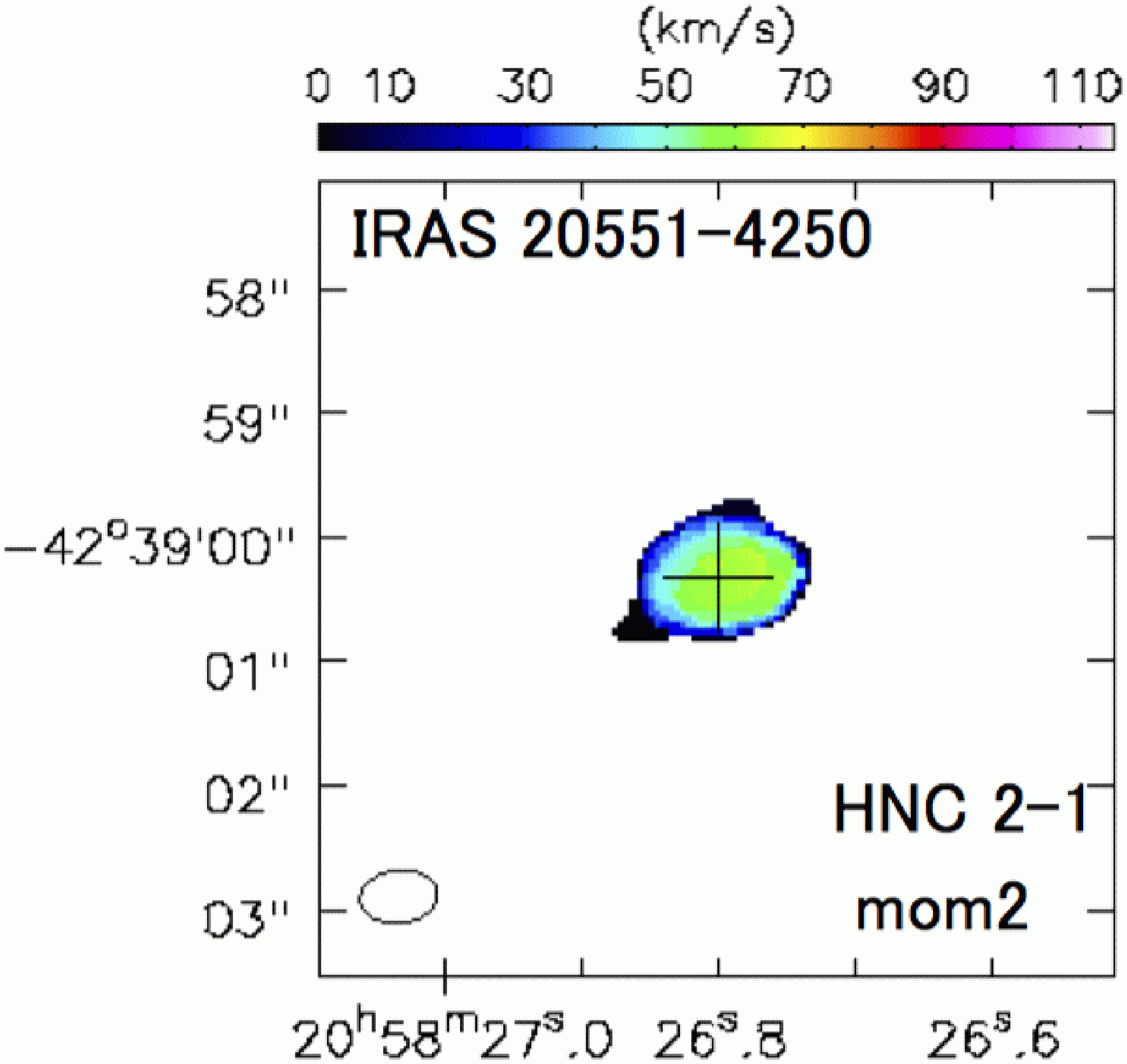} 
\includegraphics[angle=0,scale=.195]{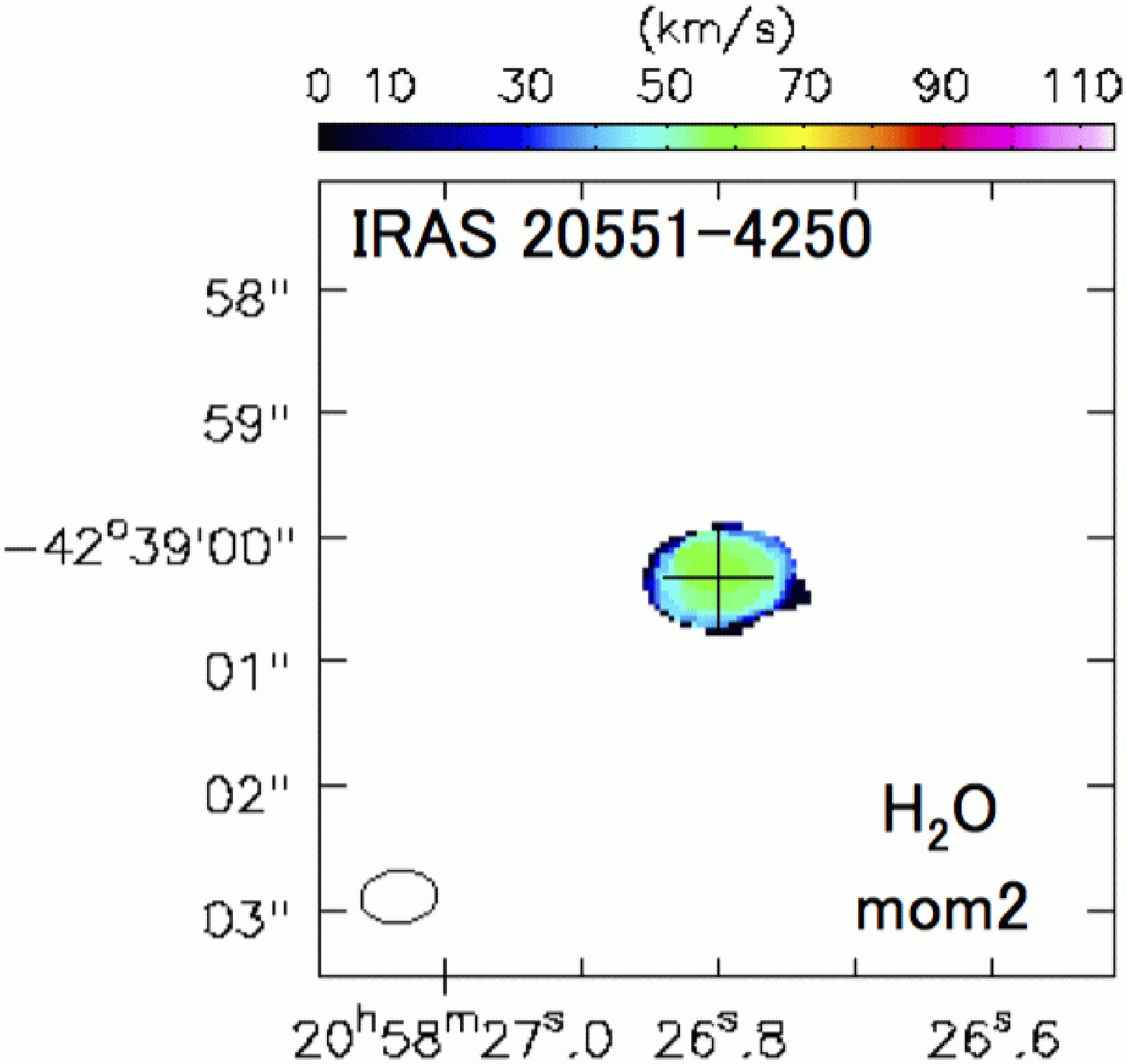} \\
\includegraphics[angle=0,scale=.195]{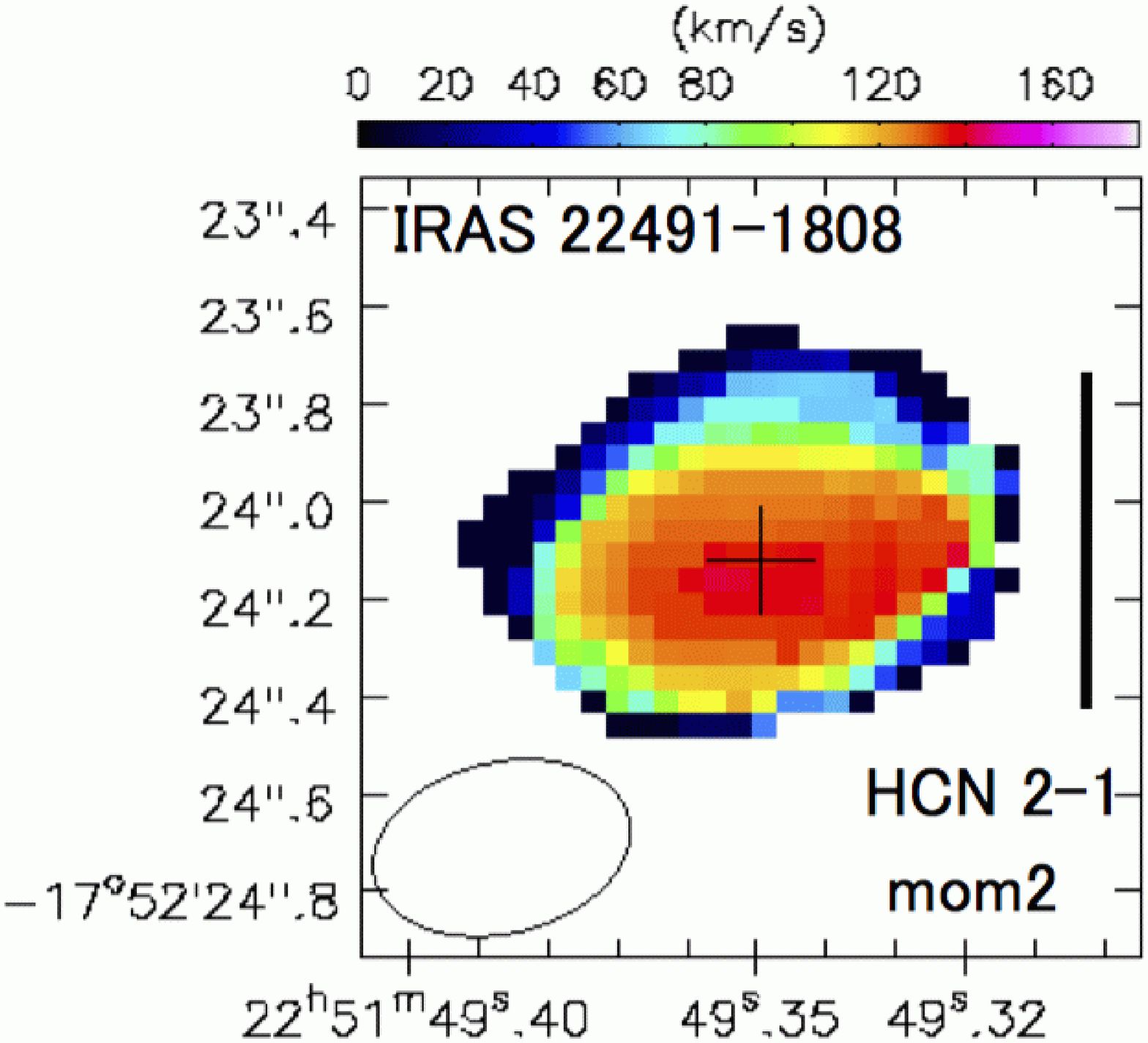} 
\includegraphics[angle=0,scale=.195]{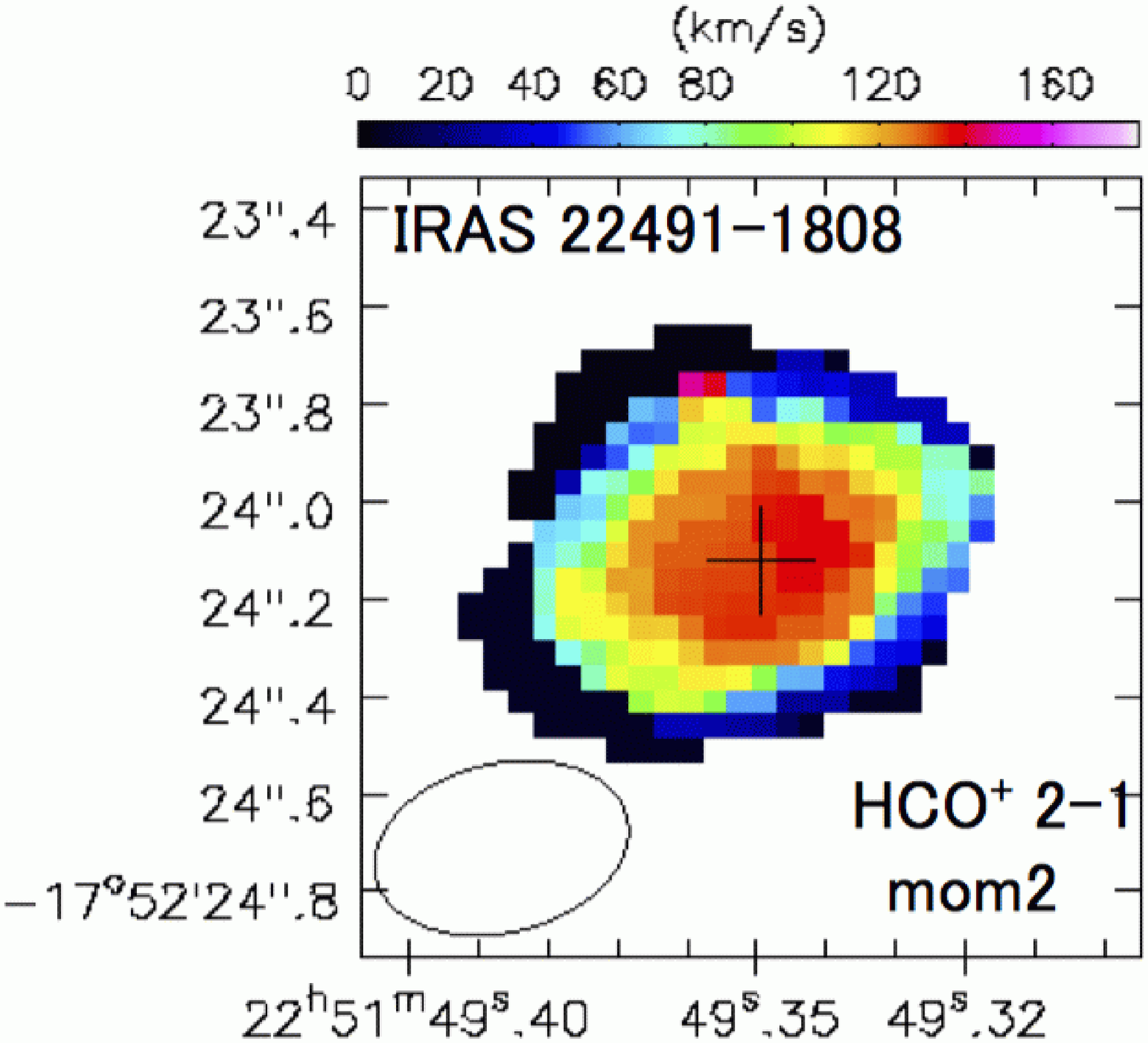} 
\includegraphics[angle=0,scale=.195]{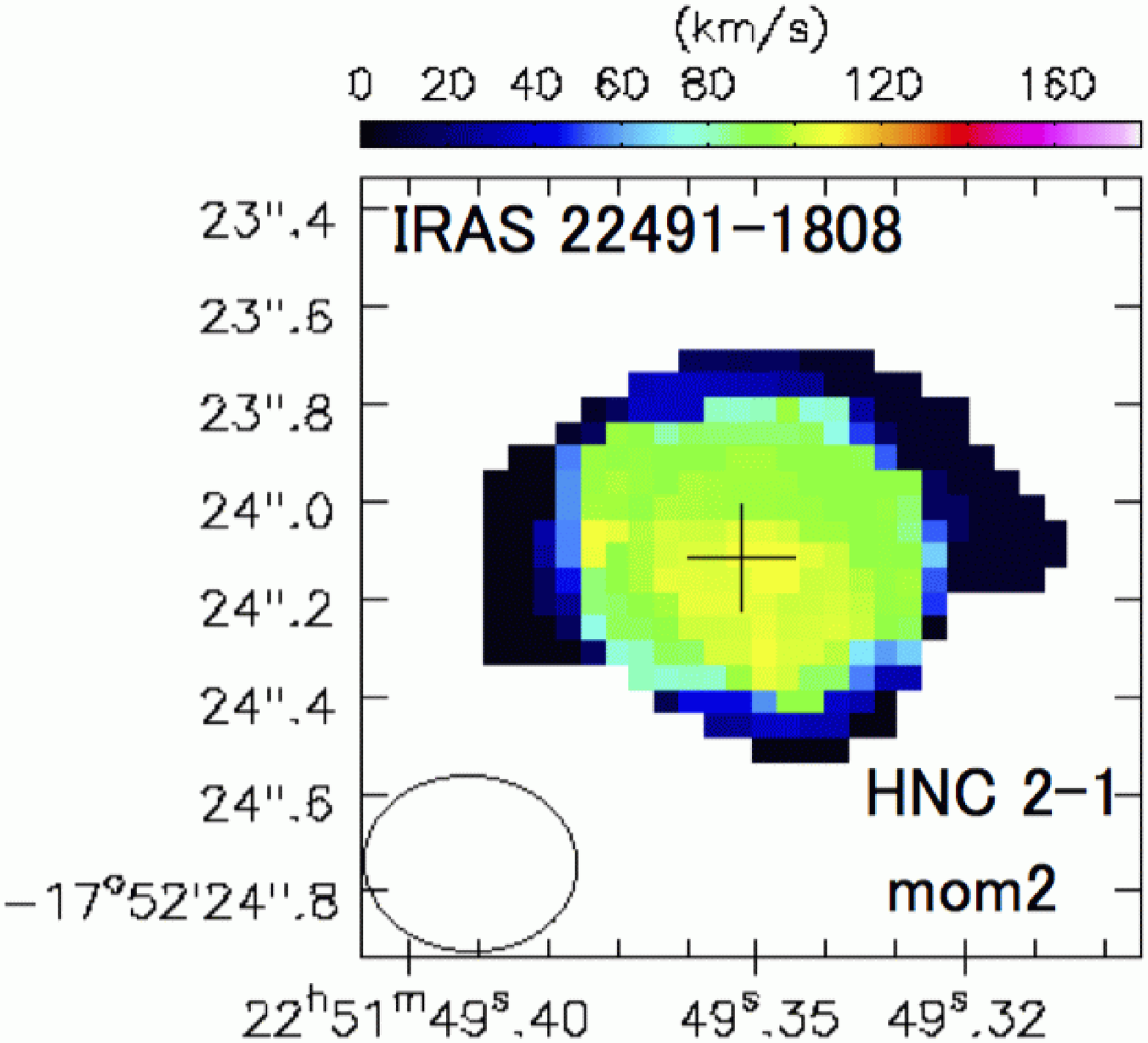} 
\includegraphics[angle=0,scale=.195]{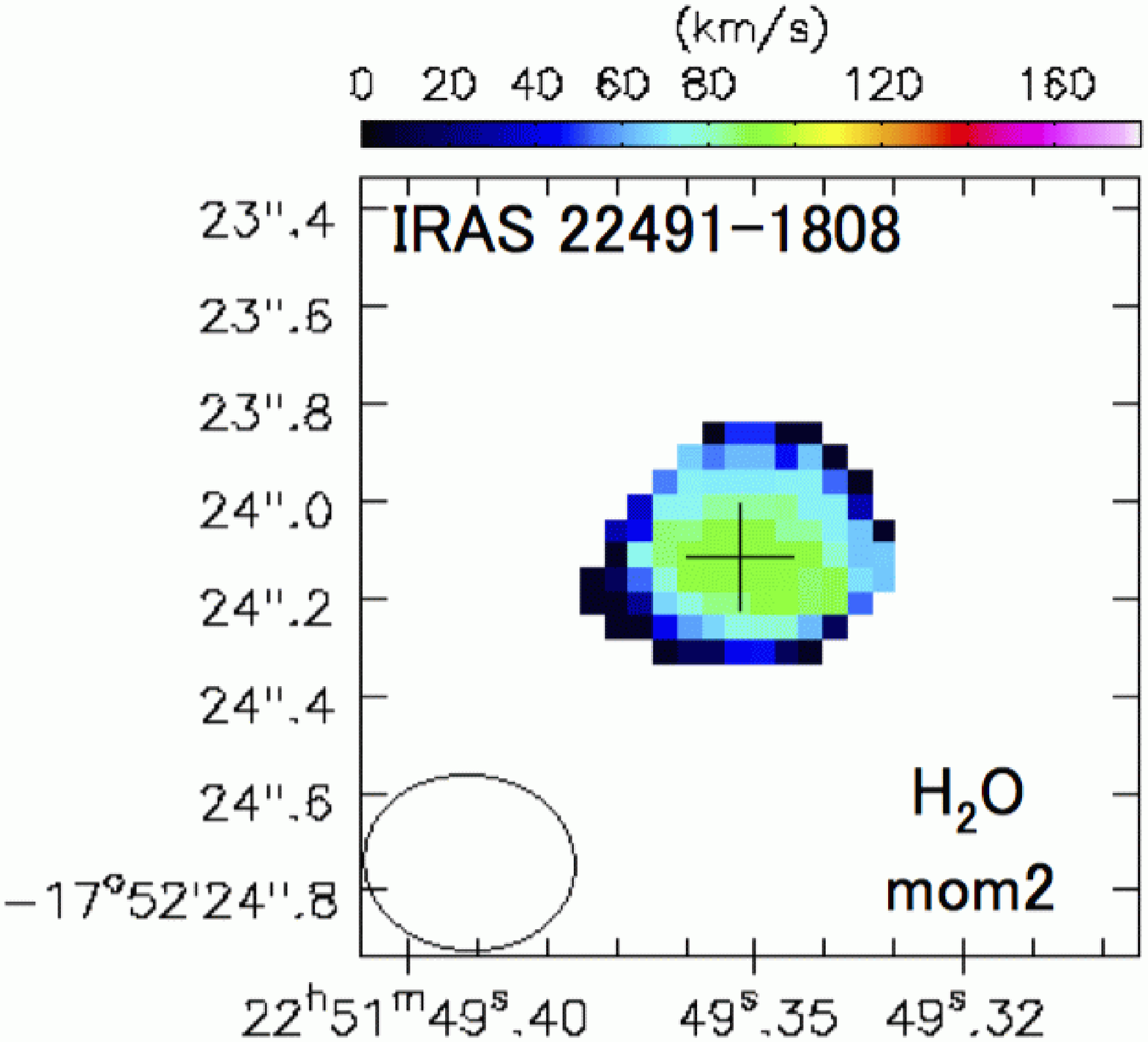} \\
\end{center}
\vspace{0.5cm}
\caption{
Intensity-weighted velocity dispersion (moment 2) maps of the 
HCN J=2--1 {\it (Left)}, 
HCO$^{+}$ J=2--1 {\it (Second left)}, HNC J=2--1 {\it (Second right)}, 
and 183 GHz H$_{2}$O {\it (Right)} lines for sources with sufficiently 
bright emission.
Continuum peak position is shown as a cross.
The length of the vertical black solid bar at the right side of 
HCN J=2--1 data {\it (Left)} corresponds to 1 kpc.
IRAS 12127$–$1412 (all four lines), IRAS 13509$+$0442 (all four lines), 
NGC 1614 (183 GHz H$_{2}$O), and 
IRAS 12112$+$0305 SW (HNC J=2--1 and 183 GHz H$_{2}$O) are not included 
as in Figure 4.
An appropriate cutoff ($\sim$2$\sigma$) is applied when we make 
these moment 2 maps so that they are not dominated by noise.
}
\end{figure}

Since probed physical scale in the original beam spectra is 
largely different among different (U)LIRGs (Table 3), we compare 
molecular emission line fluxes with the same physical scale 
for further discussion.
We choose 1 kpc, because we can investigate dense molecular emission line 
properties at energetically dominant ULIRGs' nuclei
\citep[e.g.,][]{soi00,dia10,ima11,per21}, with minimum contaminations from
spatially extended ($\gtrsim$a few kpc) star-formation in the host galaxies. 
We modify the beam to 1 kpc for all molecular lines in ULIRGs if
their beam sizes are smaller than 1 kpc, using the CASA task
``imsmooth'', and
then extract 1 kpc beam-sized spectra, which are overplotted 
as red dotted lines in Figure 3.
This is the case for all ULIRGs except IRAS 06035$-$7102, 
IRAS 12112$+$0305, and IRAS 15250$+$3609, for which we set the beam
size to 1.6 kpc, 1.2 kpc, and 1.2 kpc, respectively, because the beam
size of some lines is larger than 1 kpc.
Hereafter, we denote these spectra with 1--1.6 kpc beam sizes 
as ``$\sim$1kpc-beam'' spectra. 
For the LIRG NGC 1614, we modify the beam size of all molecular lines 
to 200 pc and extracted 200pc-beam spectra at four bright continuum peak 
positions, which are shown as red dotted lines in Figure 2a--h. 
We also create 1kpc-beam spectra for NGC 1614 (Figure 2i--j) 
for comparison with other ULIRGs.

As expected for the presence of spatially resolved components,
emission line peak fluxes usually increase with increasing 
beam sizes for all (U)LIRGs. 
Flux density at the continuum peak position in $\sim$1kpc-beam
data is tabulated in Table 5.
The continuum flux density significantly increases from the original 
beam data (Table 3) in many sources.

\begin{deluxetable}{l|cc}
\tabletypesize{\scriptsize}
\tablecaption{Continuum Peak Flux in $\sim$1kpc-beam Data \label{tbl-5}} 
\tablewidth{0pt}
\tablehead{
\colhead{} & \multicolumn{2}{c}{Peak [mJy/beam]} \\ 
\colhead{Object} & \colhead{J21a} & \colhead{J21b} \\
\colhead{(1)} & \colhead{(2)} & \colhead{(3)} 
}
\startdata 
NGC 1614 (1 kpc, 3$\farcs$1) & 10 (7.5$\sigma$) & 11 (5.1$\sigma$) \\
IRAS 06035$-$7102 (1.6 kpc, 1$\farcs$1) & 1.1 (10$\sigma$) & 1.2 (10$\sigma$) \\
IRAS 08572$+$3915 (1 kpc, 0$\farcs$90) & 1.8 (17$\sigma$) & 1.9 (17$\sigma$) \\
IRAS 12112$+$0305 NE (1.2 kpc, 0$\farcs$87) & 3.4 (31$\sigma$) & 3.7 (17$\sigma$) \\
IRAS 12112$+$0305 SW (1.2 kpc, 0$\farcs$87) & 0.39 (3.6$\sigma$) & 0.41 (1.9$\sigma$) \\
IRAS 12127$-$1412 (1 kpc, 0$\farcs$43) & 1.0 (27$\sigma$) & 1.1 (32$\sigma$) \\
IRAS 13509$+$0442 (1 kpc, 0$\farcs$42) & 0.47 (12$\sigma$) & 0.40 (10$\sigma$) \\
IRAS 15250$+$3609 (1.2 kpc, 1$\farcs$1) & 5.4 (21$\sigma$) & 6.6 (29$\sigma$) \\ 
Superantennae (1 kpc, 0$\farcs$85) & 3.9 (14$\sigma$) & 4.3 (23$\sigma$) \\ 
IRAS 20551$-$4250 (1 kpc, 1$\farcs$2) & 2.6 (45$\sigma$) & 2.8 (33$\sigma$) \\
IRAS 22491$-$1808 (1 kpc, 0$\farcs$69) & 2.1 (20$\sigma$) & 2.6 (16$\sigma$) \\
\enddata

\tablecomments{
Col.(1): Object name.
Beam size in kpc and arcsec is shown in parentheses.
Col.(2) and (3): Flux density (in mJy beam$^{-1}$) at the
continuum peak position 
in $\sim$1kpc-beam data of J21a (taken with HCN J=2--1 and HCO$^{+}$ J=2--1) 
and J21b (obtained with HNC J=2--1 and 183 GHz H$_{2}$O), respectively. 
Detection significance relative to the  noise level is shown 
in parentheses.
The continuum peak position spatially agrees with that listed 
in Table 3 (column 4).
}

\end{deluxetable}

Gaussian fits are applied to estimate the molecular emission line flux 
from nuclear $\sim$1 kpc regions in individual (U)LIRGs 
(and 200 pc regions for NGC 1614). 
We apply a single Gaussian fit as long as the observed 
emission line profile is approximated by a single peaked profile 
with small deviation.
However, some molecular emission lines in certain ULIRGs display 
double-peaked profiles with central dips and clearly deviate
from a single Gaussian profile (e.g., HCO$^{+}$ J=2--1 line of IRAS
12112$+$0305 NE in Figure 3e). 
For these, we apply two Gaussian fits (two emission components).
If the two Gaussian fits better trace the observed emission line 
profile than the single Gaussian fit, we adopt the former flux, which
is slightly ($\gtrsim$a few \% to $\sim$10\%) smaller than the latter.
Adopted Gaussian fits of individual emission lines with possible 
detection signatures in the $\sim$1kpc-beam spectra for all (U)LIRGs
(and 200pc-beam spectra for NGC 1614) are displayed in Appendix B.
Table 6 summarizes the Gaussian-fit velocity-integrated fluxes, 
which will be used to discuss molecular line flux ratios, 
unless otherwise stated for certain faint emission lines for which the
peak values or 3$\sigma$ upper limits in $\sim$1kpc-beam moment 0 maps
are used. 

\startlongtable
\begin{deluxetable}{ll|cccc}
\tabletypesize{\scriptsize}
\tablecaption{Gaussian Fit of Emission Lines \label{tbl-6}} 
\tablewidth{0pt}
\tablehead{
\colhead{Object} & \colhead{Line} &  
\multicolumn{4}{c}{Gaussian fit} \\  
\colhead{} & \colhead{} & \colhead{Velocity}
& \colhead{Peak flux density} & \colhead{FWHM} & \colhead{Integrated flux} \\ 
\colhead{} & \colhead{} & \colhead{[km s$^{-1}$]} & \colhead{[mJy]} 
& \colhead{[km s$^{-1}$]} &
\colhead{[Jy km s$^{-1}$]} \\  
\colhead{(1)} & \colhead{(2)} & \colhead{(3)} & \colhead{(4)} & 
\colhead{(5)} & \colhead{(6)} 
}
\startdata 
NGC 1614 & HCN J=2--1 & 4765$\pm$4 & 27$\pm$1 & 243$\pm$10 & 6.8$\pm$0.4 \\
(1 kpc, 3$\farcs$1) & HCO$^{+}$ J=2--1 & 4769$\pm$4 & 47$\pm$1 & 252$\pm$8 & 12$\pm$1 \\
 & HNC J=2--1 & 4766$\pm$8 & 16$\pm$1 & 222$\pm$17 & 3.7$\pm$0.4 \\
\hline
NGC 1614 W & HCN J=2--1 & 4763$\pm$3 & 6.1$\pm$0.3 & 112$\pm$7 & 0.72$\pm$0.06 \\ 
(200 pc, 0$\farcs$62) & HCO$^{+}$ J=2--1 & 4761$\pm$2 & 11$\pm$1 & 114$\pm$5 & 1.3$\pm$0.1 \\
 & HNC J=2--1 & 4744$\pm$5 & 5.3$\pm$0.8 & 77$\pm$13 & 0.42$\pm$0.10 \\
\hline
NGC 1614 N & HCN J=2--1 & 4833$\pm$2 & 7.3$\pm$0.4 & 93$\pm$6 & 0.72$\pm$0.05 \\
(200 pc, 0$\farcs$62) & HCO$^{+}$ J=2--1 & 4833$\pm$2 & 11$\pm$1 & 109$\pm$5 & 1.2$\pm$0.1 \\
 & HNC J=2--1 & 4842$\pm$4 & 5.8$\pm$0.8 & 57$\pm$9 & 0.34$\pm$0.07  \\
\hline
NGC 1614 S & HCN J=2--1 & 4643$\pm$5 & 5.1$\pm$0.4 & 109$\pm$13 & 0.58$\pm$0.08 \\ 
(200 pc, 0$\farcs$62) & HCO$^{+}$ J=2--1 & 4639$\pm$2 & 8.6$\pm$0.4 & 97$\pm$8 & 0.88$\pm$0.08 \\
 & HNC J=2--1 & 4638$\pm$6 & 3.8$\pm$0.4 & 104$\pm$18 & 0.41$\pm$0.08 \\
\hline
NGC 1614 E & HCN J=2--1 & 4708$\pm$5 & 3.8$\pm$0.3 & 124$\pm$15 & 0.49$\pm$0.07 \\
(200 pc, 0$\farcs$62) & HCO$^{+}$ J=2--1 & 4700$\pm$2 & 7.2$\pm$0.3 & 120$\pm$6 & 0.90$\pm$0.06 \\
 & HNC J=2--1 & 4744$\pm$13 & 2.7$\pm$0.6 & 112$\pm$35 & 0.32$\pm$0.12 \\
\hline
IRAS 06035$-$7102 & HCN J=2--1 & 23857$\pm$5 & 7.8$\pm$0.2 & 371$\pm$11 & 2.9$\pm$0.1 \\
(1.6 kpc, 1$\farcs$1) & HCO$^{+}$ J=2--1 & 23867$\pm$5 & 8.9$\pm$0.2 & 404$\pm$13 & 3.6$\pm$0.1 \\
 & HNC J=2--1 & 23832$\pm$11 & 4.6$\pm$0.3 & 356$\pm$25 & 1.6$\pm$0.1 \\
 & 183 GHz H$_{2}$O & 23898$\pm$29 & 2.0$\pm$0.2 & 424$\pm$77 & 0.83$\pm$0.18 \\ \hline
IRAS 08572$+$3915 & HCN J=2--1 & 17485$\pm$11 & 3.6$\pm$0.2 & 383$\pm$27 & 1.4$\pm$0.1 \\
(1 kpc, 0$\farcs$90) & HCO$^{+}$ J=2--1 & 17388$\pm$27, 17563$\pm$31 &
4.9, 4.7 (fix) \tablenotemark{A} & 162$\pm$23, 185$\pm$15 & 1.7$\pm$0.2 \\
 & HNC J=2--1 & 17481$\pm$16 & 1.9$\pm$0.2 & 315$\pm$33 & 0.61$\pm$0.09 \\
 & 183 GHz H$_{2}$O & 17351$\pm$18, 17558$\pm$15 & 2.2$\pm$0.3, 2.6$\pm$0.3 
& 133$\pm$44, 174$\pm$36 & 0.75$\pm$0.15 \\ \hline
IRAS 12112$+$0305 NE & HCN J=2--1 & 21684$\pm$20, 21942$\pm$13 & 9.2$\pm$0.6, 10$\pm$1 & 289$\pm$35, 243$\pm$22 & 5.1$\pm$0.5 \\
(1.2 kpc, 0$\farcs$87) & HCO$^{+}$ J=2--1 & 21659$\pm$7, 21957$\pm$7 & 6.8$\pm$0.4, 7.6$\pm$0.4 & 198$\pm$17, 233$\pm$22 & 3.1$\pm$0.2 \\
 & HNC J=2--1 & 21804$\pm$5 & 12$\pm$1 & 349$\pm$13 & 4.0$\pm$0.2 \\
 & 183 GHz H$_{2}$O & 21807$\pm$5 & 7.8$\pm$0.3 & 268$\pm$14 & 2.1$\pm$0.1 \\ \hline
IRAS 12112$+$0305 SW & HCN J=2--1 & 21987$\pm$18 & 1.7$\pm$0.2 & 267$\pm$44 
& 0.45$\pm$0.10 \\
(1.2 kpc, 0$\farcs$87) & HCO$^{+}$ J=2--1 & 21975$\pm$15 & 2.9$\pm$0.3 & 338$\pm$42 & 0.97$\pm$0.15 \\
 & HNC J=2--1 & 21917$\pm$15, 22077$\pm$16 & 1.8$\pm$0.5, 1.9$\pm$0.5 
& 104$\pm$41, 73$\pm$48 & 0.32$\pm$0.13\\
\hline
IRAS 12127$-$1412 & HCN J=2--1 & 39966$\pm$25 & 1.5$\pm$0.2 & 483$\pm$63 & 
0.70$\pm$0.12 \\
(1 kpc, 0$\farcs$43) & HCO$^{+}$ J=2--1 & 39867$\pm$38, 40146$\pm$16 & 1.4$\pm$0.2, 1.4$\pm$0.5 & 360$\pm$111, 93 (fix) & 0.60$\pm$0.17 \\
 & HNC J=2--1 & 39871$\pm$55, 40141$\pm$33 & 1.2$\pm$0.2, 1.5$\pm$0.4 & 376$\pm$89, 132$\pm$75 & 0.61$\pm$0.17 \\
 & 183 GHz H$_{2}$O & 40036$\pm$74 & 0.75$\pm$0.55 & 500$\pm$193 & 0.35$\pm$0.29 \\ \hline
IRAS 13509$+$0442 & HCN J=2--1 & 40935$\pm$15 & 2.2$\pm$0.2 & 286$\pm$40 
& 0.59$\pm$0.11 \\
(1 kpc, 0$\farcs$42) & HCO$^{+}$ J=2--1 & 40914$\pm$20 & 2.0$\pm$0.2 & 307$\pm$52 & 0.57$\pm$0.12 \\
 & HNC J=2--1 & 40957$\pm$16 & 1.5$\pm$0.2 & 221$\pm$40 & 0.32$\pm$0.07 \\
 & 183 GHz H$_{2}$O & 41047$\pm$134 & 0.45$\pm$0.12 & 799$\pm$482 & 0.33$\pm$0.22 \\ \hline
IRAS 15250$+$3609 & HCN J=2--1 & 16582$\pm$4 & 12$\pm$1 & 274$\pm$10 & 
3.4$\pm$0.2 \\
(1.2 kpc, 1$\farcs$1) & HCO$^{+}$ J=2--1 & 16564$\pm$7 & 7.1$\pm$0.5 & 216$\pm$17 & 1.6$\pm$0.2 \\
 & HNC J=2--1 & 16591$\pm$7 & 13$\pm$1 & 295$\pm$18 & 3.7$\pm$0.3 \\
 & 183 GHz H$_{2}$O & 16581$\pm$5 & 11$\pm$1 & 313$\pm$12 & 3.4$\pm$0.2 \\ \hline
Superantennae & HCN J=2--1 & 18539$\pm$19 & 5.5$\pm$0.3 & 878$\pm$54 & 4.9$\pm$0.4 \\
(1 kpc, 0$\farcs$85) & HCO$^{+}$ J=2--1 & 18505$\pm$21 & 4.3$\pm$0.3 & 665$\pm$52 & 2.9$\pm$0.3 \\
 & HNC J=2--1 & 18504$\pm$22 & 2.9$\pm$0.2 & 661$\pm$57 & 1.9$\pm$0.2 \\
 & 183 GHz H$_{2}$O & 18215$\pm$15, 18754$\pm$20 & 5.8$\pm$0.3, 5.5$\pm$0.2 & 391$\pm$33, 517$\pm$44 & 5.1$\pm$0.3 \\ \hline
IRAS 20551$-$4250 & HCN J=2--1 & 12890$\pm$1 & 23$\pm$1 & 202$\pm$2 & 4.8$\pm$0.1 \\ 
(1 kpc, 1$\farcs$2) & HCO$^{+}$ J=2--1 & 12887$\pm$1 & 34$\pm$1 & 214$\pm$3 & 7.4$\pm$0.1 \\
 & HNC J=2--1 & 12892$\pm$2 & 12$\pm$1 & 182$\pm$5 & 2.3$\pm$0.1 \\
 & 183 GHz H$_{2}$O & 12887$\pm$2 & 7.9$\pm$0.2 & 175$\pm$5 & 1.4$\pm$0.1 \\ \hline
IRAS 22491$-$1808 & HCN J=2--1 & 23317$\pm$5 & 12$\pm$1 & 434$\pm$11 & 5.1$\pm$0.2 \\
(1 kpc, 0$\farcs$69) & HCO$^{+}$ J=2--1 & 23285$\pm$7 & 7.2$\pm$0.2 & 466$\pm$16 & 3.3$\pm$0.2 \\
 & HNC J=2--1 & 23315$\pm$9 & 9.3$\pm$0.5 & 349$\pm$20 & 3.2$\pm$0.2 \\
 & 183 GHz H$_{2}$O & 23323$\pm$9 & 4.7$\pm$0.3 & 354$\pm$23 & 1.6$\pm$0.1 \\ \hline
\enddata

\tablenotetext{A}{Fixed to the best fit value.}

\tablecomments{Col.(1): Object name.
Beam size in kpc and arcsec is shown in parentheses.
Col.(2): Line.
Cols.(3)--(6): Gaussian fit of emission line in a $\sim$1kpc-beam 
spectrum at the continuum peak position.
For NGC 1614, that in a 200pc-beam spectrum is also shown.
Col.(3): Optical local standard of rest (LSR) velocity (v$_{\rm opt}$) 
of emission line peak (in km s$^{-1}$). 
Col.(4): Peak flux density (in mJy). 
Col.(5): Observed full width at half maximum (FWHM) (in km s$^{-1}$).
Col.(6): Gaussian-fit velocity-integrated flux (in Jy km s$^{-1}$). 
The Gaussian fit for the 183 GHz H$_{2}$O line is not attempted 
for NGC 1614 and IRAS 12112$+$0305 SW, 
because there is no emission signature in each spectrum (Figures 2 and 3).
}

\end{deluxetable}

We also attempt to estimate the fluxes of dense molecular 
(HCN, HCO$^{+}$, and HNC J=2--1) and 183 GHz H$_{2}$O emission lines, 
based on Gaussian fits in 2kpc-beam spectra (Appendix B).
We find that the flux increase from $\sim$1 kpc to 2kpc-beam data, 
is as high as $\sim$80\% in the starburst-dominated ULIRG IRAS
13509$+$0442, but significantly smaller in the remaining ULIRGs 
(Appendix B). 
Namely, we recover the bulk of ULIRGs' nuclear dense molecular
and 183 GHz H$_{2}$O line emission in the $\sim$1kpc-beam data.
A further increase of the beam size does not significantly increase dense 
molecular emission line signals, but increases noise a lot, 
making the detection of faint emission lines difficult.
Thus, we adopt quantities derived from $\sim$1kpc-beam data for our 
discussion of (U)LIRGs' nuclei.

There are serendipitously detected emission lines. 
The HC$_{3}$N J=20--19 line at $\nu_{\rm rest}$ = 181.945 GHz was covered 
in the J21b spectra of all ten (U)LIRGs and was clearly detected in four 
ULIRGs (IRAS 12112$+$0305 NE, IRAS 15250$+$3609, IRAS 20551$-$4250, 
and IRAS 22491$–$1808 in Figure 3f, 3n, 3r, and 3t, respectively).
 
Signatures of the HCN-VIB and HNC-VIB J=2--1 emission lines are recognizable 
in the spectra of a few (U)LIRGs in Figures 2 and 3.
We apply Gaussian fits in the $\sim$1kpc-beam spectra (and
200pc-beam spectra for NGC 1614) as well as create moment 0 maps 
with original beam.  
We claim significant ($>$3$\sigma$) detection as an isolated emission peak, 
of the HCN-VIB J=2--1 line in two ULIRGs (IRAS 15250$+$3609 and 
IRAS 20551$-$4250 in Figure 3m and 3u, respectively) and 
of the HNC-VIB J=2--1 line in three (U)LIRGs 
(NGC 1614 S, IRAS 15250$+$3609, and IRAS 22491$-$1808 in Figure 2f, 
3n, and 3t, respectively).
Table 7 summarizes the observed properties of these detected 
HC$_{3}$N J=20--19 and VIB emission lines.
The HCO$^{+}$-VIB J=2--1 emission line is not detected in any 
spectra (Figures 2 and 3).

\begin{deluxetable}{lc|c|cccc}
\tabletypesize{\scriptsize}
\tablecaption{Observed Properties of Other Possibly Detected Faint Emission 
Lines \label{tbl-7}} 
\tablewidth{0pt}
\tablehead{
\colhead{Object} & \colhead{Line} & \colhead{Moment 0 peak} & 
\multicolumn{4}{c}{Gaussian fit ($\sim$1 kpc beam) \tablenotemark{A}} \\  
\colhead{} & \colhead{} & \colhead{(original beam)} & \colhead{Velocity} & 
\colhead{Peak} & \colhead{FWHM} & \colhead{Flux} \\ 
\colhead{} & \colhead{} & \colhead{[Jy beam$^{-1}$ km s$^{-1}$]} & 
\colhead{[km s$^{-1}$]} & \colhead{[mJy]} & \colhead{[km s$^{-1}$]} &
\colhead{[Jy km s$^{-1}$]} \\  
\colhead{(1)} & \colhead{(2)} & \colhead{(3)} & \colhead{(4)} & 
\colhead{(5)} & \colhead{(6)} & \colhead{(7)} 
}
\startdata 
NGC 1614 S (200 pc) \tablenotemark{A} & HNC-VIB J=2--1 & 0.14
(3.6$\sigma$) & 4609$\pm$23 & 1.7$\pm$0.5 & 123 (fix)
\tablenotemark{B} & 0.22$\pm$0.06 \\ 
IRAS 12112$+$0305 NE (1.2 kpc) \tablenotemark{A} & HC$_{3}$N J=20--19
& 0.28 (5.9$\sigma$) & 21806$\pm$22 & 1.6$\pm$0.3 & 278$\pm$71  &
0.44$\pm$0.14 \\ 
 & HCN-VIB J=2--1 & 0.24 (4.9$\sigma$) \tablenotemark{C} & --- & --- &
--- & --- \\ 
IRAS 15250$+$3609 (1.2 kpc) \tablenotemark{A} & HC$_{3}$N J=20--19 & 0.62 (12$\sigma$) & 16615$\pm$10 & 3.2$\pm$0.2& 272$\pm$25  & 0.89$\pm$0.11 \\
 & HCN-VIB J=2--1 & 0.21 (3.8$\sigma$) \tablenotemark{D} & 16542$\pm$19 & 1.7$\pm$0.7 & 127 (fix) \tablenotemark{B} & 0.22$\pm$0.09 \\
   & HNC-VIB J=2--1 & 0.18 (5.3$\sigma$) & 16542$\pm$26 & 1.3$\pm$0.6 & 120$\pm$62  & 0.16$\pm$0.11 \\
IRAS 20551$-$4250 \tablenotemark{E} & HC$_{3}$N J=20--19 & 0.22 (5.6$\sigma$) & 12891$\pm$9 & 1.5$\pm$0.2 & 165$\pm$23 & 0.25$\pm$0.04 \\
 & HCN-VIB J=2--1 & 0.10 (6.6$\sigma$) & 12851$\pm$34 & 0.64$\pm$0.12 & 202$\pm$62 & 0.13$\pm$0.05 \\
 & SO 5(4)--4(3) & 0.40 (18$\sigma$) & 12892$\pm$4 & 2.5$\pm$0.1 & 174$\pm$11 & 0.44$\pm$0.04 \\
 & HOC$^{+}$ J=2--1 & 0.072 (5.3$\sigma$) & 12887$\pm$26 & 0.41$\pm$0.11 & 203$\pm$67 & 0.085$\pm$0.036 \\
IRAS 22491$-$1808 &  HC$_{3}$N J=20--19 & 0.75 (12$\sigma$) & 23302$\pm$14 & 3.0$\pm$0.2 & 355$\pm$38 & 1.1$\pm$0.1 \\
 & HCN-VIB J=2--1 & 0.16 (3.7$\sigma$) \tablenotemark{C} & --- & --- & --- & --- \\
 & HNC-VIB J=2--1 & 0.13 (4.8$\sigma$) & 23239$\pm$39 & 0.96$\pm$0.52 & 164 (fix) \tablenotemark{B} & 0.16$\pm$0.08 \\
\hline 
NGC 1614 & CS J=4--3 & 1.9 (5.1$\sigma$) & 4774$\pm$20 & 6.5$\pm$1.0 
& 249$\pm$38 & 1.7$\pm$0.4 \\
IRAS 06035$-$7102 & CS J=4--3 & 0.72 (6.9$\sigma$) & 23850$\pm$26 & 2.4$\pm$0.3 & 417$\pm$72 & 1.0$\pm$0.2 \\
IRAS 12112$+$0305 NE (1.2 kpc) \tablenotemark{A} & HC$_{3}$N J=21--20 & 0.55 (3.8$\sigma$) & 21668$\pm$60 & 1.7$\pm$0.5 & 299 (fix) \tablenotemark{B} & 0.50$\pm$0.16 \\
IRAS 20551$-$4250 & CS J=4--3 & 1.5 (24$\sigma$) & 12894$\pm$2 & 12$\pm$1 & 175$\pm$5 & 2.1$\pm$0.1 \\
IRAS 22491$-$1808 & HC$_{3}$N J=21--20 & 0.52 (7.1$\sigma$) & 23297$\pm$29 & 2.0$\pm$0.4 & 318$\pm$60 & 0.63$\pm$0.18 \\
\hline 
\enddata

\tablenotetext{A}{200 pc beam for NGC 1614 S, and 1.2 kpc beam
for IRAS 12112$+$0305 and IRAS 15250$+$3609.
For the remaining sources, 1 kpc beam size is adopted, including IRAS
06035$-$7102 (see also Figure 6b).
} 

\tablenotetext{B}{Fixed to the best fit value.}

\tablenotetext{C}{Signals of the emission tail at the lower frequency 
side of the HCO$^{+}$ J=2--1 line 
(a horizontal solid bar in Figures 3e and 3s) are integrated.
These values are only approximate for the HCN-VIB J=2--1 emission line 
flux, because we choose the integrated frequency range to 
minimize the contaminations from the nearby bright HCO$^{+}$ J=2--1 
emission line, and thus not all velocity components are likely to be covered, 
particularly for IRAS 22491$-$1808.}

\tablenotetext{D}{Only signals marked with the horizontal solid bar in 
Figure 3m are integrated.
Other possible emission components at $\nu_{\rm obs}$ = 168.6--168.75 GHz 
at even lower frequencies are not included, because they may originate from 
other components (e.g., redshifted HCO$^{+}$ J=2--1 emission 
in the P Cygni profile of outflow \citep{ima18b}).}

\tablenotetext{E}{There may be some emission signature at the expected 
frequency of the HNC-VIB J=2--1 line in Figure 3r, but its velocity profile 
is largely different from other detected bright emission lines.
We do not argue that the HNC-VIB J=2--1 emission line is clearly detected 
in our data, because certain unidentified faint emission line may also 
contribute to the observed emission signature.}

\tablecomments{Col.(1): Object name.
Col.(2): Line.
Data listed above the horizontal solid line are molecular lines detected 
in Figures 2 and 3 within the spectral windows that cover the primarily 
targeted molecular lines, and those below the line are molecular lines 
detected in other spectral windows in Figure 6. 
Only lines with significant detection ($>$3$\sigma$), either with the
Gaussian fit with all parameters set as free and/or moment 0 map, 
are listed.
Col.(3): Peak flux (in Jy beam$^{-1}$ km s$^{-1}$) in the moment 0 map of 
original beam. 
Detection significance relative to the rms noise (1$\sigma$) in the 
moment 0 map is shown in parentheses. 
The peak position spatially agrees with simultaneously obtained continuum 
peak position within peak determination uncertainty.
For the CS J=4--3 and HC$_{3}$N J=21--20 lines below the horizontal 
solid line, data of IRAS 12112$+$0305 NE and IRAS 22491$-$1808 were 
obtained during J21a observations, while the remaining data were obtained 
during J21b observations. 
Synthesized beam sizes are comparable to those of individual J21a or J21b 
data listed in Table 3 (column 6).
Cols.(4)--(7): Gaussian fit of emission line 
in $\sim$1kpc-beam spectrum at the continuum peak position. 
For NGC 1614 S, the beam size is 200 pc.
Col.(4): Optical LSR velocity (v$_{\rm opt}$) of emission line peak 
(in km s$^{-1}$). 
Col.(5): Peak flux density (in mJy). 
Col.(6): Observed FWHM (in km s$^{-1}$).
Col.(7): Gaussian-fit velocity-integrated flux (in Jy km s$^{-1}$).
}

\end{deluxetable}

In the spectrum of IRAS 20551–4250 (Figures 3q and 3u), at the high 
frequency side of the HCO$^{+}$ J=2--1 line, an emission line signature exists,
which we identify as SO 5(4)--4(3) ($\nu_{\rm rest}$ = 178.605 GHz). 
Moreover, the HOC$^{+}$ J=2--1 emission line ($\nu_{\rm rest}$ = 178.972 GHz) is 
detected (Figure 3u).
We create moment 0 maps (with original beam) of both the lines
and confirmed clear signals at the continuum peak position of IRAS
20551$-$4250. 
Gaussian fitting results in the 1kpc-beam spectrum as well as
peak flux values in the moment 0 maps of these lines are presented in Table 7. 
Our Gaussian fits of other possible emission-like features 
provide no significant detection ($<$3$\sigma$).

In spectral windows other than J21a and J21b, 
the CS J=4--3 ($\nu_{\rm rest}$ = 195.954 GHz) and HC$_{3}$N J=21--20 
($\nu_{\rm rest}$ = 191.040 GHz) lines are included in certain fraction of 
(U)LIRGs and are detected in a few (U)LIRGs.
Figure 6 displays these spectra.
The observed properties of these detected emission lines are also 
summarized in Table 7.
Both CS J=4--3 and HC$_{3}$N J $>$ 10 lines have high critical densities 
of $>$10$^{5}$ cm$^{-3}$ \citep{shi15}, and thus can be regarded as 
additional dense molecular gas tracers.
 
\begin{figure}
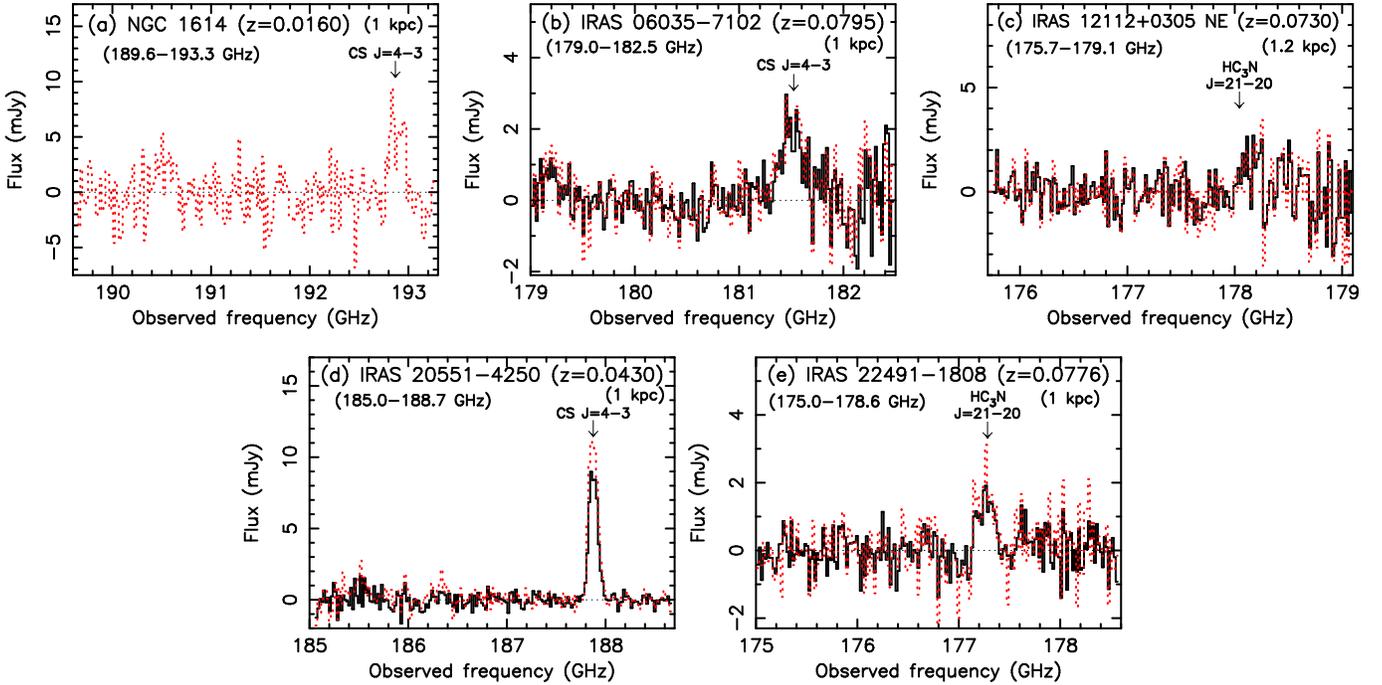

\begin{center}
\includegraphics[angle=-90,scale=.26]{f6a.eps} 
\hspace{0.05cm}
\includegraphics[angle=-90,scale=.26]{f6b.eps} 
\hspace{0.05cm}
\includegraphics[angle=-90,scale=.26]{f6c.eps} \\ 
\vspace{0.3cm}
\includegraphics[angle=-90,scale=.26]{f6d.eps} 
\includegraphics[angle=-90,scale=.26]{f6e.eps} 
\end{center}
\caption{Serendipitously detected emission lines in spectral windows other 
than those 
covering the primarily targeted lines (HCN, HCO$^{+}$, HNC, and H$_{2}$O). 
The black solid and red dotted line is an original-beam 
and a $\sim$1kpc-beam spectrum at the continuum peak, respectively.
{\it (a)}: CS J=4--3 emission line ($\nu_{\rm rest}$ = 195.954 GHz) 
in the 189.6--193.3 GHz spectrum of NGC 1614.
{\it (b)}: CS J=4--3 emission line in the 179.0--182.5 GHz spectrum 
of IRAS 06035$-$7102.
{\it (c)}: HC$_{3}$N J=21--20 emission line ($\nu_{\rm rest}$ = 191.040 GHz) 
in the 175.7--179.1 GHz spectrum of IRAS 12112$+$0305 NE.
{\it (d)}: CS J=4--3 emission line in the 185.0--188.7 GHz spectrum 
of IRAS 20551$-$4250.
{\it (e)}: HC$_{3}$N J=21--20 emission line in the 175.0--178.6 GHz 
spectrum of IRAS 22491$-$1908.
For (b) IRAS 06035$–$7102, the beam size of the CS J=4--3 line in the J21b 
data is smaller than 1 kpc (Table 3); therefore, the red dotted line spectrum 
is extracted from 1kpc-beam data (not 1.6 kpc, as shown in Figure 3), 
in order not to decrease the detection significance of this faint 
CS J=4--3 emission line.
(Gaussian-fit velocity-integrated flux of CS J=4--3 emission line agrees 
within 4\% between 1 kpc and 1.6 kpc beam.)
}
\end{figure}

\section{Discussion} 

\subsection{HCN to HCO$^{+}$ J=2--1 Flux Ratios}

From Table 6, we find that the observed HCN J=2--1 flux is significantly
higher than the HCO$^{+}$ J=2--1 flux in several ULIRGs classified as
AGN important ($\S$2; IRAS 12112$+$0305 NE, IRAS 15250$+$3609, 
the Superantennae, and IRAS 22491$-$1808), while no such trend is
seen in the three starbust-classified galaxy nuclei, NGC 1614, IRAS
12112$+$0305 SW, and IRAS 13509$+$0442 ($\S$2), where the observed 
HCO$^{+}$ flux is higher than or at least comparable to the observed 
HCN J=2--1 flux. 
Elevated HCN emission, relative to HCO$^{+}$ emission, has been
observed in a high fraction of, if not all, luminous AGNs at J=1--0 
\citep[e.g.,][]{koh05,ima07b,kri08,ima09a,pri15},
J=3--2 \citep[e.g.,][]{ima16a,ima16c,ima18a,ima19}, and 
J=4--3 \citep[e.g.,][]{gar14,ima14,vit14,izu15,izu16,ima18b}.
A similar trend is seen also at J=2--1.
However, not all AGN-important ULIRGs clearly show elevated HCN J=2--1
emission, relative to HCO$^{+}$ J=2--1 
(e.g., IRAS 06035$-$7102, IRAS 08572$+$3915, IRAS 12127$-$1412, and 
IRAS 20551$-$4250 in Table 6), as previously seen at other 
J-transition lines \citep[e.g.,][]{ima19,pri20}. 
Further detailed discussion about how the observed HCN to HCO$^{+}$
flux ratios change among J=2--1, J=3--2, and J=4--3, will be presented by 
Imanishi et al. (2021; in preparation), to better understand the 
properties of nuclear dense molecular gas in nearby (U)LIRGs 
with and without luminous AGN signatures.

\subsection{183 GHz H$_{2}$O Emission}

\subsubsection{Flux Comparison}

In the moment 0 maps in Figure 1, the 183 GHz
H$_{2}$O emission line is not clearly detected ($<$4$\sigma$) in the three 
starbust-classified galaxy nuclei (NGC 1614, IRAS 12112$+$0305 SW, and 
IRAS 13509$+$0442), whereas it is detected with higher significance  
($>$4.5$\sigma$) in the AGN-important ULIRGs' nuclei (Table 4, column 5).
Gaussian fits of the $\sim$1kpc-beam spectra also provide
significant ($>$3$\sigma$) H$_{2}$O detection in all AGN-important ULIRGs
except IRAS 12127$-$1412, but do not ($<$3$\sigma$) for the three
starburst nuclei (Table 6, column 6).
As briefly described in $\S$1, theoretically, the 183 GHz H$_{2}$O emission 
line can be extremely luminous in warm and dense molecular gas in the immediate 
vicinity of a luminous AGN, if maser amplification caused by population 
inversion occurs \citep[e.g.,][]{deg77,neu91,yat97,mal02}.
The strong 183 GHz H$_{2}$O megamaser ($>$10L$_{\odot}$) emission was 
detected in a few very nearby ($<$20 Mpc) AGNs 
\citep{hum05,hum16}.
In the AGN-hosting ULIRG, the Superantennae at $z =$ 0.0617 
($\sim$270 Mpc), the detected bright 183 GHz H$_{2}$O emission 
line was argued to primarily originate from the megamaser 
phenomena in AGN-illuminated gas at the very center of galaxy nucleus
($<<$1 kpc), 
based on substantially elevated flux (luminosity) relative to, and 
spatially more compact distribution than, other dense molecular gas 
tracers (HCN, HCO$^{+}$, and HNC J=2--1 emission lines) \citep{ima21}.
Furthermore, in such gas near an AGN, it is predicted that the 
H$_{2}$O abundance can be enhanced by X-ray radiation 
\citep[e.g.,][]{neu94,meij12} and/or H$_{2}$O evaporation from dust grain 
mantles into gas phase \citep[e.g.,][]{gon10}.
In fact, high H$_{2}$O abundance in AGNs has been observationally supported 
\citep[e.g.,][]{gon10,gon12,liu17}.
Hence, even if the maser amplification does not work as effectively 
as in the Superantennae, the 183 GHz H$_{2}$O emission in AGN-important ULIRGs 
can be elevated because of the enhanced H$_{2}$O abundance.

We investigate the 183 GHz H$_{2}$O emission line flux, in comparison 
with other dense molecular gas tracers, to determine whether the elevated 
183 GHz H$_{2}$O emission, previously found in the Superantennae, is also 
observed in other (U)LIRGs.
Figures 7a and 7b compare, respectively, the 183 GHz H$_{2}$O to 
HCO$^{+}$ J=2--1 
and the 183 GHz H$_{2}$O to HNC J=2--1 flux ratio (abscissa), 
with HCN to HCO$^{+}$ J=2--1 flux ratio (ordinate).
In Figure 7a, two sources with distinctly high 183 GHz H$_{2}$O to 
HCO$^{+}$ J=2--1 flux ratios show high HCN to HCO$^{+}$ J=2--1
flux ratios with $>$1.6.
In Figure 7c, the 183 GHz H$_{2}$O to HCN J=2--1 flux ratio is 
compared with the HNC to HCN J=2--1 flux ratio.
The Superantennae is distinguishable as the 183 GHz H$_{2}$O emission 
is elevated relative to all the HCO$^{+}$, HNC, and HCN J=2--1 emission 
lines.
IRAS 15250$+$3609 also shows elevated 183 GHz H$_{2}$O emission, 
with respect to the HCO$^{+}$ and HCN J=2--1 lines, but not 
relative to the HNC J=2--1.
The signatures of the elevated 183 GHz H$_{2}$O emission are 
much weaker in the remaining (U)LIRGs.

\begin{figure}
\begin{center}
\includegraphics[angle=-90,scale=.249]{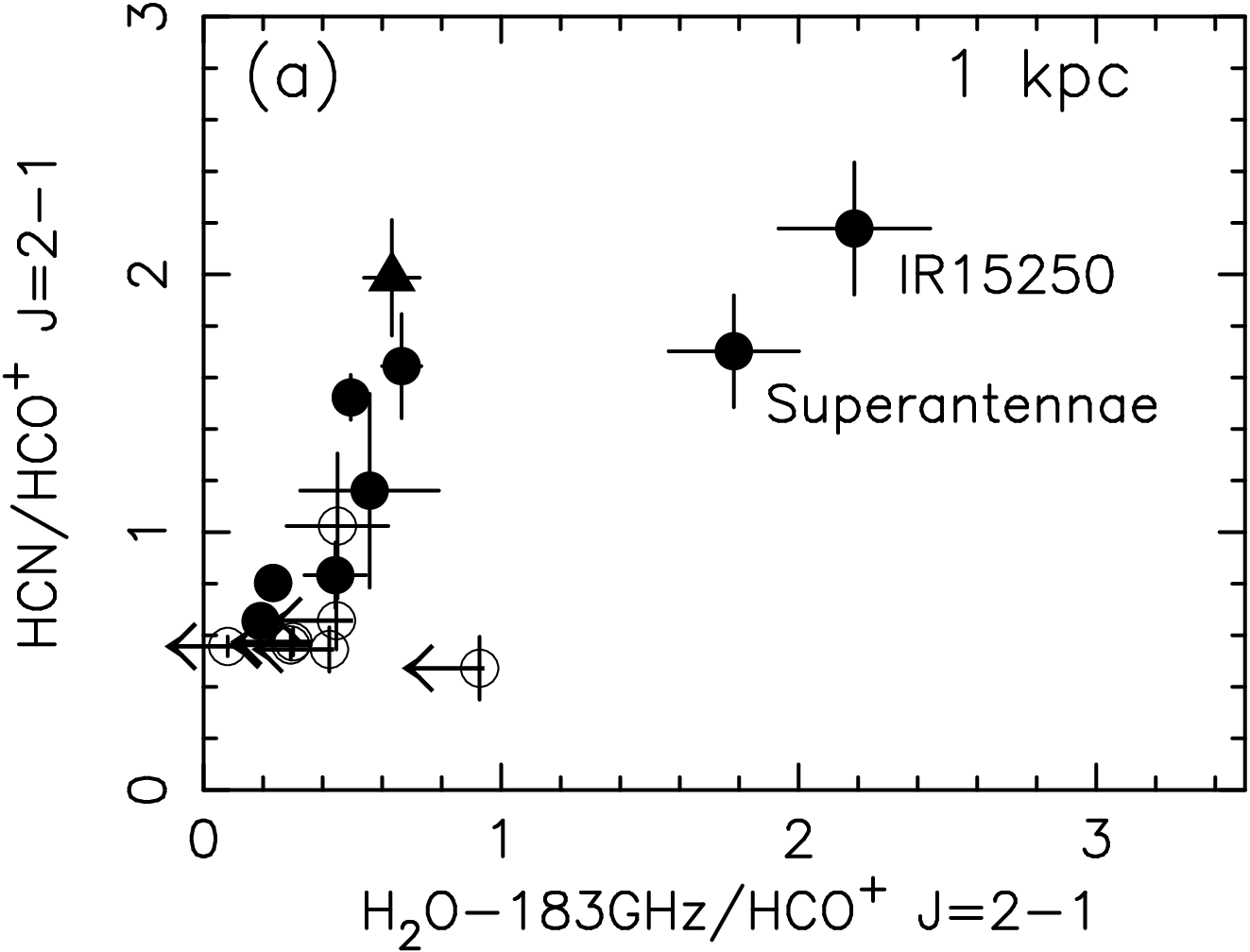} 
\hspace{0.3cm}
\includegraphics[angle=-90,scale=.249]{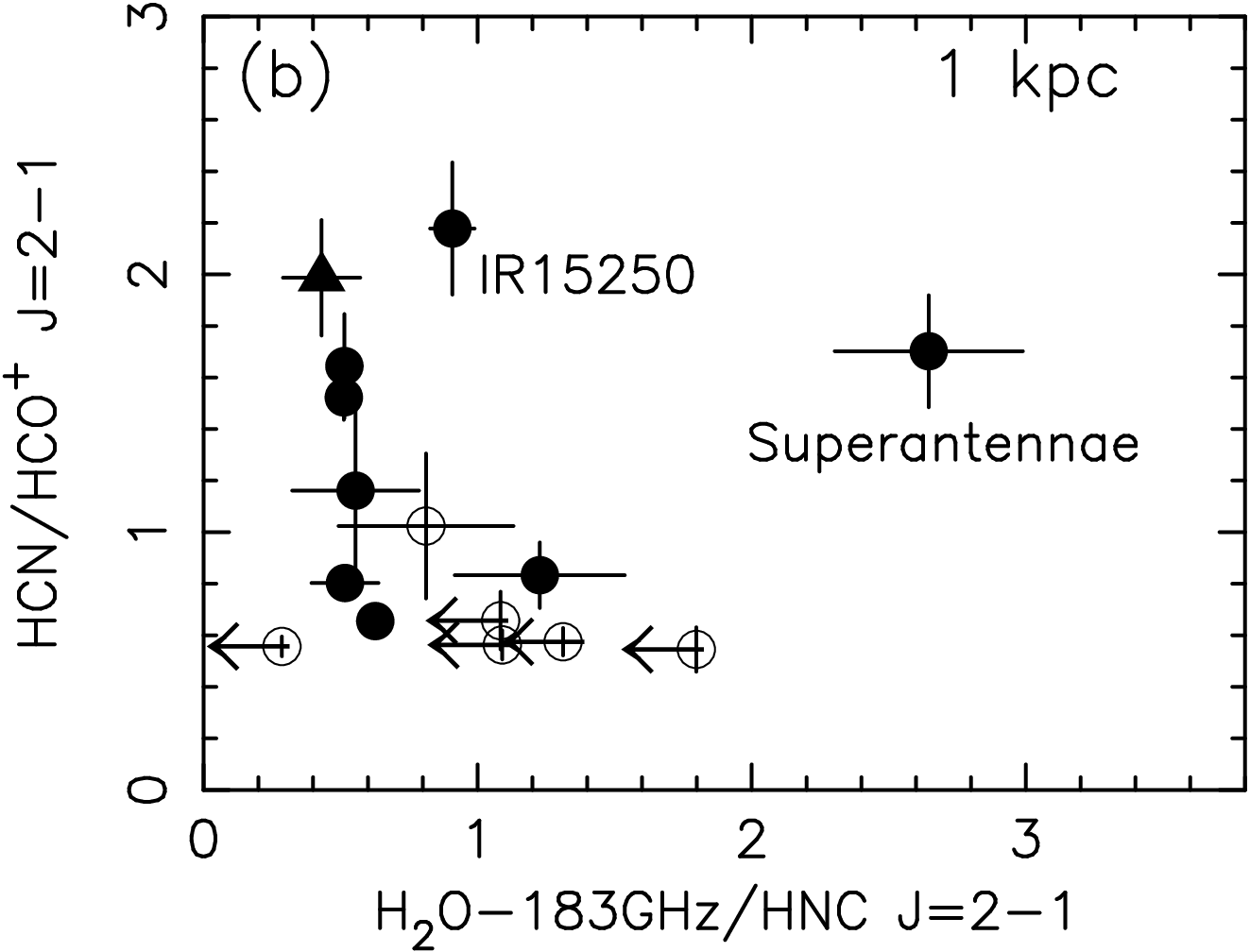} 
\hspace{0.3cm}
\includegraphics[angle=-90,scale=.249]{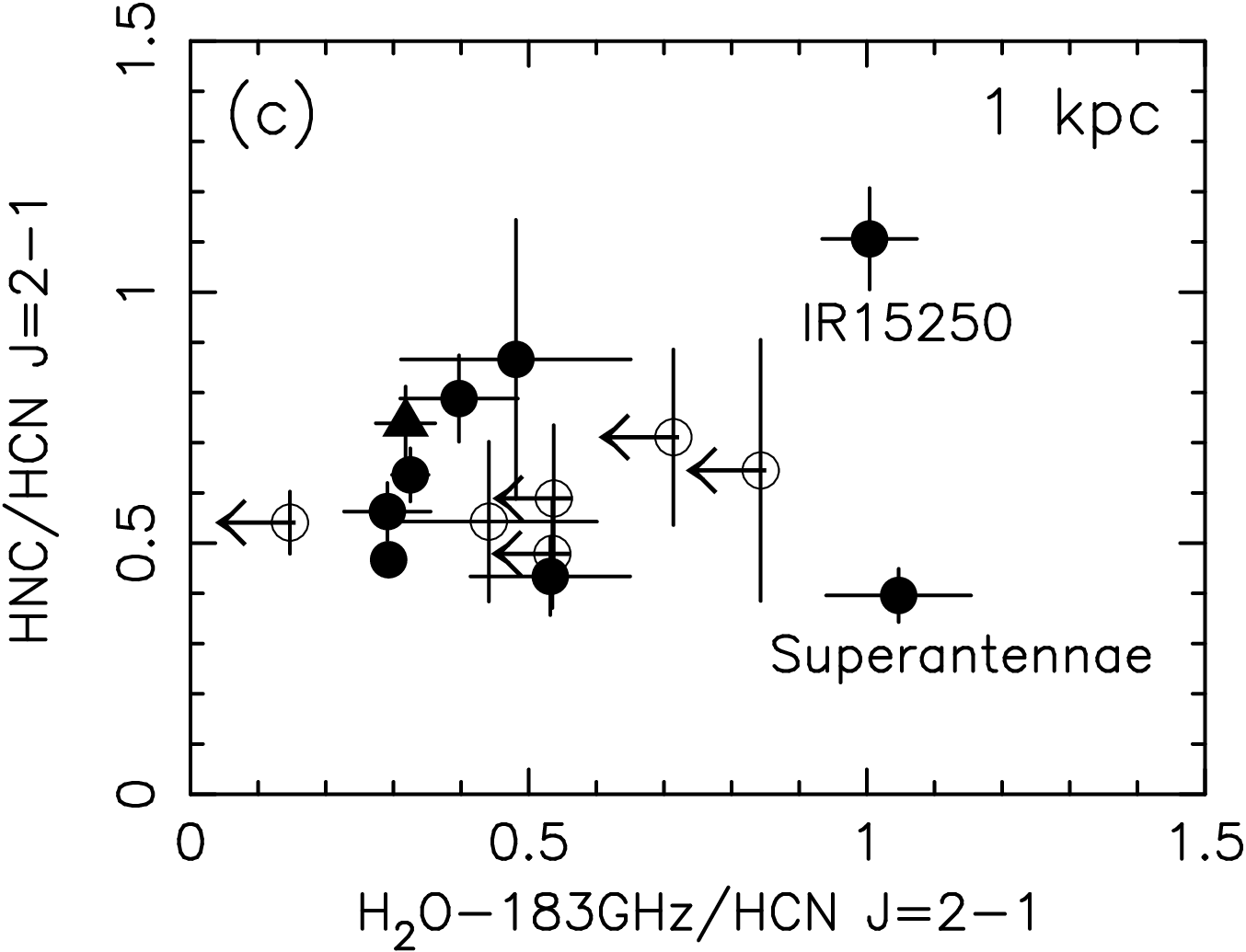}  
\end{center}
\caption{
Emission line flux ratio in $\sim$1kpc-beam data and 200pc-beam
data for NGC 1614). 
{\it (a)} 183 GHz H$_{2}$O to HCO$^{+}$ J=2--1 (abscissa) vs. 
HCN J=2--1 to HCO$^{+}$ J=2--1 flux ratio (ordinate). 
{\it (b)} 183 GHz H$_{2}$O to HNC J=2--1 (abscissa) vs. 
HCN J=2--1 to HCO$^{+}$ J=2--1 flux ratio (ordinate).
{\it (c)} 183 GHz H$_{2}$O to HCN J=2--1 (abscissa) vs. 
HNC J=2--1 to HCN J=2--1 flux ratio (ordinate).
Our data are shown as circles. 
Filled circles are AGN-important ULIRGs and
open circles are three starburst-dominated (U)LIRGs (NGC 1614, 
IRAS 12112$+$0305 SW, and IRAS 13509$+$0442).
NGC 1614 has five data points (200pc-beam data at four positions and 
1kpc-beam data).
Data of the nearby ULIRG, Arp 220 ($z =$ 0.018), measured with $\sim$13 kpc 
beam 
\citep{gal16}, are displayed as filled triangles for reference. 
The Superantennae and IRAS 15250$+$3609 (two sources exhibiting distinct 
183 GHz H$_{2}$O emission excess in certain plots) are indicated as 
Superantennae and IR15250, respectively. 
The Gaussian-fit velocity-integrated fluxes in Table 6 are used 
for bright emission lines with significant ($>>$3$\sigma$) detection. 
For IRAS 12127$-$1412 and IRAS 13509$+$0442, although some emission 
signature of the 183 GHz H$_{2}$O line in the spectra (Figure 3) is observed, 
the Gaussian fit detection significance is $<$3$\sigma$ (Table 6).
We adopt peak flux values of $\sim$1kpc-beam moment 0 maps with 
0.34 (3.2$\sigma$) and 0.26 (3.2$\sigma$) [Jy km s$^{-1}$] (Appendix C) 
as the 183 GHz H$_{2}$O emission line flux for IRAS 12127$-$1412 and 
IRAS 13509$+$0442, respectively.
For NGC 1614 (four continuum positions) and IRAS 12112$+$0305 SW, 
we adopt 3$\sigma$ upper limits in 200pc-beam and 
$\sim$1kpc-beam moment 0 maps ($<$0.36 and $<$0.76 Jy km s$^{-1}$,
respectively) as the 183 GHz H$_{2}$O emission line flux.
The HNC J=2--1 emission line for IRAS 12112$+$0305 SW is also 
$<$3$\sigma$ detection in the Gaussian fit in the 
$\sim$1kpc-beam spectrum (Table 6), 
as well as in a $\sim$1kpc-beam moment 0 map ($<$0.47 Jy km s$^{-1}$).
In {\it (b)} and {\it (c)}, IRAS 12112$+$0305 SW is not plotted because both 
the HNC J=2--1 and 183 GHz H$_{2}$O emission line fluxes are upper limits.
}
\end{figure}

\subsubsection{Compactness}

We search for the presence of spatially unresolved 
($<<$1 kpc), compact H$_{2}$O emission and measure how its fraction, 
relative to the total nuclear ($\sim$1 kpc) emission, is higher than the 
other three dense molecular lines, using several indicators.
In Figure 3, it is clear that the peak flux increase from the original 
beam to the $\sim$1 kpc beam, is much smaller for the 183 GHz H$_{2}$O line 
than the HNC J=2--1 line in the spectra of 
(e.g.,) IRAS 20551$-$4250 and IRAS 22491$-$1808 
(Figures 3r and 3t), suggesting that the fraction of 
spatially unresolved components is higher for the 183 GHz H$_{2}$O in some 
ULIRGs.

We (1) quantitatively compare the flux increase of the H$_{2}$O and
HNC J=2--1 emission lines from original beam to $\sim$1kpc-beam, based
on Gaussian fit in the spectra. 
We find that the flux increase of the H$_{2}$O is significantly 
smaller than that of HNC J=2--1 in two ULIRGs 
(the Superantennae and IRAS 20551$-$4250) 
(Appendix B, columns 3 and 4 of Table 11). 
The presence of spatially unresolved compact H$_{2}$O emission
can explain the observed trend in these ULIRGs.
However, as HNC abundance can be low in AGN-illuminated warm molecular gas 
at the center of a galaxy nucleus \citep{ima20}, it may be possible 
that the observed higher HNC J=2--1 flux increase is simply due to 
the suppression of HNC J=2--1 emission at the innermost region, 
rather than the presence of compact H$_{2}$O emission.
To overcome this ambiguity, we combine the H$_{2}$O data with the HCN and 
HCO$^{+}$ J=2--1 line data as well (different synthesized beams from
H$_{2}$O) and (2) check whether the flux increase from
$\sim$1kpc-beam to 2kpc-beam is  substantially smaller for the
H$_{2}$O emission line than that of the HCN and HCO$^{+}$ J=2--1
emission lines. 
This trend is found in three ULIRGs (IRAS 12112$+$0305 NE, 
the Superantennae, and IRAS 20551$-$4250) (Appendix B, columns 4 and 5
of Table 11).

We also (3) compare the measured emission size of the H$_{2}$O with 
those of other dense molecular gas tracers, by applying the CASA 
task ``imfit'' to the original beam moment 0 maps (Figure 1).
These results are summarized in Table 8.
Because the beam sizes are not the same between J21a and J21b data, we 
first compare the deconvolved intrinsic size of H$_{2}$O and HNC J=2--1.
The Superantennae and IRAS 20551$-$4250 show substantially smaller 
size for the H$_{2}$O than the HNC J=2--1. 
For these two ULIRGs, the intrinsic size of H$_{2}$O is also much 
smaller than those of HCN and HCO$^{+}$ J=2--1.

\begin{deluxetable}{l|ccc|ccc}
\tabletypesize{\scriptsize}
\tablecaption{Emission Size \label{tbl-8}} 
\tablewidth{0pt}
\tablehead{
\colhead{} & \multicolumn{3}{|c|}{J21a} & \multicolumn{3}{c}{J21b} \\ 
\colhead{Object} & \colhead{HCN J=2--1} & \colhead{HCO$^{+}$ J=2--1} & 
\colhead{J21a continuum} & \colhead{HNC J=2--1} & \colhead{183 GHz H$_{2}$O} & 
\colhead{J21b continuum} \\ 
\colhead{[mas]} & \colhead{[mas]} & \colhead{[mas]} & \colhead{[mas]} & 
\colhead{[mas]} & \colhead{[mas]} & \colhead{[mas]} \\
\colhead{(1)} & \colhead{(2)} & \colhead{(3)} & \colhead{(4)} & 
\colhead{(5)} & \colhead{(6)} & \colhead{(7)} 
}
\startdata 
IRAS 06035$-$7102 & 587$\pm$107, 540$\pm$140 & 703$\pm$72, 523$\pm$50 
& 705$\pm$186, 613$\pm$236 & 352$\pm$77, 248$\pm$184 
& 420$\pm$174, 194$\pm$90 & 312$\pm$75, 296$\pm$86 \\
IRAS 08572$+$3915 & 397$\pm$114, 213$\pm$180 & 478$\pm$106, 360$\pm$62 
& 318$\pm$65, 287$\pm$124 & --- & --- & 238$\pm$86, 208$\pm$155 \\
IRAS 12112$+$0305 NE & 552$\pm$57, 337$\pm$101 & 687$\pm$80, 462$\pm$135 
& 283$\pm$72, 186$\pm$156 & 479$\pm$105, 428$\pm$122 & --- & 380$\pm$94, 228$\pm$90 \\
IRAS 12127$-$1412 & 290$\pm$88, 46$\pm$69 & 333$\pm$96, 172$\pm$53 
& 151$\pm$22, 96$\pm$68 & 357$\pm$116, 185$\pm$104 & 759$\pm$349, 239$\pm$150 
& 162$\pm$40, 97$\pm$92 \\
IRAS 13509$+$0442 & 491$\pm$139, 392$\pm$119 & 659$\pm$164, 356$\pm$104 
& 481$\pm$75, 367$\pm$57 & 695$\pm$184, 376$\pm$118 & 655$\pm$276, 327$\pm$155 
& 580$\pm$97, 473$\pm$82 \\
IRAS 15250$+$3609 & 338$\pm$114, 330$\pm$145 & 597$\pm$183, 432$\pm$341 
& 445$\pm$85, 375$\pm$80 & 467$\pm$85, 234$\pm$72 & 357$\pm$85, 220$\pm$52 
& 388$\pm$71, 226$\pm$30  \\ 
Superantennae & 838$\pm$123, 446$\pm$65 & 1040$\pm$141, 466$\pm$61 
& 348$\pm$71, 284$\pm$61 & 563$\pm$132, 483$\pm$137 & 171$\pm$61, 134$\pm$46 
& 280$\pm$39, 245$\pm$30  \\ 
IRAS 20551$-$4250 & 479$\pm$59, 373$\pm$58 & 552$\pm$80, 417$\pm$95 
& 367$\pm$56, 346$\pm$74 & 382$\pm$28, 264$\pm$16 & 203$\pm$45, 135$\pm$40 
& 304$\pm$32, 236$\pm$20 \\
IRAS 22491$-$1808 & 277$\pm$36, 231$\pm$28 & 433$\pm$80, 337$\pm$64 
& 217$\pm$49, 131$\pm$41 & 285$\pm$34, 223$\pm$29 & 251$\pm$78, 157$\pm$92 
& 363$\pm$41, 160$\pm$37 \\
\enddata

\tablecomments{
Col.(1): Object name.
Cols.(2)--(7): Deconvolved size in milli-acrsecond (mas) in the original 
beam moment 0 map (Figure 1) 
derived from the CASA task ``imfit''.
Value in the major and minor axis is shown as first and second, respectively.
Because it is very difficult to constrain the deconvolved intrinsic 
size to be considerably (a factor of $>>$2--3) smaller than the beam size 
(Table 3, column 6), in the case of extremely compact emission, 
the derived value can still be larger than the actual value.
The mark ``---'' is added when the size is not sufficiently constrained 
owing to limited S/N ratios.
The position angle (PA) is not shown because its uncertainty is large, 
except in the case of the Superantennae whose PA is
$\sim$150$^{\circ}$ (east of north)
with $<$10$^{\circ}$ uncertainty for HCN, HCO$^{+}$, and HNC J=2--1 
(roughly aligned to the blueshifted and redshifted direction in 
Figure 4), and 70$\pm$80$^{\circ}$ (large uncertainty) for the 183 GHz
H$_{2}$O line.  
Col.(2): HCN J=2--1.
Col.(3): HCO$^{+}$ J=2--1.
Col.(4): J21a continuum.
Col.(5): HNC J=2--1.
Col.(6): 183 GHz H$_{2}$O.
Col.(7): J21b continuum.
}

\end{deluxetable}

We (4) create H$_{2}$O to HCO$^{+}$ J=2--1 flux ratio maps, after 
matching beam size, and find that four ULIRGs 
(IRAS 12112$+$0305 NE, the Superantennae, IRAS 20551$-$4250, and 
IRAS 22491$-$1808) show a higher ratio at the continuum peak 
(putative AGN position) than in spatially extended regions.
These maps are shown in Figure 8.
However, compared to the Superantennae, the central-concentrations of 
the elevated H$_{2}$O emission regions are weaker in the remaining
three ULIRGs.  

\begin{figure}
\begin{center}
\includegraphics[angle=0,scale=.28]{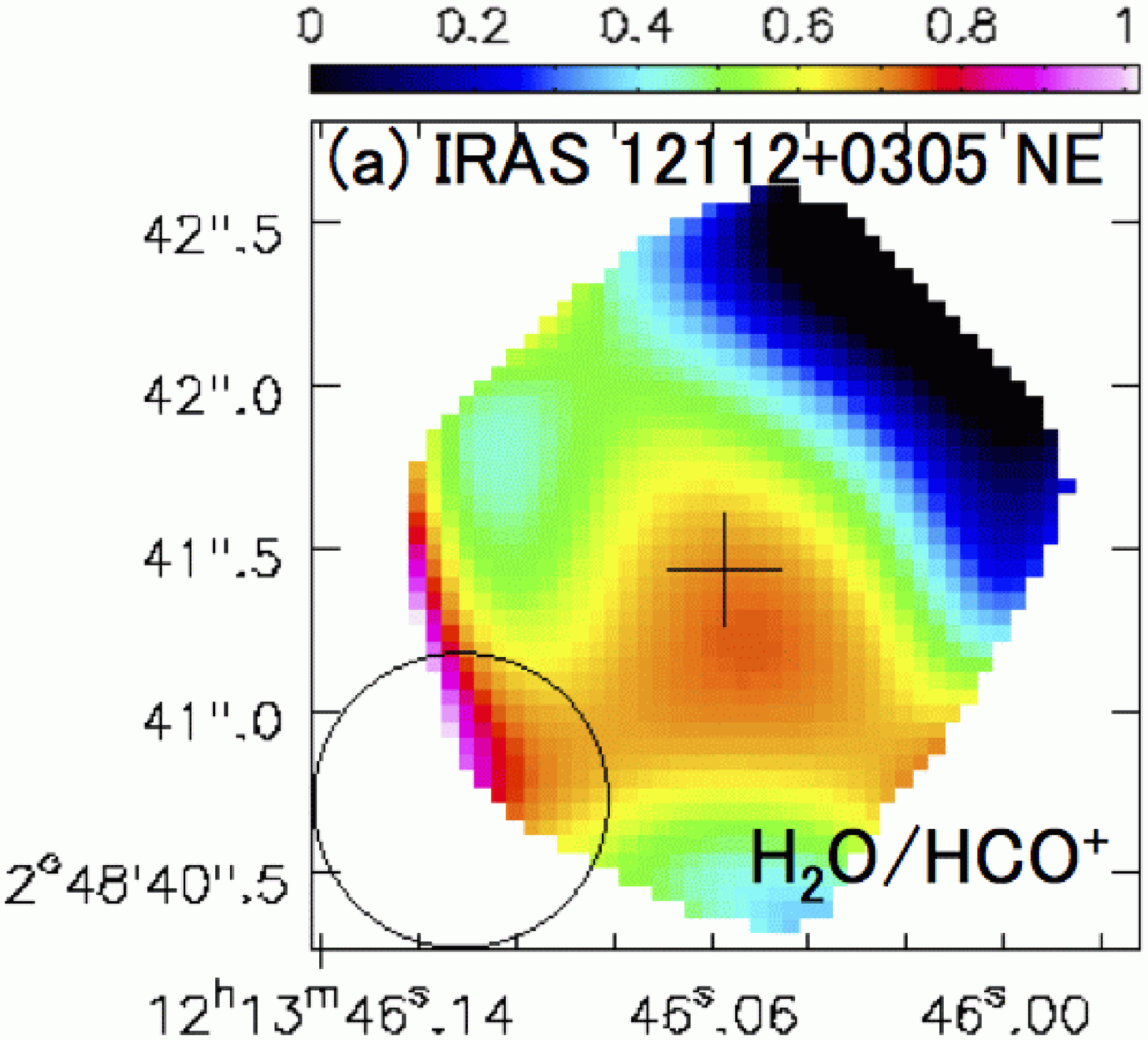} 
\hspace{0.05cm}
\includegraphics[angle=0,scale=.28]{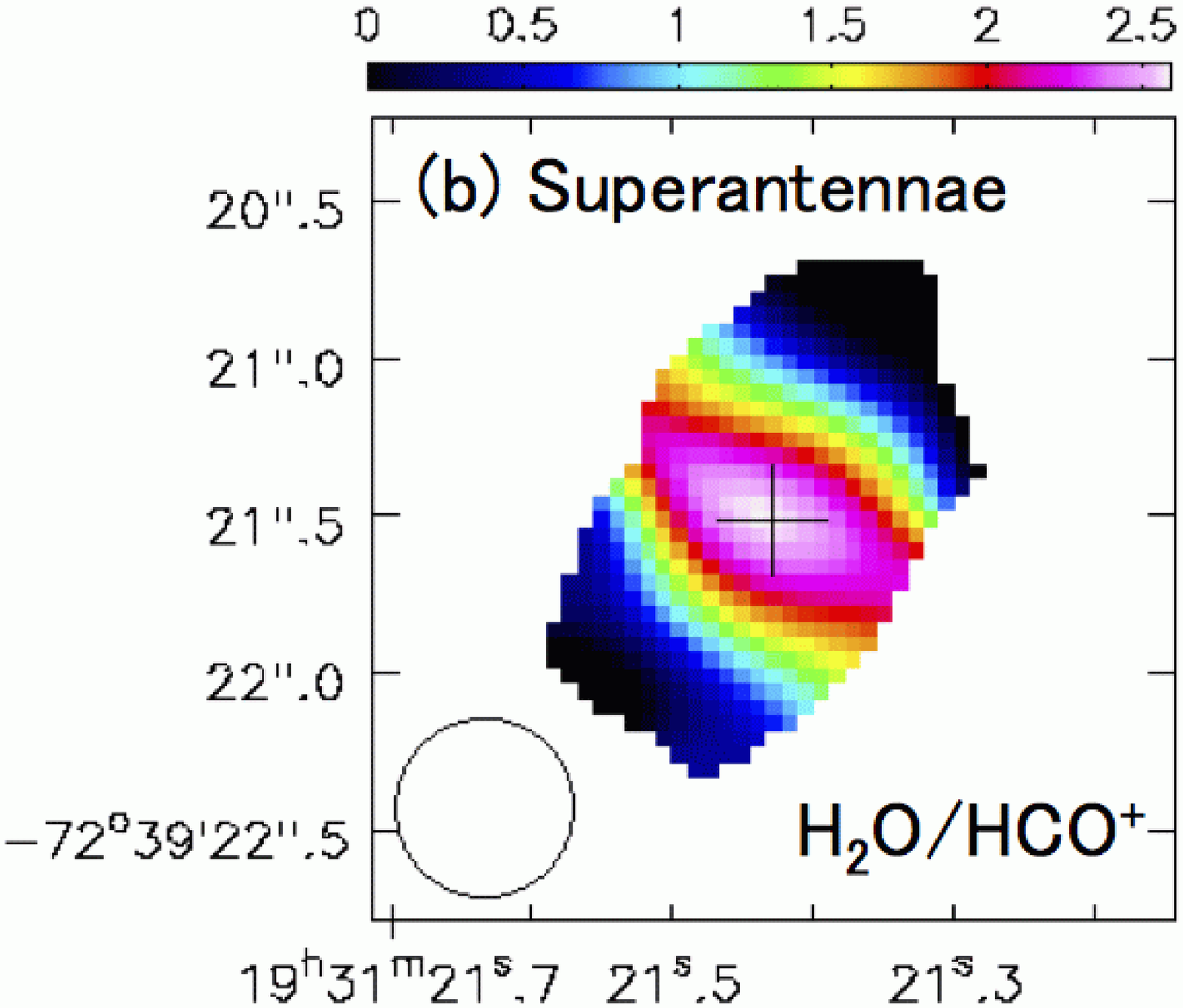} \\
\includegraphics[angle=0,scale=.28]{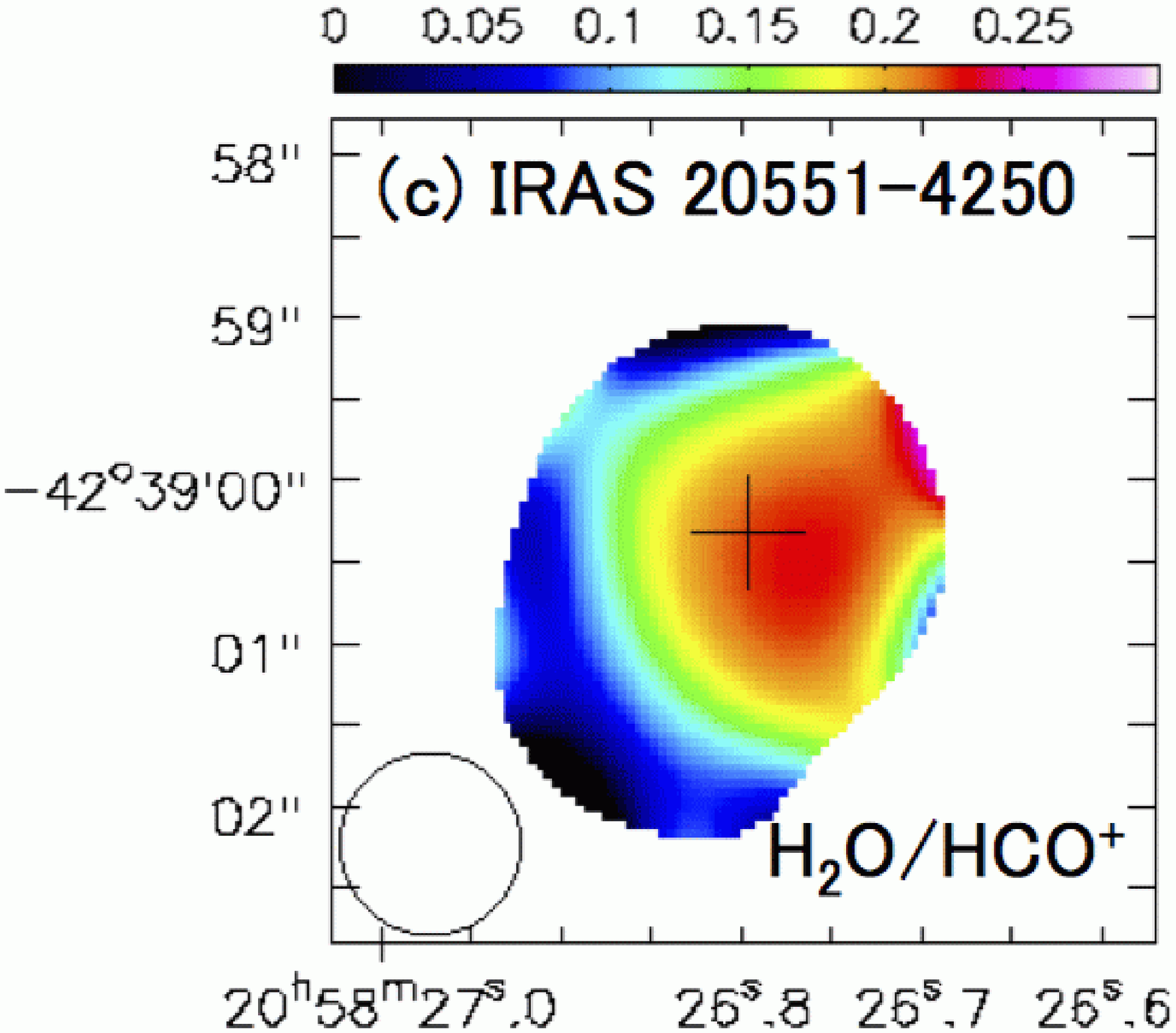} 
\hspace{0.05cm}
\includegraphics[angle=0,scale=.28]{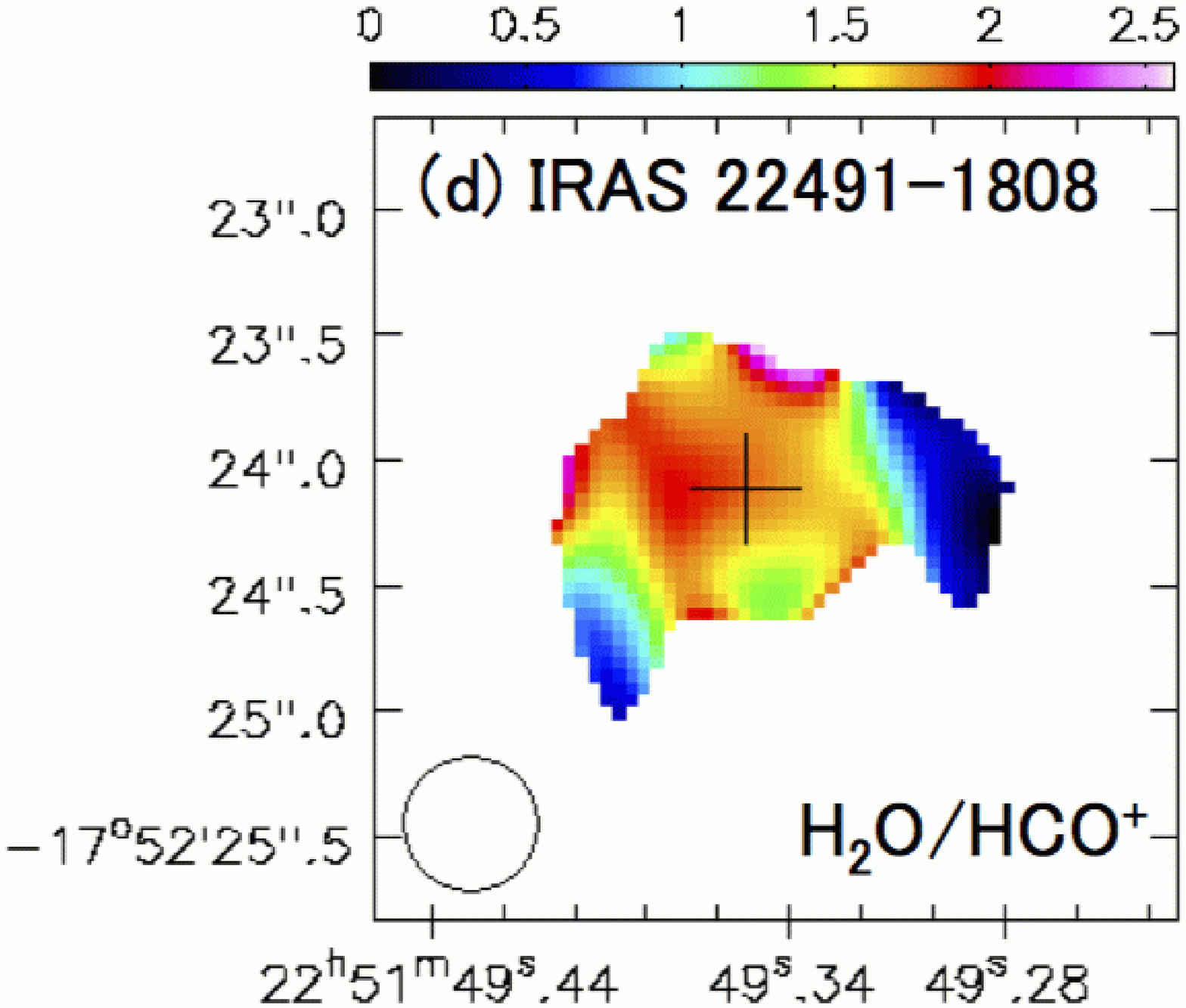} \\
\end{center}
\caption{
Ratio of 183 GHz H$_{2}$O to HCO$^{+}$ J=2--1 flux measured 
in Jy km s$^{-1}$
of {\it (a)} IRAS 12112$+$0305 NE, {\it (b)} the Superantennae, 
{\it (c)} IRAS 20551$-$4250, and {\it (d)} IRAS 22491$-$1808, 
with matched circular beam of 
0$\farcs$91, 0$\farcs$57, 1$\farcs$1 and 0$\farcs$53, respectively. 
Continuum peak position is shown as a cross.
An appropriate cutoff ($\sim$2$\sigma$) for HCO$^{+}$ J=2--1 (denominator) 
is applied to prevent the ratio map being dominated by noise.
}
\end{figure}

\subsubsection{Dynamics}

We also try to use dynamical information to constrain the
presence of compact H$_{2}$O emission.
We (5) compare the properties of the H$_{2}$O emission with those of the 
other dense 
molecular tracers in the moment 1 maps because spatially unresolved ($<<$1 kpc) 
H$_{2}$O emission may be dynamically decoupled from nuclear ($\sim$1 kpc) 
dense molecular gas, as observed in the Superantennae \citep{ima21}.
Besides the Superantennae, in IRAS 08572$+$3915, IRAS
12112$+$0305 NE, and IRAS 22491$-$1808, blueshifted and redshifted
motions identified in the HCN, HCO$^{+}$, and HNC J=2--1 lines along
similar directions, are not clearly seen in the H$_{2}$O line (Figure 4). 
However, since H$_{2}$O emission is fainter than the other dense
molecular lines and outer regions are not sufficiently probed in the
moment 1 maps (Figure 4), further discussion of the possible dynamical
difference is difficult.
In the moment 2 maps (Figure 5), dynamically distinct features of the
H$_{2}$O, compared to the other three dense molecular gas tracers, are
not clearly observed, except the Superantennae \citep{ima21}.

\subsubsection{Summary of the 183 GHz H$_{2}$O Emission}

In summary, based on any of the above methods (1)--(5), besides
the Superantennae, it is suggested that spatially more compact
H$_{2}$O emission than the nuclear ($\sim$1 kpc) HCN, HCO$^{+}$, and
HNC J=2--1 emission, may be present in three ULIRGs 
(IRAS 12112$+$0305 NE, IRAS 20551$-$4250, and IRAS 22491$-$1808).
Interestingly, despite the very high 183 GHz H$_{2}$O to HCO$^{+}$
J=2--1 and 183 GHz H$_{2}$O to HCN J=2--1 flux ratios (Figure 7), IRAS
15250$+$3609 has no obvious signature of compact H$_{2}$O emission in
the above methods.  
The compact H$_{2}$O emission in the three ULIRGs can be of
AGN-magamaser origin and/or thermal (non-maser) emission from the
H$_{2}$O-abundant AGN vicinity. 
Even if AGN megamaser emission is present, its contribution to the 
observed nuclear ($\sim$1 kpc) H$_{2}$O flux is much smaller than 
that of the Superantennae, because of much weaker signatures of 
the compact H$_{2}$O emission. 
Except the Superantennae, the bulk of the observed 183 GHz 
H$_{2}$O luminosity comes from entire nuclear regions ($\sim$1 kpc),
most likely thermal emission or possibly ensemble of stellar maser
emission \citep{kon17}. 

Among ten observed (U)LIRGs, eight ULIRGs are diagnosed to
contain luminous obscured AGNs ($\S$2).
The detection rate of very luminous AGN-origin megamaser 183 GHz
H$_{2}$O emission is low (1/8 = 12.5\%).
The detection rates of 22 GHz H$_{2}$O 6$_{1,6}$--5$_{2,3}$ megamaser 
emission in previously conducted extensive surveys are also low 
($\lesssim$10\%) for active galaxies 
\citep[e.g.,][]{bra97,hag02,gre03,bra04,hen05}, but can increase 
(10--50\%) if only highly obscured AGNs are observed
\citep[e.g.,][]{kon06,yam17,cas19,kuo20,pan20}. 
Observations of a larger number of nearby (U)LIRGs that are
diagnosed to host luminous obscured AGNs, are needed to better
constrain how the detection rates of very luminous 183 GHz H$_{2}$O
megamaser emission differ from those of 22 GHz H$_{2}$O 
magamaser emission. 

\subsection{Emission Line Luminosity and Nuclear Dense Molecular Gas Mass}

We use the following formulas to convert molecular emission line flux to 
luminosity \citep{sol05},
\begin{eqnarray}
\left(\frac{L_{line}}{\rm L_{\odot}}\right) = 1.04 \times 10^{-3} \left(\frac{\nu_{rest}}{\rm GHz}\right) (1+z)^{-1} \left(\frac{D_{L}}{\rm Mpc}\right)^2 \left(\frac{S \Delta V}{\rm Jy\ km\ s^{-1}}\right) 
\end{eqnarray}
and
\begin{eqnarray}
\left(\frac{L'_{line}}{\rm K\ km\ s^{-1}\ pc^{2}}\right) = 3.25 \times 10^{7} \left(\frac{\nu_{rest}}{\rm GHz}\right)^{-2} (1+z)^{-1} \left(\frac{D_{L}}{\rm Mpc}\right)^2 \left(\frac{S \Delta V}{\rm Jy\ km\ s^{-1}}\right), 
\end{eqnarray}
where S$\Delta$V is Gaussian-fit velocity-integrated emission line flux 
and D$_{L}$ is luminosity distance.
The derived luminosity of the HCN, HCO$^{+}$, HNC J=2--1, and 183 GHz
H$_{2}$O emission lines from the (U)LIRGs' nuclear regions,
based on the $\sim$1kpc-beam flux data (Table 6), is summarized in Table 9
(columns 2--5).

\begin{deluxetable}{l|cccc|c}
\tabletypesize{\scriptsize}
\tablecaption{Nuclear Molecular Emission Line Luminosity and Derived 
Dense Molecular Gas Mass \label{tbl-9}} 
\tablewidth{0pt}
\tablehead{
\colhead{Object} & \colhead{HCN J=2--1} & \colhead{HCO$^{+}$ J=2--1} &
\colhead{HNC J=2--1} & \colhead{183 GHz H$_{2}$O} & \colhead{M$_{\rm dense-HCO^+}$} \\
\colhead{} & \multicolumn{4}{c}{10$^{4}$ L$_{\odot}$ (10$^{7}$ K km s$^{-1}$ pc$^{2}$)} 
& \colhead{10$^{8}$ [M$_{\odot}$]} \\       
\colhead{(1)} & \colhead{(2)} & \colhead{(3)} & \colhead{(4)} & \colhead{(5)} 
& \colhead{(6)} 
}
\startdata
NGC 1614 & 0.58$\pm$0.03 (3.3$\pm$0.2) & 1.1$\pm$0.1 (5.8$\pm$0.2) 
& 0.32$\pm$0.03 (1.7$\pm$0.2) & $<$0.084 ($<$0.43) & 1--3 \\
IRAS 06035$-$7102 & 6.2$\pm$0.2 (35$\pm$1) & 7.8$\pm$0.3 (43$\pm$2) 
& 3.6$\pm$0.3 (19$\pm$2) & 1.9$\pm$0.4 (9.4$\pm$2.1) & 8--22 \\
IRAS 08572$+$3915 & 1.6$\pm$0.1 (9.0$\pm$0.8) & 2.0$\pm$0.2 (11$\pm$1) 
& 0.71$\pm$0.11 (3.7$\pm$0.6) & 0.88$\pm$0.18 (4.5$\pm$0.9) & 2--6 \\
IRAS 12112$+$0305 NE & 9.5$\pm$2.0 (53$\pm$11) & 5.7$\pm$0.4 (31$\pm$2) 
& 7.5$\pm$0.4 (39$\pm$2) & 3.9$\pm$0.3 (20$\pm$1) & 6--16 \\
IRAS 12112$+$0305 SW & 0.83$\pm$0.17 (4.7$\pm$1.0) & 1.8$\pm$0.3 (9.9$\pm$1.5) 
& $<$0.88 ($<$4.6) & $<$1.5 ($<$7.3) & 2--5 \\
IRAS 12127$-$1412 & 4.4$\pm$0.7 (24$\pm$4) & 3.8$\pm$1.1 (21$\pm$6) 
& 3.9$\pm$1.1 (20$\pm$6) & 2.2$\pm$0.7 (11$\pm$3) & 4--11 \\
IRAS 13509$+$0442 & 3.9$\pm$0.7 (22$\pm$4) & 3.8$\pm$0.8 (21$\pm$4) 
& 2.1$\pm$0.5 (11$\pm$3) & 1.8$\pm$0.6 (8.9$\pm$2.8) & 4--11 \\
IRAS 15250$+$3609 & 3.5$\pm$0.2 (20$\pm$6) & 1.6$\pm$0.2 (8.9$\pm$0.9) 
& 4.0$\pm$0.3 (21$\pm$2) & 3.6$\pm$0.2 (18$\pm$1) & 2--4 \\
Superantennae & 6.3$\pm$0.5 (35$\pm$3) & 3.7$\pm$0.1 (21$\pm$2) 
& 2.6$\pm$0.3 (13$\pm$1) & 6.9$\pm$0.5 (35$\pm$2) & 4--11 \\
IRAS 20551$-$4250 & 3.0$\pm$0.1 (17$\pm$1) & 4.6$\pm$0.1 (25$\pm$1) 
& 1.4$\pm$0.1 (7.5$\pm$0.3) & 0.91$\pm$0.03 (4.6$\pm$0.2) & 5--13 \\
IRAS 22491$-$1808 & 10$\pm$1 (59$\pm$2) & 6.9$\pm$0.3 (38$\pm$2) 
& 6.8$\pm$0.5 (36$\pm$3) & 3.5$\pm$0.3 (18$\pm$1) & 8--19 \\
\enddata

\tablecomments{
Col.(1): Object name. 
Cols.(2)--(5): Molecular emission line luminosity in units of 
(10$^{4}$ L$_{\odot}$), 
derived from $\sim$1kpc-beam data.
That in units of (10$^{7}$ K km s$^{-1}$ pc$^{2}$) is
shown in parentheses. 
Col.(2): HCN J=2--1.
Col.(3): HCO$^{+}$ J=2--1.
Col.(4): HNC J=2--1.
Col.(5): 183 GHz H$_{2}$O.
Col.(6): Dense molecular gas mass derived from the HCO$^{+}$ J=2--1 luminosity, 
assuming that the HCO$^{+}$ emission is thermalized and optically thick at 
J=2--1 and J=1--0, and using the conversion factor of 
2--5 (M$_{\odot}$ [K km s$^{-1}$ pc$^{2}$]$^{-1}$) from HCO$^{+}$ 
J=1--0 luminosity to dense H$_{2}$ mass \citep{ler17}.
}

\end{deluxetable}

Figure 9 shows the comparison of nuclear ($\sim$1 kpc) dense 
molecular line luminosity with infrared (8--1000 $\mu$m) luminosity 
from the entire (U)LIRG regions measured with IRAS's large ($>$30$''$) 
apertures.
In Figures 9a and 9b, the infrared luminosity positively correlates with 
the nuclear HCN J=2--1 and HCO$^{+}$ J=2--1 emission line luminosity. 
The previously derived relations at J=4--3 \citep{zha14,tan18} are overplotted 
for reference. 
These relations were derived by observing galaxies with a wide (six
orders of magnitude) infrared luminosity range \citep{zha14,tan18} and
were largely determined by nearby ULIRGs which dominate the high
luminosity part. 
Although the HCN and HCO$^{+}$ J=4--3 line data of nearby ULIRGs 
\citep{zha14,tan18} were taken with large apertures of single
dish telescopes, we regard that the bulk of the observed J=4--3
luminosities come from nuclear regions, because 
(1) nearby ULIRGs are usually energetically
dominated by compact ($\sim$1 kpc) nuclear regions, with small
contributions from spatially extended ($\gtrsim$a few kpc) star-forming
regions in the host galaxies \citep[e.g.,][]{soi00,dia10,ima11,per21},
and (2) when ALMA high-spatial-resolution data are available, the HCN and
HCO$^{+}$ J=4--3 emission of nearby (U)LIRGs are confirmed to be
spatially compact \citep{ima18b}.
Our J=2--1 data roughly follow the updated J=4--3 relation by \citet{tan18}.
If the HCN and HCO$^{+}$ are almost thermally excited at up to J=4 and 
optically thick, the ratio of molecular line luminosity 
(in units of K km s$^{-1}$ pc$^{2}$) to infrared luminosity 
(in units of L$_{\odot}$) is expected to be comparable for J=4--3 and J=2--1.
Our data suggest that this is the case for warm and dense molecular gas 
in nearby (U)LIRGs' nuclei.
Figures 9c and 9d show the comparison of infrared luminosity
with HNC J=2--1 and 183 GHz H$_{2}$O luminosity, respectively.

\begin{figure}
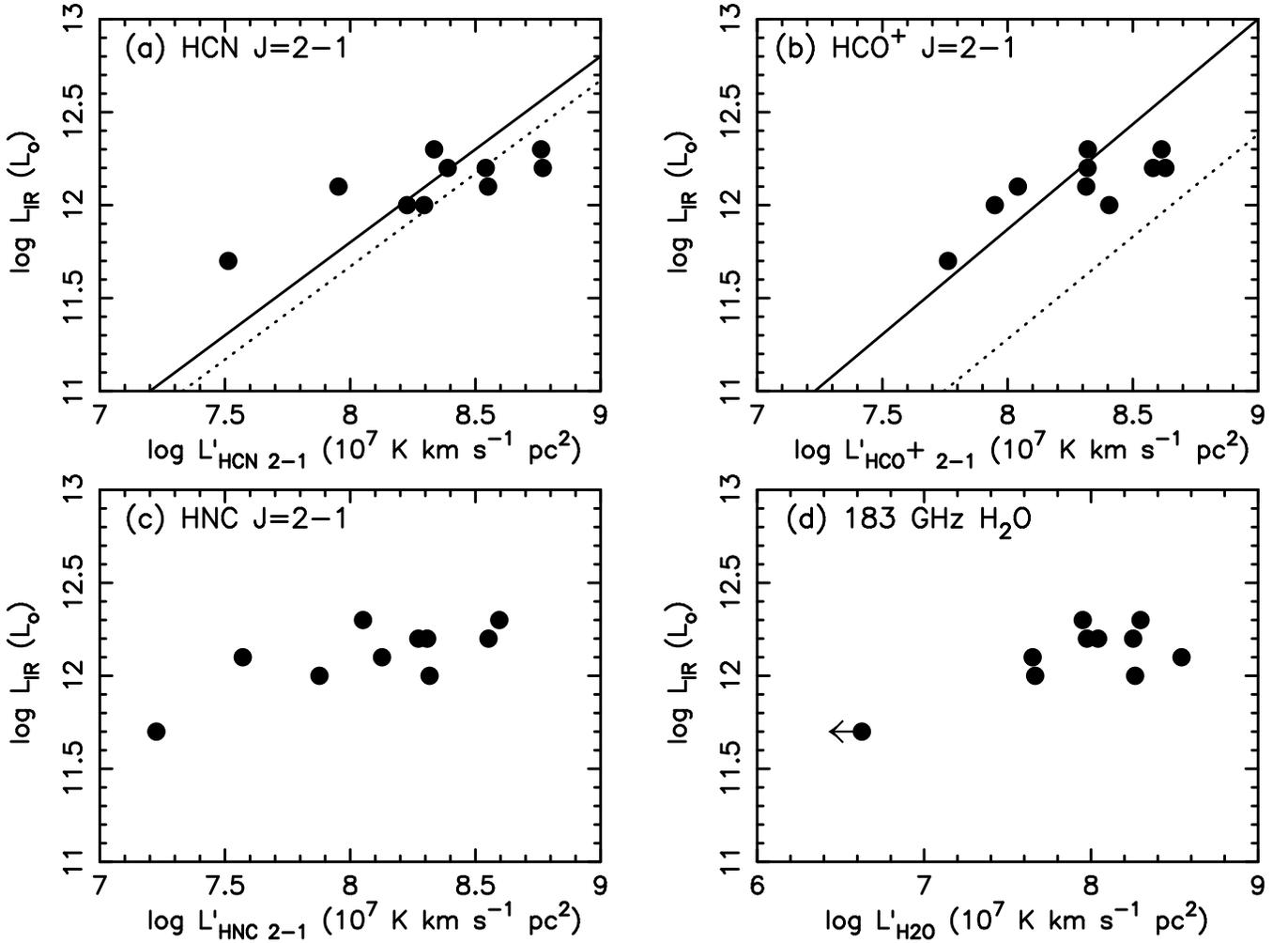

\begin{center}
\includegraphics[angle=-90,scale=.38]{f9a.eps} 
\hspace{0.5cm}
\includegraphics[angle=-90,scale=.38]{f9b.eps} \\
\includegraphics[angle=-90,scale=.38]{f9c.eps} 
\hspace{0.5cm}
\includegraphics[angle=-90,scale=.38]{f9d.eps} 
\end{center}
\caption{
Comparison of molecular emission line and infrared (8--1000 $\mu$m) luminosity.
The abscissa is {\it (a)} HCN J=2--1, {\it (b)} HCO$^{+}$ J=2--1, 
{\it (c)} HNC J=2--1, and {\it (d)} 183 GHz H$_{2}$O emission line luminosity 
in units of (K km s$^{-1}$ pc$^{2}$) measured with $\sim$1kpc-beam ALMA data.
The ordinate is infrared luminosity measured with the IRAS's 
large ($>$30$''$) apertures.
The solid lines in (a) and (b) are the latest best-fit lines derived for the 
HCN J=4--3 (log$L$$_{\rm IR}$ = 1.00log$L$$'$$_{\rm HCN(4-3)}$ + 3.80) and
HCO$^{+}$ J=4--3 (log$L$$_{\rm IR}$ = 1.13log$L$$'$$_{\rm HCO^{+}(4-3)}$ + 2.83) 
for various types of sources with wide (six orders of magnitude) 
infrared luminosity range \citep{tan18}. 
The best-fit lines derived by \citet{zha14} for 
HCN J=4--3 (log$L$$_{\rm IR}$ = 1.00log$L$$'$$_{\rm HCN(4-3)}$ + 3.67) and
HCO$^{+}$ J=4--3 (log$L$$_{\rm IR}$ = 1.10log$L$$'$$_{\rm HCO^{+}(4-3)}$ + 2.48) 
are also shown as dotted lines in (a) and (b) for reference.
For IRAS 12112$+$0305 in {\it (c)} and {\it (d)}, molecular line luminosity 
detected at the primary NE nucleus is adopted, excluding the upper limit 
at the fainter SW nucleus.
}
\end{figure}

We derive nuclear dense molecular gas mass from HCO$^{+}$ J=2--1 luminosity, 
because 
(1) the slope of infrared and HCO$^{+}$ J=2--1 luminosity relation 
is the closest to unity among the four lines (Figure 9), and (2) 
HCO$^{+}$ can be a less biased dense molecular gas mass tracer than 
HCN and HNC, particularly in the vicinity of a luminous AGN
\citep{ima20}. 
Table 9 (column 6) summarizes the derived dense molecular gas mass in 
the observed (U)LIRGs' nuclei, where we adopt the conversion factor 
of 2--5 (M$_{\odot}$ [K km s$^{-1}$ pc$^{2}$]$^{-1}$) from optically thick 
HCO$^{+}$ J=1--0 line luminosity to dense H$_{2}$ mass \citep{ler17}, 
and the above validated assumption that HCO$^{+}$ emission is thermalized 
and optically thick at J=2--1 and J=1--0.
All the (U)LIRGs are estimated to contain nuclear ($\sim$1 kpc) 
dense molecular 
gas with mass of M$_{\rm dense}$ $\gtrsim$ a few $\times$ 10$^{8}$M$_{\odot}$.

We can estimate a lower limit of the depletion time (t$_{\rm dep}$) of 
the nuclear dense molecular gas by star-formation, by assuming 
(1) lower side of the derived mass in Table 9 (column 6) and 
(2) the nuclear star formation rate (SFR) of  
SFR (M$_{\odot}$ yr$^{-1}$) = 4.5 $\times$ 10$^{-44}$ L$_{\rm IR}$ (ergs s$^{-1}$) 
\citep{ken98}, where infrared (8--1000 $\mu$m) and 
far-infrared (40--500 $\mu$m) luminosities are considered comparable 
for (U)LIRGs.
This is a stringent lower limit because 
AGN activity can contribute significantly to the observed infrared
luminosity with IRAS. 
The estimated lower limit of the depletion time 
(t$_{\rm dep}$ $\equiv$ M$_{\rm dense}$/SFR) is 
$\sim$1--3 $\times$ 10$^{6}$ yr, which is shorter than that derived in 
dense molecular clouds in our Galaxy and nearby star-forming galaxies, 
$\sim$10$^{7-8}$ yr \citep[e.g.,][]{lad10,zha14,liu16,tan18,jia20}, 
but it is longer than the free-fall time of dense ($>$10$^{4}$ cm$^{-3}$) 
molecular gas, 
t$_{ff}$ $\equiv$ $\sqrt{\frac{3\pi}{32G\rho}}$ $\lesssim$ 
5 $\times$ 10$^{5}$ yr 
(G is the gravitational constant and $\rho$ is the mass volume density).
Thus, even considering only for (U)LIRGs' nuclei where dense molecular gas 
highly concentrates through galaxy mergers, it is suggested that the bulk 
($\gtrsim$50\%) of dense molecular gas is not forming stars with a free-fall 
time scale \citep{kru07}, possibly because of various feedback processes 
from AGN and starburst activities \citep[e.g.,][]{hop11,hop13,che20}.

\subsection{Vibrationally Excited Emission Lines}

The rotational J=3--2 and J=4--3 emission lines at the vibrationally excited 
v$_{2}$=1f levels of HCN and HNC (HCN-VIB and HNC-VIB, respectively) were
detected in active galaxies, mostly those diagnosed as containing 
luminous obscured AGNs 
\citep[e.g.,][]{sak10,ima13b,aal15a,aal15b,cos15,mart16,ima16b,ima16c,ima18b,fal19,fal21,sak21}.
It is widely believed that infrared radiative pumping is responsible for 
vibrationally exciting HCN and HNC, by absorbing mid-infrared $\sim$14 $\mu$m 
and $\sim$21.5 $\mu$m photons, respectively, because the excitation 
energy level ($>$650 K) is too high to collisionally excite 
\citep[e.g.,][]{aal95,aal07b,sak10}.
Because a luminous AGN usually emits a strong 3--15 $\mu$m continuum 
originating from hot ($>$100 K) dust in the AGN vicinity, the HCN-VIB line 
can be emitted efficiently and its luminosity, relative to vibrational 
ground emission line and/or total infrared (8--1000 $\mu$m) luminosity, 
can be higher in AGNs than in starbursts.
Because HNC can be vibrationally excited by longer wavelength infrared 
photons ($\sim$21.5 $\mu$m), the HNC-VIB emission lines can be moderately 
strong in starbursts as well, where the dust temperature is usually lower 
than in the AGN's vicinity \citep[e.g.,][]{and17,mar21}.
HCN-VIB emission lines with intrinsically low absolute luminosity can also 
be detected in very nearby starbursts if sensitivity is sufficient 
\citep{kri20,mar21}.

As explained in $\S$4 and summarized in Table 7, 
the HCN-VIB J=2--1 emission line is significantly detected
($>$3$\sigma$) as an isolated peak in IRAS 15250$+$3609 and IRAS
20551$-$4250. 
In addition to these two ULIRGs, 
both IRAS 12112$+$0305 NE and IRAS 22491$-$1808 display excess emission at 
the lower frequency side of the HCO$^{+}$ J=2--1 line, close to the 
expected frequency of the HCN-VIB J=2--1 line (Figures 3e and 3s).
For both IRAS 12112$+$0305 NE and IRAS 22491$-$1808, a similar
excess was found at the lower frequency  
part of the HCO$^{+}$ J=3--2 and J=4--3 lines, which was interpreted 
as HCN-VIB J=3--2 and J=4--3 emission line contributions, respectively 
\citep{ima16c,ima18b}.
Thus, the excess components of IRAS 12112$+$0305 NE and IRAS 22491$-$1808 
in Figure 3e and 3s, respectively, may originate from the
HCN-VIB J=2--1 emission line.  
We create moment 0 maps of the excess components (with original beam) and 
confirm detection with $>$3.5$\sigma$ (Table 7).
The HNC-VIB J=2--1 emission line is also significantly detected 
($>$3$\sigma$) in an original beam moment 0 map for NGC 1614 S, IRAS
15250$+$3609, and IRAS 22491$-$1808 (Table 7).

As the number of $\sim$21.5 $\mu$m photons is typically much larger 
than that of $\sim$14 $\mu$m (rest-frame) in (U)LIRGs 
\citep[e.g.,][]{arm07,ima07a,vei09,ima09b,ima10a,her11}, 
HNC can be vibrationally excited with a higher infrared radiative
pumping rate than HCN. 
The emission line luminosity of HNC-VIB is expected to be nearly an order of 
magnitude higher than that of HCN-VIB, {\it if} the number of molecules 
illuminated by the mid-infrared radiation in the close proximity to 
energy sources (AGN and/or young stars) is comparable \citep{ima16b}. 
However, the HNC-VIB emission lines are not as bright as expected 
even in the case of detection in Figure 3 for J=2--1, as well as 
for J=3--2 and J=4--3 \citep{ima18b}.
A plausible scenario is that HNC abundance is low at the innermost 
($<<$1 kpc) obscuring material exposed to strong mid-infrared 
radiation, even if the emission line luminosity from, and abundance in, 
the entire nuclear region ($\sim$1 kpc) is roughly comparable between 
HCN and HNC.
In fact, it is well known from observations of Galactic sources that 
HNC abundance is very low in high radiation density environments around 
luminous energy sources \citep[e.g.,][]{sch92,hir98,gra14,bub19}.
In the recent ALMA observations of the very nearby ($\sim$14 Mpc) galaxy 
NGC 1068, it has been clearly demonstrated that HNC abundance is much lower 
than HCN in the vicinity of a luminous AGN \citep{ima20}.
A plausible explanation for the observed weaker-than-expected HNC-VIB 
emission is that an HNC abundance is depressed 
substantially in the mid-infrared-radiation-illuminated, innermost  
high-temperature molecular gas around the luminous energy sources in (U)LIRGs.

HCO$^{+}$ can also be vibrationally excited by absorbing mid-infrared 
$\sim$12 $\mu$m photons, and the vibrational excitation rate was estimated 
to be roughly comparable to HCN in typical infrared spectra of (U)LIRGs 
\citep{ima16b}.
However, the HCO$^{+}$-VIB line detection was never reported in emission 
in any (U)LIRGs, and only very recently, \citet{kam20} have reported the first 
detection of the HCO$^{+}$-VIB J=4--3 line in {\it absorption} against 
strong background continuum emission in the very nearby ($\sim$17 Mpc) 
radio-loud AGN NGC 1052, with its absorption peak and equivalent 
width $\sim$2.5 times smaller than those of the HCN-VIB J=4--3 line.
Considering that the HCO$^{+}$ J=2--1 flux at the vibrational ground level (v=0)
is comparable within a factor of $\sim$2 to, or sometimes higher than, 
the HCN J=2--1 flux in nearby (U)LIRGs' nuclei (Figure 7a,b), 
if the same fraction of HCO$^{+}$ and HCN are exposed to mid-infrared 
radiation, the HCO$^{+}$-VIB J=2--1 emission line should have been detected 
at least in certain (U)LIRGs with clearly detected HCN-VIB emission.
The considerably weaker HCO$^{+}$-VIB emission than HCN-VIB in (U)LIRGs 
is also very difficult to explain unless the HCO$^{+}$ abundance is 
substantially lower 
than HCN at the innermost mid-infrared-radiation-illuminated region around 
the luminous AGNs at the very center of galaxy nuclei
\citep[e.g.,][]{mal96,pap07,har10}.

\subsection{Other Fainter Dense Molecular Gas Tracers}

\subsubsection{HC$_{3}$N}

The HC$_{3}$N J=20--19 line was covered in the spectra of all the ten observed 
(U)LIRGs, and was clearly detected in four ULIRGs (IRAS 12112$+$0305 NE, 
IRAS 15250$+$3609, IRAS 20551$-$4250, and IRAS 22491$-$1808), 
as explained in $\S$4 and summarized in Table 7.
HC$_{3}$N J=21--20 data were also obtained in some sources and were detected 
in two ULIRGs (IRAS 12112$+$0305 NE and IRAS 22491$-$1808) in Figure 6
and Table 7.

Various rotational J-transition emission lines of HC$_{3}$N were detected in 
active galaxies 
\citep[e.g.,][]{aal02,wan04,aal07a,ala11,lin11,mei11,ala15,cos15,jia17,ric21}.
HC$_{3}$N lines are considered dense ($>$10$^{4}$ cm$^{-3}$) molecular 
gas tracers with fainter flux than HCN, HCO$^{+}$, and HNC 
\citep[e.g.,][]{wan04,ala11,mei11}.
Figure 10a plots  HC$_{3}$N J=20--19 and J=21--20 to HCN, HCO$^{+}$,
and HNC J=2--1 flux ratios, which are $\lesssim$0.6 
(i.e., HC$_{3}$N lines are fainter) for all the
HC$_{3}$N-detected sources. 
 
\begin{figure}
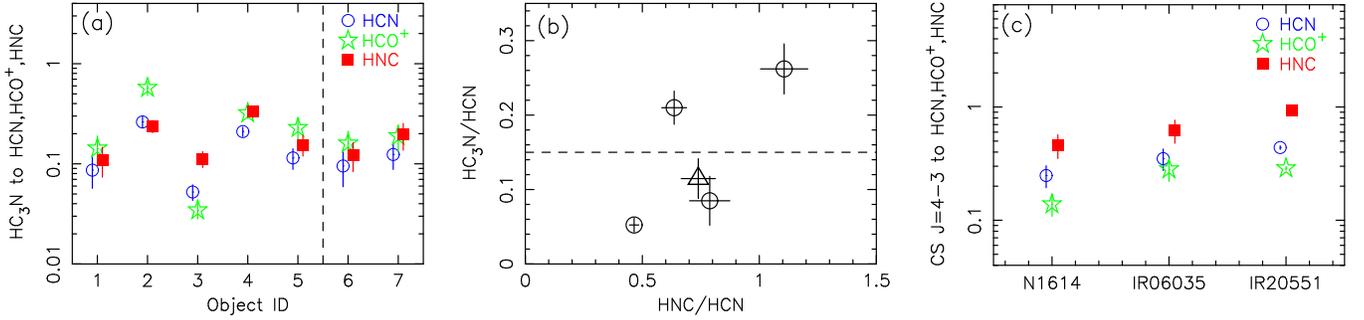

\begin{center}
\includegraphics[angle=-90,scale=.25]{f10a.eps} 
\hspace{0.2cm}
\includegraphics[angle=-90,scale=.25]{f10b.eps} 
\hspace{0.2cm}
\includegraphics[angle=-90,scale=.25]{f10c.eps} 
\end{center}
\caption{
{\it (a)}: Flux ratio of HC$_{3}$N, relative to HCN J=2--1 (blue open
circle), HCO$^{+}$ J=2--1 (green open star) and HNC J=2--1 (red filled
square) in the ordinate.
ID=1--5 (the left side of the vertical dashed line) are meant for the
HC$_{3}$N J=20--19 line, and ID=6--7 (the right side) are meant for
the HC$_{3}$N J=21--20 line. 
1: IRAS 12112$+$0305 NE,  
2: IRAS 15250$+$3609,
3: IRAS 20551$-$4250,
4: IRAS 22491$-$1808,
5: Arp 220 \citep{gal16}, 
6: IRAS 12112$+$0305 NE, and 
7: IRAS 22491$-$1808.
Each flux is measured in [Jy km s$^{-1}$] 
with $\sim$1 kpc beam for all ULIRGs, except Arp 220 
(ID=5) whose flux measurement was made with $\sim$13 kpc beam \citep{gal16}. 
{\it (b)}: Comparison of HNC J=2--1 to HCN J=2--1 flux ratio (abscissa) and 
HC$_{3}$N J=20--19 to HCN J=2--1 flux ratio (ordinate), 
measured with $\sim$1 kpc beam, for four ULIRGs with significant
detection of the faint HC$_{3}$N J=20--19 emission line in Table 10 
(open circle).
We also plot Arp 220 whose flux measurement was made with $\sim$13 kpc
beam \citep{gal16}, for reference (open triangle).
The horizontal dashed line indicates the flux ratio of 0.15 in the
ordinate (see $\S$5.5.1).
{\it (c)}: Flux ratio of CS J=4--3, relative to HCN J=2--1 (blue open
circle), HCO$^{+}$ J=2--1 (green open star) and HNC J=2--1 (red filled
square) in the ordinate. 
Each flux is measured in [Jy km s$^{-1}$] with $\sim$1 kpc beam. 
In the abscissa, ``N1614'', ``IR06035'', and ``IR20551'' means NGC
1614, IRAS 06035$-$7102, and IRAS 20551$-$4250, respectively.
In the abscissa of {\it (a)} and {\it (c)}, the horizontal positions of 
HCN, HCO$^{+}$, and HNC data for each object are slightly displaced for 
presentation.
}
\end{figure}

HC$_{3}$N can be easily destroyed by UV radiation, and thus can be 
emitted strongly in UV-shielded regions at certain distance from 
the central energy sources \citep{lin11,mei11}.
Luminous HC$_{3}$N emission lines, relative to other dense 
molecular tracers, were detected in galaxies with high column density of 
obscuring material around energy sources \citep{aal07a,lin11}.
\citet{lin11} defined HC$_{3}$N-luminous galaxies as those with 
HC$_{3}$N J=10--9 to HCN J=1--0 flux ratio of $>$0.15, which constitutes 
less than one-third of the observed galaxies.
We compare the HC$_{3}$N J=20--19 and HCN J=2--1 fluxes (Tables 6 and 7), 
and find that IRAS 15250$+$3609 and IRAS 22491$-$1808 show HC$_{3}$N J=20--19 
to HCN J=2--1 flux ratios of $>$0.15, whereas IRAS 12112$+$0305 NE and 
IRAS 20551$-$4250 do not (Table 10, column 2).
The upper excitation energy level is E$_{\rm u}$ $\sim$ 92 K for 
HC$_{3}$N J=20--19 and E$_{\rm u}$ $\sim$ 13 K for HCN J=2--1.
If both lines are thermalized and optically thick, 
the HC$_{3}$N J=20--19 to HCN J=2--1 flux ratios are expected to be 
comparable to the HC$_{3}$N J=10--9 to HCN J=1--0 flux ratios, 
because the rest frequency ($\nu_{\rm rest}$) of HC$_{3}$N J=20--19 and 
HCN J=2--1 is approximately twice that of HC$_{3}$N J=10--9 and HCN J=1--0, 
respectively, and the emission flux (in Jy km s$^{-1}$) increases 
with $\nu_{\rm rest}$$^{2}$.
However, given the much higher excitation energy of HC$_{3}$N J=20--19 
(E$_{\rm u}$ $\sim$ 92 K) than HC$_{3}$N J=10--9 (E$_{\rm u}$ $\sim$ 24 K),
HCN J=2--1 (E$_{\rm u}$ $\sim$ 13 K), and HCN J=1--0 (E$_{\rm u}$ $\sim$ 4 K),
if HC$_{3}$N is only sub-thermally excited at J=20, HC$_{3}$N J=20--19 
to HCN J=2--1 flux ratio can be smaller than that of HC$_{3}$N J=10--9 to 
HCN J=1--0 flux ratio.
Thus, IRAS 15250$+$3609 and IRAS 22491$-$1808 are safely classified as 
HC$_{3}$N-luminous galaxies \citep{lin11}, and even IRAS 12112$+$0305 NE 
and IRAS 20551$-$4250 could be in this classification.
Because ULIRGs' nuclear energy sources are usually highly obscured 
and UV-shielded regions are expected to develop, the HC$_{3}$N emission can 
be bright in ULIRGs.

\begin{deluxetable}{l|ccc|ccc}
\tabletypesize{\scriptsize}
\tablecaption{Flux Ratio of HC$_{3}$N to HCN, HCO$^{+}$, and HNC Lines 
\label{tbl-10}} 
\tablewidth{0pt}
\tablehead{
\colhead{Object} & \colhead{$\frac{\rm HC_{3}N\ J=20-19}{\rm HCN\ J=2-1}$} 
& \colhead{$\frac{\rm HC_{3}N\ J=20-19}{\rm HCO^{+}\ J=2-1}$} 
& \colhead{$\frac{\rm HC_{3}N\ J=20-19}{\rm HNC\ J=2-1}$}
& \colhead{$\frac{\rm HC_{3}N\ J=21-20}{\rm HCN\ J=2-1}$} 
& \colhead{$\frac{\rm HC_{3}N\ J=21-20}{\rm HCO^{+}\ J=2-1}$} 
& \colhead{$\frac{\rm HC_{3}N\ J=21-20}{\rm HNC\ J=2-1}$} \\
\colhead{(1)} & \colhead{(2)} & \colhead{(3)} & \colhead{(4)} & \colhead{(5)} 
& \colhead{(6)} & \colhead{(7)} 
}
\startdata
IRAS 12112$+$0305 NE  & 0.085$\pm$0.033 & 0.14$\pm$0.05 & 0.11$\pm$0.04 
& 0.095$\pm$0.036 & 0.16$\pm$0.05 & 0.12$\pm$0.04 \\
IRAS 15250$+$3609  & 0.26$\pm$0.03 & 0.57$\pm$0.09 & 0.24$\pm$0.03 
& --- & --- & --- \\
IRAS 20551$-$4250 & 0.052$\pm$0.009 & 0.034$\pm$0.006 & 0.11$\pm$0.02 
& --- & --- & --- \\
IRAS 22491$-$1808 & 0.21$\pm$0.02 & 0.32$\pm$0.04 & 0.33$\pm$0.04 
& 0.12$\pm$0.04 & 0.19$\pm$0.06 & 0.20$\pm$0.06 \\ \hline
Arp 220 ($\sim$13 kpc) & 0.11$\pm$0.03 & 0.23$\pm$0.05 & 0.16$\pm$0.04
& --- & --- & --- \\ 
\enddata

\tablecomments{
Col.(1): Object name. 
Cols.(2)--(7): Ratio of Gaussian-fit velocity-integrated emission line 
flux in (Jy km s$^{-1}$) derived from $\sim$1kpc-beam spectra.
Arp 220 data measured with $\sim$13 kpc beam \citep{gal16} 
are also added.
Col.(2): HC$_{3}$N J=20--19 to HCN J=2--1 flux ratio. 
Col.(3): HC$_{3}$N J=20--19 to HCO$^{+}$ J=2--1 flux ratio. 
Col.(4): HC$_{3}$N J=20--19 to HNC J=2--1 flux ratio. 
Col.(5): HC$_{3}$N J=21--20 to HCN J=2--1 flux ratio. 
Col.(6): HC$_{3}$N J=21--20 to HCO$^{+}$ J=2--1 flux ratio. 
Col.(7): HC$_{3}$N J=21--20 to HNC J=2--1 flux ratio.
}

\end{deluxetable}

Not only HC$_{3}$N, but also HNC emission can be strong in UV-shielded 
regions ($\S$5.4).
In regions which are UV shielded, but are strongly exposed to infrared 
20--45 $\mu$m radiation, HC$_{3}$N can be vibrationally excited by infrared 
radiative pumping, and thus rotational J-level excitation at vibrational 
ground level (v=0) can be largely affected through back decay
\citep{cos10,ric21}. 
Because this is also the case for HNC ($\S$5.4), certain flux correlation 
between HC$_{3}$N and HNC may be expected.
In Figure 10a and Table 10, while the HC$_{3}$N J=20--19 to HNC
J=2--1 flux ratio is 0.11--0.33 (a factor of $\sim$3), the
distribution of the HC$_{3}$N J=20--19 to HCN J=2--1 and 
HC$_{3}$N J=20--19 to HCO$^{+}$ J=2--1 flux ratio is much wider 
(a factor of $>$5 and $>$15, respectively.)
\citet{lin11} argued that HNC to HCN flux ratio 
and HC$_{3}$N to HCN flux ratio are correlated.
Figure 10b compares the flux ratios for four ULIRGs with significant 
HC$_{3}$N J=20--19 line detection and Arp 220 (Table 10). 
For this limited ULIRG sample, HC$_{3}$N-luminous galaxies tend to be
HNC-luminous, when normalized by HCN.
Both Figures 10a and 10b support the above expected flux correlation
between HC$_{3}$N and HNC emission.

\subsubsection{CS}

The CS J=4--3 line was covered in certain sources, and it was clearly detected 
in NGC 1614, IRAS 06035$-$7102, and IRAS 20551$-$4250 (Figure 6). 
The derived fluxes based on the Gaussian fits of $\sim$1kpc-beam spectra 
are shown in Table 7.
CS lines are considered as moderately bright, dense ($>$10$^{4}$ cm$^{-3}$) 
molecular 
gas tracers \citep{shi15}, but usually fainter than HCN, HCO$^{+}$, and 
HNC \citep[e.g.,][]{hel93,pag95,wan04,ala15,mei15,kri20,mar21,sak21}.
The CS J=4--3 to HCN, HCO$^{+}$, and HNC J=2--1 flux ratios are 
plotted in Figure 10c.
The ratios are 0.1--1 (i.e., CS lines are fainter), and slightly
higher in two AGN-hosting ULIRGs (IRAS 06035$-$7102 and IRAS
20551$-$4250) than in the starburst-classified LIRG NGC 1614. 
In the literature, \citet{izu16} argued that AGNs tend to show 
lower CS J=7--6 to HCN J=4--3 flux ratios than starbursts, while 
no systematic difference of the CS to HCN, HCO$^{+}$, and HNC flux ratios 
was reported between AGNs and starbursts \citep{ala15}.
Possible AGN effects to CS emission are not clear with currently
available data.

\section{Summary} 

We conducted ALMA band 4--5 ($\sim$2 mm) observations of nine 
ultraluminous infrared galaxies (ULIRGs; L$_{\rm IR}$ $>$ 10$^{12}$L$_{\odot}$) 
at $z =$ 0.04--0.14, and one starburst-dominated luminous infrared galaxy, 
NGC 1614 (LIRG; L$_{\rm IR}$ $=$ 10$^{11.7}$L$_{\odot}$) at $z =$ 0.016, 
with the aim of investigating the properties of dense molecular gas tracers 
(HCN, HCO$^{+}$, and HNC J=2--1) and 183 GHz H$_{2}$O 3$_{1,3}$--2$_{2,0}$ 
lines.
The following main results were obtained.

\begin{enumerate}

\item 
Continuum at $\sim$2 mm, HCN J=2--1, HCO$^{+}$ J=2--1, and HNC J=2--1 
emission lines were clearly detected in all sources in our 
sub-arcsec-resolution ($\lesssim$1$''$) data.
The 183 GHz H$_{2}$O emission lines were also clearly detected in
almost all ULIRGs that were classified as AGN important, but 
not in the three starburst-classified (U)LIRGs' nuclei (NGC 1614, IRAS
12112$+$0305 SW, and IRAS 13509$+$0442).

\item 
We found significantly higher HCN to HCO${+}$ J=2--1 flux
ratios in a high fraction of, but not all, AGN-important ULIRGs 
than in starburst-classified sources. 

\item 
We compared the molecular emission line flux ratios and found that 
two ULIRGs with elevated 183 GHz H$_{2}$O emission, relative to
HCO$^{+}$ J=2--1, tend to show elevated HCN J=2--1 emission. 
However, for the remaining (U)LIRGs, the observed 
183 GHz H$_{2}$O to HCO$^{+}$ J=2--1 flux ratios are not strongly
correlated with the observed HCN J=2--1 to HCO$^{+}$ J=2--1 flux ratios.

\item 
Besides the Superantennae that was reported to display strong 
signatures of compact, extremely luminous 183 GHz H$_{2}$O megamaser
emission in AGN-illuminated molecular gas at the very center 
($<<$1 kpc) of the galaxy nucleus \citep{ima21}, data of four
other ULIRGs (IRAS 12112$+$0305 NE, IRAS
15250$+$3609, IRAS 20551$-$4250, and IRAS 22491$-$1808) suggested an
elevated flux or the presence of a spatially unresolved compact
component of the 183 GHz H$_{2}$O emission. 
However, the bulk of the observed H$_{2}$O emission in the four
ULIRGs originates from the entire nuclear regions ($\sim$1 kpc), with
limited contributions from possible AGN-origin megamaser phenomena.

\item 
The infrared (8--1000 $\mu$m) luminosity positively correlates with 
nuclear ($\sim$1 kpc) HCN J=2--1 and HCO$^{+}$ J=2--1 luminosity. 
The correlation roughly follows the previously established
luminosity ratios between infrared and J=4--3 lines of HCN and HCO$^{+}$.
Using HCO$^{+}$ J=2--1 luminosity as the least biased tracer of 
dense molecular gas mass (M$_{\rm dense}$) at nuclear ($\sim$1 kpc) 
regions, we derived the mass to be 
M$_{\rm dense}$ $\gtrsim$ a few $\times$ 10$^{8}$M$_{\odot}$ for all (U)LIRGs.
In addition, we estimated that the depletion time of the nuclear dense 
molecular gas by star-formation is $\gtrsim$1 $\times$ 10$^{6}$ yr, 
significantly longer than the free-fall time of dense molecular gas
even considering the (U)LIRGs' nuclear regions only. 

\item 
Signatures of vibrationally excited v$_{2}$=1f HCN and HNC emission lines 
(HCN-VIB and HNC-VIB) were detected in four and three (U)LIRGs, respectively. 
However, those of HCO$^{+}$-VIB were not found in any (U)LIRGs.
The detection rate and estimated flux of HCO$^{+}$-VIB and HNC-VIB were 
much smaller than those expected from (1) the comparable observed fluxes 
of HCN, HCO$^{+}$, and HNC at the vibrational ground level (v=0), and 
(2) calculated rate of vibrational excitation by infrared radiative pumping, 
if the same proportion of these molecules are exposed to mid-infrared photons.
We suggested that at the innermost strongly-mid-infrared-exposed 
obscuring material around the luminous AGNs, the abundance of HCO$^{+}$ and 
HNC is considerably smaller than that of HCN.

\item 
Another fainter dense molecular gas tracer, HC$_{3}$N J=20--19 
line, was detected in four ULIRGs with bright HCN, HCO$^{+}$, and HNC J=2--1 
emission lines.
CS J=4--3 and HC$_{3}$N J=21--20 lines were included in the spectra of 
certain fraction of (U)LIRGs and their emission signatures were identified 
in three and two (U)LIRGs, respectively.
We found that HC$_{3}$N J=20--19 luminous sources tend to be HNC J=2--1 
luminous, when normalized by HCN J=2--1 emission.
This conforms to our expectations because both HC$_{3}$N and HNC emission 
lines are thought to largely originate from UV-shielded regions at certain 
distance from central energy sources.

\item 
Two continuum emission sources were serendipitously detected 
from our ALMA band 4--5 ($\sim$2 mm) observations of ten (U)LIRGs.
One is at $\sim$8$''$ north of IRAS 13509$+$0442 and after combining with 
previously obtained ALMA data at 0.9--1.3 mm, we interpreted it as an 
infrared luminous dusty galaxy at $z >$ 1, based on the increasing flux with 
decreasing wavelength from 2 mm to 0.9 mm and detection of one emission line 
signature. 
The other is at $\sim$8$''$ south of the Superantennae and regarded 
as a blazar, based on a flat continuum spectrum at 0.8--2 mm and 
no emission line signatures.
Detecting two serendipitous 1--2 mm continuum sources with flux of 
$\gtrsim$0.8 mJy in two out of ten observed (U)LIRG fields appears to be 
many, compared to the surface density of such sources ($\lesssim$0.03 
expected for each ALMA field at 1--2 mm within $\sim$10$''$ from the center). 

\end{enumerate}

We have now J=4--3, 3--2, and 2--1 data of HCN, HCO$^{+}$, and HNC for
the observed ten (U)LIRGs.
The combination of these line data and non-LTE modeling will better
constrain the properties of dense molecular gas
and further elucidate the nature of these (U)LIRGs' nuclei 
(M. Imanishi 2021; in preparation).

\acknowledgments

We thank the anonymous referee for his/her valuable comment which 
helped improve the clarity of this manuscript.
This paper made use of the following ALMA data:
ADS/JAO.ALMA\#2017.1.00022.S and \#2017.1.00023.S.
ALMA is a partnership of ESO (representing its member states), NSF (USA) 
and NINS (Japan), together with NRC (Canada), NSC and ASIAA
(Taiwan), and KASI (Republic of Korea), in cooperation with the Republic
of Chile. The Joint ALMA Observatory is operated by ESO, AUI/NRAO, and
NAOJ. 
M.I., T.I., and S.B. are supproted by JP21K03632, JP20K14531, and 
JP19J00892, respectively. 
Data analysis was in part carried out on the open use data analysis
computer system at the Astronomy Data Center, ADC, of the National
Astronomical Observatory of Japan. 
This research has made use of NASA's Astrophysics Data System and the
NASA/IPAC Extragalactic Database (NED) which is operated by the Jet
Propulsion Laboratory, California Institute of Technology, under
contract with the National Aeronautics and Space Administration. 

\vspace{5mm}
\facilities{ALMA}

\appendix

\section{Serendipitously Detected Continuum Sources}

A serendipitously detected continuum emitting source was identified 
in the fields of view of IRAS 13509$+$0442 and the Superantennae data.
Figure 11 displays the continuum image of the source 
at $\sim$8$''$ north of IRAS 13509$+$0442 ($\sim$20 kpc at the distance 
of this ULIRG) {\it (Left)} 
and at $\sim$8$''$ south of the Superantennae ($\sim$10 kpc at the distance 
of this ULIRG) {\it (Right)}, whose peak ICRS coordinate is 
(13$^{h}$53$^{m}$31.7$^{s}$, $+$04$^{\circ}$28$'$13$''$) and 
(19$^{h}$31$^{m}$21.1$^{s}$, $-$72$^{\circ}$39$'$29$''$), respectively.

\begin{figure}
\begin{center}
\includegraphics[angle=0,scale=.35]{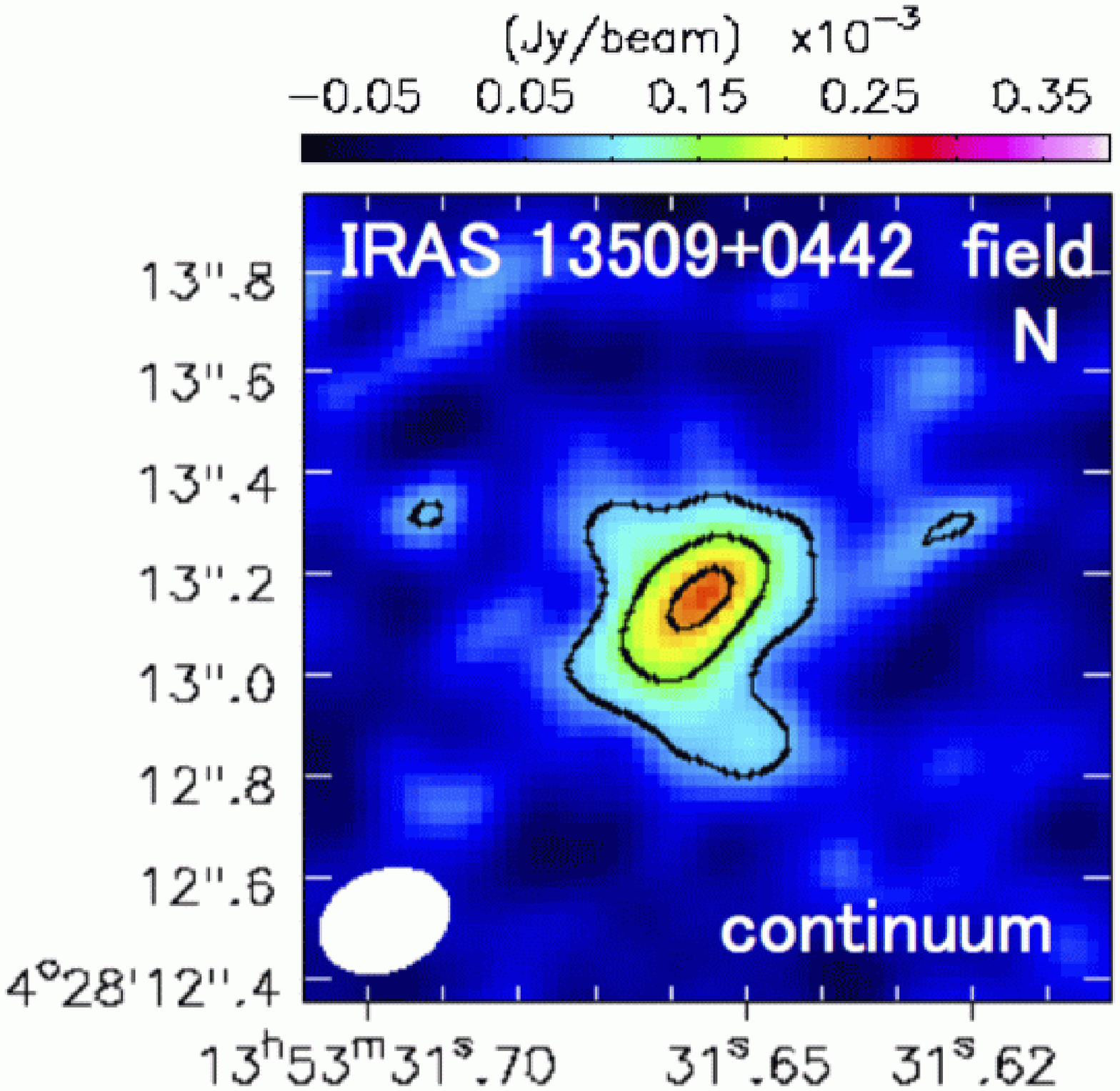} 
\hspace{0.8cm}
\includegraphics[angle=0,scale=.35]{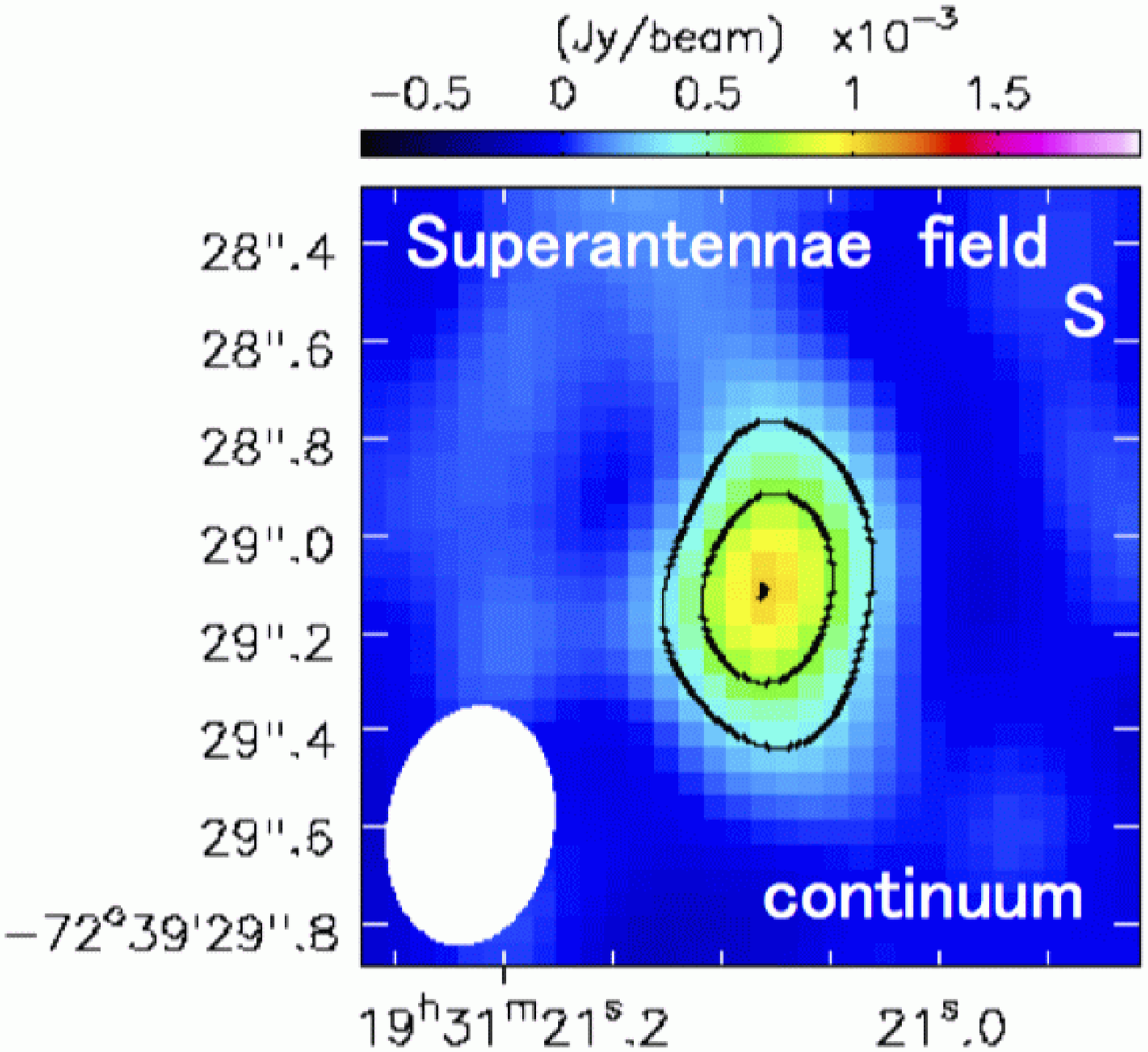} 
\end{center}
\caption{
Continuum image of serendipitously detected object at $\sim$8$''$ north 
of IRAS 13509$+$0442 {\it (Left)} and at $\sim$8$''$ south of the 
Superantennae {\it (Right)}.
In the former and latter images, continuum J21a and J21b data with 
original beam are displayed, respectively, because the 
detection significance is higher than the other continuum data. 
Contours are 3$\sigma$, 6$\sigma$, and 9$\sigma$ for continuum emission 
in both the images.
The continuum image of the left panel was created after excluding a 
possible emission line detected in this object in Figure 12.
}
\end{figure}

For the northern source of IRAS 13509$+$0442, continuum detection 
($>$25$\sigma$) was reported in ALMA band 6 (211--275 GHz) 
and 7 (275--373 GHz) data \citep{ima16c,ima18b}, but no emission line 
signature was found in the obtained spectra. 
We extracted the spectra at the continuum peak position in our newly obtained 
band 4 data.
An emission line signature with a peak flux of $>$1 mJy was detected 
at $\nu_{\rm obs}$ $\sim$ 156.34 GHz in the 154.7--158.2 GHz original beam 
spectrum in J21a data 
(Figure 12a).
This emission line is located close to the edge of the two overlapped 
spectral windows and was detected similarly in the spectra of 
both spectral windows (Figure 12b), suggesting that this emission line 
is real.
If we assume that this emission line signature originates from a bright 
CO J=2--1 ($\nu_{\rm rest}$ = 230.538 GHz) line, then the redshift of 
this source becomes $z \sim$ 0.47.
In this case, CO J=3--2 ($\nu_{\rm rest}$ = 345.796 GHz) 
and CO J=4--3 ($\nu_{\rm rest}$ = 461.041 GHz) lines are redshifted to 
the observed frequency of $\nu_{\rm obs}$ $\sim$ 235.2 and 313.6 GHz, 
respectively, both of which 
were covered in the previously taken ALMA band 6 and 7 spectra 
\citep{ima16c,ima18b}.
However, no clear emission lines were observed at these frequencies.
Previous observations have shown that the majority of 1--2 mm continuum 
sources with flux of a few mJy are at $z >$ 0.5 \citep{bri17}.
The non-detection of optical and near-infrared counterparts at the 
position of this northern continuum source \citep{kim02,ima18b} also makes 
it unlikely that this source is only at $z \sim$ 0.47. 

\begin{figure}
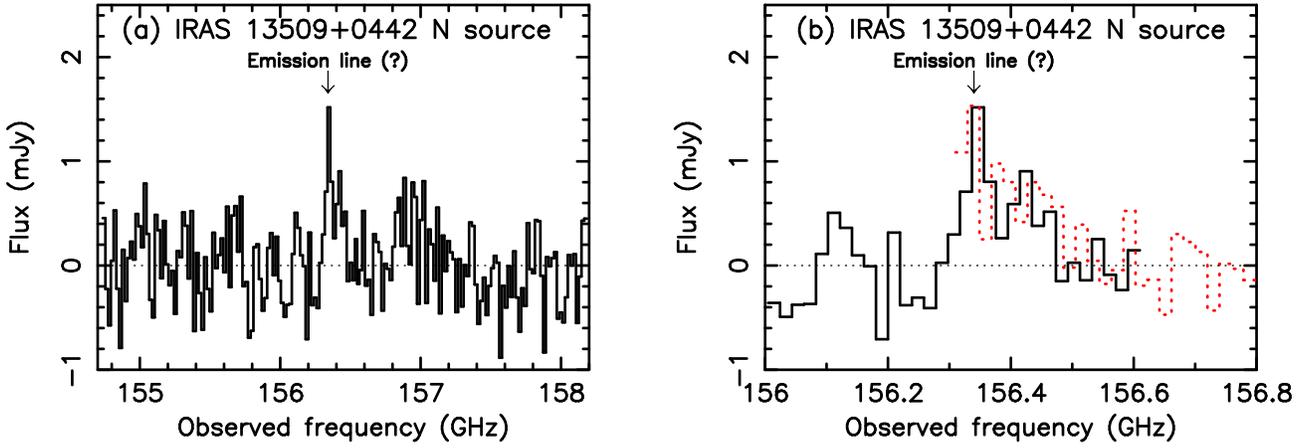

\begin{center}
\includegraphics[angle=-90,scale=.35]{f12a.eps} 
\hspace{0.8cm}
\includegraphics[angle=-90,scale=.35]{f12b.eps} \\
\end{center}
\caption{
{\it (a)}: Original beam (0$\farcs$27 $\times$ 0$\farcs$19) 
spectrum at 154.7--158.2 GHz of the northern continuum source detected within 
IRAS 13509$+$0442 field of view.
An emission line signature is detected at the observed frequency 
$\nu_{\rm obs}$ $\sim$ 156.34 GHz, indicated by a downward arrow 
with the note of ``Emission line (?)''.
{\it (b)}: Magnified view of the possible emission line features.
Spectrum in one spectral window is shown as a black solid line and that 
in another spectral window is shown as a red dotted line.
In {\it (a)} and {\it (b)}, the abscissa is the observed frequency in GHz, and 
the ordinate is flux density in mJy.
The horizontal black thin dotted line indicates the zero-flux level.
}
\end{figure}

We thus consider other possibilities in which the emission line signature 
at $\nu_{\rm obs}$ $\sim$ 156.34 GHz originates from 
CO J=3--2 ($\nu_{\rm rest}$ = 345.796 GHz) at $z \sim$ 1.21, or 
CO J=4--3 ($\nu_{\rm rest}$ = 461.041 GHz) at $z \sim$ 1.94, or 
CO J=5--4 ($\nu_{\rm rest}$ = 576.268 GHz) at $z \sim$ 2.69, or 
CO J=6--5 ($\nu_{\rm rest}$ = 691.473 GHz) at $z \sim$ 3.42.
In either case, one or two higher-J CO lines fall into the 
spectral coverages of the previously obtained ALMA band 6 and 7 data 
\citep{ima16c,ima18b}, but no clear emission line signatures were 
observed at the expected frequencies.
This is not surprising if higher-J CO emission lines are faint due to 
sub-thermal excitation \citep{car13}.
We consider that identification of the emission line signature as an even 
higher-J CO line from a source 
at higher redshift is less likely because the bulk of 1--2 mm continuum 
sources with flux of a few mJy are at $z <$ 4 \citep[e.g.,][]{bri17}.

This source appears spatially extended (Figure 11).
In fact, the continuum flux increases by a factor of $\sim$2.5 from the 
original 
beam ($\sim$0$\farcs$3 $\times$ $\sim$0$\farcs$2) to $\sim$0$\farcs$9 
beam in both J21a and J21b data, suggesting that the continuum emission 
is spatially extended with $\gtrsim$0$\farcs$3 or $\gtrsim$a few kpc 
at $z =$ 1--4. 
The CASA task ``imfit'' also confirms a factor of $>$1.5 larger deconvolved 
image size in both major and minor axis than the beam size. 
We measure the total continuum flux using the CASA task ``imfit'' 
and display the flux at 0.9--2 mm (150--330 GHz) in Figure 13a.
The continuum flux increases with decreasing wavelength, which is 
expected for the Rayleigh–Jeans part of dust thermal radiation.
An infrared luminous dusty galaxy at $z =$ 1--4 can satisfy this.
Detection of other emission lines, particularly lower-J CO lines 
which are less affected by sub-thermal excitation \citep{car13}, 
is necessary to securely identify the redshift.

\begin{figure}
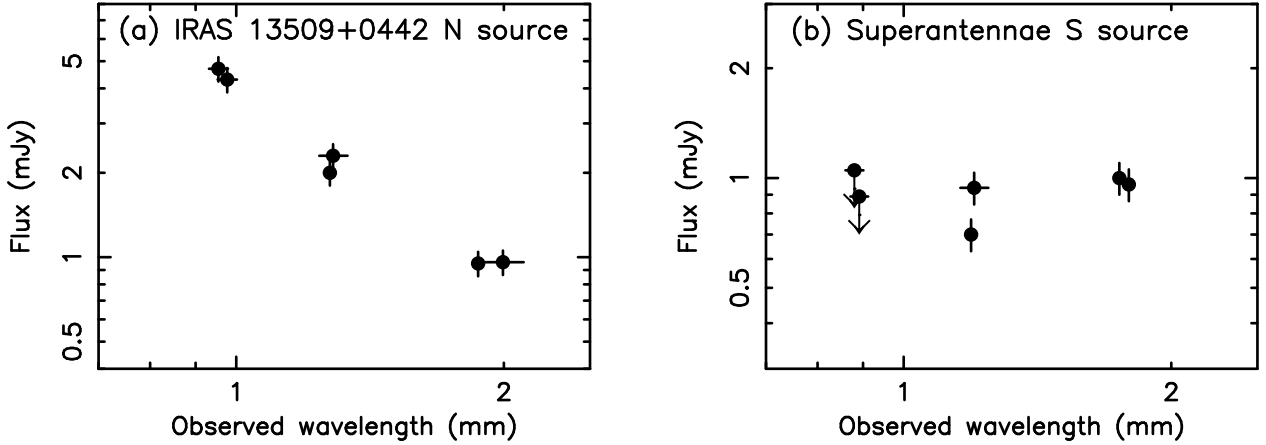

\begin{center}
\includegraphics[angle=-90,scale=.35]{f13a.eps} 
\hspace{0.8cm}
\includegraphics[angle=-90,scale=.35]{f13b.eps} 
\end{center}
\caption{
{\it (a)}: Continuum flux at 0.9--2 mm of the serendipitously detected 
source at $\sim$8$''$ north of IRAS 13509$+$0442.
The total area-integrated continuum flux measured with $\sim$0$\farcs$9 
circular beam data is plotted, because this source is spatially resolved. 
{\it (b)}: Continuum flux at 0.8--2 mm of the serendipitously detected 
source at $\sim$8$''$ south of the Superantennae. 
Peak flux in $\sim$0$\farcs$9 circular beam data is plotted, 
because this source is spatially compact.
The 3$\sigma $ upper limits are plotted for band 7 data ($\sim$0.9 mm).
In both {\it (a)} and {\it (b)}, the abscissa is the observed wavelength 
in millimeter (mm) and the ordinate is flux density in mJy.
Absolute flux calibration uncertainty of maximum $\sim$10\% in individual 
ALMA observations is considered.
}
\end{figure}

For the southern source of the Superantennae, continuum emission was 
detected with 4--8$\sigma$ in our previously obtained ALMA band 6 data 
at $\nu_{\rm obs}$ $\sim$ 250 GHz ($\sim$1.2 mm), but was not 
discussed \citep{ima16c,ima18b}, 
because band 7 continuum detection at $\nu_{\rm obs}$ $\sim$ 330 GHz 
($\sim$0.9 mm) was only marginal 
($\sim$3.5$\sigma$ in $\sim$0$\farcs$7 $\times$ $\sim$0$\farcs$5 
beam data).
Now that band 5 continuum emission at $\nu_{\rm obs}$ $\sim$ 170 GHz is 
detected ($\sim$9$\sigma$; Figure 11), we can investigate 
the nature of this source in more detail.
No emission line signature was observed in any of the ALMA band 5, 6, and 7 
spectra.
No spatial extension was detected using the CASA task ``imfit''. 
The 0.8--2 mm continuum flux of this source is shown in Figure 13b.
It is relatively flat or slightly decreasing with decreasing wavelength.
This can be produced by an AGN core, whose emission is beamed 
toward our line of sight (i.e, blazar).
X-ray detection as a hard source by Chandra observations \citep{wan16} 
and the non-detection of any emission line in ALMA spectra can naturally 
be explained by this blazar scenario.

In summary, we serendipitously detected two continuum sources with flux 
of $\gtrsim$0.8 mJy at 1--2 mm from our ALMA band 4, 5, and 6 
observations of ten (U)LIRGs. 
The surface density of such continuum sources is 
$\sim$10$^{3}$ deg$^{-2}$ \citep[e.g.,][]{hat18,gon20} or $\lesssim$0.03 per 
each ALMA observing field at 1--2 mm within $\sim$10$''$ radius from the 
center. 
Our detection rate of two such sources in ten ALMA observing fields 
is very high.

\section{Gaussian Fits of Emission Lines}

Figures 14 and 15 show Gaussian fits of individual emission lines listed in
Table 6 and serendipitously detected faint emission lines in Table 7 
(column 4--7), respectively.
Table 11 tabulates Gaussian-fit velocity-integrated fluxes of dense 
molecular (HCN, HCO$^{+}$, and HNC J=2--1) and 183 GHz H$_{2}$O
emission lines, measured with original, $\sim$1 kpc, and 2 kpc beam,
for the nine observed ULIRGs. 

\begin{figure*}
\begin{center}
\hspace*{-4.5cm}
\includegraphics[angle=-90,scale=.17]{f14a.eps} 
\hspace{0.3cm}
\includegraphics[angle=-90,scale=.17]{f14b.eps} 
\hspace{0.3cm}
\includegraphics[angle=-90,scale=.17]{f14c.eps} \\
\hspace*{-4.5cm}
\includegraphics[angle=-90,scale=.17]{f14d.eps} 
\hspace{0.3cm}
\includegraphics[angle=-90,scale=.17]{f14e.eps} 
\hspace{0.3cm}
\includegraphics[angle=-90,scale=.17]{f14f.eps} \\
\hspace*{-4.5cm}
\includegraphics[angle=-90,scale=.17]{f14g.eps} 
\hspace{0.3cm}
\includegraphics[angle=-90,scale=.17]{f14h.eps} 
\hspace{0.3cm}
\includegraphics[angle=-90,scale=.17]{f14i.eps} \\
\hspace*{-4.5cm}
\includegraphics[angle=-90,scale=.17]{f14j.eps} 
\hspace{0.3cm}
\includegraphics[angle=-90,scale=.17]{f14k.eps} 
\hspace{0.3cm}
\includegraphics[angle=-90,scale=.17]{f14l.eps} \\
\hspace*{-4.5cm}
\includegraphics[angle=-90,scale=.17]{f14m.eps} 
\hspace{0.3cm}
\includegraphics[angle=-90,scale=.17]{f14n.eps} 
\hspace{0.3cm}
\includegraphics[angle=-90,scale=.17]{f14o.eps} \\
\includegraphics[angle=-90,scale=.17]{f14p.eps} 
\hspace{0.3cm}
\includegraphics[angle=-90,scale=.17]{f14q.eps} 
\hspace{0.3cm}
\includegraphics[angle=-90,scale=.17]{f14r.eps} 
\hspace{0.3cm}
\includegraphics[angle=-90,scale=.17]{f14s.eps} \\
\vspace{0.3cm}
\includegraphics[angle=-90,scale=.17]{f14t.eps} 
\hspace{0.3cm}
\includegraphics[angle=-90,scale=.17]{f14u.eps} 
\hspace{0.3cm}
\includegraphics[angle=-90,scale=.17]{f14v.eps} 
\hspace{0.3cm}
\includegraphics[angle=-90,scale=.17]{f14w.eps} \\
\vspace{0.3cm}
\includegraphics[angle=-90,scale=.17]{f14x.eps} 
\hspace{0.3cm}
\includegraphics[angle=-90,scale=.17]{f14y.eps} 
\hspace{0.3cm}
\includegraphics[angle=-90,scale=.17]{f14z.eps} 
\hspace{0.3cm}
\includegraphics[angle=-90,scale=.17]{f14aa.eps} \\
\end{center}
\end{figure*}


\begin{figure*}
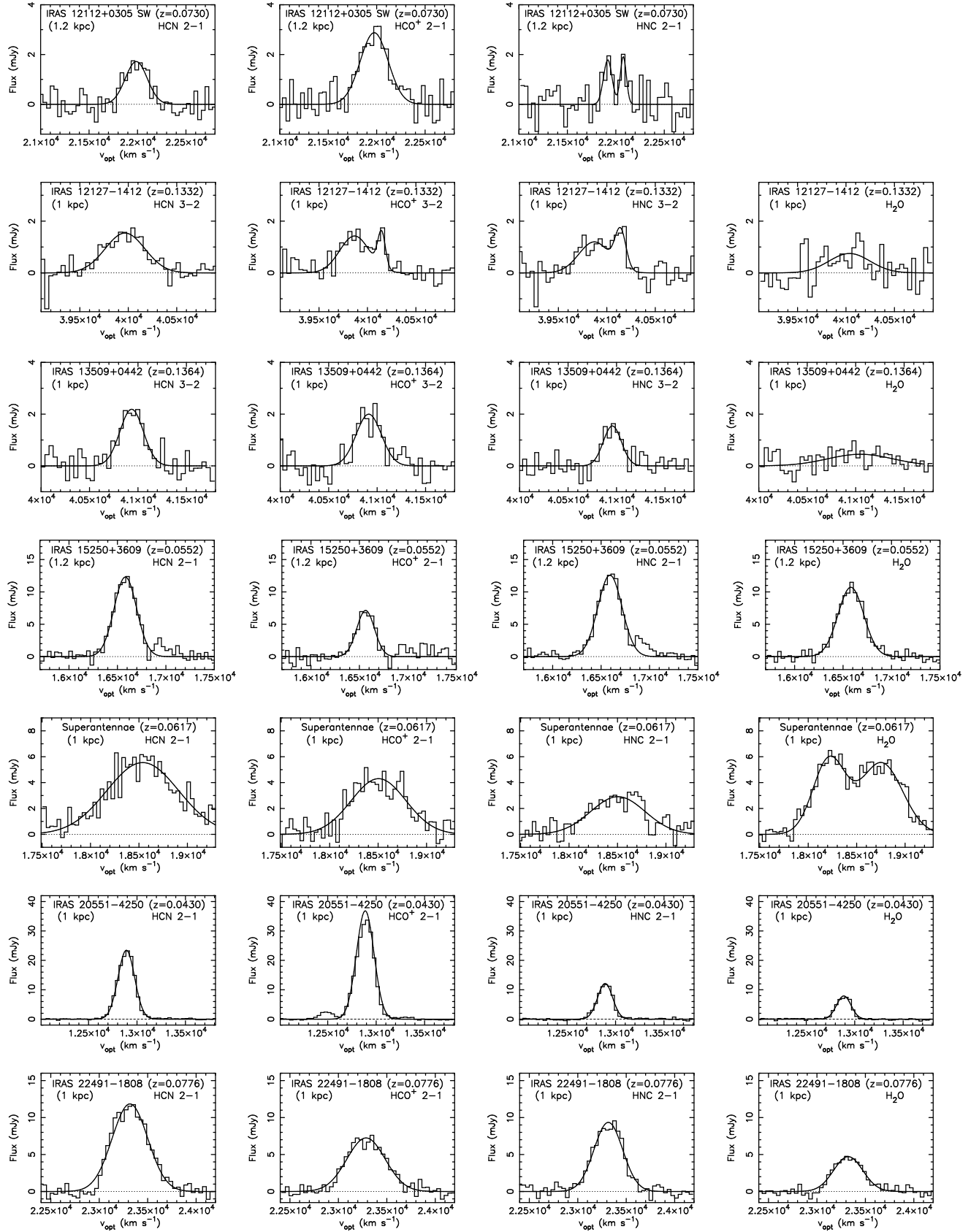

\begin{center}
\hspace*{-4.5cm}
\includegraphics[angle=-90,scale=.17]{f14ab.eps} 
\hspace{0.3cm}
\includegraphics[angle=-90,scale=.17]{f14ac.eps} 
\hspace{0.3cm}
\includegraphics[angle=-90,scale=.17]{f14ad.eps} \\
\vspace{0.3cm}
\includegraphics[angle=-90,scale=.17]{f14ae.eps} 
\hspace{0.3cm}
\includegraphics[angle=-90,scale=.17]{f14af.eps} 
\hspace{0.3cm}
\includegraphics[angle=-90,scale=.17]{f14ag.eps} 
\hspace{0.3cm}
\includegraphics[angle=-90,scale=.17]{f14ah.eps} \\
\vspace{0.3cm}
\includegraphics[angle=-90,scale=.17]{f14ai.eps} 
\hspace{0.3cm}
\includegraphics[angle=-90,scale=.17]{f14aj.eps} 
\hspace{0.3cm}
\includegraphics[angle=-90,scale=.17]{f14ak.eps} 
\hspace{0.3cm}
\includegraphics[angle=-90,scale=.17]{f14al.eps} \\
\vspace{0.3cm}
\hspace{0.35cm}
\includegraphics[angle=-90,scale=.17]{f14am.eps} 
\hspace{-0.03cm}
\includegraphics[angle=-90,scale=.17]{f14an.eps} 
\hspace{-0.03cm}
\includegraphics[angle=-90,scale=.17]{f14ao.eps} 
\hspace{-0.03cm}
\includegraphics[angle=-90,scale=.17]{f14ap.eps} \\
\vspace{0.3cm}
\includegraphics[angle=-90,scale=.17]{f14aq.eps} 
\hspace{0.3cm}
\includegraphics[angle=-90,scale=.17]{f14ar.eps} 
\hspace{0.3cm}
\includegraphics[angle=-90,scale=.17]{f14as.eps} 
\hspace{0.3cm}
\includegraphics[angle=-90,scale=.17]{f14at.eps} \\
\vspace{0.3cm}
\includegraphics[angle=-90,scale=.17]{f14au.eps} 
\hspace{0.3cm}
\includegraphics[angle=-90,scale=.17]{f14av.eps} 
\hspace{0.3cm}
\includegraphics[angle=-90,scale=.17]{f14aw.eps} 
\hspace{0.3cm}
\includegraphics[angle=-90,scale=.17]{f14ax.eps} \\
\vspace{0.3cm}
\includegraphics[angle=-90,scale=.17]{f14ay.eps} 
\hspace{0.3cm}
\includegraphics[angle=-90,scale=.17]{f14az.eps} 
\hspace{0.3cm}
\includegraphics[angle=-90,scale=.17]{f14ba.eps} 
\hspace{0.3cm}
\includegraphics[angle=-90,scale=.17]{f14bb.eps} \\
\end{center}
\caption{
Adopted Gaussian fits (solid curved lines) of the emission
lines in the $\sim$1kpc-beam spectra (and 200pc-beam spectra for NGC
1614), tabulated in Table 6. 
The abscissa is optical LSR velocity in km s$^{-1}$ and the ordinate 
is flux density in mJy.
The horizontal black thin dotted line indicates the zero flux 
level.
}
\end{figure*}

\begin{figure*}
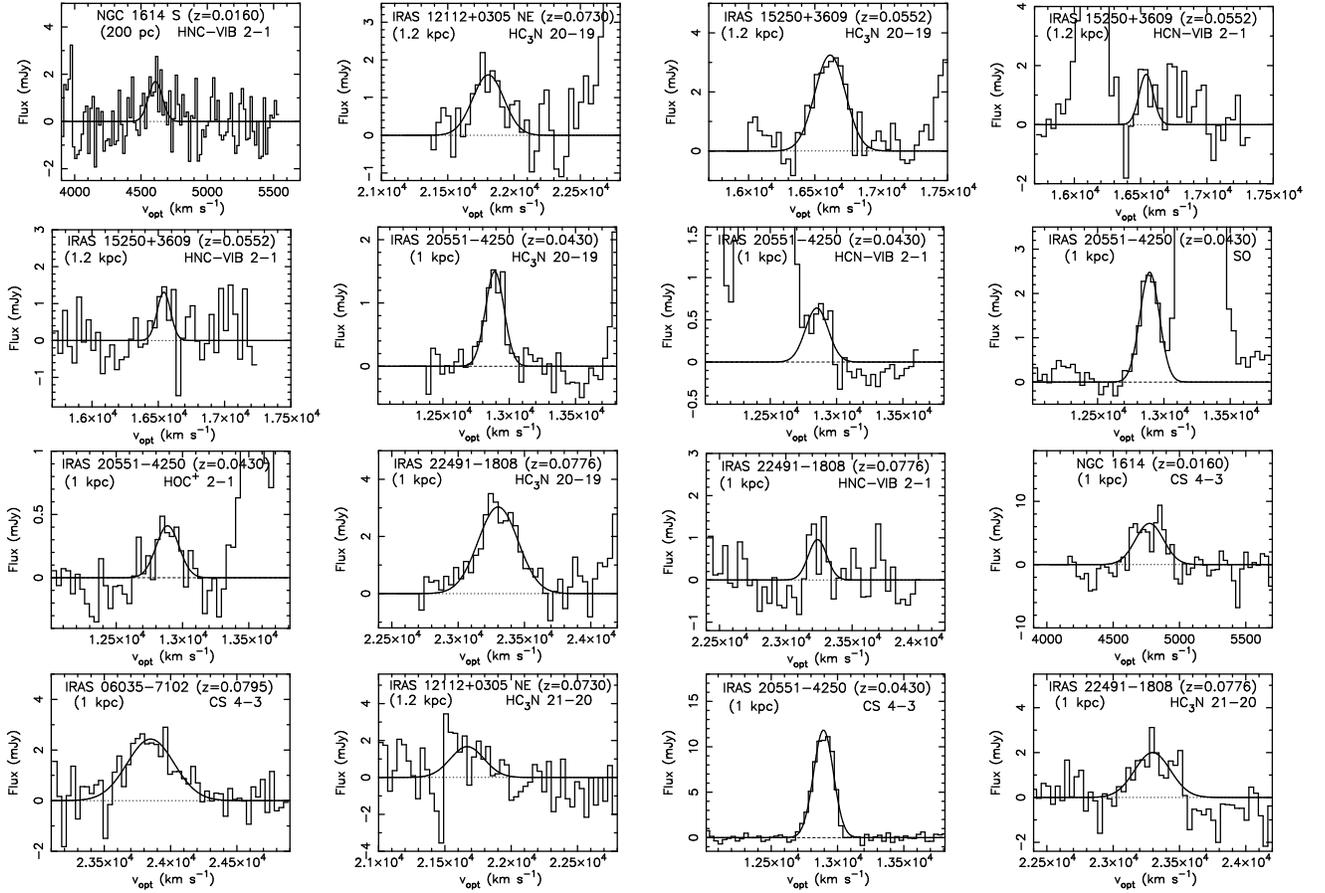

\begin{center}
\hspace{0.4cm}
\includegraphics[angle=-90,scale=.17]{f15a.eps} 
\hspace{0.2cm}
\includegraphics[angle=-90,scale=.17]{f15b.eps} 
\hspace{0.3cm}
\includegraphics[angle=-90,scale=.17]{f15c.eps} 
\hspace{-0.1cm}
\includegraphics[angle=-90,scale=.17]{f15d.eps}\\
\includegraphics[angle=-90,scale=.17]{f15e.eps} 
\hspace{-0.1cm}
\includegraphics[angle=-90,scale=.17]{f15f.eps} 
\hspace{0.3cm}
\includegraphics[angle=-90,scale=.17]{f15g.eps} 
\hspace{0.3cm}
\includegraphics[angle=-90,scale=.17]{f15h.eps} \\
\includegraphics[angle=-90,scale=.17]{f15i.eps} 
\hspace{0.3cm}
\includegraphics[angle=-90,scale=.17]{f15j.eps} 
\hspace{0.3cm}
\includegraphics[angle=-90,scale=.17]{f15k.eps} 
\hspace{0.3cm}
\includegraphics[angle=-90,scale=.17]{f15l.eps} \\
\includegraphics[angle=-90,scale=.17]{f15m.eps} 
\hspace{0.3cm}
\includegraphics[angle=-90,scale=.17]{f15n.eps} 
\hspace{0.3cm}
\includegraphics[angle=-90,scale=.17]{f15o.eps} 
\hspace{0.3cm}
\includegraphics[angle=-90,scale=.17]{f15p.eps} \\
\end{center}
\caption{
Adopted Gaussian fits (solid curved lines) of the
serendipitously detected faint emission lines in the $\sim$1kpc-beam
spectra (and 200pc-beam spectrum for NGC 1614), tabulated in Table 7
(columns 4--7). 
The abscissa is optical LSR velocity in km s$^{-1}$ and the ordinate 
is flux density in mJy.
The horizontal black thin dotted line indicates the zero flux 
level.
}
\end{figure*}

\begin{deluxetable*}{ll|ccc}
\tabletypesize{\scriptsize}
\tablecaption{Gaussian Fit Flux of Molecular Emission Lines with
Various Beam Sizes for ULIRGs  
\label{tbl-11}} 
\tablewidth{0pt}
\tablehead{
\colhead{} & \colhead{} & 
\multicolumn{3}{c}{Integrated Flux [Jy km s$^{-1}$]} \\ 
\colhead{Object} & \colhead{Line} &
\colhead{Original Beam} & \colhead{$\sim$1 kpc} & \colhead{2 kpc} \\  
\colhead{(1)} & \colhead{(2)} & \colhead{(3)} & \colhead{(4)} & 
\colhead{(5)} 
}
\startdata 
IRAS 06035$-$7102  & HCN J=2--1 & 2.7$\pm$0.1 ($-$5\%) & 2.9$\pm$0.1
(1.6 kpc) & 3.0$\pm$0.1 ($+$5.4\%) \\ 
 & HCO$^{+}$ J=2--1 & 3.4$\pm$0.1 ($-$4\%) & 3.6$\pm$0.1 (1.6 kpc) &
3.8$\pm$0.2 ($+$6.3\%) \\ 
 & HNC J=2--1 & 1.1$\pm$0.1 ($-$32\%) & 1.6$\pm$0.1 (1.6 kpc) &
1.7$\pm$0.2 ($+$4\%) \\ 
 & 183 GHz H$_{2}$O & 0.54$\pm$0.09 ($-$35\%) & 0.83$\pm$0.18 (1.6
kpc) & 0.92$\pm$0.24 ($+$10\%) \\ \hline 
IRAS 08572$+$3915 & HCN J=2--1 & 1.2$\pm$0.1 ($-$18\%) & 1.4$\pm$0.1 &
1.7$\pm$0.2 ($+$23\%) \\
 & HCO$^{+}$ J=2--1 & 1.4$\pm$0.1 ($-$15\%) & 1.7$\pm$0.2 & 
1.9$\pm$0.3 ($+$12\%)\\
 & HNC J=2--1 & 0.41$\pm$0.07 ($-$33\%) & 0.61$\pm$0.09 &
0.75$\pm$0.13 ($+$24\%) \\
 & 183 GHz H$_{2}$O & 0.62$\pm$0.08 ($-$17\%) & 0.75$\pm$0.15 &
0.82$\pm$0.30 ($+$10\%)\\ \hline 
IRAS 12112$+$0305 NE & HCN J=2--1 & 5.2$\pm$0.2 ($-$0\%) & 5.1$\pm$0.5
(1.2 kpc) & 6.1$\pm$0.3 ($+$19\%) \\
 & HCO$^{+}$ J=2--1 & 2.8$\pm$0.2 ($-$9\%) & 3.1$\pm$0.2 (1.2 kpc) &
3.7$\pm$0.3 ($+$20\%) \\
 & HNC J=2--1 & 3.7$\pm$0.2 ($-$8\%) & 4.0$\pm$0.2 (1.2 kpc) &
4.8$\pm$0.3 ($+$19\%) \\
 & 183 GHz H$_{2}$O & 2.0$\pm$0.1 ($-$4\%) &  2.1$\pm$0.1 (1.2 kpc) &
2.1$\pm$0.2 ($+$4\%) \\ \hline
IRAS 12112$+$0305 SW & HCN J=2--1 & 0.41$\pm$0.10 ($-$10\%) & 0.45$\pm$0.10 
(1.2 kpc) & 0.50$\pm$0.13 ($+$9\%) \\
 & HCO$^{+}$ J=2--1 & 0.81$\pm$0.11 ($-$16\%) & 0.97$\pm$0.15 (1.2 kpc) &
0.97$\pm$0.17 ($+$0\%) \\
 & HNC J=2--1 & 0.28$\pm$0.12 ($-$14\%) & 0.32$\pm$0.13 (1.2 kpc) & 
--- \tablenotemark{A} \\ \hline
IRAS 12127$-$1412 & HCN J=2--1 & 0.48$\pm$0.09 ($-$31\%) & 0.70$\pm$0.12 &
0.94$\pm$0.21 ($+$34\%) \\
 & HCO$^{+}$ J=2--1 & 0.37$\pm$0.11 ($-$39\%) & 0.60$\pm$0.17 &
0.90$\pm$0.27 ($+$49\%) \\
 & HNC J=2--1 & 0.36$\pm$0.08 ($-$40\%) & 0.61$\pm$0.17 &
0.93$\pm$0.35 ($+$54\%) \\
 & 183 GHz H$_{2}$O & 0.22$\pm$0.10 ($-$39\%) & 0.35$\pm$0.29 & 
--- \tablenotemark{A} \\ \hline
IRAS 13509$+$0442 & HCN J=2--1 & 0.32$\pm$0.07 ($-$46\%) &
0.59$\pm$0.11 & 1.0$\pm$0.2 ($+$77\%) \\
 & HCO$^{+}$ J=2--1 & 0.29$\pm$0.10 ($-$50\%) & 0.57$\pm$0.12 &
1.0$\pm$0.2 ($+$77\%) \\
 & HNC J=2--1 & 0.17$\pm$0.05 ($-$47\%) & 0.32$\pm$0.07 &
0.57$\pm$0.14 ($+$78\%) \\
 & 183 GHz H$_{2}$O & --- \tablenotemark{A} & 0.33$\pm$0.22 & 
--- \tablenotemark{A} \\ \hline
IRAS 15250$+$3609 & HCN J=2--1 & 3.3$\pm$0.2 ($-$4\%) & 3.4$\pm$0.2
(1.2 kpc) & 3.6$\pm$0.2 ($+$7\%) \\
 & HCO$^{+}$ J=2--1 & 1.4$\pm$0.1 ($-$12\%) & 1.6$\pm$0.2 (1.2 kpc) &
1.8$\pm$0.2 ($+$18\%) \\ 
 & HNC J=2--1 & 2.9$\pm$0.1 ($-$21\%) & 3.7$\pm$0.3 (1.2 kpc) & 
4.5$\pm$0.4 ($+$19\%) \\ 
 & 183 GHz H$_{2}$O & 2.9$\pm$0.1 ($-$16\%) & 3.4$\pm$0.2 (1.2 kpc) & 
3.9$\pm$0.3 ($+$15\%)\\ \hline 
Superantennae & HCN J=2--1 & 3.1$\pm$0.2 ($-$36\%) & 4.9$\pm$0.4 &
7.7$\pm$0.7 ($+$58\%) \\
 & HCO$^{+}$ J=2--1 & 2.0$\pm$0.2 ($-31$\%) & 2.9$\pm$0.3 &
4.7$\pm$0.6 ($+$63\%) \\
 & HNC J=2--1 & 1.2$\pm$0.2 ($-$40\%) & 1.9$\pm$0.2 & 
2.7$\pm$0.4 ($+$42\%) \\
 & 183 GHz H$_{2}$O & 4.6$\pm$0.3 ($-$10\%) & 5.1$\pm$0.3 &
5.5$\pm$0.7 ($+$8\%) \\ \hline
IRAS 20551$-$4250 & HCN J=2--1 & 4.6$\pm$0.1 ($-$5\%) & 4.8$\pm$0.1 &
5.4$\pm$0.1 ($+$11\%) \\ 
 & HCO$^{+}$ J=2--1 & 6.9$\pm$0.1 ($-$6\%) & 7.4$\pm$0.1 & 
8.5$\pm$0.2 ($+$16\%) \\
 & HNC J=2--1 & 1.8$\pm$0.1 ($-$21\%) & 2.3$\pm$0.1 & 
2.5$\pm$0.1 ($+$10\%) \\
 & 183 GHz H$_{2}$O & 1.3$\pm$0.1 ($-$7\%) & 1.4$\pm$0.1 & 
1.5$\pm$0.1 ($+$3\%) \\ \hline
IRAS 22491$-$1808 & HCN J=2--1 & 4.3$\pm$0.1 ($-$14\%) & 5.1$\pm$0.2 &
5.8$\pm$0.3 ($+$14\%) \\
 & HCO$^{+}$ J=2--1 & 3.1$\pm$0.4 ($-$8\%) & 3.3$\pm$0.2 & 
4.3$\pm$0.3 ($+$29\%) \\
 & HNC J=2--1 & 2.7$\pm$0.1 ($-$17\%) & 3.2$\pm$0.2 & 
3.9$\pm$0.4 ($+$20\%) \\
 & 183 GHz H$_{2}$O & 1.5$\pm$0.1 ($-$9\%) & 1.6$\pm$0.1 & 
2.0$\pm$0.3 ($+$22\%) \\ \hline
\enddata

\tablenotetext{A}{Detection is not significant for meaningful estimate.} 

\tablecomments{
Col.(1): Object name.
Col.(2): Line.
Cols.(3)--(5): Gaussian-fit velocity-integrated flux (in Jy km s$^{-1}$). 
For each line of each object, we adopt the same Gaussian (one or two 
components). 
Col.(3): Original beam size (Table 3, column 6).
Col.(4): $\sim$1 kpc beam. 1 kpc beam for all ULIRGs, except IRAS
06035$-$7102, IRAS 12112$+$0305, and IRAS 15250$+$3609, for which 1.6
kpc, 1.2 kpc, and 1.2 kpc beam is used, respectively.
Col.(5): 2 kpc beam.
In Col.(3) and (5), flux increase, relative to the $\sim$1 kpc
beam measurement, is shown in parentheses.
Negative value means flux decrease.
}
\end{deluxetable*}

\section{Moment 0 Maps of the 183 GHz H$_{2}$O Emission Line for 
IRAS 12127$-$1412 and IRAS 13509$+$0442.}

Figure 16 displays moment 0 maps of the 183 GHz H$_{2}$O emission
line, with 1kpc circular beam, for IRAS 12127$-$1412 and IRAS 13509$+$0442.
Detection significance of this H$_{2}$O line in these two sources is 
$<$3$\sigma$ in the Gaussian fit in the 1kpc-beam spectra (Table 6), 
but $>$3$\sigma$ in the moment 0 maps. 
The peak values in the moment 0 maps are used in Figure 7. 

\begin{figure}
\begin{center}
\includegraphics[angle=0,scale=.28]{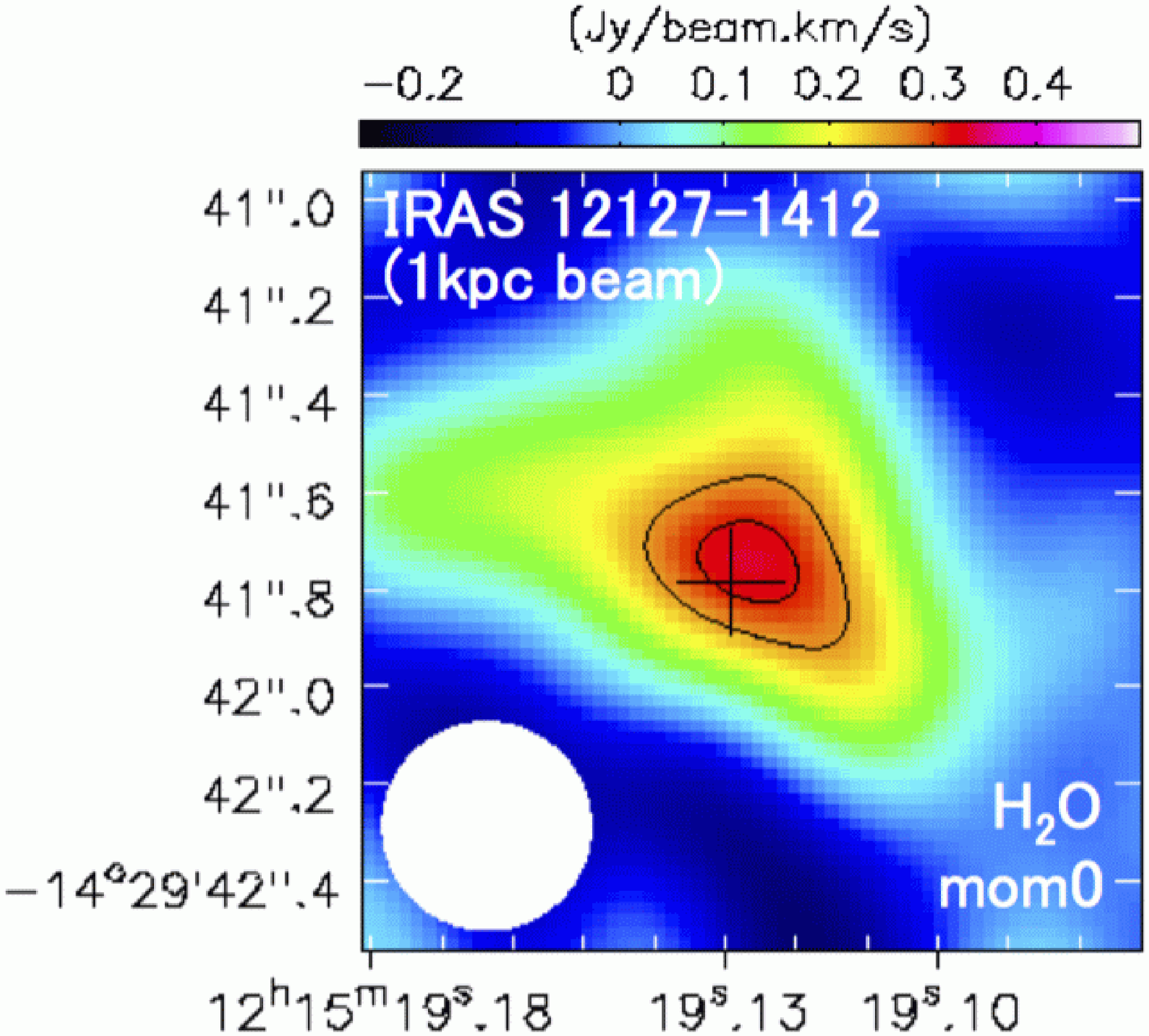} 
\hspace{0.2cm}
\includegraphics[angle=0,scale=.28]{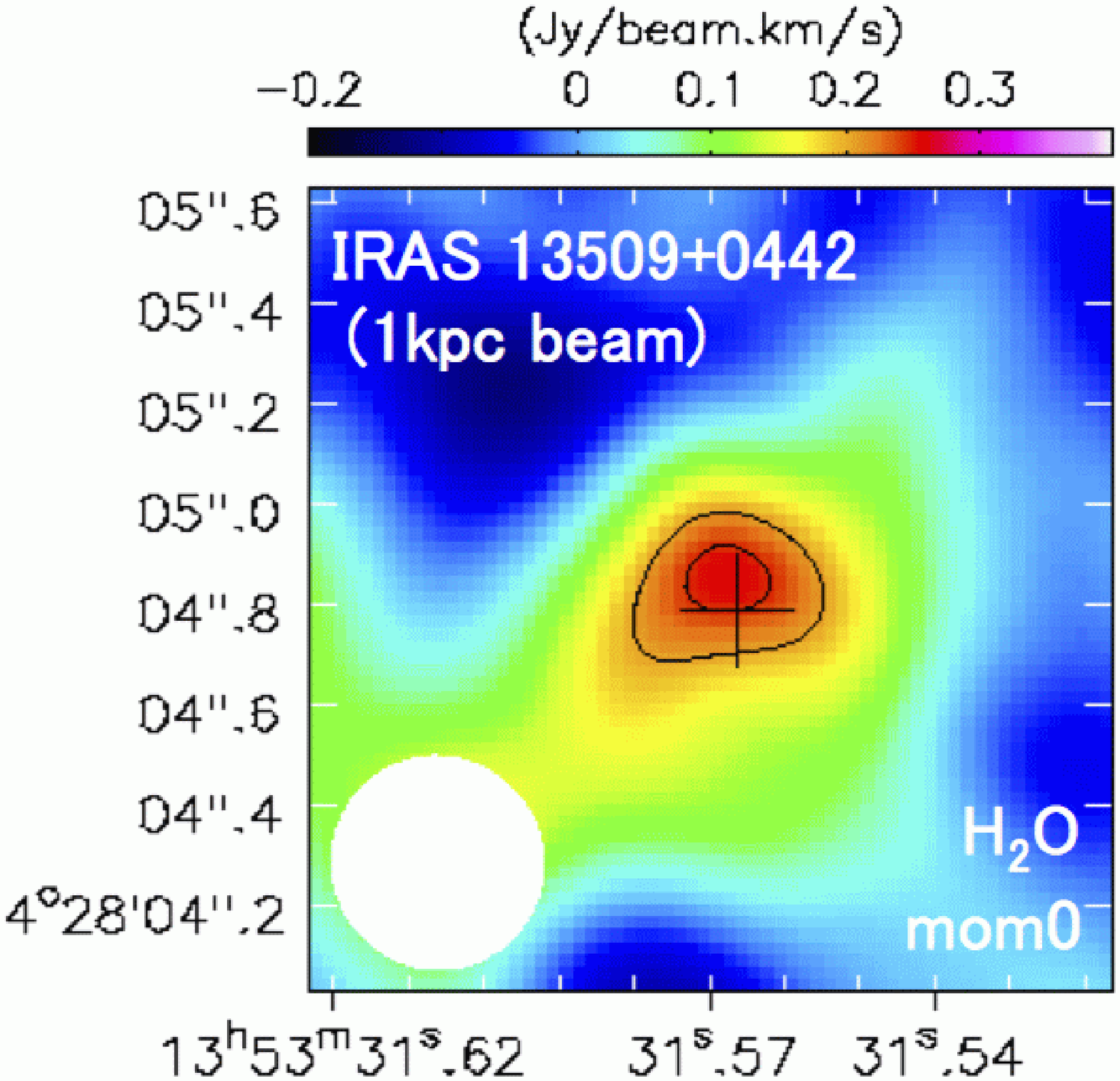} \\
\end{center}
\vspace{0.5cm}
\caption{
Moment 0 map of the 183 GHz H$_{2}$O emission line created from
1kpc-beam data. {\it (Left)}: IRAS 12127$-$1412. 
{\it (Right)}: IRAS 13509$+$0442.
Peak flux of the H$_{2}$O emission is 0.34 and 0.26 
[Jy beam$^{-1}$ km s$^{-1}$]  
for IRAS 12127$-$1412 and IRAS 13509$+$0442, respectively.
Contours are 2.5$\sigma$ and 3$\sigma$, where the rms noise is 
0.10 and 0.082 [Jy beam$^{-1}$ km s$^{-1}$] for IRAS 12127$-$1412 and 
IRAS 13509$+$0442, respectively.
The detection significance at the H$_{2}$O emission peak is 3.2$\sigma$ for 
both sources. 
Continuum peak position is shown as a cross.
Circular 1 kpc beam is shown as a filled circle in the lower-left part.
}
\end{figure}


\end{document}